\documentclass[10pt,b5paper,twoside,english]{myphdthesis}

\usepackage{ThAV} 

\usepackage[utf8]{inputenc} 
\usepackage{palatino} 
\usepackage{babel} 
\usepackage{lettrine} 

\usepackage{epigraph}

\let\originalepigraph\epigraph 
\renewcommand\epigraph[2]{\originalepigraph{\textit{#1}}{\textsc{#2}}}

\makeatletter
\renewcommand{\@dotsep}{10000} 
\makeatother
\usepackage[nottoc]{tocbibind} 

\usepackage{deluxetable}  

\graphicspath{{./Figures/}} 
\usepackage{subfig}
\usepackage{tikz} 
\usepackage{pgfplots} 
\pgfplotsset{compat=1.17}

\usepackage{amsmath,amsfonts,amssymb,mathtools,mathrsfs,amsthm}
\usepackage{latexsym}

\DeclareMathAlphabet\mathbfcal{OMS}{cmsy}{b}{n}
\renewcommand\Re{\operatorname{Re}}
\renewcommand\Im{\operatorname{Im}}
\newcommand{\ernst}{\mathcal{E}}

\newcommand{\tr}{\mathrm{Tr}}
\newcommand{\diag}{\mathrm{diag}}

\usepackage{fancyhdr}
\renewcommand{\chaptermark}[1]{\markboth{#1}{}}
\renewcommand{\sectionmark}[1]{\markright{\thesection\ #1}}

\renewcommand{\headrulewidth}{0.2pt}
\renewcommand{\footrulewidth}{0pt}

\fancypagestyle{plain}{
\fancyhf{} 
\fancyfoot[C]{\bfseries \thepage} 
\renewcommand{\headrulewidth}{0pt}
\renewcommand{\footrulewidth}{0pt}}

\begin{document}

\frontmatter
\setcounter{page}{1}

\maketitle

\chapter*{List of Publications}
\thispagestyle{plain}



\noindent
Marco Astorino and Adriano Vigan\`o \\
\emph{Binary black hole system at equilibrium} \\
Physics Letters B \textbf{820}, 136506 (2021)
\vspace{0.1cm} \\

\noindent
Marco Astorino and Adriano Vigan\`o \\
\emph{Many accelerating distorted black holes} \\
The European Physical Journal C \textbf{81}, 891 (2021)
\vspace{0.1cm} \\

\noindent
Marco Astorino, Roberto Emparan and Adriano Vigan\`o \\
\emph{Bubbles of nothing in binary black holes and black rings, and viceversa} \\
Journal of High Energy Physics \textbf{07}, 007 (2022)
\vspace{0.1cm} \\

\noindent
Marco Astorino, Riccardo Martelli and Adriano Vigan\`o \\
\emph{Black holes in a swirling universe} \\
Physical Review D \textbf{106}, 064014 (2022)
\vspace{0.1cm} \\

\noindent
Marco Astorino and Adriano Vigan\`o \\
\emph{Charged and rotating multi-black holes in an external gravitational field} \\
The European Physical Journal C \textbf{82}, 829 (2022)

\addcontentsline{toc}{chapter}{List of Publications} 
\pagestyle{fancy}
\fancyhf{}
\renewcommand{\chaptermark}[1]{\markboth{#1}{}}
\renewcommand{\sectionmark}[1]{\markright{\thesection\ #1}}
\fancyhead[LE]{\bf \thepage}
\fancyhead[RO]{\bf \thepage}
\fancyhead[LO]{}
\fancyhead[RE]{}
\renewcommand{\headrulewidth}{0.2pt}
\renewcommand{\footrulewidth}{0pt}


\tableofcontents
\pagestyle{fancy}
\fancyhf{}
\fancyhead[LE]{\bf \thepage}
\fancyhead[RO]{\bf \thepage}
\fancyhead[LO]{}
\fancyhead[RE]{}
\renewcommand{\headrulewidth}{0.2pt}
\renewcommand{\footrulewidth}{0pt}




\mainmatter

\renewcommand{\theequation}{\arabic{chapter}.\arabic{equation}}

\renewcommand{\chaptermark}[1]{\markboth{#1}{}}
\renewcommand{\sectionmark}[1]{\markright{\thesection\ #1}}
\fancyhead[RO]{\bf \thepage}
\fancyhead[LE]{\bf \thepage}
\fancyhead[LO]{{\it Introduction}}
\fancyhead[RE]{{\it Introduction}}
\renewcommand{\headrulewidth}{0.2pt}
\renewcommand{\footrulewidth}{0pt}
\addcontentsline{toc}{chapter}{Introduction}
\chapter*{Introduction}
\thispagestyle{plain}

\setlength{\epigraphwidth}{0.5\textwidth}
\epigraph{Oh leave the Wise our measures to collate\\
One thing at least is certain, light has weight\\
One thing is certain, and the rest debate--\\
Light rays, when near the Sun, do not go straight.}%
{Arthur Eddington}

General Relativity~\cite{Einstein:1915by,Hilbert:1915tx} represents one of the most astonishing achievements of humankind, and one the most successful theories in the history of physics.
From its early successes in reproducing the perihelion precession of Mercury and the deflection of light~\cite{Dyson:1920cwa,Bertotti:2003rm,Collett:2018gpf}, to the remarkable prediction of the existence of gravitational waves, that have been recently detected~\cite{LIGOScientific:2016aoc}, the theory has been tested on different scales and an excellent agreement with the experiments has been established~\cite{Will:2014kxa}.

Among the predictions of General Relativity, black holes~\cite{Schwarzschild:1916uq} perhaps represent the most fascinating one:
they are regions of spacetime where the gravitational field is so intense that nothing can escape from it, not even light.
Such objects are quite difficult to observe, and indeed many years passed before the existence of black holes was accepted by the physics community.
In particular, the hypothesis that a supermassive black hole, dubbed Sagittarius A*, is located in the centre of our Galaxy~\cite{Parsa2017,Boehle2016,Gravity2019} has recently received a spectacular confirmation, with the first image of such a black hole shot by the Event Horizon Telescope.
Moreover, there are observations of gravitational lensing around Messier 87*~\cite{EventHorizonTelescope:2019dse,EventHorizonTelescope:2019uob,EventHorizonTelescope:2019jan,EventHorizonTelescope:2019ths,EventHorizonTelescope:2019pgp,EventHorizonTelescope:2019ggy}, a proposed black hole, that are in good agreement with the Kerr metric~\cite{Kerr:1963ud,Hod:1998dt}.

Binary black hole systems are of utmost relevance from the astrophysical perspective:
their merging processes are the major source of gravitational waves~\cite{LIGOScientific:2016aoc,LIGOScientific:2016sjg,LIGOScientific:2017bnn,LIGOScientific:2018mvr} (together with binary neutron star systems~\cite{LIGOScientific:2017vwq,LIGOScientific:2017zic,LIGOScientific:2017zic} and, on the theoretical side, they disclose the non-linear nature of General Relativity, and represent an important playground in which testing the laws of black hole mechanics and of quantum gravity.
More generally, an analytical description of multi-black hole spacetimes is of fundamental importance both for the interpretation of the measurements and for the investigation of the structure of the gravitational theory.

Of course one of the main obstacle in modelling a stationary multi-gravitational source system is the necessity of a mechanism to balance the gravitational attraction of the bodies:
otherwise the system naturally tends to collapse.
Usually the equilibrium is granted by the introduction of cosmological struts which prop up the gravitating bodies, but these one-dimensional objects must be constituted by matter that violates physically reasonable energy conditions.
Alternatively, in some cases cosmic strings of infinite length can be used to support the gravitational collapse.
In both cases these objects are symptomatic for conical singularities which plague the spacetime.

It is known that in Newtonian mechanics the (uncharged) $N$-body configurations cannot be in equilibrium for $N>1$.
Indeed, since the Newtonian gravitational force is always attractive, it is clear that separated bodies cannot be in balance.
However, the situation might be different in General Relativity:
if one considers rotating objects, then the effect of spin-spin repulsion might compensate for the gravitational attraction.
Therefore, in order to better understand the nature of the gravitational interaction in General Relativity, it is an important question as to whether an equilibrium of (physically reasonable) rotating bodies is possible~\cite{Hennig:2019knn}.

The simplest instance of such a configuration is provided by two aligned black holes in vacuum, i.e.~a double Kerr--NUT solution~\cite{Kramer1980,Neugebauer1980II}.
Many authors, over the years, addressed the problem of the equilibrium configuration for such a system~\cite{Chrusciel:2011iv,Neugebauer:2009su,Hennig:2011fp,Neugebauer:2011qb}, by refining the parametrisation in order to simplify the resulting metric and to clarify the physical significance of the parameters~\cite{Manko:2018iyn}.
However, it turned out that the double Kerr--NUT solution describes an equilibrium configuration when one of the two black holes violates an inequality between the the horizon area and the angular momentum~\cite{Hennig:2009fqe} or, equivalently, when one of the two horizons disappears, i.e.~we are left with a naked singularity.
Thus, regular stationary double black holes configurations in vacuum do not exist:
the spin-spin repulsion is not strong enough to compensate for the gravitational attraction.

The situation is slightly different in the electrovacuum case:
in Newtonian physics, we know that a collection of $N$ charged bodies can reach the equilibrium, provided that the charges are sufficiently large to balance the gravitational attraction.
In the relativistic case, however, there exist upper limits on the charges and angular momenta, such that the black hole horizons are not broken.
This implies that the existence of equilibrium configurations due to charge-charge (and spin-spin interaction) is not guaranteed \emph{a priori}.

The most known example of a static equilibrium configuration is provided by the Majumdar--Papapetrou solution~\cite{Papapetrou1945,Majumdar:1947eu}, which describes a collection of an arbitrary number of extremal Reissner--Nordstr\"om black holes:
the interesting feature of this solution is that the black holes are not constrained on an axis, but can be located in any position in the plane.
The Majumdar--Papapetrou spacetime was extended to the presence of a positive cosmological constant in~\cite{Kastor:1992nn}, and then to the general case (which also includes a negative cosmological constant) in~\cite{Chimento:2013pka}.
These solutions, despite being very interesting, contain extremal black holes (i.e.~vanishing surface gravity), which represent limiting configurations that cannot be realised in Nature, as shown by Thorne~\cite{Thorne:1974ve}.
Thus, one may wonder if it is possible to construct \emph{non-extremal} and stable multi-black hole configurations in the electrovacuum case.

In the case of static, non-extremal multi-black hole configurations, the answer is negative:
it was shown~\cite{Chrusciel:2006pc} that a static multi-black hole solution must reduce to the Majumdar--Papapetrou metric.
The stationary case is more complicated, and a definitive answer is not known:
some recent papers that tackle this problem are~\cite{Hennig:2019knn,Alekseev:2019kcf,Cabrera-Munguia:2021whe}.

Turning back to the vacuum case, we showed that it is not possible to construct asymptotically flat multi-black hole configurations that are at equilibrium.
Thus, it is natural to wonder if it is possible to relax the boundary conditions and to consider more general backgrounds (different than Minkowski), such that the embedding of multi-black hole systems can reach an equilibrium configuration.
This is the main goal of this Thesis:
we will see, in the next Chapters, that there exist non-trivial backgrounds (given by external multipolar gravitational fields and by Kaluza--Klein bubbles) that allow for a multi-black hole configuration that can achieve an equilibrium state, provided that the parameters of the background (and of the holes) are fine tuned in order to remove the conical singularities.

However, as one can imagine, the construction of a multi-black hole solution is a highly non-trivial task:
because of the non-linearity of the Einstein equations, one can not simply superimpose many black hole solutions in order to obtain a new solution.
Because of these difficulties, we will rely on the application of some powerful \emph{solution generating techniques}, i.e.~techniques that have been developed to construct exact solutions of the Einstein equations without the direct integration of the equations of motion.
Since the new solutions that we will present in this Thesis are based on such techniques, we will dedicate some space to the explanation of the methods.

More precisely we begin, in Chapter~\ref{chap:gentech}, by introducing the solution generating techniques:
after an overview (and a history) of the many techniques that have been developed over the years, we explain in some detail the two techniques that allow us to construct the desired multi-black hole solutions, i.e.~the Ernst formalism and the inverse scattering method.
The inverse scattering method will be our main tool in Chapters~\ref{chap:extfield} and~\ref{chap:bubble}, while we will take advantage of the Ernst transformations in Chapter~\ref{chap:swirl}.

Subsequently, Chapter~\ref{chap:extfield} deals with multi-black holes embedded in an external gravitational field:
such a field is given by a multipolar expansion and, thus, represents a generic static and axisymmetric gravitational field.
We will see that, by choosing appropriately the multipole parameters of the field, it is possible to obtain an equilibrium configuration in many interesting cases, like a collection of collinear static black holes or a chain of accelerating black holes.

Chapter~\ref{chap:bubble} considers another interesting example of a regularising background:
the expanding Kaluza--Klein bubbles, also known as ``bubbles of nothing''.
These are interesting solutions of vacuum General Relativity, that are also connected to the vacuum stability, that resemble the expanding de Sitter spacetime.
Such an expanding behaviour provides the force necessary to balance the gravitational attraction among the black holes, and hence to reach the equilibrium.

Finally, Chapter~\ref{chap:swirl} takes a different perspective:
it is not focused on a multi-black hole solution, but rather it explores the action of a symmetry transformation of the Ernst equations (unexplored until now), which produces a new solution.
Such a solution represents a black hole embedded in a rotating (or ``swirling'', as we dubbed it) universe, which possesses many interesting features.
Even if we limit ourselves to constructing single-black hole solutions in Chapter~\ref{chap:swirl}, we discuss the possibility of implementing the swirling background in order to enforce the spin-spin configuration explained above, and reach an equilibrium configuration in a double-Kerr spacetime.

\renewcommand{\chaptermark}[1]{\markboth{#1}{}}
\renewcommand{\sectionmark}[1]{\markright{\thesection\ #1}}
\fancyhead[RO]{\bf \thepage}
\fancyhead[LE]{\bf \thepage}
\fancyhead[LO]{{\it \leftmark}}
\fancyhead[RE]{\it \rightmark}
\renewcommand{\headrulewidth}{0.2pt}
\renewcommand{\footrulewidth}{0pt}

\chapter{Solution generating techniques}
\label{chap:gentech}
\thispagestyle{plain}

Einstein equations represent a set of highly non-linear partial differential equations:
thus, it is extremely difficult to find exact analytical solutions just resting on ans\"atze.
For this reason, many people have developed, over the years, methods and techniques that allow one to construct exact solution \emph{without} the direct integration of the equations of motion.
The basic idea behind these methods is to take a known solution, that is usually called a \emph{seed}, and then to apply to it a transformation or a map in order to produce a new, non-equivalent solution of the equations of motion, by overcoming the integration of the equations themselves.
We begin this chapter by presenting a short review of the history of the solution generating techniques, in such a way to provide to the reader a perspective of the many methods that have been developed over the years, and to not limit ourselves only to the ones that we will use in this Thesis.
The historical account we will present is largely based on the one given by Alekseev in~\cite{Alekseev:2010mx}.

The history of solution generating techniques began in the mid-60s, when a class of non-linear partial differential equations, which were called completely integrable, were discovered~\cite{Gardner:1967wc}:
that class was represented by the famous Korteweg--de Vries equation.
Such equations admit various solution generating methods for the explicit construction of infinite hierarchies of solutions with an arbitrary number of parameters and for the non-linear superposition of fields.
This class is extremely important, because many fundamental equations of mathematics and physics were found, upon two-dimensional reduction, to belong to it\footnote{In fact, the integrability we are talking about was found to hold only for the symmetry reduced Einstein equations, i.e.~for spacetimes which admit $D-2$ commuting Killing vectors in $D$ dimensions.}.

The discoveries in the field of integrability led to the hope that Einstein equations could be integrable as well.
One of the very first examples was provided by Ernst~\cite{ErnstI,ErnstII}, who showed, both for the vacuum and the electrovacuum case, that the equations of motion can be written in a very compact and elegant way, such that some symmetry transformations that leave the theory invariant can be unraveled.
Subsequently, Geroch conjectured~\cite{Geroch:1972yt} that four-dimensional vacuum Einstein equations with two commuting Killing vectors admit an infinite-dimensional group of internal symmetries which allows one to obtain any solution starting from Minkowski spacetime.

Later, Kinnersley and Chitre~\cite{Kinnersley:1977pg,Kinnersley:1977ph,Kinnersley:1978pz,Kinnersley1978IV} inspected the internal symmetries of stationary and axisymmetric Einstein--Maxwell equations, and indeed they discovered an infinite-dimensional algebra by constructing its representation in terms of potentials which characterise any solution.
Shortly after, Maison~\cite{Maison:1978es} found that vacuum gravity can be related to a linear eigenvalue problem in the form of Lax, i.e.~that it is possible to reduce the solutions of non-linear Einstein equations to a sequence of linear problems.

The period of conjectures about integrability of Einstein gravity came to end in 1978, when Belinski and Zakharov~\cite{Belinsky1978}, by means of their \emph{inverse scattering method}, constructed an over-determined linear system with a complex spectral parameter, whose integrability conditions are equivalent to the two-dimensional reduced vacuum Einstein equations.
Thus, the original non-linear problem was reformulated in terms of an equivalent spectral problem.
This also implied the discovery of gravitational solitons and the formulation of the spectral problem in the language of a classical matrix Riemann problem.
The procedure was then completed in~\cite{Belinsky:1979mh}, where the authors found an expression for the conformal factor.
The electrovacuum version of the inverse scattering method does not admit a straightforward generalisation:
Alekseev~\cite{Alekseev1980} proved that, in the presence of the electromagnetic field, the solitons must possess complex (but not real) poles.

In the same year, Harrison~\cite{Harrison1978} constructed the so-called \emph{B\"acklund transformations} for vacuum Einstein equations, by using an approach due to Estabrook and Wahlquist known as ``prolongation structures'' for non-linear equations~\cite{Estabrook:1976fr}.
Almost simultaneously, Neugebauer~\cite{Neugebauer:1979iw} obtained the B\"acklund transformations following a different approach, that was further developed in~\cite{Neugebauer1980I,Neugebauer1980II}.
Finally, Julia~\cite{Julia:1982gx,Julia:1982tm} proved that the infinite-dimensional symmetries found by Geroch and Kinnersley are nothing but Kac--Moody symmetries, together with a wide treatment of the symmetries of generalised $\sigma$-models and supergravities.
In~\cite{Breitenlohner:1986um} the corresponding infinite-dimensional Geroch group was studied in detail.

These last developments were followed by a series of papers by Hauser and Ernst~\cite{Hauser:1979zz,Hauser:1979zza,Hauser1980I,Hauser1980II}, who proposed the integral equation method in order to exponentiate the infinitesimal symmetries of Kinnersley and Chitre and to obtain the finite solutions generating transformations, that culminated in the proof of the Geroch conjecture mentioned above~\cite{Hauser1981}.
A modification of the Hauser and Ernst formalism was then proposed by Sibgatullin~\cite{Sibgatullin1984,Sibgatullin:1991qr} in order to make its applicability simpler, and to make use only of the seed data on the axis of symmetry.

The monodromy transform approach proposed by Alekseev~\cite{Alekseev:1982hy,Alekseev1987,Alekseev1989,Alekseev:1999ph} started with the formulation of a spectral problem, like the inverse scattering method, but it also defined a monodromy data as a set of coordinate-independent functions of a complex parameter that characterises any local solution of vacuum or electrovacuum gravity.

All the developments we have described up to now are related to vacuum or electrovacuum gravity;
however, there are situations in which the presence of matter fields still makes the reduced Einstein equations integrable.

A very simple example is given by gravity coupled to a perfect fluid with stiff matter equation of state, where the inverse scattering method still works in the very same way as the vacuum case~\cite{Belinsky:1979wi}.
This case is equivalent to gravity plus a minimally coupled scalar field, which is particularly interesting because, as shown by Bekenstein~\cite{Bekenstein:1975ts}, it is possible to map the minimally coupled theory into the conformally coupled one:
thus, the generation methods can be used in the conformally coupled theory, where black holes and wormholes are known to exist~\cite{Astorino:2014mda,Barcelo:1999hq}.

The Einstein--Maxwel-dilaton system, which can be obtained from five-dimensional pure gravity upon Kaluza--Klein reduction, enjoys the symmetries of the vacuum case~\cite{Belinsky:1979mh}.
Symmetries are present even in the case of a dilatonic coupling different from the Kaluza--Klein one~\cite{Galtsov:1995mb,Galtsov:1995va}.
Regarding scalar fields, also the dilaton-axion systems are known to possess an integrable structure~\cite{Galtsov:1994pd,Garcia:1995qz,Galtsov:1996qko,Galtsov:1997jrl}.

Higher-dimensional gravity benefits of generating techniques as well, at least in some particular cases.
An extension of the inverse scattering method to higher dimensions in vacuum, which manages the problem of the metric normalisation, has been proposed in~\cite{Pomeransky:2005sj,Pomeransky:2006bd}, and it was lately applied to construct many interesting solutions which generalise the Myers--Perry black holes~\cite{Myers:1986un}, black rings~\cite{Emparan:2001wn,Iguchi:2006rd,Tomizawa:2005wv,Tomizawa:2006vp,Mishima:2005id,Chen:2012zb,Iguchi:2011qi},
black Saturn~\cite{Elvang:2007rd}, black di-rings~\cite{Iguchi:2007is,Evslin:2007fv}, bycicling black rings~\cite{Elvang:2007hs} and others~\cite{Rocha:2011vv,Emparan:2006mm,Emparan:2008eg}.
The Einstein--Maxwell reduced system represents a $\sigma$-model under a special ansatz, as shown in~\cite{Yazadjiev:2006hw}.

Supergravity theories, in the ungauged case, are known to admit the construction of a Belinski--Zakharov spectral problem, as shown in the simple case of the dilaton-axion system in~\cite{Bakas:1996dz,Bakas:1996gs}.
In the case of more complicated theories, part of the symmetry reduced equations takes the form of a $\sigma$-model.
The existence of a $\sigma$-model is a strong evidence of the integrability properties of a theory, and indeed it is usually possible to construct an associated linear spectral problem.
However, the existence of a spectral problem does not lead itself to a method for generating new solutions, because the space of solutions of the linear system is typically larger than the space of solutions of the original non-linear equations:
some reduction constraints must be imposed in order to provide the equivalence between the two systems.
An example is provided by 5D supergravities, whose integrability was established in~\cite{Bouchareb:2007ax,Galtsov:2008pbf,Galtsov:2008bmt,Galtsov:2008jjb}, and for which the inverse scattering method was constructed in~\cite{Figueras:2009mc}.
The inverse scattering procedure was also applied to some supergravity models in~\cite{Katsimpouri:2012ky,Katsimpouri:2013wka,Katsimpouri:2014ara}.
Studies on the integrability of the heterotic string theory were pursued in~\cite{Alekseev:2004zz,Alekseev:2008gh}.

One can observe, as a consequence of the historical perspective that we have presented, that many generation techniques exist, and thus there are many choices that can be made when dealing with the construction of exact solutions.
Each of the methods mentioned above comes with pros and cons, and their usefulness depends on the specific system that one is studying.
However, they are all restricted to the existence of a certain number of Killing vectors in the class of spacetime under consideration, more precisely $D-2$, (where $D$ is the spacetime dimension):
for our purposes, the class of metrics that we will be interested in are the \emph{stationary} and \emph{axisymmetric} spacetimes\footnote{Let us notice that, for $D>4$, a slightly different definition of stationary and axisymmetric spacetime than the $D=4$ case, is used~\cite{Godazgar:2009fi}.}.
In this Thesis, we will focus on the Ernst formalism (described in Sec.~\ref{sec:ernst}) and on the inverse scattering method (described in Sec.~\ref{sec:ism}).

The Ernst formalism~\cite{ErnstI,ErnstII} is based on a clever rewriting of the Einstein--Maxwell equations, which is suitable to make explicit the symmetries of the equations themselves:
the system is studied in an effective three-dimensional space, thus the Ernst construction operates a sort of dimensional reduction from four to three dimensions.
The Ernst equations, which represent the aforementioned rewriting, are a couple of complex equations written in terms of two complex potentials, dubbed $\ernst$ and $\Phi$.
Such potentials are defined in terms of the relevant components of the metric and of the Maxwell field, and thus they are in one-to-one correspondence with the fields of the theory:
given the metric and the Maxwell 1-form, the Ernst potentials are uniquely determined, and viceversa.

The Ernst equations allow us to reveal (a part of) the symmetries of the Einstein--Maxwell system:
such symmetries are encoded into some transformations that the Ernst potentials undergo, which leave the theory unchanged.
Thus, such transformations map a solution into another one, without changing the theory.
We will see that most of the symmetry maps of the Ernst equations are nothing but gauge transformations;
however, there are two maps, the \emph{Ehlers transformation}~\cite{Ehlers:1957zz} and the \emph{Harrison transformation}~\cite{Harrison1968} that act non-trivially on a seed solution.
We should specify that these maps actually represent four different transformations, basically because it is possible to Wick-rotate them in order to obtain non-equivalent behaviours.

More precisely, the Ehlers map allows one to add the NUT parameter to a given seed solution~\cite{Reina1975}, or, after a Wick-rotation, to add a rotation parameter that we will study extensively in Chapter~\ref{chap:swirl}.
On the other hand, the Harrison map adds a dyonic charge to a seed or, in its Wick-rotated version, an external electromagnetic field of Melvin type.
The second usage, in particular, was popularised by Ernst himself, who made use of it in order to regularise the charged C-metric~\cite{ErnstMelvin}.

The inverse scattering method~\cite{Belinsky1978}, that we will study in its original vacuum version, is based on the integrability properties of the Einstein equations.
It works, in contrast with the Ernst formalism, on an effective two-dimensional system, and it is thus a dimensional reduction from four to two dimensions.
The key intuition of Belinski and Zakharov is to rewrite the the gravitational equations as first order matrix equations, and then to recognise them as the integrability conditions of an over-determined system of matrix equations related to an
eigenvalue problem for some linear differential operator.
The eigenvalue problem is linear, and for this reason it is easier to solve than the original non-linear problem represented by the Einstein equations:
the strategy of the approach is to solve the eigenvalue problem associated to a seed solution, and then use it to algebraically build a new solution, without the direct integration of the equations of motion.
The fundamental building blocks needed in the construction of a new solution are the \emph{solitons}, that we will define later in Sec.~\ref{sec:ism}.

Because of the reduction to two dimensions, the inverse scattering method unravels more symmetries than the Ernst equations, and indeed it allows to add not only the NUT parameter to a seed, but also mass, angular momentum and acceleration.
Furthermore, and this is our main motivation in using it, the inverse scattering procedure allows us to easily construct multi-black hole spacetimes:
once one recognises that a pair of the aforementioned solitons gives rise to a black hole, then it suffices to add many pairs of solitons to obtain many black holes.
This modular structure is extremely powerful, since it allows us to keep under control the parametrisation of the solution under consideration and to construct a multi-source solution by using a non-linear superposition principle.

The rest of the Chapter is devoted to an accurate description of the Ernst formalism and of the inverse scattering method, since they will play a central role in the construction of the new solutions that we will present in the subsequent Chapters.

\section{Ernst equations}
\label{sec:ernst}

In this section we will derive the Ernst equations for the Einsten--Maxwell theory and unravel their symmetries, in order to find non-trivial transformations that map a solution of the theory into another one.
We will follow the original Ernst's papers~\cite{ErnstI,ErnstII}.

We begin by considering the Einstein--Maxwell action
\begin{equation}
S = \int d^4x \sqrt{-g} \biggl( R - \frac{1}{4} F_{\mu\nu} F^{\mu\nu} \biggr) \,,
\end{equation}
which gives rise to the Maxwell equations
\begin{equation}
\nabla_\mu F^{\mu\nu} = 0 \,,
\end{equation}
and to the Einstein equations with electromagnetic source
\begin{equation}
G_{\mu\nu} = 8\pi T_{\mu\nu} \,,
\end{equation}
that can also purposefully rewritten, by taking the trace, as
\begin{equation}
R_{\mu\nu} = 8\pi \biggl( T_{\mu\nu} - \frac{1}{2} T g_{\mu\nu} \biggr) \,.
\end{equation}
We have defined the electromagnetic stress-energy tensor as
\begin{equation}
T_{\mu\nu} = \frac{1}{4\pi} \biggl( F_{\mu\rho} F_\nu^{\;\;\rho} - \frac{1}{4} F_{\rho\sigma} F^{\rho\sigma} g_{\mu\nu} \biggr) \,.
\end{equation}
Being the stress-energy tensor traceless, $T=0$, the Einstein equations are just given by
\begin{equation}
\label{einstein-ricci}
R_{\mu\nu} = 8\pi T_{\mu\nu} \,.
\end{equation}
We have to consider an ansatz for the electromagnetic field and for the metric.
We are interested in stationary and axisymmetric fields~\cite{Heusler1996}:
we recall that a spacetime is \emph{stationary} when it admits a nowhere vanishing timelike Killing field, and it is \emph{axisymmetric} when it is invariant under the action  of the 1-parameter group $SO(2)$ and the fixed point set under the group is non-empty.

One can prove~\cite{Heusler1996} that a metric with such symmetries is given by the Lewis--Weyl--Papapetrou ansatz\footnote{Actually, such a form of the metric is legit when the trace of the Ricci tensor over the manifold orthogonal to the Killing fields is zero~\cite{Heusler1996}.}
\begin{equation}
\label{electric-metric}
{ds}^2 =
-f (dt - \omega d\phi)^2
+ f^{-1} \bigl[ e^{2\gamma} \bigl( {d\rho}^2 + {dz}^2 \bigr)
+ \rho^2 {d\phi}^2 \bigr] \,,
\end{equation}
where the functions $f$, $\omega$, $\gamma$ depend only on $(\rho,z)$.
The coordinates $(t,\phi,\rho,z)$ are called \emph{Weyl} or \emph{cylindrical} coordinates.
We will see that they will prove very useful to express the equations of motion in a suitable way.
We also choose an ansatz for the Maxwell potential\footnote{The ansatz is coherent with the Lewis--Weyl--Papapetrou symmetries, however it is not the most general stationary and axisymmetric choice.
A discussion of the constraints on the Maxwell field can be found in~\cite{Heusler1996}.}
\begin{equation}
\label{electric-maxwell}
A = A_t dt + A_\phi d\phi \,,
\end{equation}
where $A_t$ and $A_\phi$ depend only on $(\rho,z)$.

Since we are working in cylindrical coordinates $(\rho,\phi,z)$, and our functions depend on $(\rho,z)$ only, it will be convenient to express our equations in three-dimensional Euclidean space notation, thus introducing quantities like the axis unit vectors $(\hat{\rho},\hat{\phi},\hat{z})$, and the gradient, divergence, curl and Laplacian, that we report here for convenience
\begin{subequations}
\begin{align}
\nabla f & = \frac{\partial f}{\partial\rho} \hat{\rho}
+ \frac{1}{\rho} \frac{\partial f}{\partial\phi} \hat{\phi}
+ \frac{\partial f}{\partial z} \hat{z} \,, \\
\nabla \cdot \vec{V} & = \frac{1}{\rho} \frac{\partial (\rho V_\rho)}{\partial\rho}
+ \frac{1}{\rho} \frac{\partial V_\phi}{\partial\phi}
+ \frac{\partial V_z}{\partial z} \,, \\
\nabla \times \vec{V} & = \biggl( \frac{1}{\rho} \frac{\partial V_z}{\partial\phi} - \frac{\partial V_\phi}{\partial z} \biggr) \hat{\rho}
+ \biggl( \frac{\partial V_\rho}{\partial z} - \frac{\partial V_z}{\partial\rho} \biggr) \hat{\phi}
+ \frac{1}{\rho} \biggl( \frac{\partial (\rho V_\phi)}{\partial\rho} - \frac{\partial V_\rho}{\partial\phi} \biggr) \hat{z} \,, \\
\nabla^2 f & = \frac{1}{\rho} \frac{\partial}{\partial\rho} \biggl(\rho\frac{\partial f}{{\partial\rho}}\biggr)
+ \frac{1}{\rho^2} \frac{\partial^2 f}{{\partial\phi}^2}
+ \frac{\partial^2 f}{{\partial z}^2} \,.
\end{align}
\end{subequations}

\subsection{Einstein--Maxwell equations}

Let us define the Maxwell equations
\begin{equation}
M^\nu \coloneqq \nabla_\mu F^{\mu\nu} \,.
\end{equation}
We notice that the $\phi$-component can be written as
\begin{equation}
\label{m-phi}
\begin{split}
M^\phi & = e^{-2\gamma}f \biggl[ \rho^{-2} f \bigl( \nabla^2 A_\phi + \omega \nabla^2 A_t \bigr)
- 2 \rho^{-3} f \biggl( \frac{\partial A_\phi}{\partial\rho} + \omega \frac{\partial A_t}{\partial\rho} \biggr) \\
&\quad + \rho^{-2} f \nabla\omega \cdot \nabla A_t
+ \rho^{-2} \nabla f \cdot \bigl( \nabla A_\phi + \omega \nabla A_t \bigr) \biggr] \equiv 0 \,;
\end{split}
\end{equation}
one can show that Eq.~\eqref{m-phi} is equivalent to
\begin{equation}
\label{ernst-max1}
\nabla \cdot \bigl[
\rho^{-2} f \bigl( \nabla A_\phi + \omega \nabla A_t \bigr) \bigr] = 0 \,.
\end{equation}
Analogously, we can manipulate the $t$-component of the Maxwell equations
\begin{equation}
\begin{split}
M^t & = e^{-2\gamma}f \biggl[
\rho^{-2} f \omega \bigl( \nabla^2 A_\phi + \omega \nabla^2 A_t \bigr) - f^{-1} \nabla^2 A_t
+ \rho^{-2} \omega \nabla f \cdot \bigl( \nabla A_\phi + \omega \nabla A_t \bigr) \\
&\quad + f^{-2} \nabla f \cdot \nabla A_t
+ \rho^{-2} f \nabla\omega \cdot \bigl( \nabla A_\phi + 2\omega \nabla A_t \bigr)
- 2 \rho^{-3} f \omega \biggl( \frac{\partial A_\phi}{\partial\rho} + \omega \frac{\partial A_t}{\partial\rho} \biggr) \biggr] \equiv 0 \,,
\end{split}
\end{equation}
which can be written as
\begin{equation}
\label{ernst-max2}
\nabla \cdot \bigl[ -f^{-1} \nabla A_t + \rho^{-2} f \omega
\bigl( \nabla A_\phi + \omega \nabla A_t \bigr) \bigr] = 0 \,.
\end{equation}
Eqs.~\eqref{ernst-max1} and~\eqref{ernst-max2} are equivalent to the Maxwell equations.

Now we turn to the gravitational equations:
we make use of the Einstein equations in the form~\eqref{einstein-ricci}, and define
\begin{equation}
E_{\mu\nu} \coloneqq
R_{\mu\nu} - 8\pi T_{\mu\nu} \,.
\end{equation}
We consider the $tt$-component
\begin{equation}
\begin{split}
E_{tt} & =
-\frac{e^{-2\gamma}}{2\rho^2} \biggl\{
2\rho^{-2} f^3 \bigl[ (\nabla A_\phi)^2 + \omega^2 (\nabla A_t)^2 + 2\omega \nabla A_\phi \cdot \nabla A_t \bigr]
+ 2 f (\nabla A_t)^2 \\
&\quad + (\nabla f)^2 - \rho^{-2} f^4 (\nabla\omega)^2
- f \nabla^2 f \biggr\} \equiv 0 \,,
\end{split}
\end{equation}
that can also be written as
\begin{equation}
\label{ernst-einstein1}
f \nabla^2 f =
(\nabla f)^2 - \rho^{-2} f^4 (\nabla\omega)^2
+ 2 f (\nabla A_t)^2
+ 2\rho^{-2} f^3 \bigl( \nabla A_\phi + \omega \nabla A_t \bigr)^2 \,,
\end{equation}
and the $t\phi$-component
\begin{equation}
\begin{split}
E_{t\phi} & =
-\frac{e^{-2\gamma}}{2\rho^2} \biggl\{
2 f^3 \omega \bigl[ (\nabla A_\phi)^2 + \omega^2 (\nabla A_t)^2 + 2\omega \nabla A_\phi \cdot \nabla A_t \bigr]
+ \rho^2 \omega (\nabla f)^2 - f^4\omega (\nabla\omega)^2 \\
&\quad - \rho f \bigl[ 2\rho \bigl( 2 \nabla A_\phi \cdot \nabla A_t + \nabla f \cdot \nabla \omega \bigr)
+ \omega (2\rho (\nabla A_t)^2 + \rho \nabla^2 f \bigr) \bigr] \\
&\quad - \rho f^2 \biggl( \rho \nabla^2\omega - 2\frac{\partial\omega}{\partial\rho} \biggl) \biggr\}
\equiv 0 \,.
\end{split}
\end{equation}
Now we combine
\begin{equation}
\begin{split}
\omega E_{tt} + E_{t\phi} & =
-\frac{fe^{-2\gamma}}{2\rho^2} \biggl[
4\rho^{-2} \nabla A_\phi \cdot \nabla A_t
+ 4\rho^{-2} \omega (\nabla A_t)^2
+ 2\rho^{-2} \nabla f \cdot \nabla\omega \\
&\quad + \rho^{-2} f \nabla^2 \omega
-2 \rho^{-3} f \frac{\partial\omega}{\partial\rho} \biggr]
\equiv 0 \,,
\end{split}
\end{equation}
which is equivalent to
\begin{equation}
\label{ernst-einstein2}
\nabla \cdot \bigl[
\rho^{-2} f^2 \nabla\omega
+ 4\rho^{-2} f A_t \bigl( \nabla A_\phi + \omega \nabla A_t \bigr) \bigr] = 0 \,.
\end{equation}
The other non-trivial Einstein equations define $\gamma$ by quadratures:
such equations are obtained from $E_{\rho z}$ and $E_{\rho\rho} - E_{zz}$, and they are
\begin{equation}
\begin{split}
& \rho \partial_z\gamma
+ \rho^2 f^{-1} \partial_zA_t\partial_\rho A_t
- 2f (\omega\partial_zA_t-\partial_z A_{\varphi})(\omega\partial_\rho A_t-\partial_\rho A_\varphi) \\
&\quad - \frac{1}{2} \rho^2 f^{-2} \partial_z f\partial_\rho f
+ \frac{1}{2} f^2\partial_z\omega\partial_\rho\omega = 0 \,,
\end{split}
\end{equation}
and
\begin{equation}
\begin{split}
&
\rho \partial_\rho\gamma
+ \rho^2 f^{-1}(\partial_\rho A_t)^2-(\partial_z A_t)^2 \\
&\quad + f \Bigl\{
(\partial_zA_\varphi)^2
- (\partial_\rho A_\varphi)^2
+ \omega^2\Bigl[(\partial_zA_t)^2 - (\partial_\rho A_t)^2
\Bigr]
+ 2\omega(\partial_\rho A_t\partial_\rho A_\varphi-\partial_zA_t\partial_zA_\varphi)
\Bigr\} \\
&\quad + \rho^2 f^{-2}\Bigl[(\partial_z f)^2 - (\partial_\rho f)^2\Bigr]
+ f^2 \Bigl[(\partial_z\omega)^2 - (\partial_\rho \omega)^2\Bigr] = 0 \,.
\end{split}
\end{equation}
Once the other metric functions are known, $\gamma$ can be found by integrating the latter equations.

\subsection{Ernst potentials}

Now we can elaborate on the equations of motion~\eqref{ernst-max1},~\eqref{ernst-max2},~\eqref{ernst-einstein1} and~\eqref{ernst-einstein2}.
We begin by noticing that~\eqref{ernst-max1} contains a total divergence:
this means that we can introduce a potential associated to the vector whose divergence is zero.
Let us recall that, for any function $h(\rho,z)$ sufficiently well behaved, it holds
\begin{equation}
\label{potential}
\nabla \cdot \bigl( \rho^{-1} \hat{\phi} \times \nabla h \bigr) = 0 \,,
\end{equation}
as it can be verified by a direct computation.
This means that we can introduce a so-called twisted potential $\tilde{A}_\phi$, such that
\begin{equation}
\label{max-potential}
\hat{\phi} \times \nabla \tilde{A}_\phi \coloneqq
\rho^{-1} f \bigl( \nabla A_\phi + \omega \nabla A_t \bigr) \,,
\end{equation}
which, thanks to~\eqref{potential}, implies Eq.~\eqref{ernst-max1}.
Now we multiply the last equation by $\hat{\phi}\times$
\begin{equation}
\rho^{-1} \hat{\phi} \times \nabla A_\phi =
-f^{-1} \nabla \tilde{A}_\phi - \rho^{-1} \omega \hat{\phi} \times \nabla A_t \,,
\end{equation}
and then apply $\nabla\cdot$, to obtain
\begin{equation}
\nabla\cdot \bigl( f^{-1} \nabla \tilde{A}_\phi + \rho^{-1} \omega \hat{\phi} \times \nabla A_t \bigr)
= \nabla\cdot \bigl( \rho^{-1} \hat{\phi} \times \nabla A_\phi \bigr) \overset{\eqref{potential}}{=} 0 \,.
\end{equation}
This latter equation substitutes~\eqref{ernst-max1} as an equation of motion.

We also notice that, implementing the definition of the potential~\eqref{max-potential}, Eq.~\eqref{ernst-max2} is written as
\begin{equation}
\nabla \cdot \bigl( -f^{-1} \nabla A_t + \rho^{-1} \omega \hat{\phi} \times \nabla \tilde{A}_\phi \bigr) = 0 \,.
\end{equation}
We have then shown that the Maxwell equations are equivalent, via the potential $\tilde{A}_\phi$, to the system
\begin{subequations}
\label{ernst-system1}
\begin{align}
\nabla\cdot \bigl( f^{-1} \nabla \tilde{A}_\phi + \rho^{-1} \omega \hat{\phi} \times \nabla A_t \bigr) & = 0 \,, \\
\nabla \cdot \bigl( -f^{-1} \nabla A_t + \rho^{-1} \omega \hat{\phi} \times \nabla \tilde{A}_\phi \bigr) & = 0 \,.
\end{align}
\end{subequations}
The two potentials $A_t$, $\tilde{A}_\phi$ can be effectively packed in a complex potential
\begin{equation}
\Phi \coloneqq A_t + i \tilde{A}_\phi \,,
\end{equation}
so that Eqs.~\eqref{ernst-system1} are written as a single complex equation, namely
\begin{equation}
\nabla\cdot \bigl( f^{-1} \nabla \Phi + i \rho^{-1} \omega \hat{\phi} \times \nabla \Phi \bigr) = 0 \,.
\end{equation}
Now we turn to the Einstein equations~\eqref{ernst-einstein1} and~\eqref{ernst-einstein2}:
firstly, we note that Eq.~\eqref{potential}, with $h=A_t\tilde{A}_\phi$, gives
\begin{equation}
\nabla\cdot \bigl( \rho^{-1} \hat{\phi} \times A_t \nabla\tilde{A}_\phi \bigr) =
- \nabla\cdot \bigl( \rho^{-1} \hat{\phi} \times \tilde{A}_\phi \nabla A_t \bigr) \,.
\end{equation}
With this result, we can write Eq.~\eqref{ernst-einstein2} as
\begin{equation}
\label{ernst-einstein2-twist}
\nabla \cdot \bigl[
\rho^{-2} f^2 \nabla\omega
+ 2\rho^{-1} \hat{\phi}\times \Im\bigl( \Phi^* \nabla\Phi \bigr) \bigr] = 0 \,.
\end{equation}
This can be explicitly verified by expanding the product $\Im\bigl( \Phi^* \nabla\Phi \bigr)$.

As in the case of the Maxwell equations, we can introduce a twisted gravitational potential $\chi$ from equation~\eqref{ernst-einstein2-twist}, such that
\begin{equation}
\label{grav-potential}
\hat{\phi} \times \nabla\chi \coloneqq
-\rho^{-1} f^2 \nabla\omega
-2 \hat{\phi} \times \Im\bigl( \Phi^* \nabla\Phi \bigr) \,.
\end{equation}
Multiplying by $\hat{\phi}\times$ and applying $\nabla\cdot$ as before, we obtain
\begin{equation}
\nabla\cdot \bigl[
f^{-2} \nabla\chi + 2 f^{-2} \Im\bigl( \Phi^* \nabla\Phi \bigr) \bigr] = 0 \,,
\end{equation}
that substitutes Eq.~\eqref{ernst-einstein2}.
Finally, we elaborate Eq.~\eqref{ernst-einstein1} to implement the potential $\chi$, and we find
\begin{equation}
f \nabla^2 f =
(\nabla f)^2 - \bigl[ \nabla\chi + 2\Im\bigl( \Phi^* \nabla\Phi \bigr)]^2
+ 2 f \nabla\Phi \cdot \nabla\Phi^* \,.
\end{equation}
Summarising, the Einstein equations~\eqref{ernst-einstein1},~\eqref{ernst-einstein2} are replaced by the equations
\begin{subequations}
\label{ernst-system2}
\begin{gather}
\nabla\cdot \bigl[
f^{-2} \nabla\chi + 2 f^{-2} \Im\bigl( \Phi^* \nabla\Phi \bigr) \bigr] = 0 \,, \\
f \nabla^2 f =
(\nabla f)^2 - \bigl[ \nabla\chi + 2\Im\bigl( \Phi^* \nabla\Phi \bigr)]^2
+ 2 f \nabla\Phi \cdot \nabla\Phi^* \,.
\end{gather}
\end{subequations}
We introduce a gravitational complex potential
\begin{equation}
\ernst \coloneqq
f - \big|\Phi\big|^2 + i \chi \,,
\end{equation}
which, together with $\Phi$, allows us to write the systems~\eqref{ernst-system1} and~\eqref{ernst-system2} as the two complex equations
\begin{subequations}
\label{ernst-eqs}
\begin{align}
\bigl( \Re\ernst + \big|\Phi\big|^2 \bigr) \nabla^2\ernst & =
\nabla\ernst\cdot \bigl( \nabla\ernst + 2\Phi^* \nabla\Phi \bigr) \,, \\
\bigl( \Re\ernst + \big|\Phi\big|^2 \bigr) \nabla^2\Phi & = \nabla\Phi\cdot \bigl( \nabla\ernst + 2\Phi^* \nabla\Phi \bigr) \,.
\end{align}
\end{subequations}
Eqs.~\eqref{ernst-eqs} are called \emph{Ernst equations}, and $\ernst$ and $\Phi$ are known as the \emph{Ernst complex potentials}.

We can also express the equations for $\gamma$ in term of the Ernst potentials, and they are given by
\begin{align}
\begin{split}
\partial_{\rho}\gamma & =
\frac{\rho}{4(\Re\ernst + \Phi\Phi^{*})^2} \Bigl[
\bigl(\partial_{\rho}\ernst + 2\Phi^{*}\partial_{\rho}\Phi\bigr)
\bigl(\partial_{\rho}\ernst^{*} + 2\Phi\partial_{\rho}\Phi^{*}\bigr) \\
&\quad - \bigl(\partial_{z}\ernst + 2\Phi^{*}\partial_{z}\Phi\bigr)
\big(\partial_{z}\ernst^{*} + 2\Phi\partial_{z}\Phi^{*}\bigr)
\Bigr] \\
&\quad - \frac{\rho}{\Re\ernst + \Phi\Phi^{*}}
\big(\partial_{\rho}\Phi\partial_{\rho}\Phi^{*} - \partial_{z}\Phi\partial_{z}\Phi^{*}\big) \,,
\end{split}
\\
\begin{split}
\partial_{z}\gamma & =
\frac{\rho}{4(\Re\ernst + \Phi\Phi^{*})^2}
\Big[
\big(\partial_{\rho}\ernst + 2\Phi^{*}\partial_{\rho}\Phi\big)
\big(\partial_{z}\ernst^{*} + 2\Phi\partial_{z}\Phi^{*}\big) \\
&\quad + \big(\partial_{z}\ernst + 2\Phi^{*}\partial_{z}\Phi\big)
\big(\partial_{\rho}\ernst^{*} + 2\Phi\partial_{\rho}\Phi^{*}\big)
\Big] \\
&\quad - \frac{\rho}{\Re\ernst + \Phi\Phi^{*}}
\big(\partial_{\rho}\Phi\partial_{z}\Phi^{*} - \partial_{z}\Phi\partial_{\rho}\Phi^{*}\big) \,.
\end{split}
\end{align}
As before, we notice that such equations can be solved by quadratures.

We have formally reduced the problem of solving the Einstein--Maxwell equations for a stationary and axisymmetric spacetime, to the problem of solving the Ernst equations~\eqref{ernst-eqs}.
We notice that Eqs.~\eqref{ernst-eqs} represent an effective three-dimensional problem, and as such we can forget about the four-dimensional origin of the problem.
Furthermore, the writing~\eqref{ernst-eqs} allows us to study in an efficient way the symmetries of the Einstein--Maxwell equations, and to make use of such symmetries to construct new solutions starting from a seed.
We will develop this topic in the following.

\subsection{Symmetries of the Ernst equations}

One can notice that the Ernst equations~\eqref{ernst-eqs} can be derived from an effective three-dimensional action, given by
\begin{equation}
\label{ernst-action}
S = \int d^3x \,
\frac{(\nabla\ernst + 2\Phi^*\nabla\Phi)\cdot(\nabla\ernst^* + 2\Phi\nabla\Phi^*) - \nabla\Phi\cdot\nabla\Phi^*}{\Re\ernst + \big|\Phi\big|^2} \,.
\end{equation}
This is a quite useful result, because now we can find the symmetries of the Ernst equations~\eqref{ernst-eqs} just by studying the symmetries of the Ernst action~\eqref{ernst-action}\footnote{Actually, not all the symmetries of the equations of motion correspond to the symmetries of the action.
The equations of motion might enjoy more symmetries than the action does.
However, the symmetries of the action are surely symmetries of the equations of motion, and in this case they coincide}.

A smart way to study the symmetries of the action~\eqref{ernst-action} is to consider the quadratic form associated to~\eqref{ernst-action}, i.e.~the associated metric~\cite{Stephani:2003tm}.
Let us consider $\ernst$ and $\Phi$ as complex coordinates, and introduce the real coordinates $(x,y,u,v)$ as
\begin{equation}
\ernst = x + i y \,, \quad
\Phi = u + i v \,.
\end{equation}
This way, the metric associated to the action~\eqref{ernst-action} is
\begin{equation}
\label{ernst-metric}
{ds}^2 = \frac{1}{4 (u^2+v^2+x)^2}
\bigl[ {dx}^2 + {dy}^2 - 4x \bigl( {du}^2 + {dv}^2 \bigr)
- 4v dy du + 4u dy dv + 4v dx dv + 4u dx du \bigr] \,.
\end{equation}
Such a representation is particularly useful, since the Killing vectors of the metric~\eqref{ernst-metric} are equivalent to the infinitesimal generators of the symmetries of the action~\eqref{ernst-action}.
The Killing vectors are defined by the Killing equation
$\nabla_{(i}\xi_{j)} = 0$, which can be solved for the four-dimensional metric~\eqref{ernst-metric} to find
\begin{subequations}
\label{ernst-killing}
\begin{align}
\xi_1 & = 4xy \partial_x + 2(y^2-x^2) \partial_y + 2(xv+yu) \partial_z + 2(yv-xu) \partial_v \,, \\
\xi_2 & = 2(xv+yu) \partial_x + 2(yv-xu) \partial_y + (4uv-y) \partial_z + (2v^2-2u^2+x) \partial_v \,, \\
\xi_3 & = 2(xu-yv) \partial_x + 2(xv+yu) \partial_y + (z^2-v^2+x) \partial_z + (4uv+y) \partial_v \,, \\
\xi_4 & = 4x \partial_x + 4y \partial_y + 2z \partial_z + 2v \partial_v \,, \\
\xi_5 & = -v \partial_z + z \partial_v \,, \\
\xi_6 & = -2v \partial_x + 2u \partial_y + \partial_v \,, \\
\xi_7 & = 2u \partial_x + 2v \partial_y - \partial_z \,, \\
\xi_8 & = 4 \partial_y \,.
\end{align}
\end{subequations}
Thus, we find eight Killing vectors for the quadratic form~\eqref{ernst-metric}, that correspond to eight infinitesimal symmetries for the action~\eqref{ernst-action}.

We are actually interested in the finite transformations generated by the Killing vectors~\eqref{ernst-killing}:
we need to integrate the flow generated by such vectors in order to find them.
Practically, the equations that define the flow and that we have to integrate are
\begin{equation}
\frac{\partial x^i}{\partial \epsilon} = \xi^i\bigl(x^i\bigr) \,,
\end{equation}
where $x^i$ identifies the coordinates, i.e.~$x^i=(x,y,u,v)$, and $\epsilon$ is the flow parameter of the finite transformation.
More details about the connection between the infinitesimal and the finite transformations can be found in~\cite{Stephani1990}.

We can integrate the infinitesimal transformations~\eqref{ernst-killing}, and find the finite transformations~\cite{Stephani:2003tm}
\begin{subequations}
\label{ernst-group}
\begin{align}
\label{gauge1}
\ernst' & = |\lambda|^2 \ernst \,, \qquad\qquad\qquad\quad\;\;
\Phi' = \lambda \Phi \,, \\
\label{gauge2}
\ernst' & = \ernst + ib \,, \qquad\qquad\qquad\quad\,
\Phi' = \Phi \,, \\
\label{ehlers}
\ernst' & = \frac{\ernst}{1 + ic\ernst} \,, \qquad\qquad\qquad\;\,
\Phi' = \frac{\Phi}{1 + ic\ernst} \,, \\
\label{gauge3}
\ernst' & = \ernst - 2\beta^*\Phi - |\beta|^2 \,, \qquad\quad\!
\Phi' = \Phi + \beta \,, \\
\label{harrison}
\ernst' & = \frac{\ernst}{1 - 2\alpha^*\Phi - |\alpha|^2\ernst} \,, \qquad
\Phi' = \frac{\alpha\ernst + \Phi}{1 - 2\alpha^*\Phi - |\alpha|^2\ernst} \,,
\end{align}
\end{subequations}
where $\alpha$, $\beta$ and $\lambda$ are complex parameters, while $b$ and $c$ are real.

One can prove~\cite{Stephani:2003tm,2040113} that transformations~\eqref{gauge1},~\eqref{gauge2} and~\eqref{gauge3} are actually gauge transformations, i.e.~diffeomorphisms which do not change the nature of the solutions.
They can be reabsorbed by a redefinition of the coordinates, and as such they are not interesting, since they do not produce any new solutions.
The interesting transformations are given by~\eqref{ehlers} and~\eqref{harrison}, that are called \emph{Ehlers transformation}~\cite{Ehlers:1957zz} and \emph{Harrison transformation}~\cite{Harrison1968}, respectively.

The transformations~\eqref{ernst-group} form a group, that one can show to be SU(2,1)~\cite{Neugebauer1969,Kinnersley1973}.
Indeed, it is known that SU(2,1) represents the symmetry group of the Einstein--Maxwell equations.

\subsubsection{Discrete transformation}

We have analysed the symmetry transformations for the ansatz given by the metric~\eqref{electric-metric} and the Maxwell potential~\eqref{electric-maxwell}, which together represent the class of stationary and axisymmetric spacetimes.

Actually, they do not represent the unique choice of ansatz for such a class of solutions.
There is another (non-equivalent) ansatz that one can consider, which is related to the first one by a discrete double-Wick rotation
\begin{equation}
t \to i\psi \,, \quad
\phi \to i \tau \,,
\end{equation}
that produces the following metric
\begin{equation}
\label{magnetic-metric}
{ds}^2 =
f (d\psi - \omega d\tau)^2
+ f^{-1} \bigl[ e^{2\gamma} \bigl( {d\rho}^2 + {dz}^2 \bigr)
- \rho^2 {d\tau}^2 \bigr] \,,
\end{equation}
and Maxwell field
\begin{equation}
\label{magnetic-maxwell}
A = i A_\tau d\tau + i A_\psi d\psi \,.
\end{equation}
The imaginary unit in~\eqref{magnetic-maxwell} does not represent an issue, since it can be easily reabsorbed into a redefinition of the electric (or magnetic) charge.

One can reproduce the Ernst computations that we showed previously, for the metric~\eqref{magnetic-metric} and the potential~\eqref{magnetic-maxwell}.
Almost everything works as above, except for the definition of the twisted potential, that now reads
\begin{equation}
\hat{\phi} \times \nabla \tilde{A}_\tau \coloneqq
\rho^{-1} f \bigl( \nabla A_\tau + \omega \nabla A_\psi \bigr) \,,
\end{equation}
and the definition of the Ernst complex potential
\begin{equation}
\Phi \coloneqq A_\psi + i \tilde{A}_\tau \,.
\end{equation}
The roles of the time and azimuthal coordinates are exchanged, as one might expect.

One recovers again the Ernst transformations~\eqref{ernst-group}, where the Ehlers map~\eqref{ehlers} and the Harrison map~\eqref{harrison} are the only relevant transformations.
In this case, however, their action will produce different solutions with respect to the previous case, being the seed solution non-equivalent to the ansatz~\eqref{electric-metric} and~\eqref{electric-maxwell}.
This means that we have four non-trivial maps, two Ehlers and two Harrison transformations.

Thus, we will be considering two different ans\"atze for the application of the Ehlers and the Harrison transformations:
the \emph{electric} ansatz
\begin{subequations}
\label{electric-ansatz}
\begin{align}
{ds_e}^2 & =
-f (dt - \omega d\phi)^2
+ f^{-1} \bigl[ e^{2\gamma} \bigl( {d\rho}^2 + {dz}^2 \bigr)
+ \rho^2 {d\phi}^2 \bigr] \,, \\
A_e & = A_t dt + A_\phi d\phi \,,
\end{align}
\end{subequations}
and the \emph{magnetic} ansatz
\begin{subequations}
\label{magnetic-ansatz}
\begin{align}
{ds_m}^2 & =
f (d\phi - \omega dt)^2
+ f^{-1} \bigl[ e^{2\gamma} \bigl( {d\rho}^2 + {dz}^2 \bigr)
- \rho^2 {dt}^2 \bigr] \,, \\
A_m & = A_t dt + A_\phi d\phi \,.
\end{align}
\end{subequations}
The terminology ``electric'' and ``magnetic'' is not common in the literature:
we adopt it because of the way the Harrison transformation works with the different seeds~\eqref{electric-ansatz} and~\eqref{magnetic-ansatz}, when the transformation parameter is real.
In the first case, the Harrison map adds an electric charge, while in the second one it adds an external magnetic field;
then it is quite natural to use the words ``electric'' and ``magnetic''.

In the remaining part of the Section, we will analyse the effects of the non-trivial transformations given by~\eqref{ehlers} and~\eqref{harrison} for the two different kinds of metric, the electric~\eqref{electric-ansatz} and the magnetic~\eqref{magnetic-ansatz} one.

\subsection{Ehlers transformation}

We now analyse the action of the Harrison transformation~\eqref{harrison}, that we report here for convenience:
\begin{equation}
\ernst' = \frac{\ernst}{1 + ic\ernst} \,,
\qquad \Phi' = \frac{\Phi}{1 + ic\ernst} \,.
\end{equation}
As noted above, such a map will produce a different result if applied to the electric ansatz~\eqref{electric-ansatz} or the magnetic one~\eqref{magnetic-ansatz}.
We recall that the parameter $c$ is real.

The effect of the Ehlers map on the electric ansatz is to add the NUT parameter to the spacetime under consideration, as shown for the first time in~\cite{Reina1975}.
On the other hand, the action of the Ehlers transformation on the magnetic ansatz has never been investigated before, and it will be the subject of Chapter~\ref{chap:swirl}.

\subsubsection{Electric ansatz}

We start by considering the electric metric~\eqref{electric-ansatz}, and deal with the explicit example of the Schwarzschild spacetime
\begin{equation}
{ds}^2 =
-\biggl(1 - \frac{2M}{r}\biggr) {dt}^2
+ \frac{{dr}^2}{1 - 2M/r}
+ r^2 \bigl( {d\theta}^2 + \sin^2\theta {d\phi}^2 \bigr) \,.
\end{equation}
In this case, the Maxwell field is zero, so $\Phi=0$.
Being $f=1-2M/r$, the gravitational Ernst potential is given by
\begin{equation}
\ernst = 1 - \frac{2M}{r} \,.
\end{equation}
We now apply the Ehlers transformation, and obtain the new Ernst potentials
\begin{equation}
\ernst' = \frac{r^2 - 2Mr}{r^2 + c^2 (r-2M)^2}
-i \frac{c(r - 2M)^2}{r^2 + c^2 (r-2M)^2} \,, \qquad
\Phi' = 0 \,.
\end{equation}
We see that the Ehlers map does not add an electromagnetic field to a seed solution that does not possess it from the beginning.
Therefore, it maps a vacuum solution into a vacuum solution.

From the potential $\ernst$ we are able to read the new functions
\begin{equation}
f' = \frac{r^2 - 2Mr}{r^2 + c^2 (r-2M)^2} \,, \quad
\chi' = - \frac{c(r - 2M)^2}{r^2 + c^2 (r-2M)^2} \,,
\end{equation}
and to integrate $\omega'$ from the definition of the twist potential $\chi'$~\eqref{grav-potential}:
\begin{equation}
\omega' = 4cM \cos\theta \,.
\end{equation}
Regarding the function $\gamma$, it can be easily verified that the Ehlers transformation does not modify it, i.e.~$\gamma'=\gamma$.

Collecting all the information, we can write down the resulting metric
\begin{equation}
\begin{split}
ds'^2 & =
-\frac{r^2-2Mr}{r^2+c^2(r-2M)^2} \bigl(dt - 4Mc\cos\theta d\phi\bigr)^2
+ \frac{r^2+c^2(r-2M)^2}{r^2-2Mr} {dr}^2 \\
&\quad + \bigl(r^2+c^2(r-2M)^2\bigr) \bigl({d\theta}^2 + \sin^2\theta {d\phi}^2 \bigr) \,.
\end{split}
\end{equation}
Obviously, for $c=0$ we recover the Schwarzschild metric, i.e.~the seed solution.

We can perform a change of coordinates and a parameter redefinition, in order to identify the new metric.
Let us consider the change of coordinates~\footnote{The transformation of the radial coordinate is suggested by the $S^2$ element in the metric.}
\begin{equation}
R = r \sqrt{1+c^2} - \frac{2Mc^2}{\sqrt{1+c^2}} \,, \quad
T = \frac{t}{\sqrt{1+c^2}} \,,
\end{equation}
and the parameter redefinitions
\begin{equation}
M = -\frac{n}{2c} \sqrt{1+c^2} \,, \quad
c = \frac{m - \sqrt{m^2 + n^2}}{n} \,,
\end{equation}
that give the following metric:
\begin{equation}
\begin{split}
{ds'}^2 &=
-\frac{R^2-2mR-n^2}{R^2+n^2} \bigl(dT - 2n\cos\theta d\phi\bigr)^2
+ \frac{R^2+n^2}{R^2-2mR-n^2} {dR}^2 \\
&\quad + \bigl(R^2+n^2\bigr) \bigl({d\theta}^2 + \sin^2\theta {d\phi}^2 \bigr) \,.
\end{split}
\end{equation}
This solution is the well known Taub--NUT metric~\cite{Taub:1950ez,Newman:1963yy}, where $m$ is the mass and $n$ is the so-called NUT charge.
Thus, we conclude that the Ehlers transformation adds the NUT parameter to a given seed solution.
One can explicitly check, e.g., that the application of the Ehlers map to the Reissner--Nordstr\"om black hole produces the charged Taub--NUT metric.

The Taub--NUT metric has been extensively studied in the literature, starting with the famous paper by Misner~\cite{Misner:1965zz}, and recently it has been subjected to a renewed interest~\cite{Clement:2015cxa}, which led to the study of its thermodynamical properties by many authors.

\subsubsection{Magnetic ansatz}

The Ehlers transformation applied to the magnetic ansatz~\eqref{magnetic-ansatz} produces a new solution that will be investigated in Chapter~\ref{chap:swirl}.

\subsection{Harrison transformation}

The Harrison transformation is given by
\begin{equation}
\ernst' = \frac{\ernst}{1 - 2\alpha^*\Phi - |\alpha|^2\ernst} \,,
\qquad \Phi' = \frac{\alpha\ernst + \Phi}{1 - 2\alpha^*\Phi - |\alpha|^2\ernst} \,,
\end{equation}
where $\alpha$ is a complex parameter.

The effect of the Harrison map on the electric ansatz is to add a dyon, i.e.~an electric and a magnetic charge.
On the other hand, the action of the Harrison transformation on the magnetic ansatz embeds the given spacetime in the Melvin universe, which is a spacetime filled with ``uniform'' electromagnetic field.

\subsubsection{Electric ansatz}

We consider again the electric ansatz~\eqref{electric-ansatz} with the Schwarzschild metric
\begin{equation}
{ds}^2 =
-\biggl(1 - \frac{2M}{r}\biggr) {dt}^2
+ \frac{{dr}^2}{1 - 2M/r}
+ r^2 \bigl( {d\theta}^2 + \sin^2\theta {d\phi}^2 \bigr) \,.
\end{equation}
We observe, just by looking at the structure of the Harrison map, that it always produces a non-trivial electromagnetic field, even if starting with a vacuum spacetime.

Here, it is again
\begin{equation}
\ernst = 1 - \frac{2M}{r} \,,
\end{equation}
but now
\begin{equation}
\Phi' = \alpha \frac{r-2M}{r-|\alpha|^2 (r-2M)} \,,
\end{equation}
and
\begin{equation}
\ernst' = \frac{r-2M}{r-|\alpha|^2 (r-2M)} \,,
\end{equation}
from which we read (notice that $\ernst$ is real)
\begin{equation}
f' = \frac{r(r-2M)}{\bigl[ 2M|\alpha|^2 + r(1-|\alpha|^2) \bigr]^2} \,, \quad
\chi' = 0 \,.
\end{equation}
The remaining function $\gamma$ is again left unchanged, $\gamma'=\gamma$.

From the Ernst potential $\Phi'$ and by integrating the definition of the electromagnetic twist potential~\eqref{max-potential}, we find the components of the new Maxwell potential
\begin{equation}
A' = \alpha_R \frac{r-2M}{r-|\alpha|^2(r-2M)} dt
+ 2\alpha_I M \cos\theta d\phi \,,
\end{equation}
where $\alpha \coloneqq \alpha_R + i\alpha_I$.
On the other hand, the metric is
\begin{equation}
\begin{split}
{ds'}^2 & =
-\frac{r(r-2M)}{\bigl[ 2M|\alpha|^2 + r(1-|\alpha|^2) \bigr]^2} {dt}^2
+ \frac{\bigl[ 2M|\alpha|^2 + r(1-|\alpha|^2) \bigr]^2}{r(r-2M)} {dr}^2 \\
&\quad + \bigl[ 2M|\alpha|^2 + r(1-|\alpha|^2) \bigr]^2 \bigl( {d\theta}^2 + \sin^2\theta {d\phi}^2 \bigr) \,.
\end{split}
\end{equation}
Being $\chi'=0$, we found that $\omega'=0$.

We perform the change of coordinates
\begin{equation}
R = r(1-|\alpha|^2) + 2M|\alpha|^2 \,, \quad
T = \frac{t}{1-|\alpha|^2} \,,
\end{equation}
the reparametrisation
\begin{equation}
p = 2\alpha_I M \,, \quad
q = -\frac{2\alpha_R M}{1-|\alpha|^2} \,, \quad
m = M (1+|\alpha|^2) \,,
\end{equation}
and define the dyon $e^2 \coloneqq q^2+p^2$, to end up with
\begin{align}
{ds'}^2 & =
- \biggl(1 - \frac{2m}{R} + \frac{e^2}{R^2}\biggr) {dt}^2
+ \biggl(1 - \frac{2m}{R} + \frac{e^2}{R^2}\biggr)^{-1} {dR}^2
+ R^2 \bigl( {d\theta}^2 + \sin^2\theta {d\phi}^2 \bigr) \,, \\
A' & = \frac{q}{R} dt
+ p \cos\theta d\phi \,.
\end{align}
We clearly recognise the dyonic Reissner--Nordstr\"om black hole~\cite{Reissner,Nordstrom}, where $m$ is the mass, $q$ is the electric charge and $p$ is the magnetic charge.
Thus, the Harrison transformation adds a dyonic charge to a seed solution of the form~\eqref{electric-metric}.

\subsubsection{Magnetic ansatz}

We now turn to the consideration of the magnetic ansatz~\eqref{magnetic-ansatz}, again for the example of the Schwarzschild metric.
The difference from the previous case stands in the function $f=r^2\sin^2\theta$, that gives
\begin{equation}
\ernst = r^2 \sin^2\theta \,.
\end{equation}
A non-null electromagnetic field is generated via the Harrison map
\begin{equation}
\Phi' = \frac{\alpha r^2 \sin^2\theta}{1-|\alpha|^2 r^2 \sin^2\theta} \,,
\end{equation}
and
\begin{equation}
\ernst' = \frac{r^2 \sin^2\theta}{1-|\alpha|^2 r^2 \sin^2\theta} \,.
\end{equation}
We can read the Maxwell field from $\Phi'=A'_\phi+i \tilde{A}'_t$, and integrating the Maxwell twist potential~\eqref{max-potential}
\begin{equation}
A' =
2\alpha_I (r-2M) \cos\theta dt
+ \alpha_R \frac{r^2 \sin^2\theta}{1-|\alpha|^2 r^2 \sin^2\theta} d\phi \,,
\end{equation}
where again $\alpha\coloneqq\alpha_R + i\alpha_I$.

We choose $\alpha_R=-B/2$ and $\alpha_I=E/2$, so that the metric and the Maxwell field become
\begin{align}
{ds'}^2 & =
\Lambda^2 \biggl[ -\biggl(1 - \frac{2M}{r}\biggr) {dt}^2
+ \frac{{dr}^2}{1 - 2M/r} + r^2 {d\theta}^2 \biggr]
+ \Lambda^{-2} r^2 \sin^2\theta {d\phi}^2 \,, \\
A' & =
E (r-2M) \cos\theta dt
-\frac{B}{2} \Lambda^{-1} r^2 \sin^2\theta d\phi \,,
\end{align}
where
\begin{equation}
\Lambda \coloneqq 1 - \frac{E^2 + B^2}{4} r^2 \sin^2\theta \,.
\end{equation}
This solution is known as the Schwarzschild--Melvin spacetime~\cite{ErnstMelvin}, and it represents a static black hole embedded in an electromagnetic universe, i.e.~a universe filled of a ``uniform'' electric field $E$ and magnetic field $B$.
The most known version of such a solution is the purely magnetic one, with $E=0$.

The background is recovered by putting $M=0$
\begin{align}
{ds'}^2 & =
\Lambda^2 \bigl( - {dt}^2 + {dr}^2 + r^2 {d\theta}^2 \bigr)
+ \Lambda^{-2} r^2 \sin^2\theta {d\phi}^2 \,, \\
A' & =
E r \cos\theta dt
-\frac{B}{2} \Lambda^{-1} r^2 \sin^2\theta d\phi \,,
\end{align}
which is the electromagnetic generalisation of the so-called Melvin universe~\cite{Melvin:1963qx}.
Hence, the Harrison transformation, applied to a magnetic seed~\eqref{magnetic-ansatz}, adds an external electromagnetic field to the spacetime.

We point out that the Reissner--Nordstr\"om and the Melvin spacetimes are related by an analytic continuation, similarly to what happens to the electric and magnetic seeds:
that was proven by Gibbons and Wiltshire in~\cite{Gibbons:1986wg}, however not in the context of the Harrison transformation.
Such an analytic continuation was later implemented in~\cite{Astorino:2012zm} to add the cosmological constant to the Melvin spacetime.

\section{Inverse scattering method in vacuum}
\label{sec:ism}

The inverse scattering construction relies on the identification of a linear eigenvalue equation, whose integrability condition corresponds to the non-linear equations one aims to solve.
We will describe the procedure developed by Belinski and Zakharov for the Einstein equations in vacuum, so we will work with pure General Relativity with action
\begin{equation}
S = \int d^4x \sqrt{-g} \, R \,,
\end{equation}
and Einstein equations
\begin{equation}
R_{\mu\nu} = 0 \,.
\end{equation}
This section is based on the original papers by Belinski and Zakharov~\cite{Belinsky1978,Belinsky:1979mh} and on the book~\cite{Belinski:2001ph}.

\subsection{Integrable ansatz}

We work with the following metric ansatz in Weyl coordinates $(t,\phi,\rho,z)$:
\begin{equation}
{ds}^2 = f(\rho,z) \bigl( {d\rho}^2 + {dz}^2 \bigr)
+ g_{ab}(\rho,z) dx^a dx^b \,.
\end{equation}
Such a metric is the most general stationary and axisymmetric spacetime:
it is written in a form suitable for our purposes, and it contains the Lewis--Weyl--Papapetrou metric~\eqref{electric-metric} as a subcase~\cite{Heusler1996}.
Here, $f$ is a function\footnote{$f$ should not be confused with the function appearing in~\eqref{electric-metric}!} and $g$ is a $2\times2$ matrix.
The Latin indices take the values $0,1$, which correspond to $t$ and $\phi$.

The freedom in the choice of the coordinates $\rho$ and $z$ can be used, without loss of generality, to impose
\begin{equation}
\det g = -\rho^2 \,.
\end{equation}
This is a fundamental property, that must be preserved by the generation technique, as we will see in the following.

It is convenient to rewrite the vacuum Einstein equations in matrix form, in order to apply the inverse scattering formalism.
The vacuum equations naturally split into two groups, one for the matrix $g$ and the other for the function $f$.
We start by exploiting the first group.

One can easily show that the Einstein equations
\begin{equation}
R^t_{\;\;t} = 0 \,, \quad
R^\phi_{\;\;t} = 0 \,, \quad
R^t_{\;\;\phi} = 0 \,, \quad
\end{equation}
correspond, respectively, to the components $(0,0)$, $(0,1)$ and $(1,1)$ of the second order matrix equation
\begin{equation}
\label{einstein-matrix}
\partial_\rho \bigl( \rho \partial_\rho g \, g^{-1} \bigr) + \partial_z \bigl( \rho \partial_z g \, g^{-1} \bigr) = 0 \,.
\end{equation}
On the other hand, the Einstein equations
\begin{equation}
R_{\rho\rho} - R_{zz} = 0 \,, \quad
R_{\rho z} = 0 \,,
\end{equation}
determine the first order equations for $f$
\begin{subequations}
\label{einstein-f}
\begin{align}
\partial_\rho \log f & = -\frac{1}{\rho} + \frac{1}{4\rho} \tr \bigl( U^2 - V^2 \bigr) \,, \\
\partial_z \log f & = \frac{1}{2\rho} \tr \bigl( UV \bigr) \,,
\end{align}
\end{subequations}
where we have defined the $2\times2$ matrices
\begin{equation}
\label{uv}
U \coloneqq \rho \bigl(\partial_\rho g\bigr) g^{-1} \,, \quad
V \coloneqq \rho \bigl(\partial_z g\bigr) g^{-1} \,.
\end{equation}
With this first step, we managed to rewrite the Einstein equations in matrix form.
Such a form is particularly useful to recognise an integrability condition for the equations.
We explicitly notice that, as it happened in Sec.~\ref{sec:ernst}, the equations for the $(\rho,z)$ part of the metric, i.e.~for the function $f$~\eqref{einstein-f}, completely decouple from the other Einstein equations.
This implies that we can forget about $f$ and work only on $g$:
once we have found $g$, we can simply integrate Eqs.~\eqref{einstein-f} by quadratures.

\subsection{Integration scheme}

We want to translate~\eqref{einstein-matrix} into an equivalent system consisting of the relations~\eqref{uv} and two first order matrix equations for the matrices $U$ and $V$.
The first equation of such a system is obviously obtained by rewriting~\eqref{uv} in terms of $U$ and $V$:
\begin{equation}
\label{uv-eq1}
\partial_\rho U + \partial_z V = 0 \,.
\end{equation}
The second equation is obtained as the integrability condition of~\eqref{uv} with respect to $g$:
with this, we mean a differential equation for $U$ and $V$ which is identically satisfied when it is written in terms of $g$.
Thus we find
\begin{equation}
\label{uv-eq2}
\partial_z U - \partial_\rho V + \frac{1}{\rho} [U,V] + \frac{1}{\rho} V = 0 \,,
\end{equation}
where the square brackets denote the commutator.
One can indeed verify that~\eqref{uv-eq2} is satisfied for any $g$, given~\eqref{uv}.

The key step of the inverse scattering procedure consists in representing the first order Eqs.~\eqref{uv-eq1} and~\eqref{uv-eq2} as the compatibility conditions of an over-determined system of matrix equations related to an eigenvalue problem for some linear differential operator.
Such a system will depend on a complex spectral parameter $\lambda$, and the solutions of the original problem for $g$, $U$ and $V$ will be determined by the analytic structure of the eigenfunction in the complex $\lambda$-plane.

We begin by introducing the differential operators
\begin{equation}
D_1 \coloneqq \partial_z - \frac{2\lambda^2}{\lambda^2 + \rho^2} \partial_\lambda \,, \quad
D_2 \coloneqq \partial_\rho + \frac{2\lambda\rho}{\lambda^2 + \rho^2} \partial_\lambda \,,
\end{equation}
where $\lambda$ is the complex spectral parameter, independent of $\rho$ and $z$.
One can easily verify that the commutator of the operators $D_1$ and $D_2$ vanishes,
\begin{equation}
[D_1, D_2] = 0 \,.
\end{equation}
We now introduce a complex matrix $\psi(\lambda,\rho,z)$, called the generating matrix, and consider the system of equations
\begin{subequations}
\label{eigen}
\begin{align}
\label{eigen1}
D_1 \psi = \frac{\rho V - \lambda U}{\lambda^2 + \rho^2} \psi \,, \\
\label{eigen2}
D_2 \psi = \frac{\rho U + \lambda V}{\lambda^2 + \rho^2} \psi \,,
\end{align}
\end{subequations}
where $U$ and $V$ are real and do not depend on $\lambda$.

The fundamental property of the system~\eqref{eigen} is that its compatibility condition coincides exactly with Eqs.~\eqref{uv-eq1} and~\eqref{uv-eq2}.
This can be easily verified by applying $D_2$ to~\eqref{eigen1} and $D_1$ to~\eqref{eigen2}, and then by subtracting the results:
because of the commutativity of $D_1$ and $D_2$, we get zero on the left hand side, while the right hand side is a rational function of $\lambda$.
Requiring that all the coefficients of the various powers of $\lambda$ vanish, we get exactly Eqs.~\eqref{uv-eq1} and~\eqref{uv-eq2}.

We notice that the system~\eqref{eigen} gives, when $\lambda=0$,
\begin{equation}
U = \rho \bigl(\partial_\rho \psi\bigr) \psi^{-1} \,, \quad
V = \rho \bigl(\partial_z \psi\bigr) \psi^{-1} \,,
\end{equation}
which is nothing but Eq.~\eqref{uv} with $g=\psi$.
From this, we get the fundamental property
\begin{equation}
g(\rho,z) = \psi(0,\rho,z) \,,
\end{equation}
which implies that a solution of the eigensystem~\eqref{eigen} not only guarantees that the equations satisfied by $U$ and $V$ are true, but also gives a solution $g$ of~\eqref{uv}.

The integration procedure assumes that we know a ``seed'' solution $g_0$, $f_0$:
we denote $U_0$, $V_0$ and $\psi_0$ the associated matrices, solutions of~\eqref{uv},~\eqref{uv-eq1} and ~\eqref{eigen}.
We look for a solution of the form
\begin{equation}
\psi = \chi \psi_0 \,,
\end{equation}
where $\chi(\lambda,\rho,z)$ is called dressing matrix.
By inserting this ansatz into the eigensystem~\eqref{eigen}, we find the equations
\begin{subequations}
\label{eigein-chi}
\begin{align}
D_1\chi & = \frac{\rho V - \lambda U}{\lambda^2 + \rho^2} \chi
- \chi\frac{\rho V_0 - \lambda U_0}{\lambda^2 + \rho^2} \,, \\
D_2\chi & = \frac{\rho U + \lambda V}{\lambda^2 + \rho^2} \chi
- \chi\frac{\rho U_0 + \lambda V_0}{\lambda^2 + \rho^2} \,,
\end{align}
\end{subequations}
These equations are not enough to guarantee that $g$ is real and symmetric:
we have to impose some supplementary conditions, which read
\begin{equation}
\label{chi-cond1}
\chi^* (\lambda^*) = \chi (\lambda) \,, \quad
\psi^* (\lambda^*) = \psi (\lambda) \,,
\end{equation}
for the reality, and
\begin{equation}
g = \chi(\lambda) g_0 {\chi}^T(-\rho^2/\lambda) \,,
\end{equation}
for the symmetry.
The last condition is not trivial, and follows from an invariance property of the system~\eqref{eigein-chi}.
Furthermore, we require
\begin{equation}
\label{chi-cond2}
\chi(\infty) = \mathbf{1} \,,
\end{equation}
where $\mathbf{1}$ is the unit matrix, that in turn implies
\begin{equation}
g = \chi(0) g_0 \,.
\end{equation}
Thus, the problem consists of solving the system~\eqref{eigein-chi} and determining the dressing matrix $\chi$ that fulfills the supplementary conditions~\eqref{chi-cond1},~\eqref{chi-cond2}.

It is worth emphasising that the new solution $g$ must still satisfy $\det g =-\rho^2$.
It follows another condition on $\chi$:
\begin{equation}
\det \chi(0) = 1 \,.
\end{equation}
However, the best strategy is to not take into account this problem during the procedure for the construction of the solution, and simply renormalise the final result in order to obtain the correct functions.
We will call such correct functions the \emph{physical} functions.

\subsection{General \texorpdfstring{$n$}{n}-soliton solution}

The solution for the matrix $g$ that corresponds to the presence of pole singularities in the dressing matrix $\chi(\lambda,\rho,z)$, in the complex plane of the spectral parameters $\lambda$, is called the \emph{soliton solution}.
We consider the case in which the matrix $\chi$ has $n$ simple poles\footnote{The case of higher-order poles can be faced by the so-called \emph{pole fusion}.},
and is thus represented in the form
\begin{equation}
\label{chi-exp}
\chi = \mathbf{1} + \sum_{k=1}^n \frac{R_k}{\lambda - \mu_k} \,,
\end{equation}
where the matrices $R_k$ and the functions $\mu_k$ depend only on $\rho$ and $z$.
$\mu_k$ are the pole trajectories, i.e.~the positions of the poles as functions of $(\rho$, $z)$.

Now we can substitute the expansion~\eqref{chi-exp} into the eigensystem~\eqref{eigein-chi}, and impose the constraints~\eqref{chi-cond1} and~\eqref{chi-cond2}.
These equations completely determine the matrices $R_k$ and the pole trajectories $\mu_k$.
In particular, the requirement that there are no poles of order two at the points $\lambda=\mu_k$ in~\eqref{eigein-chi}, gives the differential equations
\begin{equation}
\partial_z\mu_k = -\frac{2\mu_k^2}{\mu_k^2+\rho^2} \,, \quad
\partial_\rho\mu_k = \frac{2\rho\mu_k}{\mu_k^2+\rho^2} \,.
\end{equation}
The solutions of such differential equations are the roots of the quadratic algebraic equation
\begin{equation}
\mu_k^2 + 2(z - w_k) \mu_k - \rho^2 = 0 \,,
\end{equation}
which gives the explicit expressions
\begin{subequations}
\label{solitons}
\begin{align}
\label{solitons1}
\mu_k & = \sqrt{\rho^2 + (z-w_k)^2} - (z-w_k) \,, \\
\label{solitons2}
\bar{\mu}_k & = -\sqrt{\rho^2 + (z-w_k)^2} - (z-w_k) \,,
\end{align}
\end{subequations}
where $w_k$ are arbitrary complex constants that are called \emph{poles}.
The function~\eqref{solitons1} is called a \emph{soliton}, while the function~\eqref{solitons2} is called an \emph{anti-soliton}.
The final solution is independent on the use of solitons or anti-solitons:
for definiteness, we will always make use of the solitons.

One can prove, via the Eqs.~\eqref{eigein-chi}, that the matrices $R_k$ are degenerate, and as such they can be written in the form
\begin{equation}
(R_k)_{ab} = n_a^{(k)} m_b^{(k)} \,,
\end{equation}
where $n_a^{(k)}$ and $m_a^{(k)}$ are two-components vectors.
The vectors $m_a^{(k)}$ are found from the reality conditions~\eqref{chi-cond1} by requiring that the equations are satisfied at the poles $\lambda = \mu_k$:
then one finds
\begin{equation}
\label{bz-vectors}
m_a^{(k)} = m_{0\,b}^{(k)} \bigl[ \psi_0^{-1}(\mu_k,\rho,z) \bigr]_{ba} \,,
\end{equation}
where $m_{0\,b}^{(k)}$ are arbitrary constants, and the summation over the repeated Latin indices is understood.
The vectors $m_a^{(k)}$ are usually called \emph{Belinski--Zakharov vectors} (BZ vectors).

The vectors $n_a^{(k)}$ are determined from the condition at infinity~\eqref{chi-cond2}, which gives the $n$-th order algebraic system
\begin{equation}
\sum_{l=1}^n \Gamma_{kl} n_a^{(k)} = \mu_k^{-1} m_c^{(k)} (g_0)_{ca} \,,
\end{equation}
for $k,l=1,2,\dotsc ,n$.
The symmetric matrix $\Gamma_{kl}$ is given by
\begin{equation}
\label{gamma-matrix}
\Gamma_{kl} = \frac{m_c^{(k)} (g_0)_{cb} m_b^{(l)}}{\rho^2 + \mu_k\mu_l} \,,
\end{equation}
thus the vectors $n_a^{(k)}$ are
\begin{equation}
n_a^{(k)} = \sum_{l=1}^n (\Gamma^{-1})_{lk} \mu_l^{-1} L_a^{(k)} \,,
\end{equation}
where
\begin{equation}
L_a^{(k)} = m_c^{(k)} (g_0)_{ca} \,.
\end{equation}
We are now ready to compute the matrix $g$:
recalling that
\begin{equation}
g = \psi(0) = \chi(0) \psi_0(0) = \chi(0) g_0
= \Biggl( \mathbf{1} - \sum_{k=1}^n R_k \mu_k^{-1} \Biggr) g_0 \,,
\end{equation}
and using what we have found up to now, we obtain
\begin{equation}
g_{ab} = (g_0)_{ab}
- \sum_{k,l=1}^n (\Gamma^{-1})_{kl} \frac{L_a^{(k)} L_b^{(l)}}{\mu_k\mu_l} \,.
\end{equation}
We must guarantee that the condition on the determinant is satisfied by the physical metric $g^{(ph)}$.
The computation of the determinant gives
\begin{equation}
\label{det-ph}
\det g = (-1)^n \rho^{2n} \Biggl( \prod_{k=1}^n \mu_k^{-2} \Biggr) \det g_0 \,,
\end{equation}
from which follows that the number of the solitons $n$ must be even, being $\det g_0=-\rho^2$.
Therefore, the stationary and axisymmetric solutions appear as even-soliton states.
Eq.~\eqref{einstein-matrix} must be satisfied by the physical matrix $g^{(ph)}$ as well, and one can verify that
\begin{equation}
\label{cond-ph}
g^{(ph)} = \pm \rho \sqrt{-\det g} \, g \,,
\end{equation}
does satisfy~\eqref{einstein-matrix} and $\det g^{(ph)}=-\rho^2$.
Putting together ~\eqref{det-ph} and~\eqref{cond-ph}, we finally get the physical matrix
\begin{equation}
\label{g-ph}
g^{(ph)} = \pm \rho^{-n} \Biggl( \prod_{k=1}^n \mu_k^{-2} \Biggr) g \,, \quad
\det g^{(ph)}=-\rho^2 \,.
\end{equation}
The computation of the function $f$ is quite involved:
one has to insert the expression for $g^{(ph)}$~\eqref{g-ph} into Eqs.~\eqref{einstein-f}, and in the end one gets
\begin{equation}
\label{f-ph}
f^{(ph)} =
16 C_f f_0 \rho^{-n^2/2} \Biggl( \prod_{k=1}^n \mu_k^{-2} \Biggr)^{n+1}
\Biggl[ \prod_{k>l=1}^n (\mu_k-\mu_l)^{-2} \Biggr] \det\Gamma_{kl} \,,
\end{equation}
where $C_f$ is an arbitrary gauge parameter and the factor 16 is put for convenience.

Now we have the complete recipe for constructing soliton solutions from a seed metric:
once we have chosen a seed $g_0$, $f_0$, we need to solve the linear eigensystem~\eqref{eigen} to find $\psi_0$, and then we just have to perform algebraic computations (by adding $n$ solitons to the seed) in order to find the new solution $g^{(ph)}$, $f^{(ph)}$ from~\eqref{g-ph},~\eqref{f-ph}.
This is a very powerful feature of the inverse scattering method:
once we have solved for $\psi_0$, the new metric is found by performing \emph{algebraic} computations.
This, of course, enormously reduces the difficulty of solving the equations of motions, since the Einstein equations are non-linear while the eigensystem~\eqref{eigen} is linear.

We will apply the formalism that we have presented to construct the Kerr--NUT spacetime starting with Minkowski spacetime and, as a further example, the rotating C-metric starting with Rindler spacetime.

\subsection{Kerr--NUT spacetime}

Let us consider Minkowski spacetime, in cylindrical coordinates, as a seed:
\begin{equation}
{ds}^2 = -{dt}^2 + \rho^2{d\phi}^2 + {d\rho}^2 + {dz}^2 \,.
\end{equation}
We clearly see that $g_0=\diag\bigl(-1,\rho^2\bigr)$ and $f_0=1$, and thus $\det g_0=-\rho^2$.
By Eq.~\eqref{uv} one finds that $U_0=\diag(0,2)$, while $V_0$ is zero.
From the eigensystem~\eqref{eigen} we get the solution
\begin{equation}
\psi_0 =
\begin{pmatrix}
-1 & 0 \\
0 & \rho^2 - 2\lambda z - \lambda^2 \\
\end{pmatrix}
\,,
\end{equation}
which satisfies the requirement $\psi_0(\lambda=0)=g_0$.
By using~\eqref{bz-vectors}, we find the BZ vectors
\begin{equation}
m^{(k)} = \Biggl( C_0^{(k)}, \frac{C_1^{(k)}}{\mu_k} \Biggr) \,,
\end{equation}
where $C_0^{(k)}$ and $C_1^{(k)}$ are arbitrary constants.
From~\eqref{gamma-matrix} we find
\begin{equation}
\Gamma_{kl} =
\frac{-C_0^{(k)}C_0^{(l)} + C_1^{(k)}C_1^{(l)}\mu_k^{-1}\mu_l^{-1}\rho^2}{\rho^2 + \mu_k\mu_l} \,.
\end{equation}
We can use these quantities and Eqs.~\eqref{g-ph},~\eqref{f-ph} to construct a generic $n$-soliton solution on the flat space background.

We explicitly compute the two-soliton solution on the flat background, which corresponds to the Kerr--NUT spacetime.
This means that we have two poles $\lambda=\mu_1$ and $\lambda=\mu_2$.
Firstly, we represent the constants $w_1$ and $w_2$, that appear in the solitons, as
\begin{equation}
w_1 = \tilde{z}_1 - \sigma \,, \quad
w_2 = \tilde{z}_1 + \sigma \,,
\end{equation}
where $\tilde{z}_1$ and $\sigma$ are new arbitrary constants.
$\tilde{z}_1$ is real, and is interpreted as the position of the black hole on the $z$-axis.
We now introduce the spherical-like coordinates
\begin{equation}
\rho = \sqrt{(r-m)^2-\sigma^2} \sin\theta \,, \quad
z = \tilde{z}_1 +  (r-m) \cos\theta \,,
\end{equation}
where $m$ is another (real) arbitrary constant.
In these coordinates, the solitons take the form
\begin{equation}
\mu_1 = (r-m-\sigma) (1-\cos\theta) \,, \quad
\mu_2 = (r-m+\sigma) (1-\cos\theta) \,.
\end{equation}
We impose, without loss of generality, the following conditions on the constants $C_0^{(k)}$, $C_1^{(k)}$:
\begin{subequations}
\label{constants}
\begin{align}
\label{constants1}
C_1^{(1)}C_0^{(2)} - C_0^{(1)}C_1^{(2)} & = \sigma \,, \quad
C_1^{(1)}C_0^{(2)} + C_0^{(1)}C_1^{(2)} = -m \,, \\
\label{constants2}
C_0^{(1)}C_0^{(2)} - C_1^{(1)}C_1^{(2)} & = n \,, \quad
C_0^{(1)}C_0^{(2)} + C_1^{(1)}C_1^{(2)} = a \,.
\end{align}
\end{subequations}
The first condition of~\eqref{constants1} take advantage of the non-physical arbitrariness of the normalisation
$C_a^{(k)}\to \zeta^{(k)} C_a^{(k)}$.
After such a transformation the metric $g$ does not change, hence the metric does not depend on $\zeta^{(k)}$.
The first relation of~\eqref{constants1} partially fixes this arbitrariness.
The second relation of~\eqref{constants1}, on the other hand, is just the definition of $m$.
The two conditions~\eqref{constants2} define the new constants $a$ and $n$.
Furthermore, it follows from~\eqref{constants} that
\begin{equation}
\sigma^2 = m^2 - a^2 + n^2 \,,
\end{equation}
which explicitly proves that we are introducing only three new parameters into the physical metric.

By using all the expressions that we have collected up to now, we can compute the new metric with the formulae~\eqref{g-ph} and~\eqref{f-ph}.
The line element $d\rho^2+dz^2$ is transformed, accordingly to the new spherical coordinates, as
\begin{equation}
{d\rho}^2 + {dz}^2 =
\bigl[ (r-m)^2 - \sigma^2\cos^2\theta \bigr]
\biggl[ \frac{{dr}^2}{(r-m)^2 - \sigma^2} + {d\theta}^2 \biggr] \,.
\end{equation}
The new metric is finally given, after the rotation $t\to t+2a\phi$, by
\begin{equation}
\label{kerr-nut}
\begin{split}
{ds}^2 & =
-\frac{\Delta-a^2\sin^2\theta}{\Sigma} {dt}^2
-\frac{\Delta(a\sin^2\theta+2n\cos\theta)^2 - \sin^2\theta (r^2+a^2+n^2)^2}{\Sigma} {d\phi}^2 \\
&\quad + \frac{4\Delta n\cos\theta - 4a\sin^2\theta (mr+n^2)}{\Sigma} dt d\phi
-C_f \biggl( \frac{\Sigma}{\Delta} {dr}^2 + \Sigma {d\theta}^2 \biggr) \,,
\end{split}
\end{equation}
where
\begin{equation}
\Sigma = r^2 + (n - a\cos\theta)^2 \,, \quad
\Delta = r^2 - 2mr + a^2 - n^2 \,.
\end{equation}
One recognises the Kerr--NUT metric~\cite{Plebanski:1976gy} in Boyer--Lindquist coordinates, when $C_f=-1$.
We notice that the constant $\tilde{z}_1$ disappeared in the final solution~\eqref{kerr-nut}:
this is due to the fact that the solution is invariant under translations along the $z$-axis.
The solution with horizons corresponds to the case in which $\sigma$ is real (i.e.~$m^2-a^2+n^2>0$), so that $w_1$, $w_2$ are real and the pole trajectories $\mu_1$, $\mu_2$ are real as well along $\sigma$.
If $\sigma$ is pure imaginary (i.e.~$m^2-a^2+n^2<0$), the constants $w_1$, $w_2$ and the pole trajectories $\mu_1$, $\mu_2$ are complex and conjugate to each other, and we find a solution without horizons.

The Schwarzschild metric corresponds to the choice $a=n=0$, evidently.
One can directly obtain the Schwarzschild solution from the inverse scattering procedure by choosing
\begin{equation}
C_0^{(1)} = C_1^{(2)} = 0 \,, \quad
C_0^{(2)} = C_1^{(1)} = 1 \,.
\end{equation}
Actually, such a choice for the constants always guarantees a diagonal physical metric for any seed.
The construction of the Kerr--NUT metric shows that a rotating black hole corresponds to a gravitational two-soliton solution.

\subsection{Rotating C-metric}

We start with Rindler spacetime in cylindrical coordinates~\cite{Griffiths:2009dfa}
\begin{equation}
{ds}^2 = -\mu_A {dt}^2 + \frac{\rho^2}{\mu_A}{d\phi}^2
+ \frac{\mu_A}{\rho^2 + \mu_A^2} \bigl( {d\rho}^2 + {dz}^2 \bigr) \,,
\end{equation}
where $\mu_A$ is the soliton associated to the acceleration parameter $A$.

In this case $g_0=\diag\bigl(-\mu_A,\rho^2/\mu_A\bigr)$ and $f_0=\mu_A/(\rho^2 + \mu_A^2)$, and again $\det g_0=-\rho^2$.
The eigensystem~\eqref{eigen} provides the solution
\begin{equation}
\psi_0 =
\begin{pmatrix}
\lambda - \mu_A & 0 \\
0 & \lambda + \frac{\rho^2}{\mu_A} \\
\end{pmatrix}
\,,
\end{equation}
which satisfies again the constraint $\psi_0(\lambda=0)=g_0$.
The BZ vectors are
\begin{equation}
m^{(k)} = \Biggl( \frac{C_0^{(k)}}{\mu_k-\mu_A},
C_1^{(k)} \frac{\mu_A}{\rho^2 + \mu_k\mu_A} \Biggr) \,,
\end{equation}
where $C_0^{(k)}$ and $C_1^{(k)}$ are arbitrary constants.

We now add, as in the previous Section, two solitons on top of the Rindler background by means of Eqs.~\eqref{g-ph},~\eqref{f-ph}, and explicitly show that the rotating C-metric in the form presented by Hong and Teo~\cite{Hong:2004dm} is recovered.

The change of coordinates is provided by~\cite{Harmark:2004rm} and, in order to simplify the computations, it is particularly useful to adopt the trick explained in~\cite{Chen:2012zb}:
we firstly switch to the coordinates $(u,v)$ defined by
\begin{equation}
\rho = \frac{2\kappa^2 \sqrt{(1-u^2)(v^2-1)(1+\nu u)(1+\nu v)}}{(u-v)^2} \,, \quad
z = \frac{\kappa^2 (1-uv)(2+\nu u+\nu v)}{(u-v)^2} \,,
\end{equation}
with poles parametrised as
\begin{equation}
w_1 = -\nu \kappa^2 \,, \qquad
w_2 = \nu \kappa^2 \,, \qquad
w_A = \kappa^2 \,,
\end{equation}
and the new parameters $\nu$, $\kappa$ chosen in such a way that $w_1<w_2<w_A$.
Then, we perform the M\"obius transformation
\begin{equation}
u = \frac{x+d}{1+dx} \,, \qquad
v = \frac{y+d}{1+dy} \,, \qquad
\nu = \frac{c-d}{1-cd} \,,
\end{equation}
to work with the standard C-metric coordinates $(x,y)$, and to introduce two new parameters $c$ and $d$.

We still have to parametrise the constants appearing in the BZ vectors:
we choose
\begin{subequations}
\begin{align}
C_1^{(1)}C_0^{(2)} - C_0^{(1)}C_1^{(2)} & = -\sqrt{m^2-a^2} \,, \quad
C_1^{(1)}C_0^{(2)} + C_0^{(1)}C_1^{(2)} = m \frac{1-a^2A^2}{1+a^2A^2} \,, \\
C_0^{(1)}C_0^{(2)} - C_1^{(1)}C_1^{(2)} & = \frac{2amA}{1+a^2A^2} \,, \qquad\;
C_0^{(1)}C_0^{(2)} + C_1^{(1)}C_1^{(2)} = a \,,
\end{align}
\end{subequations}
where we have introduced the mass $m$, the angular momentum $a$ and the acceleration $A$.
Finally, we set
\begin{subequations}
\begin{align}
c & = A \Bigl( m + \sqrt{m^2-a^2} \Bigr) \,, \quad
d = A \Bigl( m - \sqrt{m^2-a^2} \Bigr) \,, \\
\kappa^2 & = \frac{1-a^2A^2}{2A^2} \,, \qquad\qquad\quad\!
C_f = -\frac{1+a^2A^2}{A^6} \,.
\end{align}
\end{subequations}
With these definitions, the metric boils down to
\begin{equation}
\begin{split}
{ds}^2 & = \frac{1}{A^2(x-y)^2} \biggl[
\frac{G(y)}{1+a^2A^2x^2y^2} (dt - aA x^2 d\phi)^2
- \frac{1+a^2A^2x^2y^2}{G(y)} {dy}^2 \\
&\quad + \frac{1+a^2A^2x^2y^2}{G(x)} {dx}^2
+ \frac{G(x)}{1+a^2A^2x^2y^2} (d\phi + aA y^2 dt)^2
\biggr] \,,
\end{split}
\end{equation}
with
\begin{equation}
G(\xi) = (1 - \xi^2) (1 + r_+ A\xi) (1 + r_- A\xi) \,,
\end{equation}
and where $r_\pm$ are the usual Kerr horizons, i.e.~$r_\pm=m\pm\sqrt{m^2-a^2}$.

We did not show it explicitly, but it is possible to construct the accelerating Kerr--NUT solution by including the NUT parameter in the parametrisation of the solution constructed above, similarly to what we did for the Kerr--NUT metric, thus recovering the vacuum Pleba\'nski--Demia\'nski solution~\cite{Plebanski:1976gy}.

\subsection{Multi-black hole solutions}

We construct the double-Schwarzschild solution, also known as the Bach--Weyl metric~\cite{Bach1922} (that is a subcase of the Israel--Kahn metric~\cite{Israel1964}).
It is worth to remind to the reader that, as mentioned in the Introduction, such a solution is not regular, being affected by the presence of conical singularities.

By following the discussion in the end of the previous Sections, we just have to add four solitons to the Minkowski background, so to add two black holes to the flat background.
We choose the BZ constants as
\begin{equation}
C_0^{(1)} = C_1^{(2)} = C_0^{(3)} = C_1^{(4)} = 0 \,, \quad
C_0^{(2)} = C_1^{(1)} = C_0^{(4)} = C_1^{(3)} = 1 \,,
\end{equation}
to have a diagonal solution.
Then, by applying the inverse scattering formulae~\eqref{g-ph} and~\eqref{f-ph}, we find
\begin{subequations}
\begin{align}
g^{(ph)} & = \diag\biggl( -\frac{\mu_1\mu_3}{\mu_2\mu_4} ,
\rho^2 \frac{\mu_2\mu_4}{\mu_1\mu_3} \biggr) \,, \\
f^{(ph)} & = 16 C_f \frac{\mu_1^3\mu_2^5\mu_3^3\mu_4^5}{W_{11}W_{22}W_{33}W_{44} W_{13}^2W_{24}^2 Y_{12}^2Y_{14}^2Y_{23}^2Y_{34}^2} \,,
\end{align}
\end{subequations}
where we have defined
\begin{equation}
W_{ij} \coloneqq \rho^2 + \mu_i\mu_j \,, \quad
Y_{ij} \coloneqq \mu_i - \mu_j \,.
\end{equation}
We see that the structure of the double-black hole solution (at least at the level of the matrix $g^{(ph)}$) is very simple:
it is just an alternating series of solitons, where in the $tt$ component the odd solitons are at the numerator and the even solitons are at the denominator, while in the $\phi\phi$ component the situation is reversed.

One can easily guess the structure of the complete Israel--Kahn solution~\cite{Israel1964}, i.e.~the multi-black hole solution with $N$ black holes (that corresponds to $n=2N$ solitons):
\begin{subequations}
\begin{align}
g^{(ph)} & = \diag\Biggl(
-\frac{\prod_{k=1}^N \mu_{2k-1}}{\prod_{l=1}^N \mu_{2l}},
\rho^2 \frac{\prod_{l=1}^N \mu_{2l}}{\prod_{k=1}^N \mu_{2k-1}}
\Biggr) \,, \\
\begin{split}
f^{(ph)} & = 16C_f f_0
\Biggl( \prod_{k=1}^N \mu_{2k}^{2N+1}
\mu_{2k-1}^{2N-1} \Biggr)
\Biggl( \prod_{k=1}^{2N} \frac{1}{\rho^2+\mu_k^2} \Biggr)
\Biggl( \prod_{k=1,l=1,3,\cdots}^{2N-1} \frac{1}{(\mu_k-\mu_{k+l})^2} \Biggr) \\
&\quad\times
\Biggl( \prod_{k=1,l=2,4,\cdots}^{2N-2} \frac{1}{(\rho^2+\mu_k\mu_{k+l})^2} \Biggr) \,.
\end{split}
\end{align}
\end{subequations}
The resulting metric represents $N$ black holes aligned on the $z$-axis.
The standard \\ parametrisation for the poles $w_k$ is given by
\begin{equation}
w_1 = z_1 - m_1\,, \quad w_2 = z_1 + m_1\,, \quad \dotsc \quad
w_{2N-1} = z_N - m_N\,, \quad w_{2N} = z_N + m_N \,,
\end{equation}
with obvious ordering $w_1<w_2<\cdots<w_{2N-1}<w_{2N}$.
The parameters $z_k$ represent the positions of the black holes on the $z$-axis, while $m_k$ are the mass parameters.
The black hole horizons correspond to the regions
$w_{2k-1}<z<w_{2k}$ ($k=1,\dotsc,N$), while the complementary regions are affected by the presence of conical singularities, that guarantee the stationarity of the solution by preventing the collapse of the system.

It is conceptually easy to extend the previous discussion to the construction of the double-Kerr--NUT solution~\cite{Letelier:1998ft}:
one can adopt the parametrisation
\begin{subequations}
\begin{align}
C_1^{(1)}C_0^{(2)} - C_0^{(1)}C_1^{(2)} & = \sigma_1 \,, \quad
C_1^{(1)}C_0^{(2)} + C_0^{(1)}C_1^{(2)} = -m_1 \,, \\
C_0^{(1)}C_0^{(2)} - C_1^{(1)}C_1^{(2)} & = n_1 \,, \quad
C_0^{(1)}C_0^{(2)} + C_1^{(1)}C_1^{(2)} = a_1 \,,
\end{align}
\end{subequations}
and
\begin{subequations}
\begin{align}
C_1^{(3)}C_0^{(4)} - C_0^{(3)}C_1^{(4)} & = \sigma_2 \,, \quad
C_1^{(3)}C_0^{(4)} + C_0^{(3)}C_1^{(4)} = -m_2 \,, \\
C_0^{(3)}C_0^{(4)} - C_1^{(3)}C_1^{(4)} & = n_2 \,, \quad
C_0^{(3)}C_0^{(4)} + C_1^{(3)}C_1^{(4)} = a_2 \,.
\end{align}
\end{subequations}
We simply double the parametric choice we made for the Kerr--NUT solution.
However, it is important to notice that the various parameters we have introduced do not correspond to the physical charges, in general.
For instance, the angular momentum parameters $a_1$ and $a_2$ do not corresponds to the actual angular momenta of the black holes.
Such charges will be, in general, functions of all the parameters of the solutions.
This implies that turning off $a_1=a_2=0$ does not guarantee that the angular momenta of the black holes will be zero, since they will depend on other parameters of the theory.

\chapter{Black holes in an external gravitational field}
\label{chap:extfield}
\thispagestyle{plain}

As we argued in the Introduction, multi-black holes solutions are intriguing both from the theoretical and the phenomenological point of view.
On the theoretical side, these solutions disclose the non-linear nature of General Relativity and represent an important playground in which testing the laws of black hole mechanics.
On the experimental side, the recent remarkable observations of gravitational waves heavily rely on the interactions between two black holes in a binary system: 
thus an analytical description of such a spacetime
is of utmost relevance for the interpretation of the measurements.
Of course one of the main obstacle in modelling a stationary multi-gravitational sources system is to provide a mechanism to balance the gravitational attraction of the bodies.
Otherwise the system naturally tends to collapse.

The purpose of this Chapter is to provide an example of a background that can be used to regularise a multi-black hole spacetime.
Such a background is given by an external gravitational field:
we will show that there exists a solution of vacuum Einstein equations which represents a multipolar expansion of a generic external gravitational field.
Such a solution was discovered by Erez and Rosen a long time ago~\cite{Erez}, and it is the natural background to be used to regularise a multi-black hole spacetime.
Indeed, the gravitational background provides the physical explanation of the regularity of a double-black hole system:
the external field might be tuned in such a way to balance the gravitational attraction between the sources and, thus, to prevent the collapse.

External gravitational fields represent a natural setting for multi-black holes systems, as recent gravitational waves detection proceeding from the center of galaxies confirms.
In fact the observed astrophysical black holes are not isolated systems, as they are always embedded in external gravitational fields. 
In particular, it has been shown in~\cite{deCastro:2011zz} that the multipolar gravitational field, we will deem here, can be produced by a distribution of matter such as thin disks or rings, typical shapes of gravitational objects such as galaxies or nebulae.
Anyway, the solutions considered in this Chapter will be pure vacuum solution, without any energy-momentum tensor.
In principle a distribution of matter might be possibly considered very far away from the black holes:
in this sense our solutions can be interpreted as local models for binary or multi-black hole configurations.
In a certain sense the metrics presented in this Chapter have to be considered as the gravitational analogous of the stationary black holes in Melvin magnetic universe.
In fact, also in this latter case, the solution is a pure electrovacuum solution with no definite sources for the electromagnetic external field, therefore their feasibility remains in the proximity domain of the black bodies.
The multipolar expansion is the key ingredient that allows to circumvent the no-go theorem about the non-existence of static~\cite{Beig2009} configurations of many-body systems, with a suitable separation condition.
Being the multipolar expansion non-asimptotically flat, one of the main hypothesis of the theorem does not hold and regular multi-black hole solutions are allowed to exist.

Single black holes in external multipolar gravitational field have been pioneered, in the literature, by Doroshkevich, Zel'dovich and Novikov~\cite{NovikovZeldovich}, later studied by  Chandrasekhar~\cite{Chandrasekhar:1985kt}, and Geroch and Hartle~\cite{Geroch:1982bv};
these solutions are known in the literature as deformed black holes.
The novelty of our proposal, based on an Ernst's insight in the context of the C-metric~\cite{ErnstGeneralized}, and already implemented in the case of a single black hole~\cite{Kerns1982}, is to take advantage of the external field to sustain the black bodies and prevent their collision.
From a mathematical point of view this means that the solution can be regularised from conical singularities that usually affect multi-black hole metrics.

We start by investigating the properties of the external gravitational field and its representation as a seed for the inverse scattering procedure.
Then, we will construct an array of static black holes, which generalises the Israel--Khan solution, and explicitly show that such a solution can be regularised, by removing all the conical singularities via a tuning of the external field parameters.
Subsequently, we will specialise to a very interesting subcase, i.e.~the binary black hole system immersed in an external gravitational field, and also present the charged and rotating generalisations of that case.
Finally, we will add acceleration and construct a chain of accelerating black hole.
All of these systems can be regularised, in order to obtain legit multi-black hole spacetimes.
This Chapter is based on the papers~\cite{Astorino:2021dju,Astorino:2021boj,Astorino:2021rdg}.

\section{Background gravitational field}

We discuss the general solution to the Einstein equations in vacuum, which contemplates both the internal deformations of the source, and the contributions which come from matter far outside the source.
This general solution finds its roots in the pioneeristic work of Erez and Rosen~\cite{Erez}, and it was lately discussed and expanded in~\cite{NovikovZeldovich,Geroch:1982bv,Chandrasekhar:1985kt} to include the deformations due to an external gravitational field.

The general solution for the Weyl metric
\begin{equation}
\label{weyl}
{ds}^2 = -e^{2\psi(\rho,z)} {dt}^2 + e^{-2\psi(\rho,z)} \bigl[ e^{2\gamma(\rho,z)} ({d\rho}^2 + {dz}^2) + \rho^2 {d\phi}^2 \bigr] \, ,
\end{equation}
is given, following the conventions in~\cite{Breton:1998sr}, by
\begin{subequations}
\begin{align}
\psi & = \sum_{n=1}^{\infty} \biggl( \frac{a_n}{r^{n+1}} + b_n r^n \biggr) P_n \, , \\
\gamma & = \sum_{n,p=1}^\infty \biggl[
\frac{(n+1)(p+1) a_n a_p}{(n+p+2)r^{n+p+2}} (P_{n+1} P_{p+1} - P_n P_p)
+ \frac{np b_n b_p r^{n+p}}{n+p} (P_n P_p - P_{n-1} P_{p-1}) \biggr] \, ,
\end{align}
\end{subequations}
where $r\coloneqq\sqrt{\rho^2+z^2}$ defines the asymptotic radial coordinate and $P_n=P_n(z/r)$ is the $n$-th Legendre polynomial.
The real constants $a_n$ describe the deformations of the source, while the real parameters $b_n$ describe the external static gravitational field.

We observe that the ``internal'' part $a_n$, which is related to the deformations of the source, is asymptotically flat:
this seems to contradict Israel's theorem~\cite{Israel:1967wq}, which states that the only regular and static spacetime in vacuum is the Schwarzschild black hole.
Actually, the internal deformations lead to curvature singularities not covered by a horizon~\cite{Erez}, in agreement with the theorem.
Because of this feature, in the following we will discard the internal contributions and focus on the external ones only.

On the converse, the ``external'' part $b_n$ is not asymptotically flat.
This is in agreement with the physical interpretation:
this part of the metric represents an external gravitational field generated by a distribution of matter located at infinity.
This interpretation parallels the Melvin spacetime one, and in fact there is no stress-energy tensor here to model the matter responsible of the field.
The asymptotia is not flat because at infinity there is the source matter and the spacetime is not empty.
Actually, it can be shown that the curvature invariants, such as the Kretschmann scalar, can grow indefinitely at large distances\footnote{For the black hole model we are considering below, these large distances can be quantified in several orders of magnitude larger than the scale of the black holes. However, before reaching such distances, large values of the curvature invariants can be covered by a Killing horizon such as the one provided by the acceleration.} in some directions.
In this sense, a spacetime containing such an external gravitational field should be considered local, in the sense that the description is meaningful in the neighborhood of the black holes that one embeds in this background.
In this regard, these metrics are not different with respect to the usual single distorted black holes studied in the literature~\cite{Breton:1998sr,Abdolrahimi:2015gea}.
To have a completely physical solution, one should match the black holes immersed in the external field with an appropriate distribution of matter (such as galaxies), which possibly is asymptotic to Minkowski spacetime at infinity.
A model for matter content which is consistent with the multipolar background expansion treated here, and based on a thin ring distribution, can be found in~\cite{deCastro:2011zz}\footnote{The analysis of the sectors where the scalar curvature invariants are larger, at least for the firsts order of the multipole expansion, suggests that a ring matter distribution is the more appropriate one for these backgrounds.}.

Accelerating generalisations of the multipolar gravitational background studied in this Section naturally are endowed with a Killing horizon of Rindler type.
These accelerating horizon are typically located in between the black hole sources and the far region, and they will be studied later on.

The constants $a_n$ and $b_n$ are related to the multipole momenta of the spacetime.
The relativistic definition of the multipole momenta was given by Geroch~\cite{Geroch:1970cd}, and lately refined by Hansen~\cite{Hansen:1974zz}.
That definition applies for a stationary, axisymmetric and asymptotically flat spacetime:
the modified Ernst potential  associated to the spacetime is expanded at infinity, and the first coefficients of the expansion correspond to the multipole momenta~\cite{Fodor}.
Clearly, the internal deformations asymptote a flat spacetime, while the external ones do not.

In the following, we compute the multipole momenta associated to the above deformations, in order to clarify the intepretation of $a_n$ and $b_n$.
In particular, we calculate the internal multipole contributions by means of the standard definition, and then we propose a new approach for determining the multipoles associated to external deformations.
We shall see that the proposed definition is analogous to the usual one and gives a consistent interpretation of the external field parameters.

\subsection{Background field multipoles}

Let us start with the internal deformations.
If we turn off $b_n$, then we are left with
\begin{equation}
\psi_\text{int} = \sum_{n=1}^{\infty} \frac{a_n}{r^{n+1}} P_n \,.
\end{equation}
We define $\xi$ as a function of the Ernst potential
\begin{equation}
\ernst = \frac{1-\xi}{1+\xi} \,,
\end{equation}
that in our case reads
\begin{equation}
\xi_\text{int} \coloneqq \frac{1-e^{2\psi_\text{int}}}{1+e^{2\psi_\text{int}}} \,.
\end{equation}
We want to expand the above expression around infinity:
following~\cite{Fodor}, we bring infinity to a finite point by defining
$\bar{\rho}=\rho/(\rho^2+z^2)$, $\bar{z}=z/(\rho^2+z^2)$, and by conformally rescaling $\xi_\text{int}$:
\begin{equation}
\bar{\xi} = \frac{1}{\bar{\rho}^2 + \bar{z}^2} \,
\xi_\text{int} \,.
\end{equation}
One can prove that $\bar{\xi}$ is uniquely determined by its value on the $z$-axis:
since infinity corresponds to $\bar{\rho}=\bar{z}=0$ in the new coordinates, then we expand $\bar{\xi}$ around $\bar{z}$ for $\rho=0$
\begin{equation}
\label{int-exp}
\bar{\xi}(\bar{\rho}=0) = \sum_{j=0}^\infty M_j \bar{z}^j \, ,
\end{equation}
where $M_j$ are the expansion coefficients.
The multipole momenta are completely determined by the coefficients $M_j$, and in particular
one can show (see~\cite{Fodor} and references therein) that the first four coefficients $M_0,\dotsc,M_3$ are exactly equal to the first four multipole momenta $\mathcal{Q}_0^\text{int},\dotsc,\mathcal{Q}_3^\text{int}$.
Thus, for the internal deformations, we find
\begin{equation}
\label{poles-int}
\mathcal{Q}_0^\text{int} = M_0 = 0 \,, \quad
\mathcal{Q}_1^\text{int} = M_1 = -a_1 \,, \quad
\mathcal{Q}_2^\text{int} = M_2 = -a_2 \,, \quad
\mathcal{Q}_3^\text{int} = M_3 = -a_3 \,.
\end{equation}
We see that there is a direct correspondence between the coefficients $a_n$ and the multipole momenta, at least at the first orders.
$\mathcal{Q}_0^\text{int}$ is the monopole term, which is zero since no source (e.g.~no black hole) is present.
$\mathcal{Q}_1^\text{int}$ is the dipole, $\mathcal{Q}_2^\text{int}$ is the quadrupole and $\mathcal{Q}_3^\text{int}$ is the octupole moment.
The subsequent multipole momenta can be still computed from the expansion~\eqref{int-exp} by means of a recursive algorithm, but they will be non-trivial combinations of the coefficients $M_j$~\cite{Fodor}.
For instance, the 16-pole is given by
$\mathcal{Q}_4^\text{int} = M_4 - 1/7 M_0^2 M_2$.

The above construction can not work for the external deformations:
the asymptotia is not flat, and infinity is the place where the sources of the external field are thought to be, hence it does not make sense to expand there.
On the converse, it is meaningful to detect the effects of the deformation in the origin of the Weyl coordinates:
thus, by paralleling the Geroch--Hansen treatise, we propose to expand the modified Ernst potential in the \emph{origin} of the cylindrical coordinates.

Now we consider only the external deformations
\begin{equation}
\psi_\text{ext} = \sum_{n=1}^{\infty} b_n r^n P_n \,,
\end{equation}
with modified Ernst potential
\begin{equation}
\xi_\text{ext} = \frac{1-e^{2\psi_\text{ext}}}{1+e^{2\psi_\text{ext}}} \,.
\end{equation}
Accordingly to the above discussion, we assume that $\xi_\text{ext}$ is  completely determined on the $z$-axis, so we expand around the origin for $\rho=0$
\begin{equation}
\label{ext-exp}
\xi_\text{ext}(\rho=0) = \sum_{j=0}^\infty N_j z^j \,,
\end{equation}
where $N_j$ are the expansion coefficients.
We \emph{define} the first four multipole momenta $\mathcal{Q}_0^\text{ext},\dotsc,\mathcal{Q}_3^\text{ext}$ as the coefficients $N_0,\dotsc,N_3$, which are equal to
\begin{equation}
\mathcal{Q}_0^\text{ext} = N_0 = 0 \, , \quad
\mathcal{Q}_1^\text{ext} = N_1 = -b_1 \, , \quad
\mathcal{Q}_2^\text{ext} = N_2 = -b_2 \, , \quad
\mathcal{Q}_3^\text{ext} = N_3 = \frac{b_1^3}{3} - b_3 \,.
\end{equation}
Again, the monopole moment is zero because of the absence of a source.
We observe, contrary to~\eqref{poles-int}, that the octupole $\mathcal{Q}_3^\text{ext}$ is given by a non-trivial mixing of the constants $b_1$ and $b_3$.
This definition occurs only for the first momenta:
a definition which takes into account higher momenta can be achieved by generalising the procedure in~\cite{Fodor}.

For the time being, we content ourselves by proposing the following interpretation:
the coefficients $b_n$ are related to the multipole momenta generated by the external gravitational field, similarly to what happens for the internal deformations.
Then, the presence of the external gravitational field affects the momenta of a black hole source immersed in it.

\subsection{Seed for the inverse scattering construction}

We are interested in the external deformations only, hence we consider $a_n=0$ hereafter, and focus only on the contributions from $b_n$.
We express the external gravitational field metric in a form which is suitable for the inverse scattering procedure, i.e.
\begin{subequations}
\label{seed-extfield}
\begin{align}
g_0 & = \diag\Biggl[ -\exp\biggr(2\sum_{n=1}^{\infty} b_n r^n P_n \biggr) ,
\rho^2 \exp\biggl(-2\sum_{n=1}^{\infty} b_n r^n P_n \biggr) \Biggr] \, , \\
f_0 & = \exp\Biggl[
2 \sum_{n,p=1}^\infty \frac{np b_n b_p r^{n+p}}{n+p} \bigl(P_n P_p - P_{n-1} P_{p-1}\bigr)
- 2\sum_{n=1}^{\infty} b_n r^n P_n
\Biggr] \, ,
\end{align}
\end{subequations}
The parameters $b_n$ are related the multipole momenta of the external field, as explained above.
Metric~\eqref{seed-extfield} represents a generic static and axisymmetric gravitational field.

Since we want to add black holes on the background represented by the gravitational field, we take the metric~\eqref{seed-extfield} as a seed for the inverse scattering procedure.
Following the discussion in section~\ref{sec:ism}, we need the generating matrix $\psi_0$, which serves as starting point to build the multi-black hole solution.
The function which satisfies equations~\eqref{eigen} is~\cite{LetelierBrasil}
\begin{equation}
\label{psi0-extfield}
\psi_0(\rho,z,\lambda) =
\begin{pmatrix}
-e^{F(\rho,z,\lambda)} & 0 \\
0 & (\rho^2 -2\lambda\rho - \lambda^2) e^{-F(\rho,z,\lambda)}
\end{pmatrix}
\,,
\end{equation}
where
\begin{equation}
\begin{split}
F(\rho,z,\lambda) & = 2 \sum_{n=1}^\infty b_n
\Biggl[ \sum_{l=0}^\infty \binom{n}{l} \biggl(\frac{-\rho^2}{2 \lambda}\biggr)^l \biggl(z + \frac{\lambda}{2}\biggr)^{n-l} \\
&\quad - \sum_{l=1}^n \sum_{k=0}^{[(n-l)/2]}
\frac{(-1)^{k+l}2^{-2k-l} n! \lambda^{-l}}{k!(k+l)!(n-2k-l)!}
\rho^{2(k+l)} z^{n-2k-l} \Biggr] \,.
\end{split}
\end{equation}
Now we can construct the BZ vectors~\eqref{bz-vectors}:
we parametrise
$m_0^{(k)}=\bigl(C_0^{(k)},C_1^{(k)}\bigr)$,
where $C_0^{(k)}$, $C_1^{(k)}$ are constants that will be eventually related to the physical parameters of the solution.
The BZ vectors are thus
\begin{equation}
m^{(k)} = \biggl( -C_0^{(k)} e^{-F(\rho,z,\mu_k)}, 
\frac{C_1^{(k)}}{\mu_k} e^{F(\rho,z,\mu_k)} \biggr) \, .
\end{equation}
Depending on the value of $C_0^{(k)}$ and $C_1^{(k)}$, the spacetime will be static or stationary, as we will see in the following.

\section{Array of static black holes}

We now proceed to the generalisation of the Israel--Khan solution~\cite{Israel1964}, which represents an array of collinear Schwarzschild black holes.
The Israel--Khan metric is plagued by the presence of conical singularities which can not be removed by a fine tuning of the physical parameters\footnote{Actually, in the limit of an infinite number of collinear black holes, the conical singularities disappear and the metric is regular. See~\cite{Myers:1986rx}.}.
On the converse, we will see that the external gravitational field will furnish the force necessary to achieve the complete equilibrium among the black holes.

Given the seed metric~\eqref{seed-extfield} and the generating matrix~\eqref{psi0-extfield}, we construct a new solution by adding $2N$ solitons with constants
\begin{equation}
C_0^{(k)} =
\begin{cases}
1 & k \text{ even} \\
0 & k \text{ odd}
\end{cases}
, \qquad
C_1^{(k)} =
\begin{cases}
0 & k \text{ even} \\
1 & k \text{ odd}
\end{cases}
\,.
\end{equation}
This choice guarantees a diagonal, and hence static, metric.
Each couple of solitons adds a black hole, then the addition of $2N$ solitons gives rise to a spacetime containing $N$ black holes, whose metric is
\begin{subequations}
\label{n-bh}
\begin{align}
g_N & = \diag\Biggl[
-\frac{\prod_{k=1}^N \mu_{2k-1}}{\prod_{l=1}^N \mu_{2l}} \exp\Biggl(2{\sum_{n=1}^{\infty} b_n r^n P_n}\Biggr),
\rho^2 \frac{\prod_{l=1}^N \mu_{2l}}{\prod_{k=1}^N \mu_{2k-1}} \exp\Biggl(-2{\sum_{n=1}^{\infty} b_n r^n P_n}\Biggr)
\Biggr] \,, \\
\label{n-bh-fph}
\begin{split}
f_N & = 16C_f f_0
\Biggl( \prod_{k=1}^N \mu_{2k}^{2N+1}
\mu_{2k-1}^{2N-1} \Biggr)
\Biggl( \prod_{k=1}^{2N} \frac{1}{\rho^2+\mu_k^2} \Biggr)
\Biggl( \prod_{k=1,l=1,3,\cdots}^{2N-1} \frac{1}{(\mu_k-\mu_{k+l})^2} \Biggr) \\
&\quad\times
\Biggl( \prod_{k=1,l=2,4,\cdots}^{2N-2} \frac{1}{(\rho^2+\mu_k\mu_{k+l})^2} \Biggr)
\exp\Biggl[ 2 \sum_{k=1}^{2N} (-1)^{k+1} F(\rho,z,\mu_k) \Biggr] \,.
\end{split}
\end{align}
\end{subequations}
Metric~\eqref{n-bh} is, by construction, a solution of the vacuum Einstein equations,
and it represents a collection of $N$ Schwarzschild black holes, aligned along the $z$-axis, and immersed in the external gravitational field~\eqref{seed-extfield}.

We limit ourselves to the case of real poles $w_k$, since it represents the physically most relevant situation.
These constants are chosen with ordering
$w_1<w_2<\cdots<w_{2N-1}<w_{2N}$
and with parametrisation
\begin{equation}
w_1 = z_1 - m_1 \,, \quad w_2 = z_1 + m_1 \,, \quad \dotsc \quad
w_{2N-1} = z_N - m_N \,, \quad w_{2N} = z_N + m_N \,.
\end{equation}
The constants $m_k$ represent the black hole mass parameters, while $z_k$ are the black hole positions on the $z$-axis.

\begin{figure}
\includegraphics[scale=0.9]{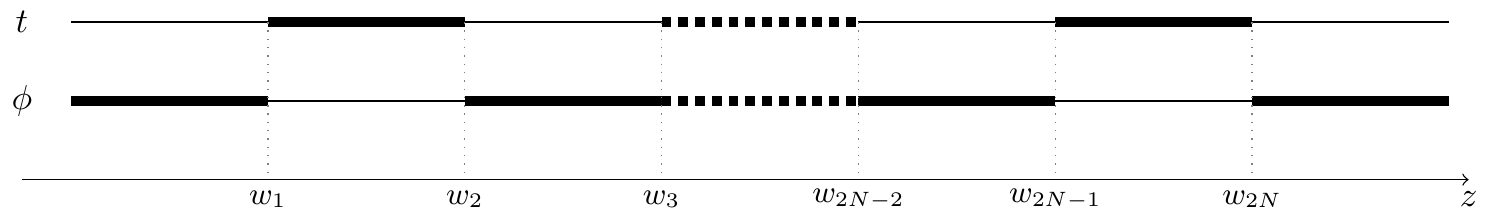}
\caption{{\small Rod diagram for the multi-black hole spacetime~\eqref{n-bh}.
The horizons correspond to the timelike rods (thick lines of the $t$ coordinate), while the conical singularities correspond to ``bolts'' where conical singularities can be avoided by imposing an appropriate periodicity on the angular coordinate.}}
\label{fig:n-rods}
\end{figure}

The black hole horizons correspond to the regions
$w_{2k-1}<z<w_{2k}$ ($k=1,\dotsc,N$), while the complementary regions are affected by the presence of conical singularities, as happens for the Israel--Khan solution (cf.~Fig.~\ref{fig:n-rods}), which can be recovered by setting $b_n=0$.
Differently from that case, our solution can be regularised, as we will show in the next subsection.

\subsection{Conical singularities and regularisation}

The infinite multipole momenta $b_n$ allow us to regularise the metric, i.e.~to remove all the conical singularities, by tuning their values.
More precisely, given $N$ black holes, there are exactly $N+1$ conical singularities:
two cosmic strings, one rear the first black hole and one ahead the last black hole, and $N-1$ struts located between the $N$ black holes.
Hence, one needs at least $N+1$ parameters to fix the singularities.

The manifold exhibits angular defects when the ratio between the length and the radius of small circles around the $z$-axis is different from $2\pi$.
Working in Weyl coordinates, a small circle around the $z$-axis has radius $R=\sqrt{g_{zz}}\rho$ and length $L=2\pi\sqrt{g_{\phi\phi}}$~\cite{Astorino:2021dju}.
The regularity condition corresponds then to
$L/(2\pi R)\to 1$ as $\rho\to 0$.
It is easy to prove that, for the static and axisymmetric metrics of the class we are considering, the above condition is equivalent to $\mathcal{P}\equiv f g_{tt}\to 1$ as $\rho\to 0$.

We choose for convenience the gauge parameter $C_f$ as
\begin{equation}
C_f = 2^{2(2N+1)} \Biggl[ \prod_{i=1}^N (w_{2i}-w_{2i-1})^2 \Biggr] \Biggl[
\prod_{k=1}^{N-1} \prod_{j=1}^{N-k}
(w_{2k-1} - w_{2k+2j})^2 (w_{2k} - w_{2k+2j-1})^2 \Biggr] \,.
\end{equation}
The quantity $\mathcal{P}=f g_{tt}$ is equal to
\begin{equation}
\mathcal{P}_k =\Biggl[ \prod_{i=1}^{2k} \prod_{j=2k+1}^{2N}
(w_j-w_i)^{2\,(-1)^{i+j+1}} \Biggr]
\exp\Biggl[ 4\sum_{n=1}^\infty b_n \sum_{j=2k+1}^{2N} (-1)^{j+1} w_j^n \Biggl] \,,
\end{equation}
between the $k$-th and $(k+1)$-th black holes (i.e.~$w_{2k}<z<w_{2k+1}$), for $1 \leq k < N$.
In the region $z<w_1$ we find
\begin{equation}
\label{n-P0}
\mathcal{P}_0 =
\exp\Biggl[ 4\sum_{n=1}^\infty b_n \sum_{j=1}^{2N} (-1)^{j+1} w_j^n \Biggl] \, ,
\end{equation}
while for $z>w_{2N}$ we simply have
\begin{equation}
\mathcal{P}_N = 1 \, ,
\end{equation}
thanks to our choice of $C_f$.
These expressions are the natural generalisations of the conical singularities for the Israel--Khan metric~\cite{Gregory:2020mmi}.

The above expressions provide a system of equations
$\mathcal{P}_k=1$ for $0 \leq k < N$,
which can be solved, e.g., for the parameters
$b_1,\dotsc,b_N$, with the result of fixing all the conical singularities.
Hence the solution can be made completely regular outside the black hole horizons.

\subsection{Smarr law}

We investigate the Smarr law for spacetime~\eqref{n-bh}:
to this end, we compute the total mass of the spacetime and the entropy and temperature of the black holes.

The mass is easily found by means of the Komar--Tomimatsu integral~\cite{Komar,Tomimatsu:1984pw}.
The result for the $k$-th black hole (i.e.~the black hole in the interval $w_{2k-1}<z<w_{2k}$) is
\begin{equation}
\label{n-mass}
M_k =
\alpha \int_{w_{2k-1}}^{w_{2k}} \rho g_{tt}^{-1} \partial_\rho g_{tt}
= \frac{\alpha}{2} (w_{2k}-w_{2k-1})
= \alpha m_k \,,
\end{equation}
where $\alpha$ is a constant which takes into account the proper normalisation of the timelike Killing vector, generator of the horizon, $\xi=\alpha\partial_t$ associated to~\eqref{n-bh}.

The entropy of a black hole is related to the area as $S_k=\mathcal{A}_k/4$, hence
\begin{equation}
\label{n-entropy}
S_k =
\frac{1}{4} \lim_{\rho\to0} \int_0^{2\pi} d\phi \int_{w_{2k-1}}^{w_{2k}} dz \sqrt{f g_{\phi\phi}}
= \pi m_k  W
\exp\Biggl[ 2\sum_{n=1}^\infty b_n \sum_{j=2k}^{2N} (-1)^{j+1} w_j^n \Biggl] \,,
\end{equation}
where
\begin{equation}
\log W = \lim_{\rho\to0} \log\sqrt{f g_{\phi\phi}} =
\log 2 + \sum_{i=1}^{2k-1} \sum_{j=2k}^{2N} (-1)^{i+j+1} \log|w_j-w_i| \,.
\end{equation}
The product $\sqrt{f g_{\phi\phi}}$ is independent of $z$ in the limit $\rho\to0$, and that is crucial in the derivation of~\eqref{n-entropy}.

Finally, the temperature is found via the Wick-rotated metric, and the result is
\begin{equation}
\label{n-temp}
T_k = \frac{\alpha}{2\pi} \lim_{\rho\to0} \rho^{-1} \sqrt{\frac{g_{tt}}{f}} =
\frac{\alpha}{2\pi} \lim_{\rho\to0} \frac{1}{\sqrt{f g_{\phi\phi}}} =
\frac{\alpha m_k}{2 S_k} \,.
\end{equation}
It is easily shown, by using~\eqref{n-mass},~\eqref{n-entropy} and~\eqref{n-temp}, that the Smarr law is satisfied:
\begin{equation}
\label{n-smarr}
\sum_{k=1}^N M_k = 2 \sum_{k=1}^N T_k S_k \, .
\end{equation}
We notice that the explicit value of $\alpha$ is not needed for~\eqref{n-smarr} to work, while it is relevant in the study of the thermodynamics~\cite{Astorino:2021dju}.

\subsection{Schwarzschild black hole}

The simplest non-trivial example we can consider for the complete external multipolar expansion from the general solution~\eqref{n-bh}, is clearly the single black hole configuration for $N=1$.
In that case the functions that appear in the Weyl static metric~\eqref{n-bh} take the form
\begin{subequations}
\begin{align}
g_1 & = \diag\biggl[
-\frac{\mu_1}{\mu_2} \exp\biggr(2\sum_{n=1}^{\infty} b_n r^n P_n \biggr) ,
\rho^2 \frac{\mu_2}{\mu_1} \exp\biggr(-2\sum_{n=1}^{\infty} b_n r^n P_n \biggr) \biggr] \,, \\
f_1 & = \frac{16 C_f \, f_0 \, \mu_1\mu_2^3 \, e^{2[F(\mu_1) - F(\mu_2)]}}{(\mu_1-\mu_2)^2(\rho^2+\mu_1^2)(\rho^2+\mu_2^2)} \,.
\end{align}
\end{subequations}
This spacetime represents a static black hole embedded in an external gravitational field.
The limit to the Schwarzschild metric is clear:
it is obtained by switching off all the multipoles $b_n=0$.
In order to recover the standard Schwarzschild metric in spherical coordinates, the following transformation is needed
\begin{equation}
\rho = \sqrt{r(r-2m)}\sin\theta \, , \quad
z = z_1 + (r-m) \cos \theta \, .
\end{equation}
A solution of this kind is not completely new, since it was already present in Chandrasekhar's book~\cite{Chandrasekhar:1985kt}, see also~\cite{Breton:1998sr}.
However, the form we are writing here is more general because, thanks to the extra parameter $z_1$, allows to place the black hole in any point of the $z$-axis.
In the absence of the external field the location of the black hole is irrelevant because the solution is symmetric under a finite shift of $z_1$.
But, when the external gravitational field is not zero, a translation along the $z$-axis is significant, since the relative position of the black hole with respect to the external multipolar sources has some non-trivial effects on the geometry and on the physics of the black hole.

\begin{figure}
\captionsetup[subfigure]{labelformat=empty}
\centering
\hspace{-0.1cm}
\subfloat[\centering $m=0.5$, $b_2=0.4$, $z_1=2$]{{\includegraphics[width=4.5cm]{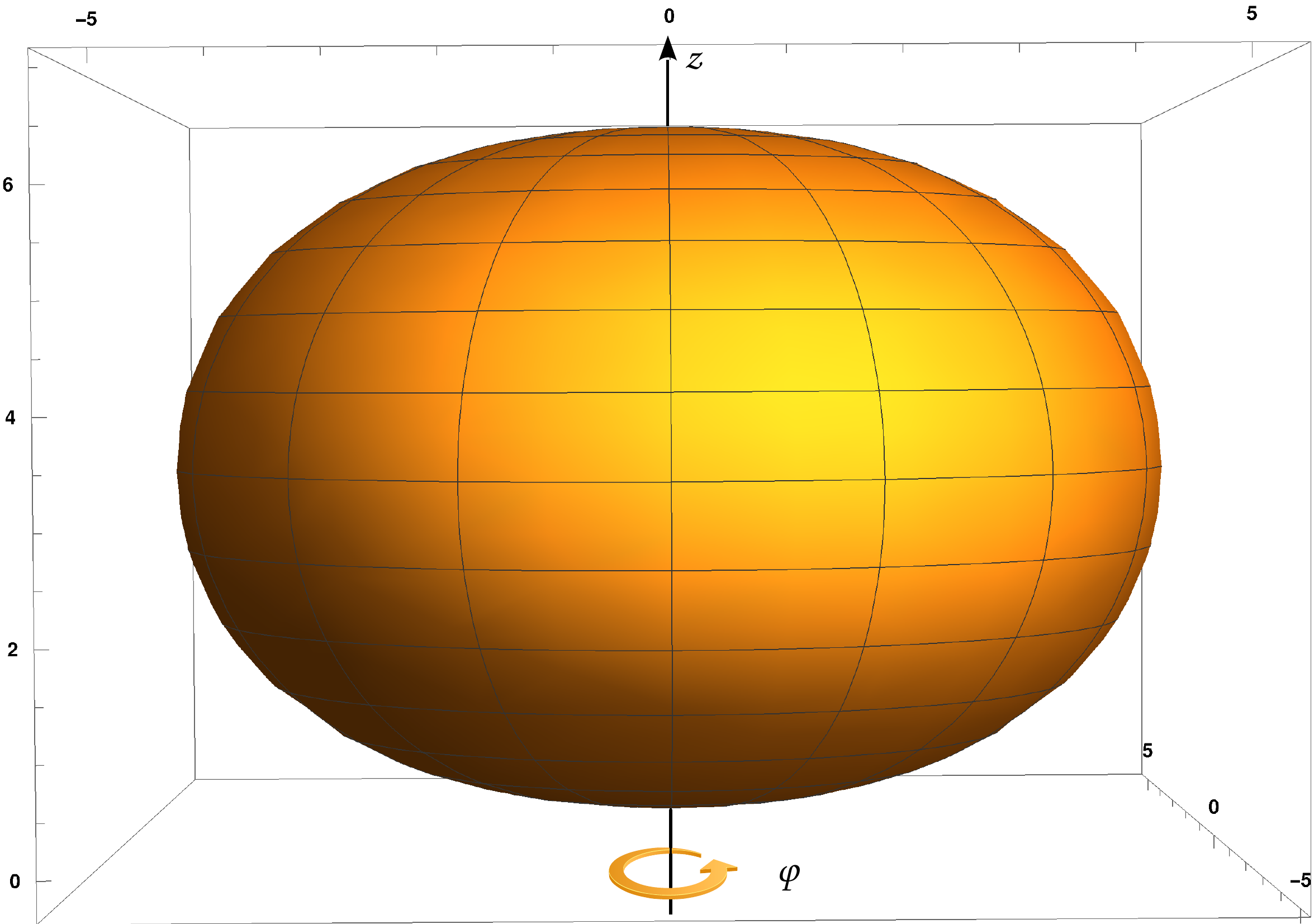}}}
\subfloat[\centering $m=1$, $b_2=-0.1$, $z_1=2$]{{\hspace{-0.3cm} \includegraphics[width=5.5cm]{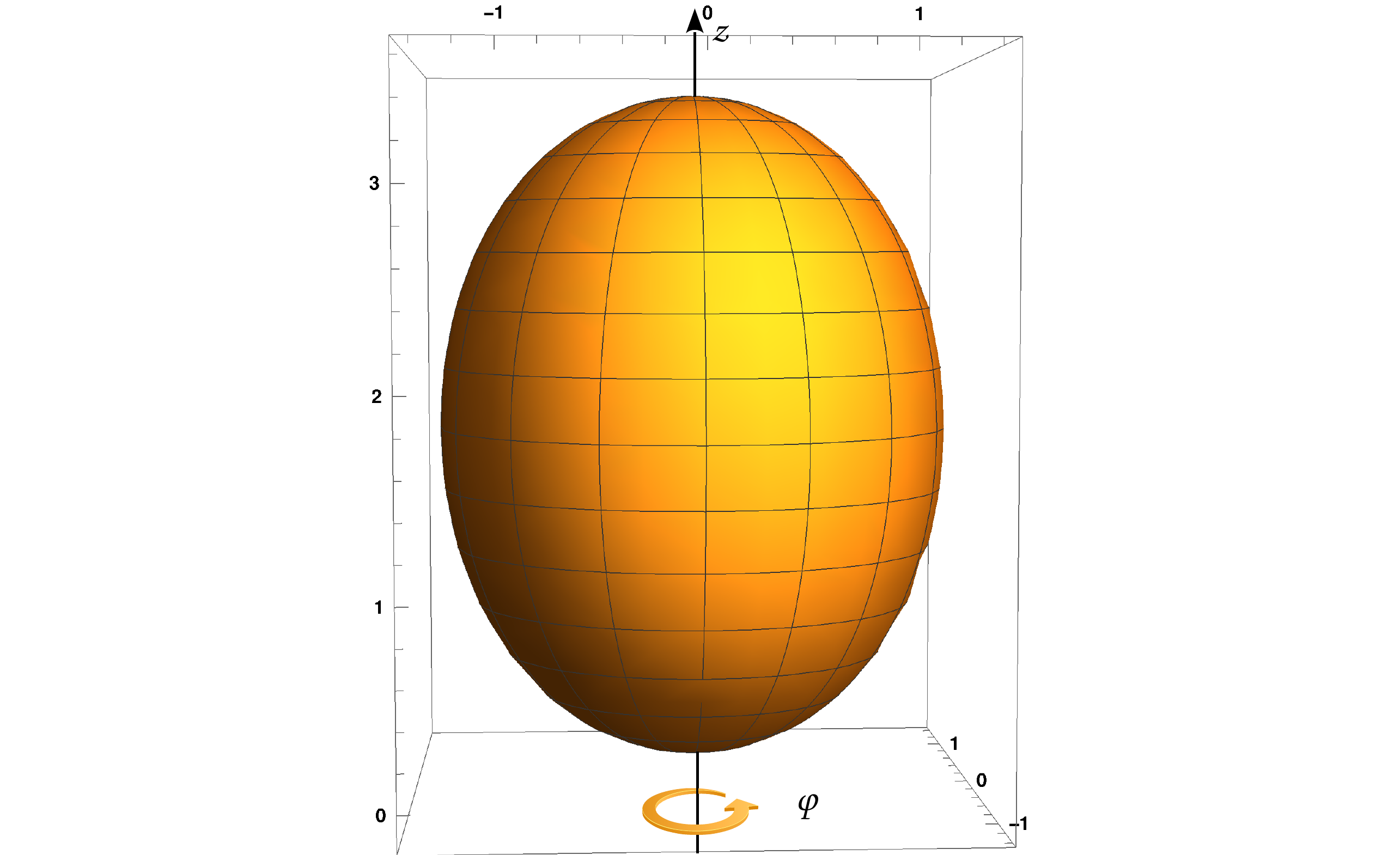}}}%
\subfloat[\hspace{-1.5cm} $m=1$, $b_2=-0.5$, $z_1=2$]{{\hspace{-1.5cm} \includegraphics[width=6.5cm]{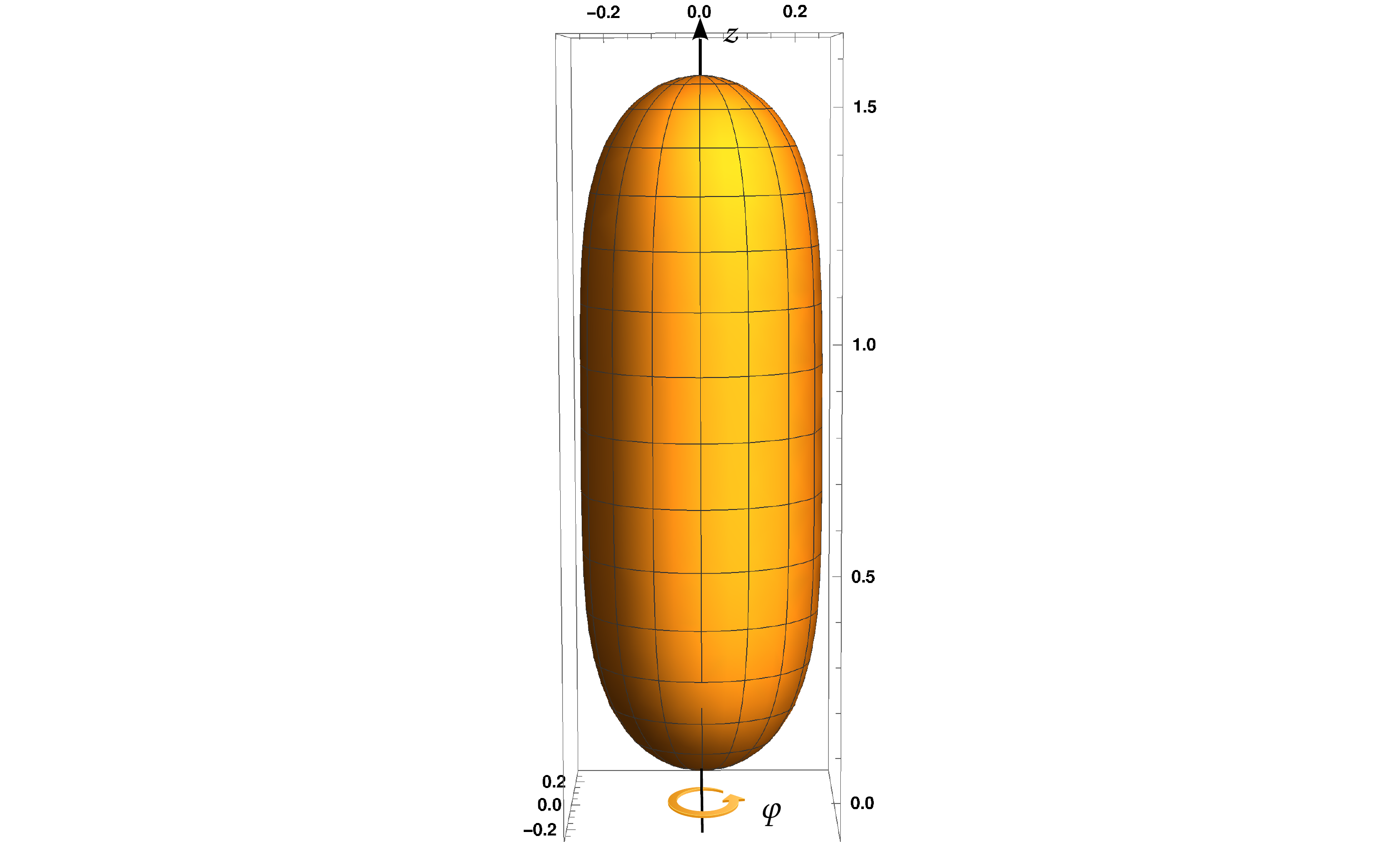}}}
\caption{\small Embeddings in Euclidean three-dimensional space $\mathbb{E}^3$ of the event horizon of single black holes distorted by the external dipolar and quadrupolar gravitational field, for three different sets of physical parameters.
$b_1$ and $C_f$ are fixed by the regularity constraint~\eqref{costraints-sh1}.}
\label{fig:bh-b1-b2}
\end{figure}

In particular the translation along the $z$-axis affects the event horizon shape.
In fact, as can be evaluated by computing the equatorial and polar circles around the event horizon, it is possible to understand how the horizon surface stretches or contracts depending on the position of the black hole and the values of the external parameters.
Some pictorial examples of the black hole horizon deformation for different external
gravitational backgrounds are given in Fig.~\ref{fig:bh-b1-b2}.

\begin{figure}
\centering 
\hspace{-3.5cm} 
\subfloat[\centering Black hole horizon with conical singularity]{{\includegraphics[width=10cm]{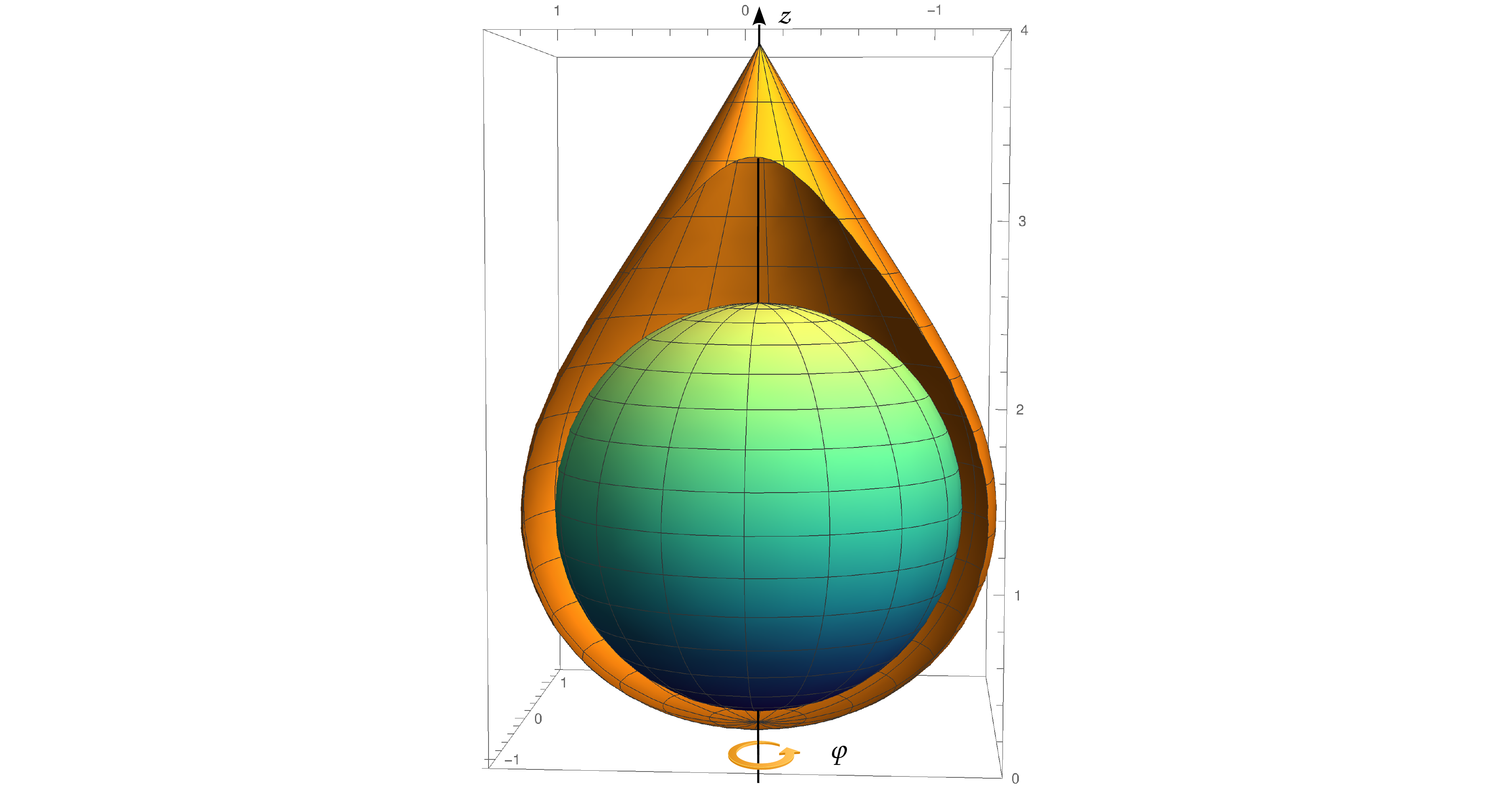}}}
\hspace{-1.5cm}
\subfloat[\centering Black hole horizon regularised on both poles]{{\includegraphics[width=5cm]{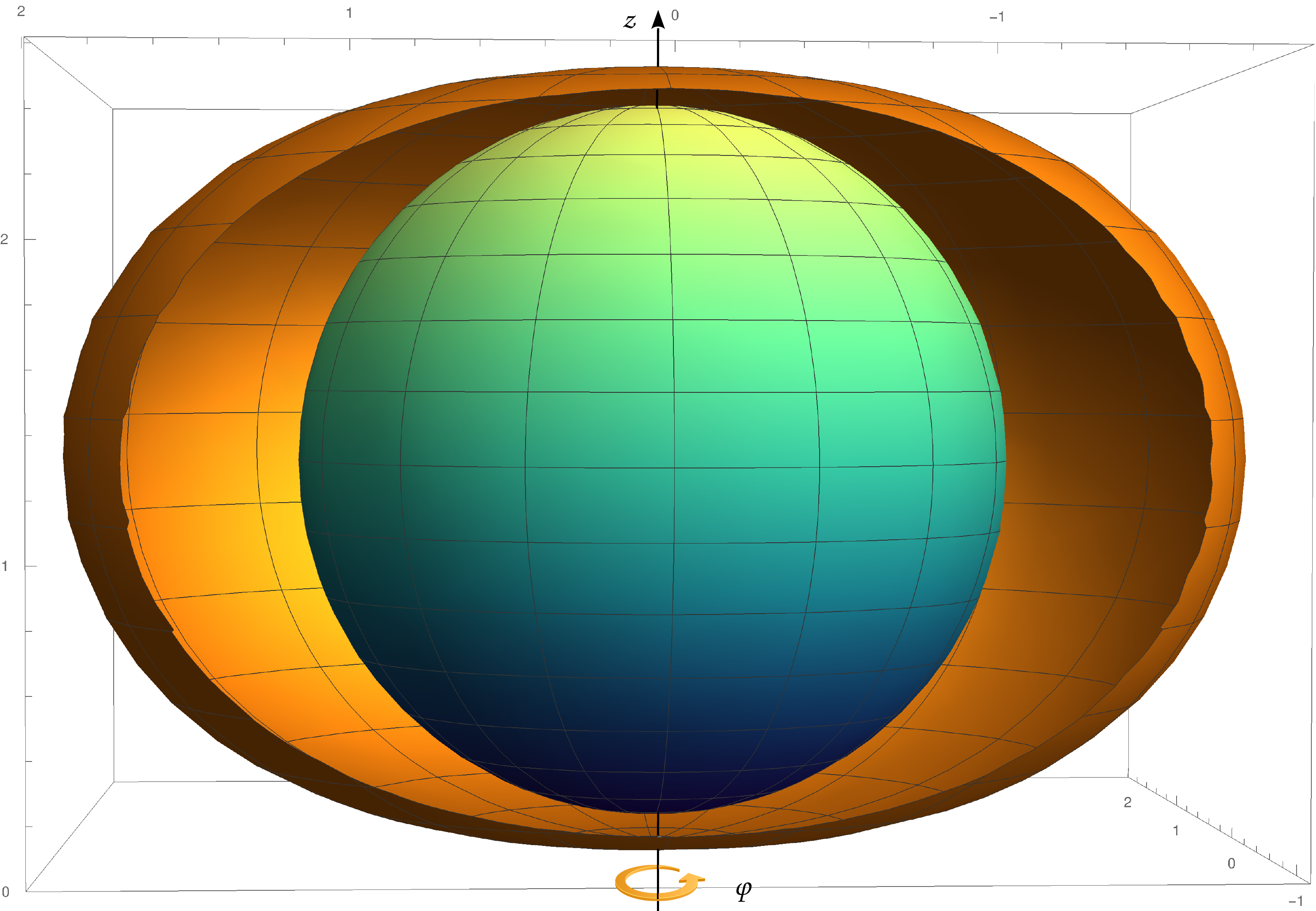}}}
\caption{\small Embedding in $\mathbb{E}^3$ of a single static black hole horizon (orange surface) immersed into a dipolar and quadrupolar external gravitational field, for $m = 0.6$, $z_1=-1.3$, $b_2=0.3$.
A section is taken to appreciate the deformation w.r.t.~the null external field case, drawn in green as a reference:
the standard spherical Schwarzschild horizon, which is everywhere covered by the horizon swollen by the presence of the external gravitational field.
The two black holes in external field differ only for the value of $b_1$, which in panel $(b)$ is chosen according to Eq.~\eqref{costraints-sh1} to remove also the second conical singularity, while in panel $(a)$ is $ b_1=0.5$.}
\label{fig:con-sing}
\end{figure}

Physically, the solution is affected by conical singularities because the arbitrary external field deforms and attempts to accelerate the black hole, which however does not contain an acceleration parameter, thus causing the appearance of the defects.
Only a fine tuned field, that is able to maintain the non-accelerating character of the solution, can avoid this pathological behaviour.

The equilibrium constraint which removes the conical defects can be loosen with respect to the one usually found in the literature~\cite{Breton:1998sr}.
The imposition usually requested in the literature, i.e~$\sum_{n=1}^\infty b_{2n+1}=0$, is not fundamental when the black hole can be adjusted coherently with the external gravitational field.
In fact, when $z_1 \neq 0$ the two regularising constraints we have to impose to avoid conical singularities become, as seen above,
\begin{equation}
\label{costraints-sh1}
C_f = \frac{(w_1-w_2)^2}{4} \,, \quad
\sum_{n=1}^\infty b_n \bigl( w_1^n - w_2^n \bigr) = 0 \,.
\end{equation}
When $b_n=0$ for $n>2$, we obtain a special subcase of the solution~\cite{Astorino:2021dju}, in the limit in which one of the black hole vanishes or where the two horizon rods merge, remaining only with a single black hole configuration.
Fig.~\ref{fig:con-sing} in this section refers, for simplicity, to this truncated expansion of the external field.
A qualitatively analogous behaviour of the black hole horizon occurs in the full multipolar expansion. 

\section{Binary system: a case study}

We consider an important subcase of the multi-black hole spacetime described in the previous section, namely the binary black hole configuration.
The importance of this subcase comes from its phenomenological relevance, since black holes usually present themselves in pairs;
moreover, they are the main sources of gravitational waves events.
It is known that gravitational waves are mainly generated in binary black hole systems, thus it is worth to construct and study regular exact solutions that might describe such configurations.

From the full solution~\eqref{n-bh} we can extract, by choosing $N=2$ and $b_n=0$ for $n>2$, the binary black hole system immersed in an external dipole-quadrupole field
\begin{subequations}
\label{binary}
\begin{align}
g & = \frac{\mu_1\mu_3}{\mu_2\mu_4}
\exp\biggl[2b_1z + 2b_2\biggl(z^2 - \frac{\rho^2}{2}\biggr)\biggr] \,, \\
\begin{split}
f & = 16C_f \frac{\mu_1^3\mu_2^5\mu_3^3\mu_4^5}{W_{11}W_{22}W_{33}W_{44}W_{13}^2W_{24}^2Y_{12}Y_{14}Y_{23}Y_{34}}
\exp\biggl[-b_1^2\rho^2 + \frac{b_2^2}{2} \bigl(\rho^2 - 8z^2\bigr)\rho^2 - 4b_1b_2 z \rho^2 \\
& \quad + 2 b_1 (-z + \mu_1 - \mu_2 + \mu_3 - \mu_4 ) \\
&\quad + b_2 \bigl(-2z^2 + \rho^2 + 4z (\mu_1 - \mu_2) + \mu_1^2 - \mu_2^2
+ (\mu_3 - \mu_4) (4z + \mu_3 + \mu_4) \bigr)\biggr] \,.
\end{split}
\end{align}
\end{subequations}
where
$W_{ij}=\rho^2+\mu_i\mu_j$,
$Y_{ij}=(\mu_i-\mu_j)^2$.
The parametrisation is inherited from the one in the previous Section, and the two event horizons extend in the regions $w_1<z<w_2$ and $w_3<z<w_4$ when $\rho=0$.

The conical singularities are removed as in the previous Section:
in this case, we can specialise the general formulas found above, and express the regularisation conditions in terms of the physical parameters
\begin{align}
C_f & = 256 m_1^2 m_2^2 (m_1+m_2+z_1-z_2)^2(m_1+m_2-z_1+z_2)^2 \,, \\
\label{b1-binary}
b_1 & = -\frac{(m_1z_1+m_2z_2)}{4m_1m_2(z_1-z_2)}
\log\biggl[\frac{(m_1-m_2+z_1-z_2)(m_1-m_2-z_1+z_2)}{(m_1+m_2+z_1-z_2)(m_1+m_2-z_1+z_2)}\biggr] \,, \\
\label{b2-binary}
b_2 & = \frac{(m_1+m_2)}{8m_1m_2(z_1-z_2)}
\log\biggl[\frac{(m_1-m_2+z_1-z_2)(m_1-m_2-z_1+z_2)}{(m_1+m_2+z_1-z_2)(m_1+m_2-z_1+z_2)}\biggr] \,.
\end{align}
It is understood that $C_f$, $b_1$, $b_2$ will assume the above values, from now on.
Note that the physical parameters characterising the black holes were left unconstrained:
this means that the physical properties of the binary black hole system are completely generic.
This fact guarantees a wide flexibility in a possible phenomenological scenario~\footnote{In fact the presented solution is not the minimal one which can be regularised.
It is sufficient to consider only one multipole term for the external  gravitational field to remove all the singularities from the double black hole configuration, at the price of fixing the position or the mass of a black hole.}.
Of course one could alternatively keep the $b_n$ generic, to model an arbitrary external gravitational background:
in that case the intrinsic black holes parameters would adjust to fit the given background.

\subsection{Near-horizon limit}

\begin{figure}
\centering
\includegraphics[scale=0.3]{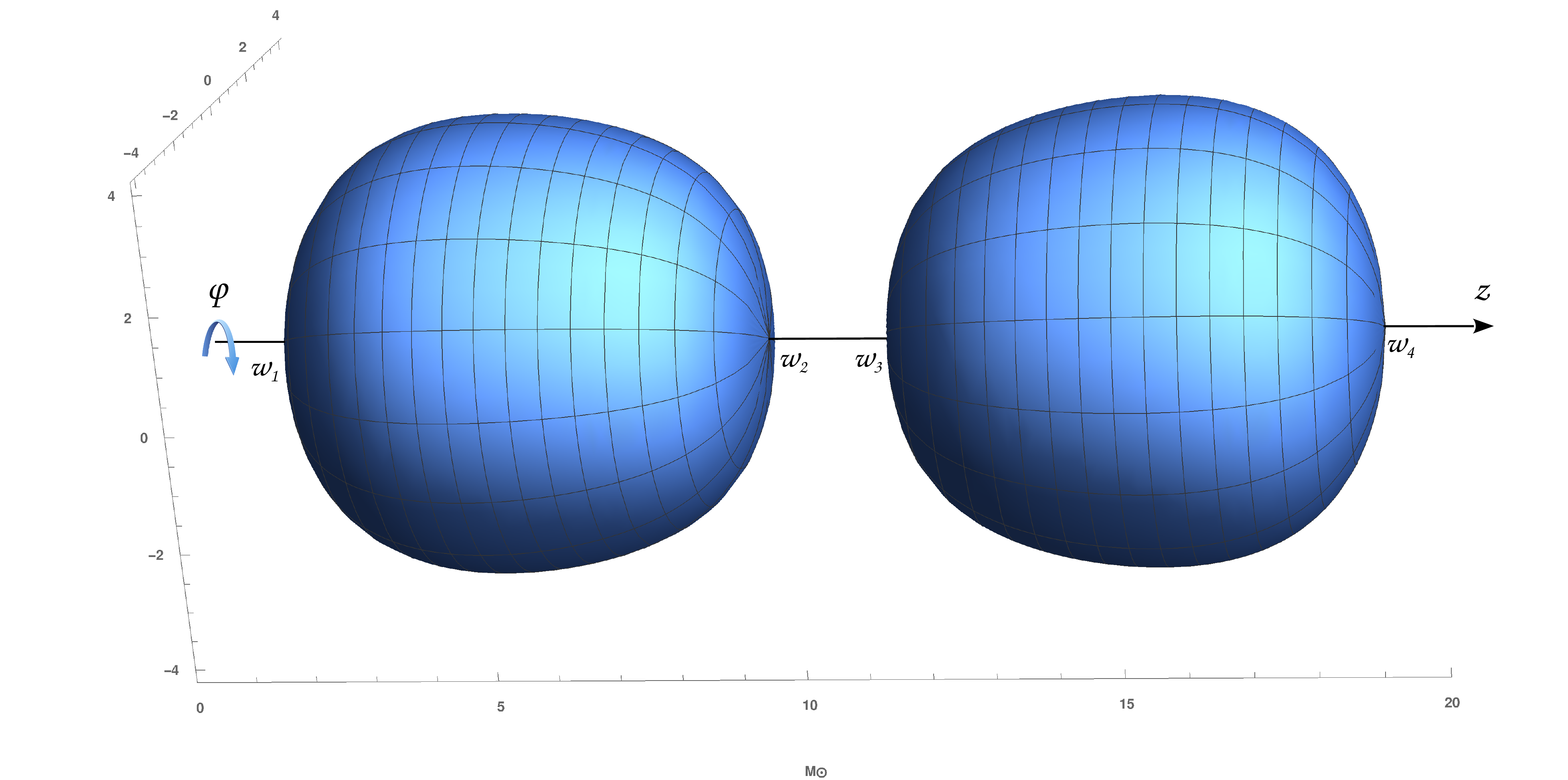}
\caption{\small Embedding diagram in $\mathbb{E}^3$ of the surfaces of the two black hole event horizons for the parametric values $z_1=5$, $z_2=15$, $m_1=4$, $m_2=4$.
This picture shows the deformation of the horizons due to both the external gravitational field and the mutual interaction between the sources.
The horizon surface is smooth because of the absence of any conical singularity.}
\label{fig:embeddingbinary}
\end{figure}

We consider the near-horizon limit of the binary metric~\eqref{binary}, in order to show that it contains two distorted Schwarzschild black holes.
We zoom in to the first black hole horizon by performing the change of coordinates
\begin{equation}
\rho = \sqrt{r(r-2m_1)}\sin\theta \,, \quad
z = z_1+(r-m_1)\cos\theta \,,
\end{equation}
and by taking the limit $r\to 2m_1$, by which the metric~\eqref{binary} boils down to
\begin{equation}
\label{near-horizon}
\begin{split}
{ds}^2 & \simeq h(\theta) \biggl[ -\biggl(1-\frac{2m_1}{r}\biggr) e^{F_1(\theta)} {dt}^2
+ \frac{D^2 e^{F_2(\theta)}}{1-2m_1/r} {dr}^2 \biggr] \\
&\quad + (2m_1)^2 \biggl[ D^2 h(\theta) e^{F_2(\theta)} {d\theta}^2
+ \frac{\sin^2\theta}{h(\theta)} e^{-F_1(\theta)} {d\phi}^2 \biggr] \,,
\end{split}
\end{equation}
where
\begin{align}
h(\theta) & = \frac{m_1\cos\theta+m_2+z_1-z_2}{m_1\cos\theta-m_2+z_1-z_2} \,, \\
F_1(\theta) & = 2 [b_1 + b_2(z_1+m_1\cos\theta)] (z_1 + m_1\cos\theta) \,, \\
\begin{split}
F_2(\theta) & = 2 b_1 (m_1\cos\theta - 2m_1 - 4m_2 - z_1) \\
&\quad + 2 b_2 (m_1^2\cos^2\theta + 2m_1z_1\cos\theta - 2m_1^2
- z_1^2 - 4m_1z_1 - 8m_2z_2) \,,
\end{split}
\end{align}
and
\begin{equation}
D = \frac{m_1+m_2-z_1+z_2}{m_1-m_2-z_1+z_2} \,.
\end{equation}
One clearly recognises the structure of a distorted Schwarzschild black hole~\cite{Geroch:1982bv}.
Actually the first black hole horizon is deformed by the presence of both the external field and the second black hole.
Indeed, when the external field and the second black hole vanish, one recovers the standard Schwarzschild metric.
Obviously, a similar description holds for the second black hole as well.
A pictorial representation of the deformation that the two horizons undergo is given in the embedding diagram of Fig.~\ref{fig:embeddingbinary}.

Another element about the geometry of the black hole horizon can be clarified by computing the length of the equatorial and polar circles in the near-horizon geometry~\eqref{near-horizon}.
The equator ($\theta=\pi/2$) length is given by
\begin{equation}
\label{equator}
L_\text{equator} = 2m_1 \int_0^{2\pi} \frac{e^{-F_1(\pi/2)/2}}{\sqrt{h(\pi/2)}} d\phi
= 4\pi m_1 \sqrt{\frac{z_1-z_2-m_2}{z_1-z_2+m_2}}
e^{-2(b_1+b_2z_1)z_1} \,,
\end{equation}
while the polar length is
\begin{equation}
\label{polar}
L_\text{polar} = 4m_1 D \int_0^\pi \sqrt{h(\theta)} e^{F_2(\theta)/2} d\theta \,.
\end{equation}
It is not possible to analytically perform the latter integral, nevertheless we can consider a numerical comparison between~\eqref{equator},~\eqref{polar} and $L_\text{Schwarzschild}=4\pi m_1$.
The result is
\begin{equation}
\label{comparison}
L_\text{equator} > L_\text{Schwarzschild} \,, \qquad
L_\text{polar} > L_\text{Schwarzschild} \,.
\end{equation}
Thus, not only a deformation along the $z$-axis occurs, but there is also an enlargement of both the equatorial and the polar circle with respect to the Schwarzschild one.
This is consistent with the behaviour of the black hole temperature, as we will see below.

\subsection{Thermodynamics}

Going back to the exact metric~\eqref{binary}, we now compute some physical quantities for that spacetime, in order to discuss the thermodynamics of the system.

The mass of the two black holes is found by means of the Komar--Tomimatsu integral~\cite{Komar,Tomimatsu:1984pw}.
The formula for the conserved mass, in the case of the static spacetime~\eqref{binary}, takes the form
\begin{equation}
M = \alpha \int_{w_{2i-1}}^{w_{2i}} dz\, \rho\, g_{tt}^{-1}\, \partial_\rho g_{tt} \big|_{\rho=0} \,,
\end{equation}
where $\alpha$ is a constant that takes into account the normalisation of the timelike Killing vector $\xi=\alpha\partial_t$, which generates the stationary symmetry.
It is well known that, in the absence of asymptotic flatness, $\alpha$ is not necessarily equal to one, as happens for black holes in AdS~\cite{Henneaux:1985tv} or Melvin~\cite{Ashtekar:2000hw,Astorino:2016hls} backgrounds.
The integration is performed over the intervals $w_1<z<w_2$ and $w_3<z<w_4$, which correspond to the black holes, and the result is
\begin{equation}
\label{mass}
M_1 = \alpha m_1 \,, \qquad
M_2 = \alpha m_2 \,.
\end{equation}
The horizons areas are found integrating over the horizon surfaces, i.e.
\begin{equation}
\mathcal{A} = \int_{0}^{2\pi} d\phi
\int_{w_{2i-1}}^{w_{2i}} dz \, \sqrt{g_{zz}g_{\phi\phi}} \Big|_{\rho=0} \,,
\end{equation}
thus
\begin{align}
\mathcal{A}_1 & =
16\pi m_1^2 \frac{m_1+m_2-z_1+z_2}{m_1-m_2-z_1+z_2}
e^{-2b_1 (2m_2+m_1+z_1)-2b_2((m_1+z_1)^2+4m_2z_2)} \,,
\\
\mathcal{A}_2 & =
16\pi m_2^2 \frac{m_1+m_2-z_1+z_2}{m_2-m_1-z_1+z_2}
e^{-2b_1(m_2+z_2)-2b_2(m_2+z_2)^2} \,.
\end{align}
The temperature is obtained from the surface gravity as $T=\kappa/(2\pi)$.
Recalling that $\kappa^2=-\frac{1}{2} (\nabla_\mu \xi_\nu)^2$,
the metric~\eqref{binary} gives rise to
\begin{equation}
\kappa^2 = -\frac{\alpha^2}{4} \frac{(\partial_z V)^2+(\partial_\rho V)^2}{fV} \bigg|_{\rho=0} \,.
\end{equation}
The temperatures are then
\begin{align}
T_1 & =
\frac{\alpha}{8\pi m_1} \frac{m_1-m_2-z_1+z_2}{m_1+m_2-z_1+z_2}
e^{2b_1 (2m_2+m_1+z_1)+2b_2((m_1+z_1)^2+4m_2z_2)} \,,
\\
T_2 & =
\frac{\alpha}{8\pi m_2} \frac{m_2-m_1-z_1+z_2}{m_1+m_2-z_1+z_2}
e^{2b_1(m_2+z_2)+2b_2(m_2+z_2)^2} \,.
\end{align}
One can verify that the same results are found via the Euclidean method~\cite{Hartle:1976tp}.
We notice that the presence of the external field lowers the black holes temperature, with respect to the Schwarzschild one, and hence the surface gravity.
A lower surface gravity means a lower gravitational ``pressure'' on the horizon, which then can swell up.
This feature is in agreement with~\eqref{comparison} and with the related observations.
Moreover, it explains how the external gravitational field acts, providing an external pressure in the region of the holes, to sustain the mutual gravitational collapse of the binary system. 

Defining the entropy as $S=\mathcal{A}/4$, the above quantities satisfy the Smarr law~\cite{Smarr:1972kt} both for the individual black holes $M_i=2 T_i S_i$ ($i=1,2$)
and for the double configuration
\begin{equation}
M_1 + M_2 = 2 T_1 S_1 + 2 T_2 S_2 \,.
\end{equation}
This result holds regardless of the value of the constant $\alpha$.
Nevertheless, a choice for $\alpha$ must be done in order to study the thermodynamics of the system.

We are interested in the first law of thermodynamics from a local point of view:
the involved quantities are evaluated on the horizons, therefore the sources at infinity (which generate the external field) are not accessible to local observers near the black holes.
Hence we will discard work terms, in the first law, due to the variation of the parameters $b_n$~\cite{Geroch:1982bv}.

We consider the system at thermal equilibrium from now on, i.e.~$T_1=T_2\equiv\bar{T}$.
This condition is satisfied by imposing $m_1=m_2$. We furthermore choose
\begin{equation}
\alpha = \sqrt{\frac{z_1-z_2-2m_1}{z_1-z_2}}
e^{-(b_1+b_2(m_1+z_2))(m_1+z_2)} \,,
\end{equation}
in order to fulfill a Christodoulou--Ruffini mass formula~\cite{Christodoulou:1972kt}, as it happens for regular metrics in which the asymptotic symmetry is different from the flat one~\cite{Astorino:2016hls,Caldarelli:1999xj,Astorino:2016ybm}.
For black hole configurations endowed with $N$ disconnected horizons the best proposal is an additive generalisation such that
\begin{equation}
\label{cristo-N}
\sum_{i=1}^{N} M_i =
\sum_{i=1}^{N} \sqrt{\frac{\mathcal{A}_i}{16 \pi}} \,.
\end{equation}
It is worth noting that $m_1=m_2$ is not the only possibility for thermal equilibrium, but it is clearly the simplest one.
Moreover, this choice guarantees the integrability of the masses~\eqref{mass}.

Defining the total mass
$\bar{M}\equiv M_1+M_2$ and the total entropy
$\bar{S}\equiv S_1 + S_2$,
the first law of thermodynamics
\begin{equation}
\label{firstlaw}
\delta\bar{M} = \bar{T} \delta\bar{S} \,,
\end{equation}
is verified.
We notice that the variation in~\eqref{firstlaw} is taken with respect to the free parameters $m_1$, $z_1$, $z_2$, in which $\alpha$ depends:
thus the presence of $\alpha$ is crucial to the first law.

We now turn to the verification of the second law of thermodynamics. At this scope we consider a process in which the initial state is described by two black holes at finite distance with total mass $\bar{M}$ and entropy $\bar{S}$,
while the final state is modeled by a single black hole of mass
\begin{equation}
M_0 = m_0 e^{-(b_1+(m_0+z_0)b_2)(m_0+z_0)} \,,
\end{equation}
and entropy
\begin{equation}
S_0 = 4\pi m_0^2
e^{2(b_1+(m_0+z_0)b_2)(m_0+z_0)} .
\end{equation}
Here $m_0$ and $z_0$ are the mass parameter and the position of the single black hole, respectively. $M_0$ and $S_0$ are computed from metric~\eqref{binary} with a single black hole.

Note that, contrary to~\cite{TOMIMATSU1984374,Breton:1998sr}, we do not assume $z_0=0$ \emph{a priori}.
The presence of the external gravitational field breaks the translation invariance along the $z$-axis, since the field acts differently on different points of the axis.
\begin{figure}
\centering
\includegraphics[scale=0.25]{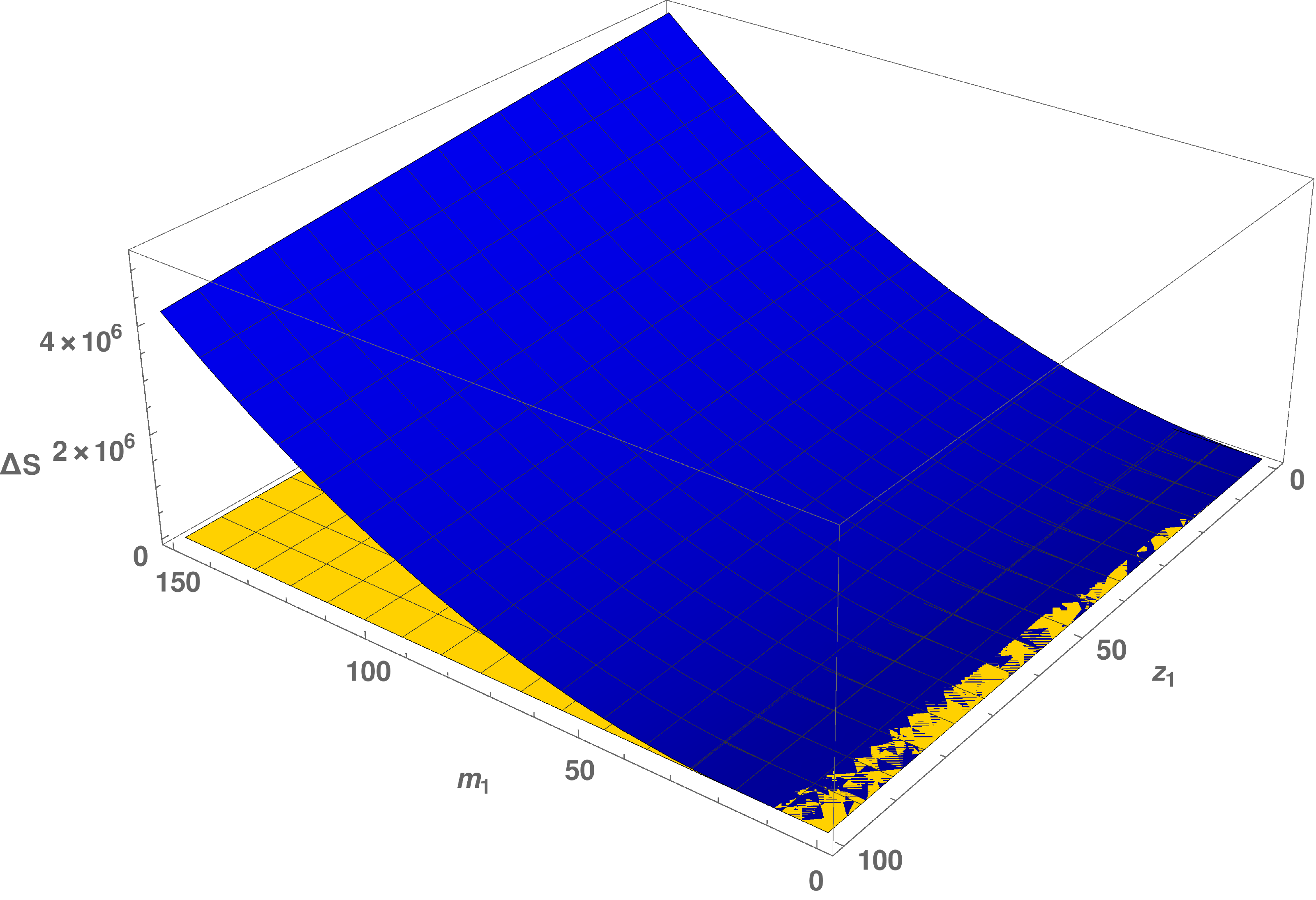}
\caption{\small The blue surface represents the entropy variation $\Delta S$, while the yellow surface is the 0 plane. The parameters vary over the values $m_1\in[0,150]$ and $z_1\in[0,100]$. Within the whole parametric range the single black hole state results more entropic than the disjoint pair.}
\label{fig:entropy}
\end{figure}
This fact is relevant, because it allows to regularise the single black hole metric with $b_1\neq0$.
In the single black hole case, the regularisation condition is
$b_1 = -2 b_2 z_0$.

In order to have a meaningful comparison we need to compare the two states at equal energy~\footnote{We are assuming that there is no energy loss due to emitted radiation.} and also their background must coincide.
Being $b_2$ unconstrained in the single black hole configuration, we just have to equate the values for $b_1$, which fixes the position of the single source at
\begin{equation}
z_0= \frac{z_1+z_2}{2} \,.
\end{equation}
Then we require $M_0=\bar{M}$,
that fixes $m_0$ and $z_2$ as
\begin{align}
m_0 & = 2 m_1 \sqrt{\frac{2m_1-z_1+z_2}{z_2-z_1}} \,, \\
z_2 & = z_1 + (\sqrt{17}-1) m_1 \,.
\end{align}
The entropy variation is given by
$\Delta S=S_0-\bar{S}$,
where the value of the parameters found above must be substituted into the expression.
The result, which is a function of $m_1$ and $z_1$, is quite involved, but it can be plotted as in Fig.~\ref{fig:entropy}.
It is clear from the plot that the function is always positive or null, i.e.
\begin{equation}
\Delta S \geq 0 \,,
\end{equation}
which verifies the second law of thermodynamics.
This is one of the few cases in which the second law for a binary system can be verified analytically;
it has been done, e.g., for the Majumdar--Papapetrou solution~\cite{Astorino:2019ljy}.

\subsection{Binary charged system}
\label{sec:charged}

A natural generalisation of the solution presented in the previous Section includes the addition of the electric charge;
thus we will construct a multi-Reissner--Nordstr\"om solution immersed in an external gravitational field.
Such a solution is the multi-black hole version of the metric presented in~\cite{Breton:1998sr}.

\subsubsection{The charging transformation}

There are several procedures to extend a stationary and axisymmetric solution of General Relativity to support an electromagnetic field.
These methods allow, for instance, to generate the Reissner--Nordstr\"om spacetime from the Schwarzschild black hole solution, such as the Harrison~\cite{Harrison1968} or the Kramer--Neugebauer~\cite{Kramer1969} transformations.
We want to present here perhaps the simplest version of this charging transformation, which maps a given static and axisymmetric vacuum solution to another static and axisymmetric electrovacuum spacetime, typically adding monopole electric charge.

Let us consider the most general static and axisymmetric metric for Einstein--Maxwell theory, the Weyl metric~\eqref{weyl}.
Suppose that the electromagnetic vector potential, compatible with the symmetries of this system, is null, i.e.~$A=0$.
Then an electrically charged solution can be generated by the following transformation on the $\psi$ function of the metric and of the electric field
\begin{subequations}
\label{hat-transf}
\begin{align}
\label{hhat}
e^{2\hat{\psi}} & = \frac{e^{2\psi}(1 - \zeta^2)^2}{(1-\zeta^2 e^{2\psi})^2}  \,, \\
\label{Ahat}
\hat{A} & = \frac{\zeta(e^{2\psi} - 1)}{1 - \zeta^2 e^{2\psi}} \, dt \,.
\end{align}
\end{subequations}
The continuous parameter $\zeta$ can be considered real and it is related to the electric charge of the spacetime.
In Appendix~\ref{app:harrison-kramer-neugebauer} we derive this transformation from the Kramer--Neugebauer one and we show how the latter is contained in the Harrison transformation, up to some gauge transformations.
There we also provide a simple example of application of the charging transformation~\eqref{hat-transf}, where the electrically charged black hole solution of Reissner--Nordstr\"om is generated from the Schwarzschild metric.

\subsubsection{Generating two distorted Reissner--Nordstr\"om black holes}
\label{double-RN}

Now we want to charge a multi-Schwarzschild solution embedded in an external gravitational field.
To keep the model as simple as possible, without constraining the physical parameters of the black hole, we consider, as a seed, the double-black hole spacetime immersed in an external gravitational field possessing dipole and quadrupole momenta only.
Note that this choice is done only for simplicity, but it could be chosen any external gravitational expansion endowed with multipoles of any order;
likewise the charging method allows one to deal easily with an arbitrary number of sources.
However we will act with the charging transformation~\eqref{hat-transf} on the solution~\eqref{n-bh} for $N=2$ and with the coefficients $b_1$ and $b_2$ different from zero only.
The resulting seed metric, which coincides with Eq.~\eqref{binary}, is
\begin{subequations}
\begin{align}
e^{2\psi} & =  \frac{\mu_1\mu_3}{\mu_2\mu_4}
\exp\biggl[2b_1z + 2b_2\biggl(z^2 - \frac{\rho^2}{2}\biggr)\biggr] \,, \\
\label{gamma}
\begin{split}
e^{2\gamma} &=  \frac{ 16C_f \, e^{2\psi} \, \mu_1^3\mu_2^5\mu_3^3\mu_4^5}{W_{11}W_{22}W_{33}W_{44}W_{13}^2W_{24}^2Y_{12}Y_{14}Y_{23}Y_{34}}
\exp\biggl\{-b_1^2\rho^2 + \frac{b_2^2}{2} \bigl(\rho^2 - 8z^2\bigr)\rho^2 - 4b_1b_2 z \rho^2 \\
&\quad + 2 b_1 (-z + \mu_1 - \mu_2 + \mu_3 - \mu_4 )
+ b_2 \bigl[-2z^2 + \rho^2 + 4z (\mu_1 - \mu_2) + \mu_1^2 - \mu_2^2 \\
&\quad + (\mu_3 - \mu_4) (4z + \mu_3 + \mu_4) \bigr] \biggr\} \,,
\end{split}
\end{align}
\end{subequations}
where again
$W_{ij}=\rho^2+\mu_i\mu_j$ and $Y_{ij}=(\mu_i-\mu_j)^2$, while the solitons are defined as in~\eqref{solitons}.
The new solution will maintain the same $\gamma$ as~\eqref{gamma} while transforming $\psi$ according to~\eqref{hhat}.

$\zeta$ is related to the electric charge of the black holes, therefore the transformed metric will represent a couple of charged black hole embedded in an external gravitational field.
In the limit of null external field $b_n=0$ and putting to zero one of the two black hole masses ($m_2=0$ for instance), we exactly recover the Reissner--Nordstr\"om solution.
Otherwise, when the external gravitational field is not present, the solution approaches the double charged masses of~\cite{Alekseev:2007gt,Manko:2007hi}.

We are interested in solutions regular outside the event horizons, therefore we have to consider the quantity $\mathcal{P}\equiv f g_{tt}$ to avoid the possible conical singularities of the charged metric.
In fact, when $\mathcal{P}$ differs from $1$, it takes into account the deficit or excess angle along the three regions of the axial axis of symmetry outside the black holes event horizons, i.e.~for $\rho=0$ and $z \in (-\infty,w_1)$, $z \in (w_2,w_3)$ and $z \in (w_4,\infty)$.
The solution is made regular from line singularities by imposing the following three constraints on the metric parameters:
\begin{align}
\label{Cf}
C_f  &= 16 (w_1-w_2)^2(w_2-w_3)^2(w_1-w_4)^2(w_3-w_4)^2 \,, \\
\label{b1}
b_1  &=  \frac{w_1^2 - w_2^2 + w_3^2 - w_4^2}{2(w_1-w_2)(w_1+w_2-w_3-w_4)(w_3-w_4)} \log \biggl[\frac{(w_1-w_3)(w_2-w_4)}{(w_2-w_3)(w_1-w_4)} \biggr] \,, \\
\label{b2}
b_2  &= -\frac{w_1 - w_2 + w_3 - w_4}{2(w_1-w_2)(w_1+w_2-w_3-w_4)(w_3-w_4)}  \log \biggl[\frac{(w_1-w_3)(w_2-w_4)}{(w_2-w_3)(w_1-w_4)} \biggr] \,.
\end{align}
While this remains formally the same regularisation constraint of the uncharged case, the physical meaning of the parameters is different, as it can be easily understood from the single source case treated in Appendix~\ref{app:harrison-kramer-neugebauer}.
In fact the two black hole horizons are located in $\rho=0$ and
$z\in(w_1,w_2)$, $z\in(w_3,w_4)$, where
\begin{equation}
\label{wi-charged}
w_1 = z_1 - \sigma_1 \,, \quad
w_2 = z_1 + \sigma_1 \,, \quad
w_3 = z_2 - \sigma_2 \,, \quad
w_4 = z_2 + \sigma_2 \,, \quad
\end{equation}
Henceforward we consider $C_f$ and $b_n$ fixed, as in Eqs.~\eqref{Cf}-\eqref{b2}, in order to assure the absence of conical singularities.
Note that the proper distance between the two event horizon surfaces converges:
\begin{equation}
\ell = \int_{w_2}^{w_3} dz\sqrt{g_{zz}} \Big|_{\rho=0} < \infty \,.
\end{equation} 
This means that the balancing condition is non-trivial and can be realised for a finite separation between the sources. 

\subsubsection{Charges and Smarr law}

The electric charge of each black hole can be computed thanks to the Gauss law~\cite{Emparan:2001bb}
\begin{equation}
\label{charge}
Q_i =
- \frac{1}{4\pi} \int_0^{2\pi} d\phi \int_{w_{2i-1}}^{w_{2i}} dz \, \rho g_{tt}^{-1} \partial_\rho A_t \big|_{\rho=0} =
\frac{2 \zeta \sigma_i}{1-\zeta^2} \,.
\end{equation}
Note that non-null results occur only in the regions which define the event horizon of the black holes, as expected.
Also note that, since the charging transformation is a one parameter transformation, it adds only an independent electric charge to the system.
Thus the free physical parameters of the solution are five: $z_i$, $\sigma_i$ and $\zeta$.
Hence the two black holes cannot vary independently their electric charge.
More general solutions involving independent electric charge parameters can be built, but with more refined generating techniques such as~\cite{Alekseev1980} (or~\cite{Sibgatullin1993}).

The mass of the charged black holes can be defined by evaluating, on their respective event horizon, the following integral\footnote{The normalisation of the timelike Killing vector here is considered unitary.
The generic normalisation factor $\alpha$ for the mass is reintroduced in Eqs.~\eqref{smarr-i},~\eqref{Mtot-Qtot}.}
\begin{equation}
\label{mass-charged}
M_i = \frac{1}{4} \int_{w_{2i-1}}^{w_{2i}} dz \rho g_{tt}^{-1}  \bigl( \partial_\rho g_{tt} - 2 A_t \partial_\rho A_t \bigr) \big|_{\rho=0} =
\frac{1 - 2 A_0 \zeta + \zeta^2}{1-\zeta^2} \sigma_i \bigg|_{A_0=0} = \frac{1 + \zeta^2}{1-\zeta^2} \sigma_i \,.
\end{equation}
Considering the mass as a local quantity, i.e.~defined close to the horizon, we can fix the gauge degree of freedom in the electric potential as $A_0=0$.
Of course other gauge fixings can be pursued, for instance requiring that the electric potential vanishes at large radial distances $\sqrt{\rho^2+z^2} \to \infty$.

From the masses and electric charges of the black holes, Eqs.~\eqref{charge} and~\eqref{mass-charged}, we can deduce, for any $\zeta \neq 1$, the value of 
\begin{equation}
\label{sigma}
\sigma_i = \sqrt{M_i^2 - Q_i^2}  \,.
\end{equation}
As expected the masses and electric charges are not independent, but they can be expressed just in terms of the parameters $M_i$ and $\zeta$.
In fact the electric charges can be written, thanks to Eqs.~\eqref{charge} and~\eqref{mass-charged}, as
\begin{equation}
\label{mass-charge}
Q_i = \frac{2\zeta}{1+\zeta^2} M_i \,.
\end{equation}
The entropy for each black hole is taken as a quarter of the event horizon area
\begin{equation}
S_i = \frac{1}{4} \int_0^{2\pi} d\phi \int_{w_{2i-1}}^{w_{2i}} dz  \sqrt{g_{zz} g_{\varphi\varphi}}\ \big|_{\rho=0} \,,
\end{equation}
which gives
\begin{subequations}
\begin{align}
S_1  & = \pi \frac{(w_2-w_1)^2(w_4-w_1)}{(w_3-w_1)(1-\zeta^2)^2} e^{-2b_1(w_2-w_3+w_4)-2b_2(w_2^2-w_3^2+w_4^2)} \,, \\
S_2 & = \pi \frac{(w_4-w_3)^2(w_4-w_1)}{(w_4-w_2)(1-\zeta^2)^2} e^{-2 w_4 (b_1+b_2w_4)} \,.
\end{align}
\end{subequations}
The temperature of the event horizons, computed as in the previous section, can be written as
\begin{equation}
T_i = \frac{\alpha \sigma_i}{2 S_i} \,.
\end{equation}
From Eq.~\eqref{mass-charge}, it is easy to see that when the masses of the two black holes coincide, $M_1=M_2$, also $Q_1=Q_2$.
Then it is possible to take $\zeta$ as in the single Reissner--Nordstr\"om case treated in Appendix~\ref{app:harrison-kramer-neugebauer}
\begin{equation}
\zeta = \frac{M_1 - \sqrt{M_1^2-Q_1^2}}{Q_1} \,.
\end{equation}
In this symmetric case one can straightforwardly check that both the temperature and the surface area of the two black holes coincide.
However the thermal equilibrium can be reached also for more general sources configurations.
Note that the event horizons become extremal in the limit $\zeta \to 1$, as in the single black hole case.

The Coulomb electric potential $\Theta$ evaluated on both the event horizons takes the value 
\begin{equation}
\Theta = - \xi^\mu A_\mu \big|_{\rho=0} =  
\alpha(\zeta-A_0) - \alpha\Theta_\infty  \,.
\end{equation}
We have now all the ingredients at our disposal to verify the Smarr law, both for the single element 
\begin{equation}
\label{smarr-i}
\alpha M_i = 2 T_i S_i - \Theta Q_i \,, 
\end{equation}
and, thus, for the double black hole configuration
\begin{equation}
M = \sum_{i=1}^2 2 T_i S_i - \Theta Q \,,
\end{equation}
where we defined 
\begin{equation}
\label{Mtot-Qtot}
M = \alpha \sum_{i=1}^2 M_i \,, \quad
Q = \sum_{i=1}^2 Q_i \,.
\end{equation}
Since distorted black holes have a preferred interpretation as local systems, we primarily focused on local quantities, basically defined on the horizon.
Nevertheless the above results hold also in the case one considers the presence of the asymptotic Coulomb potential.
The gauge freedom encoded in $A_0$ can be used to put to zero the value of the potential at large distance:
when $A_0=\zeta^{\text{sign}(b_2)}$, then  $\Theta_\infty=0$.

The above results are valid for any $\alpha$, the  normalisation parameter of the Killing vector that generate the horizon $\xi=\alpha \partial_t$. Then, in this context, $\alpha$ can practically regarded as unitary.
However, in discussing the first law of black hole thermodynamics, it is necessary to select a particular value for $\alpha$, as described in~\cite{Astorino:2021dju}, for a local point of view based on the assumption that the observers are located close to the hole and they have no access to infinity.

This charged black binary configuration is one of the few multi-black hole examples where it is concretely possible to test the second law of black holes thermodynamics, as done for the uncharged case~\cite{Astorino:2021dju} or for the Majumdar--Papapetrou black holes~\cite{Astorino:2019ljy}.
For instance, when the system is isolated, it is easy to verify that for two configurations, with the same energy and background field, the disjoint state is always less entropic than a collapsed state, which can be thought as the final state:
$S_{\odot\odot} < S_{\odot}$.
however, for different boundary conditions this charged case has a richer phase transitions scenario, from adiabatic merging to black hole brimming.

\subsubsection{Majumdar--Papapetrou limit}

Inspecting the values of the regularising parameters $b_n$ in Eqs.~\eqref{b1},~\eqref{b2}, we notice that there is a special case for which they vanish:
that happens for $w_1=w_2$ and $w_3=w_4$, that, according to Eqs.~\eqref{wi-charged},~\eqref{sigma} and~\eqref{mass-charge} corresponds to extremality, $M_i=Q_i$, or $\zeta=1$.
In that case the standard Minkowski asymptotics is retrieved.

That is not surprising, because the charged double black hole configuration presented in section~\ref{double-RN} naturally contains a particular subcase of the Majumdar--Papapetrou solution~\cite{Majumdar:1947eu,Papapetrou1945}, the one which describes two identical black holes located along the $z$-axis.
The black holes have to be identical because the charging transformation~\eqref{hat-transf} adds only one independent charge and the Majumbdar--Papapetrou solution possesses only extremal horizons.
In fact the Majumdar--Papapetrou solution is the only configuration, within the double Reissner--Nordstr\"om system~\cite{Alekseev:2007gt,Manko:2007hi}, that can reach the equilibrium by balancing the gravitational attraction thanks to the electric repulsion of the sources, without requiring any hyper-extreme event horizon, and thus preserving its black hole interpretation.

Actually, the limit to this version of the Majumdar--Papapetrou metric, describing a twin couple of (extremal) charged black holes, can be obtained easily, just considering the extremal limit of the charged solution, i.e.~$M_i=Q_i$.
It is given by the following simple form\footnote{The Majumdar--Papapetrou solution describing a binary black hole system in Weyl coordinate can be found in~\cite{Astorino:2019ljy}.}
\begin{subequations}
\begin{align}
{d\hat{s}}^{2} &= -e^{2\hat{\psi}}  {dt}^2 + e^{-2\hat{\psi}} \bigl( {d\rho}^2 + {dz}^2 + \rho^2 {d\phi} \bigr) \,, \\
\hat{A}_t & = \biggl( 1 + \frac{M_1}{\sqrt{\rho^2+(z-z_1)^2}} + \frac{M_2}{\sqrt{\rho^2+(z-z_2)^2}} \biggr)^{-1} \,,
\end{align}
\end{subequations}
where
\begin{equation}
e^{2\hat{\psi}} = \hat{A}_t^2 \,.
\end{equation}

\subsection{Binary Kerr--NUT system}
\label{sec:rotating}

We briefly discuss the rotating and NUTty generalisation of the binary black hole system immersed in an external gravitational field, that is the case $N=2$.
The generalisation to an array of rotating and NUT black holes is straightforward, at least conceptually.

We start again with the seed metric~\eqref{seed-extfield} and add four solitons (i.e.~two black holes).
However we make a different choice for the BZ constants:
we choose~\cite{Letelier:1998ft}
\begin{subequations}
\begin{align}
C_1^{(1)}C_0^{(2)} - C_0^{(1)}C_1^{(2)} & = \sigma_1 \, , \quad
C_1^{(1)}C_0^{(2)} + C_0^{(1)}C_1^{(2)} = -m_1 \, , \\
C_0^{(1)}C_0^{(2)} - C_1^{(1)}C_1^{(2)} & = n_1 \, , \quad
C_0^{(1)}C_0^{(2)} + C_1^{(1)}C_1^{(2)} = a_1 \, ,
\end{align}
\end{subequations}
and
\begin{subequations}
\begin{align}
C_1^{(3)}C_0^{(4)} - C_0^{(3)}C_1^{(4)} & = \sigma_2 \, , \quad
C_1^{(3)}C_0^{(4)} + C_0^{(3)}C_1^{(4)} = -m_2 \, , \\
C_0^{(3)}C_0^{(4)} - C_1^{(3)}C_1^{(4)} & = n_2 \, , \quad
C_0^{(3)}C_0^{(4)} + C_1^{(3)}C_1^{(4)} = a_2 \, .
\end{align}
\end{subequations}
Here $m_i$ are the mass parameters, $a_i$ are the angular momenta and $n_i$ are the NUT parameters.
We have also defined
$\sigma_i^2\equiv m_i^2 - a_i^2 + n_i^2$.
The poles $w_k$ are naturally defined as
\begin{equation}
w_1 = z_1 - \sigma_1 \,, \quad
w_2 = z_1 + \sigma_1 \,, \quad
w_3 = z_2 - \sigma_2 \,, \quad
w_4 = z_2 + \sigma_2 \,,
\end{equation}
where $z_i$ are the positions of the black holes.

The resulting metric is computed by following the inverse scattering method of Sec.~\ref{sec:ism}, but it is quite involved and we will not write it explicitly here.
The simplest form of the metric is achieved by using the bipolar coordinates
\begin{equation}
\begin{cases}
\rho = \sigma_1 \sqrt{x_1^2-1} \sqrt{1-y_1^2} \\
z = z_1 + \sigma_1 x_1 y_1
\end{cases}
\,, \quad
\begin{cases}
\rho = \sigma_2 \sqrt{x_2^2-1} \sqrt{1-y_2^2} \\
z = z_2 + \sigma_2 x_2 y_2
\end{cases}
\,.
\end{equation}
The metric regularisation from angular defects on the symmetry axis can be pursuit as in the static case above, by tuning one physical parameter of the solution for each spacelike rod.

It is not sufficient to put equal to zero all the $N$ constants $n_i$ in order to avoid Misner strings or other issues carried by the NUT charge:
in the multi-black hole case the actual NUT charge is a combination of the parameters of the solution~\cite{Alekseev:2019kcf}, hence it is not obvious how to remove the Misner singularities \emph{a priori}.
A way to compute the NUT charge is provided in~\cite{Alekseev:2019kcf,Herdeiro:2008kq}:
given the quantities
\begin{equation}
\Omega_i \coloneqq \frac{g_{t\phi}}{g_{tt}} \bigg|_{w_{i-1}<z<w_i} \,,
\end{equation}
the absence of Misner strings is guaranteed by the condition
\begin{equation}
\Omega_1 = \Omega_3 = \Omega_5 \,.
\end{equation}
Once the latter condition is satisfied, the common constant value of $\Omega$ can be reduced to zero by a shift of the time coordinate, namely $t\to t+C\phi$, where $C$ is an appropriate constant.
In particular, the NUT charge $n$ of the spacetime is usually identified as
\begin{equation}
4 n = \Omega_1 - \Omega_5 \,.
\end{equation}
The requirement of a vanishing NUT charge provides a constraint on the parameters of the solution, which then guarantees the absence of the pathologies usually associated to such a quantity.

Once the NUT charge (and the Misner strings) have been removed, it is possible to investigate the conical singularities.
In fact, the presence of a Misner strings implies the existence of a conical singularity, which can not be removed until the string is present.
The general formula for the computation of the ratio between the circumference and the radius around the $z$-axis is~\cite{Alekseev:2019kcf}
\begin{equation}
\frac{L}{2\pi R} =
\sqrt{\frac{g_{\phi\phi}}{g_{zz}\rho^2}} \to 1 \,;
\end{equation}
the expression used in the diagonal case can be applied in the stationary case as well, provided that $g_{t\phi}=O(\rho^2)$ as $\rho\to0$.
Thus, such a formula can be used to constrain the values of the dipole and quadrupole parameters, $b_1$ and $b_2$, in order to remove the conical singularities and obtain a regular spacetime.
Since the explicit expressions of the regularising parameters are quite involved, it is not possible to write them here. It is easier to check numerically that the rotating solution can be regularised as in the static cases.

In the simplest case of a dipole-quadrupole configuration, the binary system is characterised by ten parameters:
$\{ m_1,m_2,a_1,a_2,n_1,n_2,z_1,z_2,b_1,b_2 \}$.
Contrary to the standard double-Kerr case, in which there is no external field, the relevant physical quantity is not the distance between the holes $z_2-z_1$.
This happens because the external field does not act uniformly on the $z$-axis, hence the translation invariance along the axis is broken.
This means that the positions of the black holes are two independent parameters.

However, when the external field parameters are set to zero, $b_1=b_2=0$, one recovers the usual double-Kerr--NUT solution~\cite{Letelier:1998ft} (see~\cite{Manko:2018iyn} for a recent account on the equilibrium configurations).
Moreover, one can see that the positions $z_1$ and $z_2$ can be reabsorbed into a single parameter $l=z_2-z_1$.
Obviously, in such a limit the spacetime can not be regularised to give a physical solution, and the conical singularities can not be avoided.
The spacetime is then affected by the presence of struts or cosmic strings, unless one admits ``naked singularity-black hole'' or ``naked singularity-naked singularity'' configurations\footnote{Note that the ``black hole-black hole'' equilibrium configuration found in~\cite{Letelier:1998ft} is not reliable, since the conical singularity is computed in presence of a Misner string by means of a non-suitable formula for the NUTty case.}.

\section{Array of accelerating black holes}

In this Chapter, inspired by an Ernst's insight in~\cite{ErnstGeneralized}, where he was able to remove the conical singularity of an accelerating black hole thanks to the first non-trivial term (the dipole) of the multipolar expansion of the external gravitational field, we want to push forward his idea and, at the same time, incorporate multi-black hole sources.
More specifically, we want $(i)$ to generalise the regularised C-metric constructed by Ernst~\cite{ErnstGeneralized} including the full multipolar expansion of the external gravitational field and $(ii)$ to embed an arbitrary number of collinear accelerating black holes into a gravitational background.
The introduction of the acceleration in the deformed black hole scenario is not just for
sake of generality.
The acceleration parameter brings an extra Killing horizon and a conformal infinity factor, which can improve the local interpretation of these black hole systems embedded in an external gravitational field. 
The idea we pursue in this work was also pioneered by Gibbons, but by regularising with negative mass particles~\cite{Gibbons:1974zd}.

\subsection{Accelerating multipolar gravitational background}

A natural way to obtain the C-metric by means of the inverse scattering technique is to immerse a black hole in an accelerating background, as shown in Chapter~\ref{chap:gentech}.
This means that two solitons have to be added to the Rindler spacetime, which is nothing but Minkowski spacetime adapted to an accelerated observer.
The Rindler spacetime can be expressed in Weyl coordinates in the following way:
\begin{subequations}
\label{acc}
\begin{align}
g_\text{acc} & = \diag\biggl( -\mu_A , \frac{\rho^2}{\mu_A} \biggr) \, , \\
f_\text{acc} & = \frac{\mu_A}{\rho^2 + \mu_A^2} \,,
\end{align}
\end{subequations}
where $\mu_A = \sqrt{\rho^2 + (z-w_A)^2} - (z-w_A)$ is the soliton which contains the acceleration parameter of the Rindler metric.
A nice parametrisation for the constant $w_A$ is indeed
\begin{equation}
w_A = \frac{1}{2A} \,,
\end{equation}
where $A$ is the acceleration.
The addition of two solitons to~\eqref{acc} gives the standard C-metric, while more (even) solitons allow one to construct the accelerating multi-black hole metric discovered by Dowker and Thambyahpillai~\cite{Dowker:2001dg}.

We want to immerse many accelerating black holes in an external gravitational field, hence the natural background is the accelerated version of the external field background.
The external gravitational field is described by the metric
\begin{subequations}
\label{ext}
\begin{align}
g_\text{ext} & = \diag\Biggl[ -\exp\biggr(2\sum_{n=1}^{\infty} b_n r^n P_n \biggr) ,
\rho^2 \exp\biggl(-2\sum_{n=1}^{\infty} b_n r^n P_n \biggr) \Biggr] \,, \\
f_\text{ext} & = \exp\Biggl[
2 \sum_{n,p=1}^\infty \frac{np b_n b_p r^{n+p}}{n+p} \bigl(P_n P_p - P_{n-1} P_{p-1}\bigr)
- 2\sum_{n=1}^{\infty} b_n r^n P_n
\Biggr] \,,
\end{align}
\end{subequations}
where the parameters were defined at the beginning of the Chapter.

The metric which includes both the acceleration~\eqref{acc} and the external field background~\eqref{ext} is naturally given by
\begin{subequations}
\label{seed}
\begin{align}
g_0 & = \diag\Biggl[ - \mu_A \exp\biggr(2\sum_{n=1}^{\infty} b_n r^n P_n \biggr) ,
\frac{\rho^2}{\mu_A} \exp\biggl(-2\sum_{n=1}^{\infty} b_n r^n P_n \biggr) \Biggr] \,, \\
\begin{split}
f_0 & = \frac{\mu_A}{\rho^2 + \mu_A^2} \exp\Biggl[
2 \sum_{n,p=1}^\infty \frac{np b_n b_p r^{n+p}}{n+p} \bigl(P_n P_p - P_{n-1} P_{p-1}\bigr)
- 2\sum_{n=1}^{\infty} b_n r^n P_n   \\
&\quad + \frac{\rho^2 + \mu_A^2}{\mu_A} \sum_{n=1}^\infty b_n \sum_{l=0}^{n-1} w_A^{n-1-l} \, r^l P_l
\Biggr] \,,
\end{split}
\end{align}
\end{subequations}
We see that by turning off the external field (i.e.~$b_n=0$), one is left with the Rindler spacetime only.
On the converse, by removing the acceleration, in the limit $w_A\to\infty$, one recovers the background~\eqref{ext}.

Thus, we will take~\eqref{seed} as a background to construct our black hole spacetime.
Since the addition of solitons in the inverse scattering technique is equivalent to the addition of black holes to the seed spacetime~\eqref{seed}, it is quite natural to interpret the resulting metric as a collection of many black holes which are accelerating in an external gravitational field.
Following the discussion in Sec.~\ref{sec:ism}, we need the generating matrix $\Psi_0$ to build a black hole spacetime on top of the background~\eqref{seed}.
The function which satisfies equations~\eqref{eigen} generalises the one presented in~\cite{LetelierBrasil}, and it is given by
\begin{equation}
\label{psi0}
\Psi_0(\rho,z,\lambda) =
\begin{pmatrix}
- (\lambda-\mu_A) e^{F(\rho,z,\lambda)} & 0 \\
0 & (\lambda+\rho^2/\mu_A) e^{-F(\rho,z,\lambda)}
\end{pmatrix}
\,,
\end{equation}
where
\begin{equation}
\begin{split}
\label{F}
F(\rho,z,\lambda) & = 2 \sum_{n=1}^\infty b_n
\Biggl[ \sum_{l=0}^\infty \binom{n}{l} \biggl(\frac{-\rho^2}{2 \lambda}\biggr)^l \biggl(z + \frac{\lambda}{2}\biggr)^{n-l} \\
&\quad - \sum_{l=1}^n \sum_{k=0}^{[(n-l)/2]}
\frac{(-1)^{k+l}2^{-2k-l} n! \lambda^{-l}}{k!(k+l)!(n-2k-l)!}
\rho^{2(k+l)} z^{n-2k-l} \Biggr] \,.
\end{split}
\end{equation}
Now we can construct the BZ vectors~\eqref{bz-vectors}:
we parametrise
$m_0^{(k)}=\bigl(C_0^{(k)},C_1^{(k)}\bigr)$,
where $C_0^{(k)}$, $C_1^{(k)}$ are constants that will be eventually related to the physical parameters of the solution.
The BZ vectors are thus
\begin{equation}
m^{(k)} = \biggl( - \frac{C_0^{(k)}}{\mu_k-\mu_A} e^{-F(\rho,z,\mu_k)},
C_1^{(k)} \frac{\mu_A}{\rho^2 + \mu_A\mu_k} e^{F(\rho,z,\mu_k)} \biggr) \,.
\end{equation}
Depending on the value of $C_0^{(k)}$ and $C_1^{(k)}$, the spacetime will be static or stationary.

\subsection{Array of static accelerating black holes}
\label{sec:array}

We construct the generalisation of the Dowker--Thambyahpillai solution~\cite{Dowker:2001dg}, which represents an array of collinear accelerating black holes.
The Dowker--Thambyahpillai metric is characterised by the presence of conical singularities, which can not be removed by a fine tuning of the physical parameters without admitting naked singularities.

Given the accelerating background~\eqref{seed} and the generating matrix~\eqref{psi0}, we construct a new solution by adding $2N$ solitons (which correspond to $N$ black holes) with constants
\begin{equation}
\label{diagonal}
C_0^{(k)} =
\begin{cases}
1 & k \text{ even} \\
0 & k \text{ odd}
\end{cases}
\,, \qquad
C_1^{(k)} =
\begin{cases}
0 & k \text{ even} \\
1 & k \text{ odd}
\end{cases}
\,.
\end{equation}
This choice guarantees a non-rotating metric, which is the one we are interested in.
A different choice for these constants allows the inclusion of the rotation parameter $a$:
for the $k$-th pair of BZ constants one takes
\begin{subequations}
\label{rot-Cik}
\begin{align}
C_0^{(2k-1)} C_0^{(2k)} + C_1^{(2k-1)} C_1^{(2k)} & = - \sqrt{m_k^2 - a_k^2} \,, \\
C_0^{(2k-1)} C_0^{(2k)} - C_1^{(2k-1)} C_1^{(2k)} & = m_k \frac{1 - A^2 a_k^2}{1 + A^2 a_k^2} \,, \\
C_0^{(2k-1)} C_1^{(2k)} + C_1^{(2k-1)} C_0^{(2k)} & = -\frac{2A m_k a_k}{1 + A^2 a_k^2} \,, \\
C_0^{(2k-1)} C_1^{(2k)} - C_1^{(2k-1)} C_0^{(2k)} & = a_k \,.
\end{align}
\end{subequations}
In the single black hole case and with no external field, i.e.~$b_n=0$  for all $n$, the above parametrisation leads to the standard form of the rotating C-metric~\cite{Hong:2004dm}, as shown in Chapter~\ref{chap:gentech}.
Likewise, with a null external gravitational field with the choice~\eqref{rot-Cik}, in the multi-black hole case, leads to a vacuum multi-Pleba\'nski-Demia\'nski metric\footnote{We do not expand explicitly here the full expression of the multi-rotating-C-metric, neither with nor without the external gravitational field, because it is quite lengthy, but it can be straightforwardly written down with~\eqref{rot-Cik}.}.
Being interested in the phenomenological setting, we have not included the NUT parameter in the above definitions;
however, it is possible to include the NUT charge as well in the inverse scattering formalism~\cite{LetelierSolitons}.
One has to be aware that the absence of the NUT parameters in~\eqref{rot-Cik}, for the multi-black hole case, does not imply an overall zero NUT charge for the whole spacetime~\cite{Alekseev:2019kcf}, contrary to the single black hole case.

The metric resulting from the diagonal choice~\eqref{diagonal} is
\begin{subequations}
\label{n-acc}
\begin{align}
\begin{split}
g_N & = \diag\Biggl[
-\mu_A\frac{\prod_{k=1}^N \mu_{2k-1}}{\prod_{l=1}^N \mu_{2l}} \exp\Biggl(2{\sum_{n=1}^{\infty} b_n r^n P_n}\Biggr), \\
&\quad \frac{\rho^2}{\mu_A} \frac{\prod_{l=1}^N \mu_{2l}}{\prod_{k=1}^N \mu_{2k-1}} \exp\Biggl(-2{\sum_{n=1}^{\infty} b_n r^n P_n}\Biggr)
\Biggr] \,,
\end{split} \\
\begin{split}
f_N & = 16C_f \, f_0 \,
\frac{\mu_A^{2N+1}}{\rho^2+\mu_A^2}
\Biggl( \prod_{k=1}^{2N} \mu_k^{2N+1} \Biggr)
\Biggl( \prod_{k=1}^N \frac{1}{(\mu_A-\mu_{2k})^2} \Biggr)
\Biggl( \prod_{k=1}^N \frac{1}{(\rho^2+\mu_A\mu_{2k-1})^2} \Biggr) \\
&\quad\times \Biggl( \prod_{k=1}^{2N} \frac{1}{\rho^2+\mu_k^2} \Biggr)
\Biggl( \prod_{k=1,l=1,3,\cdots}^{2N-1} \frac{1}{(\mu_k-\mu_{k+l})^2} \Biggr)
\Biggl( \prod_{k=1,l=2,4,\cdots}^{2N-2} \frac{1}{(\rho^2+\mu_k\mu_{k+l})^2} \Biggr) \\
&\quad\times \exp\Biggl[ 2 \sum_{k=1}^{2N} (-1)^{k+1} F(\rho,z,\mu_k) \Biggr] \,.
\end{split}
\end{align}
\end{subequations}
Metric~\eqref{n-acc} is, by construction, a solution of the vacuum Einstein equations,
and it represents a collection of $N$ accelerating black holes, aligned along the $z$-axis, and immersed in the external gravitational field~\eqref{seed}.
Actually, as it usually happens for the C-metrics, the result~\eqref{n-acc} can be interpreted as $N$ \emph{pairs} of black holes which accelerate in two opposite directions~\cite{Griffiths:2006tk}.
However, since the black holes in each pair are causally disconnected, being on two opposite sides of the acceleration horizon and unable to communicate with each other, we restrict our attention to one of the two sides only, and we focus mostly  on the genuine $N$-black hole solution.

We consider real poles $w_k$, since they give the physically relevant situation.
These constants are chosen with ordering
$w_1<w_2<\cdots<w_{2N-1}<w_{2N}<w_A$
and with parametrisation\footnote{Parametrisation~\eqref{parametrisation} slightly differs from the standard one presented in~\cite{Hong:2003gx}, however it is coherent with \cite{Griffiths:2006tk}. The main advantage of~\eqref{parametrisation} is its close resemblance with the one used for the non-accelerating case~\cite{Astorino:2021boj} and it is more suitable for non-accelerating limits.}
\begin{equation}
\label{parametrisation}
w_1 = z_1 - m_1 \,, \;
w_2 = z_1 + m_1 \,, \; \dotsc \;
w_{2N-1} = z_N - m_N \,, \;
w_N = z_N + m_N \,, \;
w_A = \frac{1}{2A} \,.
\end{equation}
The constants $m_k$ represent the black hole mass parameters, $z_k$ are the black hole positions on the $z$-axis and $A$ is the acceleration.

\begin{figure}
\includegraphics[scale=0.9]{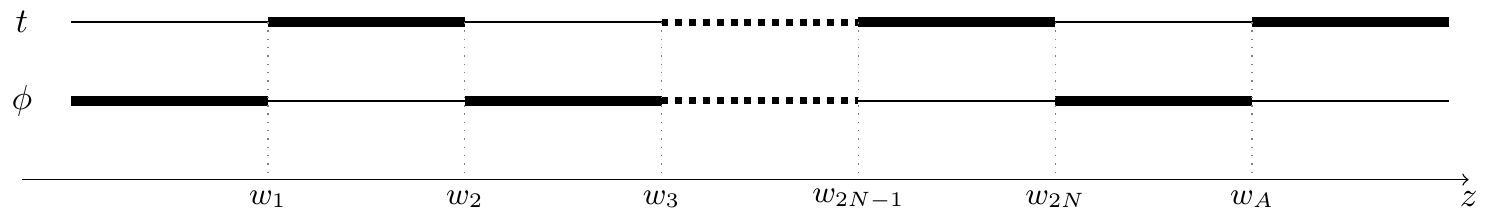}
\caption{{\small Rod diagram for the multi-black hole spacetime~\eqref{n-acc}.
The horizons correspond to the timelike rods (thick lines of the $t$ coordinate), while the conical singularities correspond to ``bolts'' where conical singularities can be avoided by imposing an appropriate periodicity on the angular coordinate.}}
\label{fig:rods}
\end{figure}

The black hole horizons correspond to the regions
$w_{2k-1}<z<w_{2k}$ ($k=1,\dotsc,N$), while the complementary regions are, in principle, affected by the presence of conical singularities, as it happens for the Dowker--Thambyahpillai metric (cf.~Fig.~\ref{fig:rods}).
The metric~\eqref{n-acc} constitutes an extension of the multi-black hole solution presented in~\cite{Astorino:2021boj}, because here we incorporate an additional acceleration horizon which corresponds to the region $z>w_A$ of the spacetime.

\subsubsection{Behaviour at infinity and acceleration horizon}

The multi-black hole solution~\eqref{n-acc} comes with the black holes curvature singularities which are covered, as usual, by the event horizons.
These singularities do not represent a problem, being the usual ones which are encountered in black hole spacetimes.

However, the external gravitational field metric~\eqref{ext} (and then~\eqref{n-acc}) may be characterised by an unbounded growth of curvature invariants at spatial infinity, which corresponds to $\sqrt{\rho^2+z^2}\to\infty$ in Weyl coordinates.
See~\cite{Abdolrahimi:2015gea} for a detailed study of possible curvature singularities in the domain of outer communication, for the distorted Kerr black hole.
This feature is due to the presence of the sources that generate the external gravitational field, and the latter are thought to be located at large distances from the horizon.
In this respect, as already remarked in~\cite{Astorino:2021dju} and~\cite{Astorino:2021boj}, this kind of metrics have to be considered \emph{local}, in the sense that the global solution would correspond to the matching between the black hole spacetime and an energy-momentum tensor which generates the external field~\cite{deCastro:2011zz}.
Hence the behaviour at infinity does not invalidate the physics in proximity of the black holes:
in fact the Smarr law, as well as the first law of thermodynamics, can be achieved for such systems.
Moreover, since these systems are regular in the neighbourhood the horizons, one can even study the second law of thermodynamics in a non-trivial setting~\cite{Astorino:2021dju}.

The novelty of the spacetime~\eqref{n-acc} presented here, is that the curvature unboundedness at infinity is covered by the acceleration horizon given by $z>w_A$.
Thus, the local interpretation of the spacetime is improved in our setting, since the metric is mostly meaningful, between the event and the acceleration horizon.
Being the singularity not directly accessible, the spacetime is completely regular in the physical regions, where observers enjoy the usual metric signature for the manifold and the local model for the distorted multi-black hole system is supposed to hold.

The above discussion is meaningful, and the the acceleration horizon covers the singularity at infinity, when spatial infinity can not be accessed by an observer that moves along a time- or light-like curve from the Lorentzian region.
In fact, that observer can not reach the acceleration horizon:
this can be understood by a simple computation, that we will exploit for the standard C-metric\footnote{The presence of the external field does not affect the argument, since it appears only with exponential terms, that do not modify the sign of the components of the metric.}.

Let us consider the standard C-metric
\begin{equation}
{ds}^2 = \frac{1}{\Omega^2} \biggl( -Q {dt}^2 + \frac{{dr}^2}{Q} + \frac{r^2}{P} {d\theta}^2 + P r^2 \sin^2\theta {d\phi}^2 \biggr) \,,
\end{equation}
where
\begin{equation}
Q = \biggl( 1 - \frac{2m}{r} \biggr) (1 - A^2 r^2) \,, \quad
P = 1 + 2 A m \cos\theta \,, \quad
\Omega = 1 + A r \cos\theta \,.
\end{equation}
In ingoing Eddington--Finkelstein coordinate
$dv = dt + dr/Q$, the metric becomes
\begin{equation}
{ds}^2 = \frac{1}{\Omega^2} \biggl( -Q {dv}^2 + 2 dv dr + \frac{r^2}{P} {d\theta}^2 + P r^2 \sin^2\theta {d\phi}^2 \biggr) \,,
\end{equation}
or
\begin{equation}
\label{acchorizon}
2 dv dr = \Omega^2 {ds}^2 + Q {dv}^2 - {d\sigma}^2 \,,
\end{equation}
where ${d\sigma}^2 = \frac{r^2}{P} {d\theta}^2 + P r^2 \sin^2\theta {d\phi}^2$.

The observer beyond the acceleration horizon ($r>1/A$) is characterised by ${ds}^2\leq0$ and $dv>0$ (future-directed worldline), so being $Q<0$ for $r>1/A$ and ${d\sigma}^2>0$ for the deformed $S^2$ line element, we find that it must be $dr<0$ for the equality~\eqref{acchorizon} to hold.
This means that the observer can not cross the acceleration horizon, since the radial coordinate is forced to decrease for $r>1/A$.

\subsubsection{Regularisation}
\label{sec:regularisation}

The spacetime exhibits conical singularities when the ratio between the length and the radius of small circles around the $z$-axis is different from $2\pi$.
A small circle around the $z$-axis has radius $R=\sqrt{g_{zz}}\rho$ and length $L=2\pi\sqrt{g_{\phi\phi}}$ in Weyl coordinates~\cite{Astorino:2021dju}.
Thus, the regularity condition is nothing but
$L/(2\pi R)\to 1$ as $\rho\to 0$.
It is easy to prove that, for a static and axisymmetric metric, such a condition is equivalent to $\mathcal{P}\equiv f g_{tt}\to 1$ as $\rho\to 0$.
In the case of our multi-black hole metric~\eqref{n-acc}, we can remove the angular defects by choosing the gauge constant $C_f$, and by tuning the external field parameters.

The constant $C_f$ is chosen as
\begin{equation}
\begin{split}
C_f & = 2^{4N} \Biggl[ \prod_{i=1}^N (w_{2i}-w_{2i-1})^2 \Biggr]
\Biggl[ \prod_{k=1}^{N-1} \prod_{j=1}^{N-k}
(w_{2k-1} - w_{2k+2j})^2 (w_{2k} - w_{2k+2j-1})^2 \Biggr] \\
&\quad\times \Biggl[ \prod_{l=1}^N (w_A-w_{2l-1})^2 \Biggr]
\exp\Biggl( -2\sum_{n=1}^\infty b_n w_A^n \Biggl)
\,.
\end{split}
\end{equation}
The quantity $\mathcal{P}=f g_{tt}$ is equal to
\begin{equation}
\label{pk}
\begin{split}
\mathcal{P}_k & =
\Biggl[ \prod_{i=k}^{N-1} \frac{(w_A-w_{2i+1})^2}{(w_A-w_{2i+2})^2} \Biggr]
\Biggl[ \prod_{i=1}^{2k} \prod_{j=2k+1}^{2N}
(w_j-w_i)^{2\,(-1)^{i+j+1}} \Biggr] \\
&\quad \times\exp\Biggl[ 4\sum_{n=1}^\infty b_n \sum_{j=2k+1}^{2N} (-1)^{j+1} w_j^n \Biggl] \,,
\end{split}
\end{equation}
between the $k$-th and $(k+1)$-th black holes (i.e.~$w_{2k}<z<w_{2k+1}$), for $1 \leq k < N$.
In the region $z<w_1$ we find
\begin{equation}
\label{p0}
\mathcal{P}_0 =
\Biggl[ \prod_{i=1}^N \frac{(w_A-w_{2i-1})^2}{(w_A-w_{2i})^2} \Biggr]
\exp\Biggl[ 4\sum_{n=1}^\infty b_n \sum_{j=1}^{2N} (-1)^{j+1} w_j^n \Biggl] \,,
\end{equation}
while for $w_{2N}<z<w_A$ we simply have
\begin{equation}
\mathcal{P}_N = 1 \,,
\end{equation}
thanks to our choice of $C_f$.

The expressions~\eqref{pk},~\eqref{p0} provide a system of equations $\mathcal{P}_k=1$, which can be solved for $b_1,\dotsc,b_N$ to completely regularise the spacetime.

\subsubsection{Smarr law}
\label{sec:smarr}

Let us derive the thermodynamic parameters which appear in the Smarr law.
Firstly, we compute the mass of the spacetime by means of the Komar--Tomimatsu integral~\cite{Komar,Tomimatsu:1984pw}.
The result for the $k$-th black hole (i.e.~the black hole in the interval $w_{2k-1}<z<w_{2k}$) is
\begin{equation}
\label{n-acc-mass}
M_k =
\alpha \int_{w_{2k-1}}^{w_{2k}} \rho g_{tt}^{-1} \partial_\rho g_{tt}
= \frac{\alpha}{2} (w_{2k}-w_{2k-1})
= \alpha m_k \, ,
\end{equation}
where $\alpha$ is a constant which takes into account the proper normalisation of the timelike Killing vector, generator of the horizon, $\xi=\alpha\partial_t$. In general $\alpha$ is not unitary for non-asymptotically flat spacetimes.

The black hole entropy is related to the area as $S_k=\mathcal{A}_k/4$, hence
\begin{equation}
\label{n-acc-entropy}
S_k =
\frac{1}{4} \lim_{\rho\to0} \int_0^{2\pi} d\phi \int_{w_{2k-1}}^{w_{2k}} dz \sqrt{f g_{\phi\phi}}
= \pi m_k  W
\exp\Biggl[ 2\sum_{n=1}^\infty b_n \sum_{j=2k}^{2N} (-1)^{j+1} w_j^n \Biggl] \,,
\end{equation}
where
\begin{equation}
\begin{split}
\log W & = \lim_{\rho\to0} \log\sqrt{f g_{\phi\phi}} = \\
&\quad \log 2
+ \sum_{i=1}^{2k-1} \sum_{j=2k}^{2N} (-1)^{i+j+1} \log|w_j-w_i|
+ \sum_{i=2k}^{2N} (-1)^{i+1} \log|w_A-w_i| \,.
\end{split}
\end{equation}
The temperature is found via the Wick-rotated metric, and the result is
\begin{equation}
\label{n-acc-temp}
T_k = \frac{\alpha}{2\pi} \lim_{\rho\to0} \rho^{-1} \sqrt{\frac{g_{tt}}{f}} =
\frac{\alpha}{2\pi} \lim_{\rho\to0} \frac{1}{\sqrt{f g_{\phi\phi}}} =
\frac{\alpha m_k}{2 S_k} \, .
\end{equation}
It is easy to show, by using~\eqref{n-acc-mass},~\eqref{n-acc-entropy} and~\eqref{n-acc-temp}, that the Smarr law is satisfied:
\begin{equation}
\label{n-acc-smarr}
\sum_{k=1}^N M_k = 2 \sum_{k=1}^N T_k S_k \,.
\end{equation}
The thermodynamics quantities just computed, can be compared with the standard ones in the absence of the external gravitational field~\cite{Gregory:2020mmi}.

\subsection{Accelerating Schwarzschild black hole}
\label{sec:c-metric}

\begin{figure}
\includegraphics[scale=0.9]{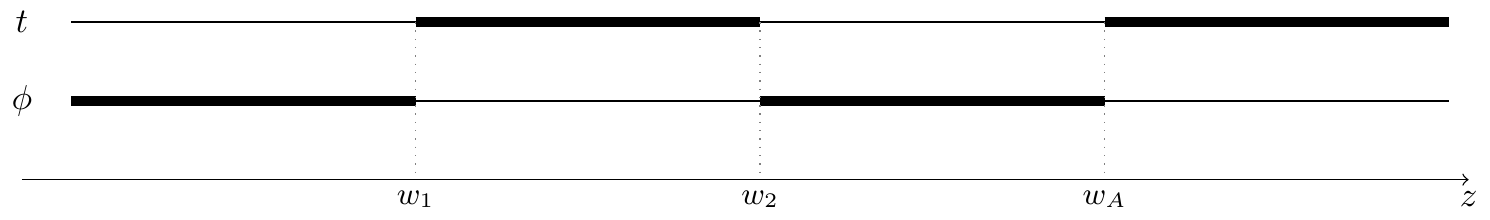}
\caption{{\small Rod diagram for the C-metric embedded in the external field~\eqref{1-acc}.
We notice that the presence of the external field does not affect the rod diagram, i.e.~the structure of the poles.
Being the rods defined by the poles $w_n$ (see~\cite{Harmark:2004rm}), we obtain the usual C-metric diagram.}}
\label{fig:c-metric}
\end{figure}

The first specialization of the multi-source metric proposed above that is worth discussing, is the single black hole case, i.e.~$N=1$.
The very first prototype of these kind of metrics were built by Ernst~\cite{ErnstGeneralized} with the aim of regularising the conical singularity of the C-metric, through the presence of an external gravitational field possessing only the first term of the external multipolar expansion, the dipole\footnote{The zeroth-order of the external multipolar expansion is just a constant that can be reabsorbed.
While the dipole term in the standard internal multipolar expansion is often, under certain assumptions, washed away thanks to a coordinate shift to the center of mass, in this external multipolar expansion this change of reference can not erase the dipole contribution.}.
The $N=1$ characterization of the metric~\eqref{n-acc} represents the full multipolar expansion of the external gravitational field with respect to the Ernst solution.
Moreover, our metric carries an extra parameter $z_1$ which describes the position of the black hole with respect to the multipoles\footnote{This differs with respect to the Ernst metric, which considered a fixed position for the black hole in the center of the coordinate system, that is for $-m/A<z<m/A$.}.
The metric is quite simple and can be written from~\eqref{n-acc} with $N=1$
\begin{subequations}
\label{1-acc}
\begin{align}
g_\text{1} & = \diag\Biggl[ - \frac{\mu_1  \mu_A}{\mu_2}  \exp\biggr(2\sum_{n=1}^{2} b_n r^n P_n \biggr) ,
\rho^2 \frac{\mu_2}{\mu_1 \mu_A} \exp\biggl(-2\sum_{n=1}^{2} b_n r^n P_n \biggr) \Biggr] \, , \\
f_\text{1} & = -\frac{16 C_f \, f_0 \, \mu_1^3 \mu_2^3 \,
e^{2[F(\mu_1) - F(\mu_2)]}}{(\mu_A-\mu_1)^2(\mu_A-\mu_2)^2(\mu_1-\mu_2)^2(\rho^2 + \mu_1^2)(\rho^2 + \mu_2^2)} \,,
\end{align}
\end{subequations}
where the sum in $F$~\eqref{F} is limited to the second term.
The rod diagram remains the same of the standard C-metric, since the poles are not affected by the presence of the external gravitational field, as can be appreciated in Fig.~\ref{fig:c-metric}.

The limit to the standard C-metric in spherical coordinates $(t,r,\theta,\varphi)$
\begin{equation}
\label{c-metric}
\begin{split}
{ds}^2 & = \frac{1}{(1 + A r \cos \theta)^2} \biggl[ -\biggl(1-\frac{2m}{r} \biggr) \bigl(1-A^2r^2\bigr) {dt}^2 + \frac{dr^2}{\bigl(1-\frac{2m}{r} \bigr)\bigl(1-A^2r^2 \bigr)} \\
&\quad + \frac{r^2 {d\theta^2}}{(1+2Am\cos\theta)} + r^2 (1+2Am\cos\theta) \sin^2 \theta {d\phi}^2 \biggr] \,,
\end{split}
\end{equation}
is obtained via the rescaling $t\to \sqrt{A}\, t$, $\phi \to \phi/\sqrt{A}$ and the following change of coordinates
\begin{align}
\rho & = \frac{\sqrt{r(r-2m)(1-A^2r^2)(1+2Am\cos\theta)}}{(1+Ar\cos\theta)^2} \sin\theta \,, \\
z & = z_1 + \frac{(Ar+\cos\theta)(r-m+Amr\cos\theta)}{(1+Ar\cos\theta)^2} \,,
\end{align}
with $z_1=0$ and $b_n=0$.
In order to obtain exactly Eq.~\eqref{c-metric}, one has to fix the poles in terms of the physical quantities
\begin{equation}
\label{acc-wi}
w_1 =  z_1 - m \, , \quad
w_2 =  z_1 + m \, , \quad
w_A = \frac{1}{2A} \, , \quad
C_f = \frac{1}{8A^3} \, .
\end{equation}
These coordinates and parametrisations are also useful for describing the accelerating and distorted black hole metric.

However, in order to remove the two conical singularities, which are generally present on the $z$-axis of the single accelerating black hole above, some constants have to be properly constrained, as explained in Sec.~\ref{sec:regularisation}. 
For simplicity, in this section we will focus on the first two terms of the multipolar expansion, thus we take $b_n=0$, $\forall n > 2$.
In this case we can explicitly write down the values of the physical parameters which regularise the metric for $z \in (z_1+m,1/2A)$ and $z\in (-\infty,z_1-m)$, respectively:
\begin{align}
\label{reg-acc-Cf}
C_f & = -\frac{(w_1-w_2)^2(w_1-w_A)^2}{8w_1^2} e^{-2w_A(b_1+b_2w_A)}
= -\frac{m + 2 A m (m-z_1)^2}{8A^2 (m-z_1)^2}
e^{-\frac{2 A b_1+b_2}{2A^2}} \,, \\
\label{reg-acc-b1}
b_1 & = \frac{2 b_2 (w_2^2-w_1^2) + \log(\frac{w_2-w_A}{w_1-w_A})}{2(w_1-w_2)} = -2 b_2 z_1 - \frac{1}{4m} \log \biggl[\frac{1-2A(m+z_1)}{1+2A(m-z_1)} \biggr] \,.
\end{align}
Note that $w_i$ have the same definition as in~\eqref{acc-wi}, while in the presence of the external gravitational field $C_f$ can be upgraded to remove the conical singularity,  as done in~\eqref{reg-acc-Cf}.
We remark that the external multipolar distortion is usually designed to model local black holes.
Nevertheless, thanks to the above regularisation, the spacetime remains completely regular in the physical regions.
In fact, between the event horizon and the accelerating horizon, where the signature is $(-+++)$, the metric is free from both conical and curvature singularities, for any finite value of the accelerating parameter $A$ satisfying the rod ordering relation $z_1-m<\frac{1}{2A}$.

In order to have a better intuition of the physics introduced by the external field, it is instructive to analyse the weak field limit of the accelerating metric~\eqref{1-acc}, that is when the black hole mass parameter $m$ is small.
In this case we can appreciate the contribution of the external multipoles on the acceleration given to an inertial observer, which we consider located in the origin of the (spherical) coordinates for simplicity. 
The four-dimensional timelike worldline of an observer with proper time $\lambda$ and constant radial $\bar{r}$ and polar coordinates $\bar{x}=\cos\bar{\theta}$ is 
\begin{equation}
y^\mu(\lambda) =
\begin{pmatrix}
\frac{1+A\bar{r}\bar{x}}{\sqrt{1-A^2\bar{r}^2}}
e^{-\frac{\bar{r}}{2(1+A\bar{r}\bar{x})} \bigl[ 2b_2\bar{r} (A \bar{r}+\bar{x})^2 +2 b_1 (A \bar{r}+\bar{x}) (1+ A \bar{r}\bar{x})^2-b_2 \bar{r} (1-A^2\bar{r}^2)(1-\bar{x}^2) \bigr]} \lambda \\
\bar{r} \\
0 \\
0 \\
\end{pmatrix}
\,.
\end{equation}
This choice fulfils the normalisation property of the four-velocity, $u_\mu u^\mu=-1$, where $u^\mu \coloneqq dy^\mu(\lambda)/d\lambda$.
The absolute value of the four-acceleration, $a^\mu \coloneqq (\nabla_\nu u^\mu)u^\nu$, for this observer is given by
\begin{equation}
\label{acc-bi}
|a| = \sqrt{a_\mu a^\mu}\  \Big|_{\bar{r}=0} =
| A -b_1 | \exp\biggl(-\frac{b_2+2Ab_1}{4A^4} \biggr) \,.
\end{equation}
Note that, because $a^\mu u_\mu=0$, $|a|$ corresponds also to the magnitude of the three-acceleration in the rest frame of the observer, the external gravitational field has a non-trivial role in the background acceleration quantified by~\eqref{acc-bi}.
In the vanishing multipoles limit, $b_n=0$, the standard C-metric acceleration $|a|=A$ is retrieved from~\eqref{acc-bi}.

The non-relativistic limit, i.e.~small values of the accelerating parameter $A \approx 0$, of the regularising condition~\eqref{reg-acc-b1}, can provide some further understanding of the multipolar deformations.
In fact in this approximation we expect to retrieve a Newtonian picture:
the force felt by a massive monopole in an uniform gravitational field is
\begin{equation}
\label{approx}
m A \approx \frac{1- e^{-4 m (b_1+2b_2 z_1)}}{4} \,.
\end{equation}
One obtains a very simple expression when the external field is weak, i.e.~$b_1\approx b_2\approx 0$:
in such a case the exponential in~\eqref{approx} can be expanded and
\begin{equation}
m A \approx m (b_1 + 2 b_2 z_1) \,.
\end{equation}
The last equation is nothing but the Newton law $\vec{F}=m\vec{a}$, hence $b_1 + 2 b_2 z_1$ is interpreted, in the Newtonian limit, as a constant external gravitational field strength.
We see that the regularisation condition has a nice and physically transparent limit, which is consistent with the analysis performed by Bonnor~\cite{Bonnor1988} on the Ernst solution.

The metric described in this section has interesting applications in the realm of black hole pair creation.
Usually, it is speculated that in the presence of a strong electromagnetic field the C-metric can describe a couple of casually disconnected and charged black holes (eventually rotating ~\cite{Astorino:2013xxa}) popping-out from vacuum and accelerating away~\cite{Gibbons:1986cq,Garfinkle:1993xk,Hawking:1994ii}.
In that picture the regularising interaction between the electric charge of the black hole and the background electromagnetic Bonnor--Melvin universe~\cite{Bonnor1954,Melvin:1963qx} reveals to be crucial.
On the other hand, the metric presented in this section provides the regularisation by means of the external gravitational field.
Therefore, in this setting the pair creation of black holes is fostered by the energy of the external gravitational field.
Thus this picture seems to be more phenomenological, because it does not require the black holes to be constitute by charged matter, an occurrence that appears outside empirical observations, at the moment.
In fact the metric~\eqref{1-acc} allows the pair creation of neutral black holes in a gravitational background.
The rate of the pair creation is proportional to the intensity of the external gravitational field, but
its computation is outside the scope of this paper and will be done elsewhere. 

The thermodynamics quantities and the Smarr law follow directly from the general multiple case presented in the Sec.~\ref{sec:smarr}.
Otherwise, the first law of black hole thermodynamics can be retrieved as a trivial specialisation of the double configuration studied in the next section, when one of the two masses of the double configuration vanishes.

\subsection{Accelerating binary system}
\label{sec:double-c-metric}

The accelerating external gravitational background provides us a second interesting opportunity to generalise the solution describing a binary black hole system at equilibrium, as presented in~\cite{Astorino:2021dju}.
This extension not only represents an enrichment of the physical model features, but it provides also a mechanism to protect the physical region of the black hole against the unbounded growth of the scalar curvature invariants at spatial infinity.
In fact, for finite values of the radial coordinate $r$, an observer will encounter the accelerating Killing horizon and conformal infinity before reaching spatial infinity, further enforcing the local nature of the model.

As explained in Sec.~\ref{sec:array}, the metric can be analytically generated for the whole multipolar series, which includes infinite independent terms, each one with its independent integration constant.
However, in this Section we work out explicitly a simple example on a truncated multipolar expansion, namely keeping only the dipole and quadrupole deformations.
Indeed these two quantities are sufficient to regularise the metric without constraining the proper physical parameters of the black holes configuration\footnote{$F$ in $f_2$ is defined according to~\eqref{F}, but only up to the second order.}.
That is because the number of constraints on the physical parameters of the metric coincides with the number of causally connected black holes, and we are now considering a double accelerating C-metric.
In this case the two blocks of the spacetime metric can be written as follows:
\begin{subequations}
\label{2-acc}
\begin{align}
g_2 & = \diag\Biggl[ - \frac{\mu_1 \mu_3 \mu_A}{\mu_2 \mu_4}  \exp\biggr(2\sum_{n=1}^{2} b_n r^n P_n \biggr) ,
\frac{\rho^2 \mu_2 \mu_4}{\mu_1 \mu_3 \mu_A} \exp\biggl(-2\sum_{n=1}^{2} b_n r^n P_n \biggr) \Biggr] \,, \\
f_2 & = \frac{f_1 \, \mu_1^2 \mu_2^2 \mu_3^5 \mu_4^5
e^{2[F(\mu_3) - F(\mu_4)]}}{Y_{23}^2Y_{A3}^2Y_{14}^2Y_{34}^2Y_{A4}^2 W_{13}W_{24}W_{33}W_{44}} \,,
\end{align}
\end{subequations}
where $f_1$ is the single-black hole value for the non-Killing elements of the metric encountered in previous Section.
We have to properly tune three parameters\footnote{Two of these parameters, $b_1$ and $b_2$, are related to physical quantities, while $C_f$ is a gauge constant of the metric.} of the solution to obtain a metric devoid of angular defects, one for each sector of the $z$-axis, between the timelike rods of Fig.~\ref{fig:rods}.
A possible choice in terms of the poles $w_i$ is
\begin{align}
\label{Cf-2}
\begin{split}
C_f & = \frac{8 e^{-2w_A(b_1+b2w_A)}}{w_1^2 w_3^2} \\
&\quad \times (w_1-w_2)^2(w_2-w_3)^2(w_1-w_4)^2(w_3-w_4)^2(w_1-w_A)^2(w_3-w_A)^2 \,,
\end{split}
\\
\label{b1-2}
b_1  &= \frac{4b_2(w_2^2-w_1^2+w_4^2-w_3) + 2\log\Bigl[\frac{(w_A-w_2)(w_A-w_4)}{(w_A-w_1)(w_A-w_3)} \Bigr]}{4(w_1-w_2+w_3-w_4)} \,, \\
\label{b2-2}
\begin{split}
b_2 & = \frac{2(w_1-w_2+w_3-w_4)\log\Bigl[\frac{(w_1-w_3)(w_2-w_4)(w_4-w_A)}{(w_2-w_3)(w_1-w_4)(w_3-w_A)} \Bigr]}{4(w_1-w_2)(w_3-w_4)(-w_1-w_2+w_3+w_4)} \\
&\quad + \frac{(w_4-w_3) \log \Bigl[\frac{(w_1-w_3)(w_2-w_4)(w_4-w_A)}{(w_2-w_3)(w_1-w_4)(w_3-w_A)}  \Bigr]}{4(w_1-w_2)(w_3-w_4)(-w_1-w_2+w_3+w_4)} \,.
\end{split}
\end{align}
Physically, this choice can be interpreted as the specific deformation of the external gravitational field multipoles required to support a generic binary configuration.
Other possible choices may have a different physical interpretation:
for instance, fixing the position of the holes through $z_i$ instead of $b_n$ to remove the conical singularities, corresponds in adjusting the mutual position of the black holes in a given multipolar configuration. 
Henceforth we will considered the value of the three parameters $b_1$, $b_2$ and $C_f$ constrained as in~\eqref{Cf-2},~\eqref{b1-2},~\eqref{b2-2} to ensure the spacetime to be free from angular defects anywhere.

The equilibrium is achieved for finite proper distance of the two black hole sources, as it can be checked from
\begin{equation}
\ell = \int_{w_2}^{w_3} dz \sqrt{g_{zz}(\rho,z)} \Big|_{\rho=0} < \infty \,.
\end{equation}

\subsubsection{Thermodynamics}

The mass of each member of the double configuration can be evaluated by integrating on their respective rod, as done for the general case~\eqref{n-mass}:
$M_i = \alpha m_i $, thus the total mass is given by $M = M_1 + M_2$.
The total entropy of the system is taken as the quarter of the two black hole surfaces, as described in Sec.~\ref{sec:smarr}:
$S = S_1 + S_2$,
where
\begin{subequations}
\begin{align}
\begin{split}
S_1 & = \frac{\pi (w_1-w_2)^2(w_1-w_4)(w_3-w_A)}{2 (w_1-w_3)(w_2-w_A)(w_5-w_A)} e^{-2b_1(w_2-w_3+w_4)-2b_2(w_2^2-w_3^2+w_4^2)} \\
& = \frac{4\pi A m_1^2
(m_1+m_2-z_1+z_2)
[1+2A(m_2-z_2)]}{(m_1-m_2-z_1+z_2)[-1+2A(m_1+z_1)][-1+2A(m_2+z_2)]} \\
&\quad \times e^{-2b_1(m_1+2m_2+z_1)-2b_2(m_1^2+2m_1z_1+z_1^2+4m_2z_2)} \,,
\end{split}
\\
\begin{split}
S_2 & = \frac{\pi (w_3-w_4)^2(w_4-w_1)}{2(w_2-w_4)(w_4-w_A)}  e^{-2w_4(b_1+b_2w_4)} \\ 
& = \frac{4\pi Am_2^2
(m_1+m_2-z_1+z_2)
e^{-2(m_2+z_2)(b_1+b_2(m_2+z_2))}}{(m_1-m_2+z_1-z_2)[-1+2A(m_2+z_2)]} \,.
\end{split}
\end{align}
\end{subequations}
The two horizon temperatures, computed as in~\eqref{n-temp}, simply result $T_i =M_i/(2S_i)$, fulfilling straightforwardly the Smarr law both for each single black source and for the binary configuration.

The first law of black hole thermodynamics can also be verified, once the normalisation of the timelike Killing vector is chosen to make the mass of the system integrable and continuously connected with the known cases, i.e.~the non-accelerating configuration and the single black hole configuration.
The Christodoulou--Ruffini mass formula~\cite{Christodoulou:1972kt} suggests the use of an integrating factor $\alpha$ such that 
\begin{equation}
\sum_{i=1}^2 M_i = \sum_{i=1}^2 \sqrt{\frac{S_i}{4\pi}} \,.
\end{equation}
It can be verified that this occurs whenever 
\begin{equation}
w_A = \frac{w_2(w_3-w_2) e^{2b_1(w_2-w_3+w_4)+2b_2(w_2^2-w_3^2+w_4^2)} + w_3(w_2-w_4) e^{2w_4(b_1+b_2w_4)}}{(w_3-w_2) e^{2b_1(w_2-w_3+w_4)+2b_2(w_2^2-w_3^2+w_4^2)} + (w_2-w_4) e^{2w_4(b_1+b_2w_4)}} \,.
\end{equation}
The resulting value of the integrating factor is
\begin{equation}
\alpha = \sqrt{\frac{w_4-w_2}{2(w_4-w_2)(w_5-w_4)}} e^{-w_4(b_1+b_2w_4)} \,.
\end{equation}
Then the first law of black hole thermodynamics holds for each member of the black hole configuration as follows
\begin{equation}
\delta M_i =  T_i \, \delta S_i \,.
\end{equation}
In the presence of thermodynamic equilibrium between the two horizons, $T_1=T_2$, which can be achieved constraining another integrating constant of the solution (e.g.~$w_4$), a first law for the whole black hole configuration can be written
\begin{equation}
\delta M = T \, \delta S \,.
\end{equation}

\subsection{Accelerating particles}
\label{sec:particles}

There exist many particle-like solutions in General Relativity that have been extensively studied over the years.
These solutions are related to the Curzon--Chazy family of metrics~\cite{Curzon,Chazy}, and they represent the gravitational field generated by point-like particles.
The particles themselves are nothing but naked singularities\footnote{The structure of these curvature singularities is quite complicated and depends on the direction one approaches them. See~\cite{Griffiths:2006tk} and references therein.}.
It is quite easy to construct a metric that contains a collection of many Curzon--Chazy particles:
these multi-particle solutions are affected by the presence of conical singularities~\cite{Einstein:1936fp}, as expected on physical grounds.

The accelerated version of two Curzon--Chazy particles was found by Bonnor and Swaminarayan~\cite{Bonnor1964}:
in such a solution the particles are accelerated by two cosmic strings\footnote{Actually, the conical singularities disappear when one of the two particles has negative mass.}.
The metric, in the case of a single accelerating particle, was later regularised, following the lines of~\cite{ErnstGeneralized}, by Bi\v c\'ak, Hoenselaers and Schmidt~\cite{Bicak1983} with the introduction of an external field, by which they were able to remove the conical singularities.
Moreover, they showed that their external field, that actually corresponds to our multipolar expansion~\eqref{ext} when $b_n=0$ for $n>1$, can be obtained by sending to infinity the second particle of the Bonnor--Swaminarayan solution.
This corroborates the idea that the field background is generated by sources located at infinity.

It is worth mentioning an alternative approach, pursued by Gibbons~\cite{Gibbons:1974zd}, who managed to regularise an accelerating black hole by means of a negative-mass Curzon--Chazy particle.

We will show that the generalisation of the Bi\v c\'ak--Hoenselaers--Schmidt solution to $N$ particles and with the generic gravitational field~\eqref{acc} can be achieved by means of an appropriate limit of our multi-black hole solution~\eqref{n-acc}.

\subsubsection{The limit to the Bonnor--Swaminarayan metric}

Let us begin by considering the limit to the Bonnor--Swaminarayan solution, i.e.~two accelerating particles with no external gravitational field.
We consider then the metric~\eqref{n-acc} with $b_n=0$ for all $n$.
To clarify the limit, we specialise to the case $N=2$, nonetheless the generalisation to any $N$ is straightforward.

It is useful to rewrite the metric~\eqref{n-acc} in the canonical Weyl form~\eqref{weyl}
\begin{equation}
\label{weyl-acc}
{ds}^2 = -e^{2\psi(\rho,z)} {dt}^2 + e^{-2\psi(\rho,z)} \bigl[ e^{2\gamma(\rho,z)} \bigl( {d\rho}^2 + {dz}^2 \bigr) + \rho^2 {d\phi}^2 \bigr] \,,
\end{equation}
and to work with the potential
\begin{equation}
\psi = \frac{1}{2} \log\biggl(\frac{\mu_1}{\mu_2}\biggr) + \frac{1}{2} \log\biggl(\frac{\mu_3}{\mu_4}\biggr) + \frac{1}{2} \log\mu_A \,.
\end{equation}
Noticing that
\begin{equation}
\frac{\mu_k}{\mu_{k+1}} = \frac{\mu_k - \bar{\mu}_{k+1}}{\mu_{k+1} - \bar{\mu}_k} =
\frac{R_k + R_{k+1} - 2m_k}{R_k + R_{k+1} + 2m_k} \,,
\end{equation}
where
\begin{equation}
R_k = \sqrt{\rho^2 + (z + m_k - z_k)^2} \,, \quad
R_{k+1} = \sqrt{\rho^2 + (z - m_k - z_k)^2} \,,
\end{equation}
we can write
\begin{equation}
\psi = \frac{1}{2} \log\biggl(\frac{R_1 + R_2 - 2m_1}{R_1 + R_2 + 2m_1}\biggr) + \frac{1}{2} \log\biggl(\frac{R_3 + R_4 - 2m_2}{R_3 + R_4 + 2m_2}\biggr) + \frac{1}{2} \log\mu_A \,.
\end{equation}
One recognises two Schwarzschild potentials (the first two terms) and the Rindler potential (the last term).
Indeed, at the level of the Weyl potential a superposition principle holds;
the non-linearity is encoded in the function $\gamma$, that we do not explicitly write here.

\begin{figure}
\includegraphics[scale=0.9]{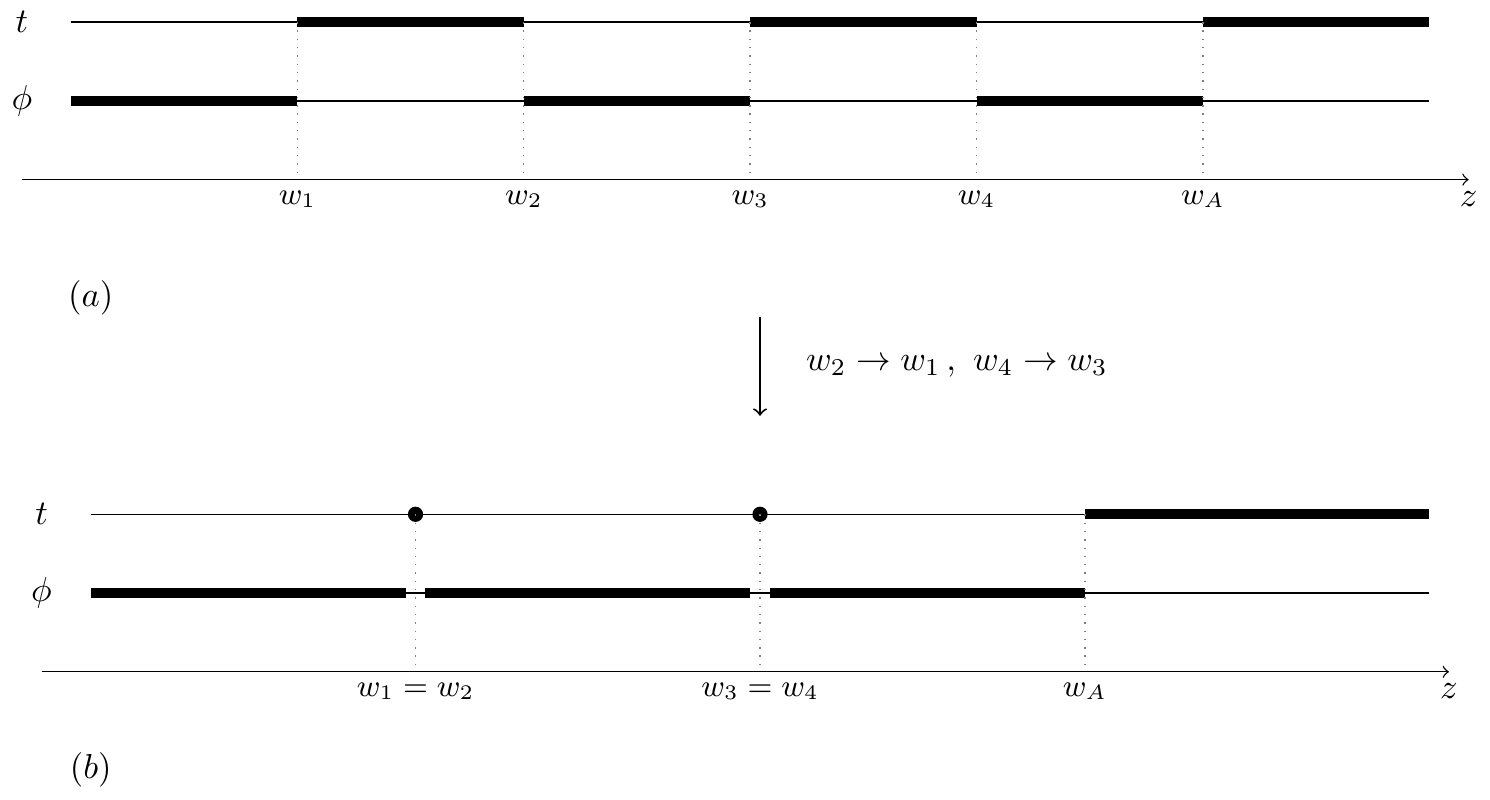}
\caption{{\small Rod diagrams for $(a)$ the double C-metric in an external field and for $(b)$ the Bonnor--Swaminarayan metric~\eqref{bonnor}.
We see that the limit considered in the main text $w_{2k}\to w_{2k-1}$ corresponds to shrinking the timelike rods representing the event horizons.
The horizons disappear through that limit, and the resulting objects are naked singularities which are represented by points in the rod diagram.}}
\label{fig:particles}
\end{figure}

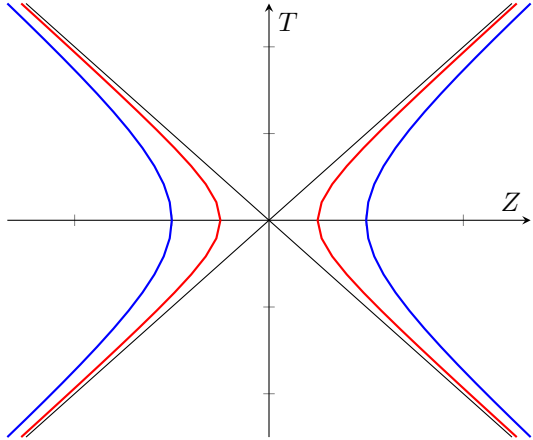
\begin{figure}
\centering
\begin{tikzpicture}

\begin{axis}[
  axis lines=middle,
  clip=false,
  xlabel=$Z$,
  ylabel=$T$,
  xticklabels=\empty,
  yticklabels=\empty,
]

\addplot[black, thin] (x,x);
\addplot[black, thin] (x,-x);

\addplot[blue, thick] (sqrt(x*x+4),x);
\addplot[blue, thick] (-sqrt(x*x+4),x);

\addplot[red, thick] (sqrt(x*x+1),x);
\addplot[red, thick] (-sqrt(x*x+1),x);

\end{axis}

\end{tikzpicture}
\caption{A spacetime diagram of the Bonnor--Swaminarayan metric in the boost-rotation coordinates of~\cite{Bicak:1999sa}, for the section $\rho=0$.
The hyperbolae represent the worldlines of two pairs of (causally disconnected) accelerating particles.}
\label{fig:hyperbolae}
\end{figure}

Now we consider the limit in which the finite timelike rods of Fig.~\ref{fig:rods} are pinched to a point, i.e.~when $w_{2k}\to w_{2k-1}$:
this is equivalent to consider $m_k\to 0$ (see Fig.~\ref{fig:particles}).
We expand for small $m_k$ to order $O\bigl(m_k^2\bigr)$, to find
\begin{equation}
\psi \approx -\frac{m_1}{\sqrt{\rho^2 + (z - z_1)^2}} - \frac{m_2}{\sqrt{\rho^2 + (z - z_2)^2}} + \frac{1}{2} \log\mu_A \,.
\end{equation}
Again, we recognise the various terms in the last expression:
the first two are Curzon--Chazy potentials and represent point-like particles, while the last term is still the Rindler one.
Then it is natural to interpret the potential as the one corresponding to two accelerating particles (see Fig.~\ref{fig:hyperbolae}).

We cast the potential in the usual Bonnor--Swaminarayan form by performing the change of coordinate
$\bar{z} = 1/(2A^2) - z$,
by which
\begin{equation}
\mu_A = \sqrt{\rho^2 + \bar{z}^2} + \bar{z} \,,
\end{equation}
and then defining the new constants
\begin{equation}
2\alpha_1^2 = \frac{2A^2}{2A^2 z_1 - 1} \,, \quad
2\alpha_2^2 = \frac{2A^2}{2A^2 z_2 - 1} \,.
\end{equation}
One finally finds
\begin{equation}
\label{bonnor}
\psi = -\frac{m_1}{\sqrt{\rho^2 + (\bar{z} - \frac{1}{2\alpha_1^2})^2}} - \frac{m_2}{\sqrt{\rho^2 + (\bar{z} - \frac{1}{2\alpha_2^2})^2}} + \frac{1}{2} \log \bigl( \sqrt{\rho^2 + \bar{z}^2} + \bar{z} \bigr) \,,
\end{equation}
which is the Bonnor--Swaminarayan potential (cf.~\cite{Bonnor1964} and~\cite{Griffiths:2009dfa}).
The generalisation to $N$ accelerating particles is easily found as
\begin{equation}
\psi = - \sum_{k=1}^N \frac{m_k}{\sqrt{\rho^2 + (\bar{z} - \frac{1}{2\alpha_k^2})^2}} + \frac{1}{2} \log \bigl( \sqrt{\rho^2 + \bar{z}^2} + \bar{z} \bigr) \,.
\end{equation}
The $\gamma$ function, which completes the Weyl metric~\eqref{weyl-acc}, is found by quadratures.

One can check that the Bonnor--Swaminarayan metric~\eqref{bonnor}, for generic values of the parameters, is affected by conical singularities.
Such singularities can be removed only when $m_2<0$:
in this case the axis is everywhere regular, except at the locations of the point particles.
More explicitly, the regularisation is achieved for~\cite{Griffiths:2009dfa}
\begin{equation}
m_1 = -m_2 = \frac{\bigl(\alpha_1^2 - \alpha_2^2)^2}{4\alpha_1^3\alpha_2^3} \,.
\end{equation}

\subsubsection{The limit to the Bi\v c\'ak--Hoenselaers--Schmidt metric}

The inclusion of the external gravitational field is now a simple matter:
this can be done following the lines of Sec.~\ref{sec:ism}, i.e.~by means if the inverse scattering method, or via the same limiting procedure of the multi-black hole metric~\eqref{n-acc} above, now with the field parameters $b_n$ turned on.
In both cases, the resulting Weyl potential is
\begin{equation}
\label{bhs}
\psi = - \sum_{k=1}^N \frac{m_k}{\sqrt{\rho^2 + (\bar{z} - \frac{1}{2\alpha_k^2})^2}} + \frac{1}{2} \log \bigl( \sqrt{\rho^2 + \bar{z}^2} + \bar{z} \bigr) + \sum_{n=1}^{\infty} b_n r^n P_n \,.
\end{equation}
The meaning of the terms in the potential is clear, and the function $\gamma$ can be found again by quadratures.
This potential specialises to the Bi\v c\'ak--Hoenselaers--Schmidt one~\cite{Bicak1983} for $N=1$ and $b_1\neq0$, $b_n=0$ for $n>1$, that reads
\begin{equation}
\psi = - \frac{m_1}{\sqrt{\rho^2 + (\bar{z} - \frac{1}{2\alpha_1^2})^2}} + \frac{1}{2} \log \bigl( \sqrt{\rho^2 + \bar{z}^2} + \bar{z} \bigr) + b_1 z \,.
\end{equation}
The limiting procedure that leads to~\eqref{bhs} does not affect the external field, hence it can be fine tuned again to support the particles attraction against the gravitational collapse and to remove the conical singularities.

It is worth mentioning that Bi\v c\'ak--Hoenselaers--Schmidt also found two accelerating particles described by internal multipole momenta~\cite{BicakMultipole}, in analogy with the Erez--Rosen metric~\cite{Erez}.
The multipole metric, initially obtained through a coalescing limit of the Bonnor--Swaminarayan solution, can be regularised everywhere on the $z$-axis (except at the two particles), thus obtaining a regular accelerating metric without the need of any external field.
Relying on the discussion presented in~\cite{Astorino:2021boj}, one can easily write down the most general potential for $N$ accelerating particles with arbitrary multipole momenta and immersed in an external gravitational field:
\begin{equation}
\psi = - \sum_{k=1}^N \frac{m_k}{\sqrt{\rho^2 + (\bar{z} - \frac{1}{2\alpha_k^2})^2}} + \frac{1}{2} \log \bigl( \sqrt{\rho^2 + \bar{z}^2} + \bar{z} \bigr) + \sum_{n=1}^{\infty} \biggl( \frac{a_n}{r^{n+1}} + b_n r^n \biggr) P_n \,.
\end{equation}
$a_n$ are the internal momenta, that describe the deformations of the point-like sources.
We do not delve into the details of this solution, because it is beyond our scope.
However, it would be interesting to explicitly write down the function $\gamma$ and to check that the conical singularities can be removed by tuning the parameters $a_n$ and $b_n$.

\section{Outlook}

In this Chapter we constructed, thanks to the inverse scattering method, a large family of new solutions which generalise the Israel--Khan metric~\cite{Israel1964} and the array of accelerating black holes~\cite{Dowker:2001dg}, by the introduction of an external gravitational field.
We were able to treat analytically the whole multipolar expansion of the gravitational background introducing a countable number of integration constants, characterising the gravitational multipoles of the external field, which are useful to remove all the conical singularities typical of the collinear multi-black hole configurations.

The external gravitational field can be interpreted as the centrifugal force field caused by the rotation of the two sources in a vacuum background, thanks to the equivalence principle of General Relativity, in a rotating frame located at the center of mass of a binary system.
While this picture can not model a full black hole merging, it might be useful to describe a metastable stationary phase of the coalescence process, when the gravitational energy radiated by the system is still negligible, and the orbits stay almost regular.

We have computed the physical charges and verified the Smarr law for various configurations.
In some particular cases, like the binary system, we also studied the thermodynamics and inspected some geometrical properties

We found that these spacetimes are relevant not only because they enrich our scarce theoretical knowledge of multi-black hole solutions or because they represent the first multi-black hole solutions which can be regularised without the need of extra fields as the electromagnetic field\footnote{Actually, in the presence of Maxwell electrodynamics, Ernst showed also how to regularise the charged C-metric, thanks to an external electromagnetic field such as the Melvin universe~\cite{ErnstMelvin}.
Even though axial magnetic fields in the center of the galaxies can be of some prominence, charged black holes are not considered plausible objects because matter in the Universe is most often neutral.}, but also because these metrics allow us to discuss some intriguing physical processes.

For example, in the case of the binary system, we were able to discuss not only the first law of thermodynamics, but also the second law:
by mimicking a merging process, i.e.~by comparing the initial and the final state of a merging process, we found that the entropy of the system is destined to increase.
This is a non-trivial result that it is difficult to check for the usual multi-black hole metrics, because of the presence of conical defects.

Further, regularised C-metrics~\footnote{We notice that the single charged C-metric can be regularised without the need of an external field~\cite{Kinnersley:1970zw}.
However, the regularisation can be achieved only in the extremal case $m=e$, which is a problematic limit for the black hole solution.} can describe the pair creation of a couple (or possibly four in case of the double C-metric) of black holes that accelerate in opposite direction remaining causally disconnected.
This process is propelled at expense of the external field, in our case the multipolar gravitational background.
The significance and the novelty of our picture is given by the fact that the accelerating black hole couple can be uncharged, a feature in line with phenomenological observation.

As a by-product of our construction, we showed how to extend the vacuum Pleba\'nski--Demia\'nski class of metrics to the rotating and accelerating multi-Kerr black holes, with or without the presence of the external gravitational field.
We are also able to detect some notable known metrics as limits of our general solution describing accelerating particles with or without the external gravitational background, such as the Bonnor--Swaminarayan and the Bi\v c\'ak--Hoenselaers--Schmidt solutions.

In general we have shown that, as the inverse scattering method predicts, in practice basically any diagonal seed can be used as a background for the solution generating technique.
In particular, this technique reveals to be useful to embed and overlap a generic number of black hole sources, possibly providing a mechanism to regularise the conical singularities that usually afflicts these metrics.
Of course, it would be interesting to explore also different backgrounds.

All these results can be extended to gravitational theories where the solution generating techniques hold, from minimally to conformally coupled scalar fields or other scalar tensor theories such as some classes of Brans--Dicke or $f(R)$ gravity.

\chapter{Black holes in an expanding bubble of nothing}
\label{chap:bubble}
\thispagestyle{plain}

We consider another example of a background which allows us to regularise a multi-black hole spacetime (again, with the aim of the inverse scattering method):
the expanding Kaluza--Klein bubbles, also known as ``bubbles of nothing''.
In the present Chapter we will also consider systems in higher spacetime dimensions, i.e.~$D>3$.

Expanding bubbles of nothing are simple but surprising solutions of gravitational theories with compact dimensions~\cite{Witten:1981gj}.
They provide channels for the non-perturbative decay of Kaluza--Klein vacua, but they are also interesting as simple time-dependent spacetimes that share many features with de Sitter cosmologies~\cite{Aharony:2002cx}.
This latter view, more than the former, will be relevant in this Chapter, where we present a suggestive new way of regarding these bubbles, and investigate their relation to some black hole systems.
Other aspects of the relation between black holes and bubbles of nothing have been studied in~\cite{Emparan:2001wk,Horowitz:2002cx,Elvang:2002br,Elvang:2004iz,Tomizawa:2007mz,Iguchi:2007xs,Kastor:2008wd,Kunz:2008rs,Yazadjiev:2009gr,Nedkova:2010gn,Kunz:2013osa}.

More specifically, we will explain that expanding bubbles of nothing are a pervasive feature of systems of black holes with multiple or non-spherical horizons\footnote{The precise notion of the topology that is required will become clearer below.}.
To demonstrate the idea, we will show that expanding bubbles of nothing arise as a limit of static black hole binaries (in four dimensions) and of black rings (in five dimensions).
These systems allow us to illustrate a general phenomenon using explicit exact solutions of vacuum gravity.
We expect that versions of all the constructions are possible in six or more dimensions, but then the solutions must be obtained numerically.
Other lesser-known kinds of bubbles, in five or more dimensions, arise from different black hole binaries and will be briefly examined.
Towards the end of the Chapter we will discuss more general configurations using topological arguments, and argue that expanding bubbles are also present in systems such as the Schwarzschild--de Sitter  and Nariai solutions.

We will also study how the expansion in bubble spacetimes acts on gravitationally interacting systems, in a manner similar to inflation in de Sitter.
We will show that bubbles in four and five dimensions admit within them black hole binaries and black rings in static (although unstable) equilibrium, their attraction being balanced against the expansion of the background spacetime.
The same mechanism is expected to work in more general situations for which exact solutions are not available.
The present Chapter is based on the paper~\cite{Astorino:2022fge}.

\section{Black holes, black rings and Kaluza--Klein bubbles}

\subsection{Expanding bubbles of nothing from black hole binaries and black rings}

The solution for an expanding bubble of nothing was originally presented in~\cite{Witten:1981gj} in the form
\begin{equation}
\label{bon1}
{ds}^2 = r^2 ( -{dT}^2 + \cosh^2 T d\Omega_n)
+ \frac{{dr}^2}{1-\frac{r_0^n}{r^n}}
+ \biggl( 1-\frac{r_0^n}{r^n} \biggr)r_0^2 {d\phi}^2 \,,
\end{equation}
where $d\Omega_n$ is the standard $n$-dimensional sphere element.
This is obtained from the Schwarzschild--Tangherlini solution in $n+3$ dimensions by rotating to imaginary values the time coordinate and one polar angle.
However, the relation between black holes and bubbles that we will discuss is of a different kind and does not involve any such rotation.
Since the coordinate $\phi$ must be periodically identified, $\phi\sim \phi + 4\pi/n$, the solution has Kaluza--Klein asymptotics, but the latter fact will also be of minor relevance for our discussion.

To understand the geometry, observe that the time-symmetric section at $T=0$ is the product of a ``cigar'' along the $(r,\phi)$ directions, and spheres $S^n$ of radius $r$.
These spheres cannot be shrunk to zero size since they reach a minimum radius at $r=r_0$\footnote{The manifold can not be extended beyond $r=r_0$, thus such a point does not represent an actual singularity of the spacetime.
Moreover, the Kretschmann scalar is well behaved at $r=r_0$.}.
The minimal sphere constitutes the bubble of nothing, and when it evolves for $T>0$, it expands in a de Sitter-like fashion.

The coordinates in~\eqref{bon1} cover the spacetime globally, but we can also write it using static-patch coordinates\footnote{\label{foot:change}The change is $t=\mathrm{arctanh}(\tanh T/\cos\chi)$, $\xi=\cosh T \sin\chi$, where $\chi$ is a polar angle of $S^n$.}, where
\begin{equation}
\label{bonstat}
{ds}^2 = r^2 \biggl[ -(1-\xi^2) {dt}^2 + \frac{{d\xi}^2}{1-\xi^2} + \xi^2 d\Omega_{n-1} \biggr]
+ \frac{{dr}^2}{1-\frac{r_0^n}{r^n}}
+ \biggl(1-\frac{r_0^n}{r^n}\biggr) r_0^2 {d\phi}^2 \,.
\end{equation}
We could set $\xi=\cos\theta$ to relate it more manifestly to the Schwarzschild--Tangherlini solution with imaginary $t$ and $\phi$, but the form above makes clearer the existence of a de Sitter-like horizon at $\xi^2=1$.
In the full spacetime, this is an infinite acceleration horizon that extends from the bubble at $r=r_0$ to infinity.
Observers who sit on the bubble midpoint between the horizons, that is, near $r=r_0$ and around $\xi=0$, find themselves partly surrounded (but not enclosed) by a horizon with topology $S^{n-1}\times \mathbb{R}^2$.
Like in de Sitter, these observers do not have access to the entire $S^n$ bubble.
They only see the half of it that remains static, while the portion of the bubble beyond the horizon expands exponentially.

\begin{figure}
\centering
\includegraphics[width=\textwidth]{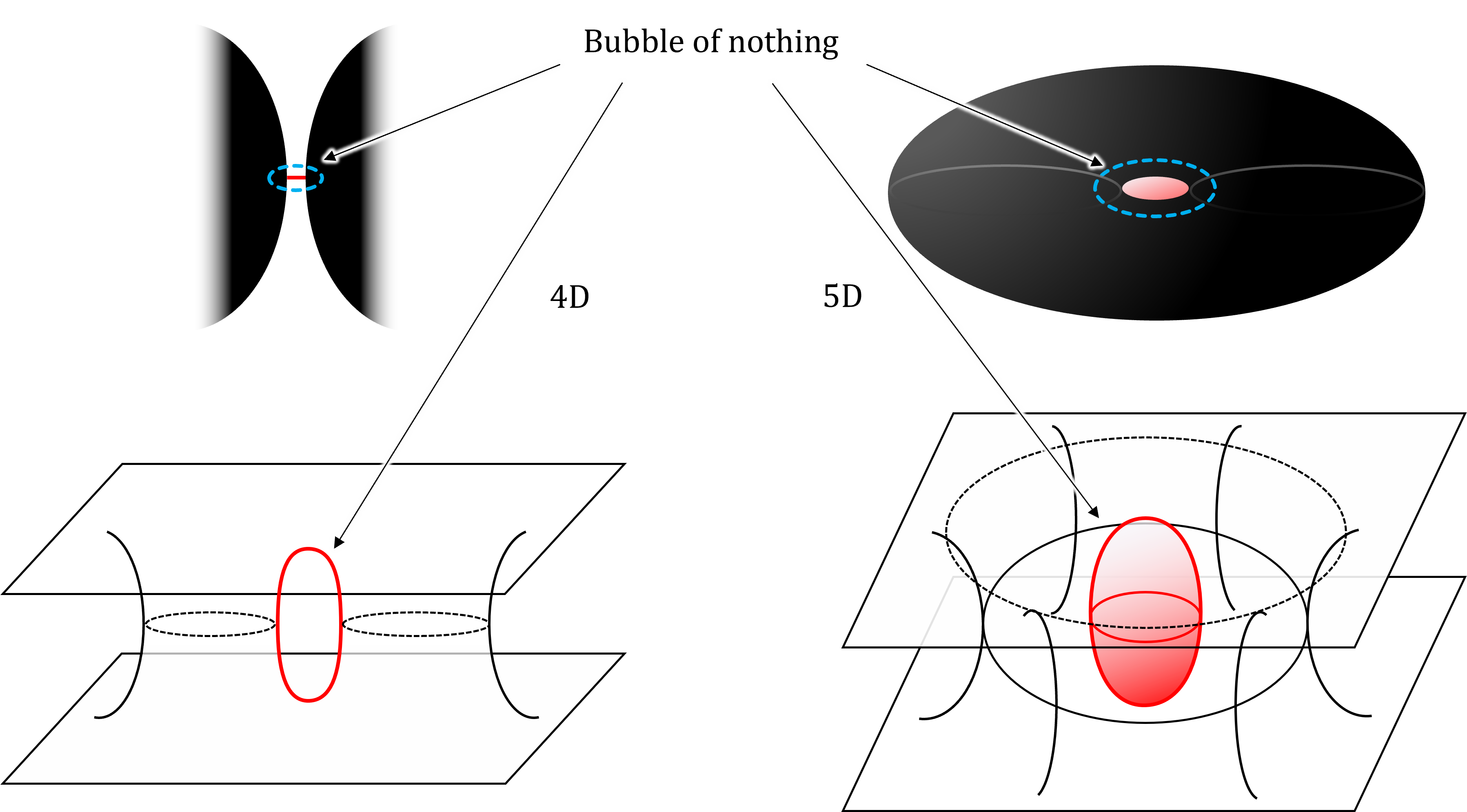}
\caption{\small Bubbles of nothing as limits of black hole systems.
The top pictures are illustrative cartoons, and the bottom ones show time-symmetric spatial sections of the maximally extended solutions.
Left top: the 4D bubble of nothing arises as the geometry in between two black holes, in the limit when their size is very large.
The horizons of the black holes correspond to acceleration horizons of the bubble.
Left bottom: the bubble is a minimal circle (in red) linking the Einstein--Rosen throats of the two black holes.
This circle  encloses ``nothing'', and its expansion occurs as the throats stretch in the black hole interiors.
The angle $\phi$ around the rotation axis is suppressed in these figures.
Right top: the 5D bubble is similarly recovered from the central region of a very fat black ring.
Right bottom: the bubble is a sphere (in red) that wraps the portion of the Einstein--Rosen bridge in the inner ``hole'' of the ring.
In the bottom figure, the ring's $S^2$ is not represented.
In the solutions we discuss, the black hole binary and the black ring are kept static by semi-infinite cosmic strings and cosmic membranes (not shown), respectively, which pull them outwards, but other means of maintaining them in equilibrium are possible.}
\label{fig:bubblebhs}
\end{figure}

Let us examine the case $n=1$ of a four-dimensional expanding bubble.
This is seldom considered when studying Kaluza-Klein spacetimes, but we will give it a new twist.
The sphere $S^{n-1}$ now consists of the two endpoints of the interval $-1\leq \xi\leq 1$, so the observer in the bubble lies between two approximately planar (for $r\approx r_0$) acceleration horizons.
Such Rindler-type horizons are known to describe the geometry near a black hole, and we will find that this interpretation is also apt for the bubble geometries~\eqref{bonstat}.
That is, we will show that the four-dimensional bubble appears as the geometry in between two black holes, when they are separated by a distance much smaller than their radius (see Fig.~\ref{fig:bubblebhs}).

One may wonder in what sense can a black hole binary contain an expanding bubble.
The answer is much the same as for the static-patch metric~\eqref{bonstat}.
When $n=1$, the bubble is a circle that links the Einstein--Rosen bridges of the black hole pair, i.e.~a minimal cycle that encloses nothing (Fig.~\ref{fig:bubblebhs}, bottom left)\footnote{Strictly speaking, the cycle $\Omega_1$ in~\eqref{bon1} need not be a compact $S^1$ (it might be a non-compact $\mathbb{R}$, that gives rise to a different topology), but we will take it to be so.
In the binary, we are identifying asymptotic regions to yield the smallest maximal analytic extension.}.
The static observer in between the two black holes is limited by the horizons to only have access to a portion of this circle, namely, the segment of the axis between the two horizons.
The rest of the circle lies beyond the horizons.
Initially, at $T=0$, this other half-circle is another segment between the two Einstein--Rosen throats.
As $T$ evolves, these throats stretch, so the portions of the circle inside the black holes expand in time.
The expansion of the bubble is then the familiar stretching that occurs in the interior of the black hole\footnote{The same effect is responsible for the growth of holographic volume complexity\cite{Susskind:2014rva}.}.
The compactification of the $\phi$ direction is a consequence of focusing on a small region around the symmetry axis, so the radius of the $\phi$ circles can only reach a finite maximum.

Exact solutions for a static configuration of a pair of black holes, kept apart by semi-infinite cosmic strings that pull on them, have been known for long~\cite{Bach1922,Israel1964}.
We will use them to explicitly exhibit the limit where they reduce to~\eqref{bonstat}.
We emphasize that there is no Wick-rotation involved in this connection:
the time and angular coordinates retain their physical meaning throughout the limit.

The five-dimensional bubble, described originally in~\cite{Witten:1981gj}, also admits a similar interpretation.
Now the acceleration horizon, with topology $S^1\times \mathbb{R}^2$, is connected.
We will find that~\eqref{bonstat} with $n=2$ arises as the limit of a black ring, with horizon topology $S^1\times S^2$, when the size of the $S^2$ is much larger than the inner rim of the ring circle.
The static coordinates only cover the hemisphere of the $S^2$ bubble that consists of the disk of the inner ``hole'' of the ring.
In global coordinates, the $S^2$ is a minimal sphere that wraps the Einstein--Rosen bridge in the inner hole of the ring.
In this case, we will use the solution, first found in~\cite{Emparan:2001wk}, for a static black ring held in place by an infinite cosmic membrane attached to the outer rim of the ring.
We expect that this construction generalizes to all $n\geq 3$, but the required solutions, with horizons of topology $S^{n-1}\times S^2$, are only known numerically~\cite{Kleihaus:2009wh,Kleihaus:2010pr}.

We will also briefly discuss how a certain type of five-dimensional black hole binary, in the limit of small separation, gives rise to a five-dimensional expanding bubble of a different kind than the $n=2$ bubble above:
the minimal cycle is not a single sphere $S^2$, but two $S^2$ that lie on orthogonal spaces and which touch each other at both North and South poles.
They compactify the spacetime on a two-torus (instead of a circle).
A more general discussion of the topology of other configurations will be presented in the concluding Section.

Let us mention that the specific mechanism that keeps the black hole pair, or the black ring, in static equilibrium is not an essential aspect of the construction.
Cosmic strings and membranes, in the form of conical deficits along the outer symmetry axes of the systems, are easy to work with, but the two black holes could also carry electric charges of opposite sign (a dihole~\cite{Emparan:1999au,Emparan:2001bb}) and be held in equilibrium by an external electric field, namely, a fluxbrane.
A similar construction is also possible for dipole black rings~\cite{Emparan:2004wy}.
As long as the black holes are not extremal, they will have bifurcation surfaces and there will be expanding bubbles, with the effects of the electric field becoming negligible in the region between the horizons, since the external fluxbrane polarizes the system so as to cancel the opposite fluxbrane-like field between the charged black holes \cite{Emparan:2001gm}.
Other equilibration methods are possible, but their differences only show up far from the gap between the horizons.
In the limit to the bubble solution, the distinctions between these geometries disappear.

Indeed, the explanation we have given should make clear, and we will elaborate further on this in the concluding Section, that, as long as the black holes in a binary have bifurcate horizons, expanding bubbles are also present in them even if the horizons are not static but dynamically merge and collapse.
But in these cases the expansion of the bubble only lasts a finite time.

\subsection{Black hole binaries and black rings inside expanding bubbles}

The previous remarks bring us to the other main subject of this Chapter, namely, the static equilibrium configurations of black hole binaries and black rings.

\begin{figure}
\centering
\includegraphics[width=.8\textwidth]{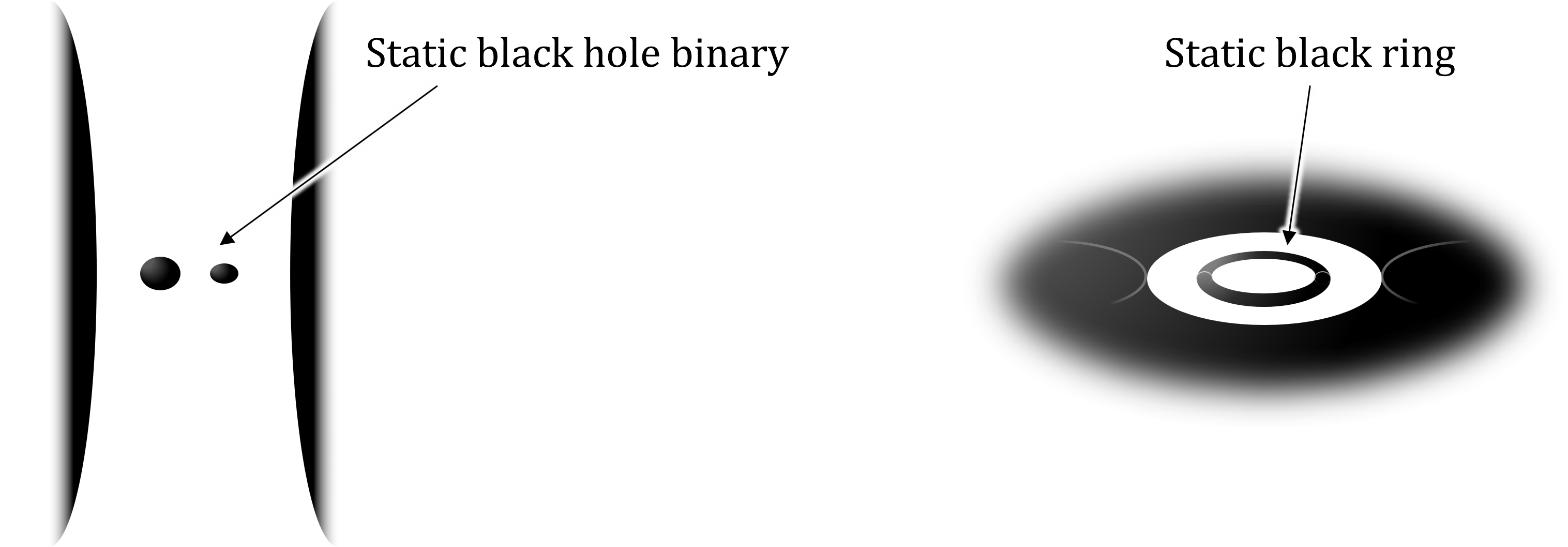}
\caption{ \small Static black hole binaries and black rings obtained by placing them inside expanding bubbles of nothing, which surround them with acceleration horizons.
The binary need not be symmetrical. These configurations can be regarded as limits of double nested static binaries (four black holes along a line), and of two coaxial black rings.}
\label{fig:bhsinbubbles}
\end{figure}

A possibility for balancing the attraction in the binary is the expansion of spacetime.
Indeed, one expects that such binaries in (unstable) equilibrium exist in the de Sitter universe, but the solutions can only be constructed approximately for very small black holes, or numerically~\cite{Dias}.
Nevertheless, the expansion in a bubble should achieve the same effect.
To prove this, we will construct exact solutions where an expanding bubble hosts a black hole binary (see Fig.~\ref{fig:bhsinbubbles} left) (possibly with different masses) in static equilibrium.
The interpretation of the expanding bubbles given above provides another explanation for why this is possible:
we can introduce a binary of two small black holes in the gap between two very large black holes, and then tune the distances between them so that the attraction in the small binary is balanced by the pull of the larger black holes.

The analogues of these configurations involving black rings in five dimensions can also be readily constructed (see Fig.~\ref{fig:bhsinbubbles} right).
We will present a solution for a static black ring inside a five-dimensional bubble of nothing, and show that it can be recovered as the limit of a concentric, static double black ring system.

All these metrics can be given in exact closed form since they are Weyl solutions, which admit a systematic construction with an arbitrary number of collinear black holes, or concentric black rings~\cite{Weyl:1917gp,Bach1922,Israel1964,Emparan:2001wk}.
The configurations are characterized by their rod structure~\cite{Emparan:2001wk,Harmark:2004rm}, which specifies the sources along the different symmetry axes.
This structure makes transparent the features of all the constructions discussed above and their limits.
Indeed, the connections between the black hole binary and the static black ring, and the corresponding expanding bubbles in four and five dimensions, have been apparent at least since the analysis in~\cite{Emparan:2001wk}.
Nevertheless, to our knowledge, this connection does not appear to be widely known, and it has not been examined in detail in the literature.

\section{Bubbles as limits of black hole binaries and black rings}

We will now show explicitly how the metrics for the bubbles of nothing in 5D and 4D are recovered as limits of static black ring and binary black hole solutions in the manner illustrated in Fig.~\ref{fig:bubblebhs}.

\subsection{From black ring to bubble}
\label{subsec:ringtobubble}

The simplest instance is the relationship between the five-dimensional bubble of nothing of~\cite{Witten:1981gj} and the static black ring of~\cite{Emparan:2001wk}.
The metric of the latter is
\begin{equation}
\begin{split}
\label{bring}
{ds}^2 & = -\frac{F(x)}{F(y)} {d\tilde t}^2
+ \frac{R^2}{(x-y)^2}\biggl[ F(x)\biggl( (y^2-1) {d\tilde\psi}^2 + \frac{F(y)}{y^2-1} {dy}^2\biggr) \\
&\quad + F(y)^2\biggl( \frac{{dx}^2}{1-x^2}+\frac{1-x^2}{F(x)}{d\tilde\phi}^2\biggr)\biggr] \,,
\end{split}
\end{equation}
with
\begin{equation}
\label{Fring}
F(\xi) = 1-\mu \xi \,.
\end{equation}
Readers unfamiliar with these $(x,y)$ coordinates are referred to~\cite{Emparan:2001wk} and~\cite{Emparan:2006mm} for a detailed explanation.
Roughly, $x\in[-1,1]$ is the cosine of the polar angle of the ring's $S^2$, and $-1/y\in (0,-1/x)$ is a radial coordinate away from these spheres.
The coordinates $\tilde\psi$ and $\tilde\phi$ are, respectively, the angle of the $S^1$ and the azimuthal angle of the $S^2$ of the black ring.
The parameter $R$ sets the scale for the size of the black ring, and varying $\mu\in[0,1)$ changes its shape from thin to fat.
The horizon lies at $y=-\infty$, and the absence of conical singularities along the $\tilde\psi$ rotation axis at $y=-1$, and in the inner disk of the ring at $x=1$, is obtained when we identify
\begin{equation}
\tilde\psi\sim \tilde\psi +2\pi\sqrt{1+\mu} \,, \qquad
\tilde\phi\sim \tilde\phi +2\pi\sqrt{1-\mu} \,.
\end{equation}
In order to make the ring very big and fat, and to blow up the inner disk region, we will take $\mu\to 1$ and $R\to\infty$ while zooming in onto $x\approx 1$.
For this purpose, we change
\begin{equation}
x = 1-\frac{r^2-r_0^2}{2R^2} \,, \qquad
\mu = 1-\frac{r_0^2}{2R^2} \,,
\end{equation}
where $r$ and $r_0$ are a new coordinate and a constant parameter that remain finite as $R\to\infty$.
In addition we introduce a coordinate $\xi$ via
\begin{equation}
y = -\frac{1+\xi^2}{1-\xi^2} \,,
\end{equation}
and rescale the Killing coordinates to have canonical normalization,
\begin{equation}
\tilde t = 2 R t \,, \qquad
\tilde\psi = \sqrt{2}\psi\,, \qquad
\tilde\phi = \frac{r_0}{\sqrt{2} R}\phi \,.
\end{equation}
Then, in the limit $R\to\infty$, the metric~\eqref{bring} becomes
\begin{equation}
\label{5dbubble}
{ds}^2 \to
r^2\biggl( -(1-\xi^2) {dt}^2 
+\frac{{d\xi}^2}{1-\xi^2} + \xi^2 {d\psi}^2 \biggr) 
+ \frac{{dr}^2}{1-\frac{r_0^2}{r^2}}
+ \biggl(1-\frac{r_0^2}{r^2}\biggr)r_0^2 {d\phi}^2 \,,
\end{equation}
which is indeed the same as the metric~\eqref{bonstat} of the bubble of nothing for $n=2$.

Observe that the $\phi$ circles in~\eqref{5dbubble} cannot reach arbitrarily large sizes but become a compact direction at infinity.
This is a consequence of focusing on the region close to the disk at $x=1$, which limits the growth of these circles.

One might wonder whether rotating black rings, with the rotation adjusted to balance the tension and gravitational self-attraction, have a limit to the bubble of nothing.
The answer is no:
in the limit where the rotating ring becomes very fat, it approaches a singular, horizonless solution instead of the non-singular geometry~\eqref{5dbubble}.

\subsection{Weyl metrics and rod structures}

All other solutions in this chapter will be presented as vacuum Weyl metrics, using cylindrical coordinates 
\begin{equation}
\label{static-metric}
{ds}^2 = f(\rho,z) \bigl({d\rho}^2 + {dz}^2\bigr) + g_{ab}(\rho,z) {dx}^a {dx}^b \,.
\end{equation}
We have already presented such a metric in Sec.~\ref{sec:ism}, nevertheless we will provide a broader overview in the context of higher-dimensional spacetimes.
For more complete expositions, we refer to~\cite{Belinski:2001ph,Emparan:2001wk,Emparan:2008eg,Harmark:2004rm}.
The main feature of~\eqref{static-metric} is the presence of $D-2$ Killing coordinates $x^a$:
in four dimensions they are $(t,\phi)$, and in five dimensions they include an additional angle, $(t,\phi,\psi)$.

For all the solutions presented here, the metric $g_{ab}(\rho,z)$ along the Killing directions will be diagonal.
Static and axisymmetric solutions can then be systematically constructed by specifying a set of rod-like sources along the $z$-axis for the three-dimensional Newtonian potentials associated to the metric functions $g_{ab}$;
they are not physical rods, but coordinate singularities in the axis $\rho=0$ of the Weyl metrics.
Their importance derives from the fact that, given the rod distribution, the form of $g_{ab}(\rho,z)$ directly follows from a simple algebraic construction. 
Subsequently, $f(\rho,z)$ can be obtained by a line integral in the case of diagonal metrics, and more generally by the inverse scattering method~\cite{Belinski:2001ph}.
The rods (with linear density $1/2$) are specified along each direction $x^a$, in such a way that at every value of $z$ there is a rod along one and only one of the directions.

The rod structure provides an easy diagrammatic way to interpret static (or more generally stationary) axisymmetric solutions. 
On a rod along a direction $x^a$, the corresponding Killing vector has a fixed point set.
When we have an angular Killing vector, such as $\partial_{\phi}$ or $\partial_{\psi}$, then the corresponding circles shrink to zero size at the rod, and the periodicity of the angle must be appropriately chosen in order to avoid conical singularities.
The regularity condition on $x^a\sim x^a + \Delta x^a$ at any given rod is
\begin{equation}
\label{nocone}
\Delta x^a = 2\pi \lim_{\rho\to 0}\rho \sqrt{\frac{f}{g_{aa}}} \,.
\end{equation}
When the Killing vector is timelike $\partial_t$, the rod represents a horizon, and through Euclidean continuation $t\to i\tau$, Eq.~\eqref{nocone} gives its associated temperature $T=\Delta\tau ^{-1}$.
If the rod is finite, it defines an event horizon, while infinite rods are generically associated to accelerating horizons, such as Rindler ones.

The topology of the solutions can also be inferred from the rod structure.
If there is a rod along a direction $x^a$, the other directions $x^b$ are fibered along the corresponding portion of the axis.
At a point where rods along $x^a$ and $x^b$ meet, the two fibers shrink to zero.
As a result, the solutions have a ``bubbling'' structure.

To illustrate these features we will describe the simplest examples that are relevant to us here.

\subsubsection{Four-dimensional solutions}

\begin{figure}
\includegraphics[scale=0.84]{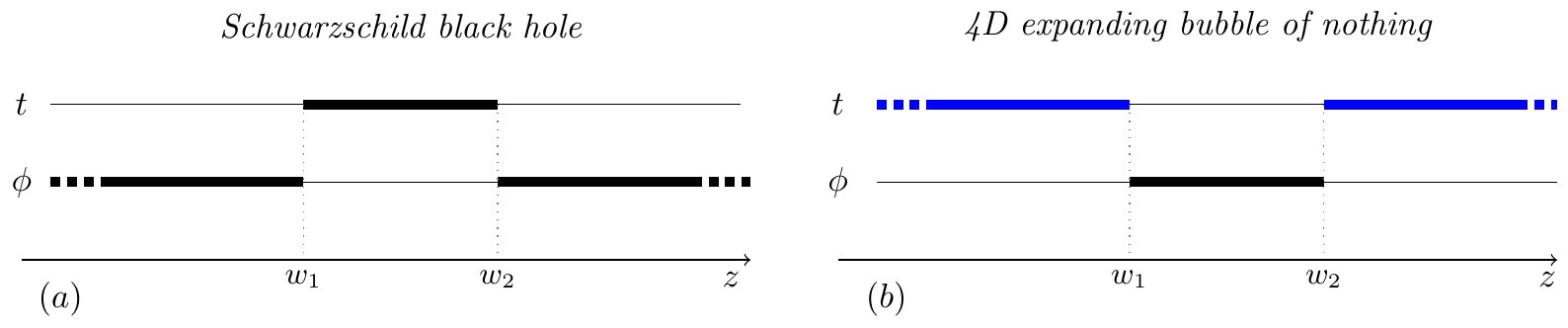}
\caption{\small $(a)$ rod diagram for the Schwarzschild black hole.
The finite timelike rod defines the black hole horizon (a sphere $S^2$, from the fibration of the $\phi$ parallel circles over the segment $w_1<z<w_2$).
Exchanging $t\to\phi$ gives $(b)$: rod diagram of the expanding bubble of nothing.
The semi-infinite timelike rods represent the bubble acceleration horizons (two of them, each with topology $\mathbb{R}^2$).
Here and in the following figures, acceleration horizons are pictured in blue.}
\label{fig:1bh-1bubble}
\end{figure}

The Schwarzschild black hole rod structure is given by a finite timelike rod and two semi-infinite spacelike rods (see Fig.~\ref{fig:1bh-1bubble}(a)).
The four-dimensional bubble of nothing is the double Wick-rotated version of the Schwarzschild metric, thus its rod structure is consistently given by exchanging the $t$ and $\phi$ rods of the previous solution (see Fig.~\ref{fig:1bh-1bubble}(b)).
We see from their respective timelike rods that in the Schwarzschild solution the horizon is finite, while the bubble of nothing possesses two infinite acceleration horizons.

The corresponding metrics are given by
\begin{align}
\label{gschbub}
g_{ab}^\text{Schwarz}dx^a dx^b & =
-\frac{\mu_1}{\mu_2} {dt}^2
+ \rho^2 \frac{\mu_2}{\mu_1} {d\phi}^2 \,, \\
g_{ab}^\text{bubble}dx^a dx^b & =
-\rho^2\frac{\mu_2}{\mu_1} {dt}^2
+ \frac{\mu_1}{\mu_2} {d\phi}^2 \,,
\end{align}
and, in both cases,
\begin{equation}
\label{fschbub}
f = C_f \frac{ 4\mu_1\mu_2^3}{\mu_{12}W_{11}W_{22}} \,.
\end{equation}
Here, and in the following, we introduce
\begin{equation}
\label{muij-Wij}
\mu_i = w_i-z+\sqrt{\rho^2 + (z-w_i)^2} \,, \qquad 
\mu_{ij} = (\mu_{i}-\mu_{j})^2 \,, \qquad
W_{ij} = \rho^2 + \mu_i\mu_j \,.
\end{equation}
The parameters $w_i$, chosen in increasing order, specify the rod endpoints, and they define the physical properties of the metric:
the position of the horizons, the size and mass of the black holes, and the rotation axes.
The parameter $C_f$ is an arbitrary gauge constant. It corresponds to a rescaling of $\rho$ and $z$, and it can be chosen, without loss of generality, to fix the normalization of one of the Killing directions, for instance, setting the periodicity of one of the angles to any prescribed value, such as canonical periodicity $2\pi$.

It is now straightforward to verify that taking
\begin{equation}
w_1 = z_0 - \frac{r_0}2 \,, \qquad
w_2 = z_0 + \frac{r_0}2 \,, \qquad
C_f=r_0^2 \,,
\end{equation}
and defining
\begin{equation}
\rho = \sqrt{r(r-r_0)}\sin\theta \,, \qquad
z = z_0 + \biggl( r-\frac{r_0}2\biggr) \cos\theta \,,
\end{equation}
in~\eqref{gschbub} and~\eqref{fschbub}, we recover
\begin{align}
g_{ab}^\text{Schwarz}dx^a dx^b & = -\biggl( 1-\frac{r_0}{r}\biggr) {dt}^2
+ r^2\sin^2\theta {d\phi}^2 \,, \\
\label{4dbub}
g_{ab}^\text{bubble} dx^a dx^b & = -r^2\sin^2\theta {dt}^2
+ \biggl( 1-\frac{r_0}{r}\biggr) {d\phi}^2 \,,
\end{align}
and
\begin{equation}
\label{f-4d}
f(\rho,z) \bigl({d\rho}^2 + {dz}^2\bigr) =
\frac{{dr}^2}{1-\frac{r_0}{r}}
+ r^2 {d\theta}^2 \,,
\end{equation}
which are the conventional forms of the Schwarzschild and bubble solutions (up to possible constant rescalings of the Killing coordinates $t$ and $\phi$). They are obviously equivalent under $t\leftrightarrow\phi$.
The form of the bubble of nothing in~\eqref{bonstat} is recovered by rescaling $\phi$ by $r_0$\footnote{We could have achieved this by adequately choosing $C_f$, but, in general, we will not take $\phi$ to be canonically normalized with $\phi\sim \phi+2\pi$, but rather its periodicity will be suitably adjusted.}, and setting $\cos\theta=\xi$.

Finally, observe that if in either of the solutions we send one of the rod endpoints, $w_1$ or $w_2$, to infinity while keeping the other fixed, then we recover the geometry of Rindler space, with an infinite acceleration horizon.
The Minkowski spacetime can be obtained when both the poles are simultaneously pushed infinitely far away in opposite directions, i.e.~$w_1 \to -\infty $ and $w_2 \to \infty$.

\subsubsection{Five-dimensional solutions}

\begin{figure}
\includegraphics[scale=0.84]{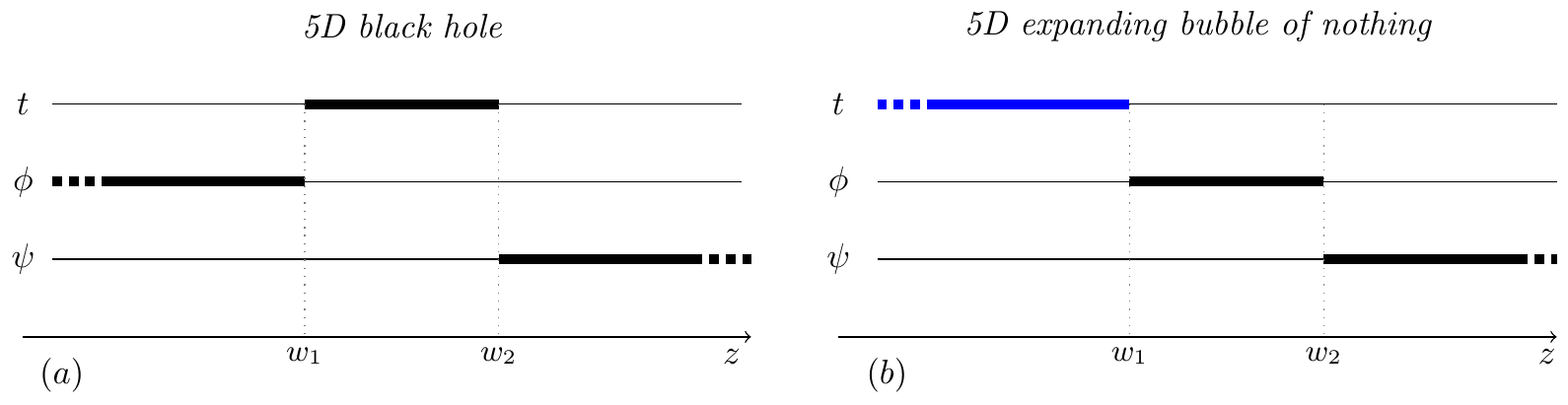}
\caption{\small $(a)$:
rod diagram for the 5D Tangherlini black hole.
The finite timelike rod defines the black hole horizon (a sphere $S^3$, fibering $\phi$ and $\psi$ circles over $w_1<z<w_2$).
Exchanging $t \leftrightarrow \phi$ gives $(b)$:
rod diagram of the five-dimensional expanding bubble of nothing.
The timelike semi-infinite rod represents the bubble acceleration horizon (which is connected, with topology $S^1\times \mathbb{R}^2$:
the $\psi$ circles are trivially fibered over $-\infty<z<w_1$).}
\label{fig:5D-1bh-1bubble}
\end{figure}

The previous analysis has a straightforward counterpart in five dimensions.
The rod structures of the Schwarzschild--Tangherlini black hole and the five-dimensional expanding bubble are given in Fig.~\ref{fig:5D-1bh-1bubble}, which makes evident that they are related by a double-Wick rotation that effectively exchanges $t$ and $\phi$.

These rod structures dictate that
\begin{align}
\label{g_5D_bh_bubble}
g_{ab}^\text{Tang} dx^a dx^b & =
-\frac{\mu_1}{\mu_2} {dt}^2
+ \frac{\rho^2}{\mu_1} {d\phi}^2
+ \mu_2 {d\psi}^2 \,, \\
\label{g5dbubble}
g_{ab}^\text{5D-bubble} dx^a dx^b & =
-\frac{\rho^2}{\mu_1} {dt}^2
+ \frac{\mu_1}{\mu_2} {d\phi}^2
+ \mu_2 {d\psi}^2 \,,
\end{align}
while $f(\rho,z)$ is again identical for both spacetimes
\begin{equation}
f = C_f \, \frac{\mu_2 W_{12}}{W_{11}W_{22}} \,.
\end{equation}
To express the Tangherlini black hole in spherical coordinates
\begin{equation}
{ds}^2 = - \biggl(1-\frac{r_0^2}{r^2} \biggr) {dt}^2 + \frac{{dr}^2}{1-\frac{r_0^2}{r^2}}
+ r^2 {d\theta}^2
+ r^2 \sin^2{\theta} {d\phi}^2
+ r^2 \cos^2{\theta} {d\psi}^2 \,,
\end{equation}
we choose
\begin{equation}
w_1 = z_0 - \frac{r_0^2}{4} \,, \qquad
w_2 = z_0 + \frac{r_0^2}{4} \,, \qquad
C_f =1 \,,
\end{equation}
and change
\begin{equation} \label{rho-z-tangherlini}
\rho = \frac{r}{2} \sqrt{r^2-r_0^2} \sin{2\theta} \,, \qquad
z = z_0 + \frac{1}{4} \bigl(2r^2-r_0^2 \bigr)\cos{2\theta} \,.
\end{equation}
Similarly, the five-dimensional expanding bubble in spherical coordinates takes the form
\begin{equation}
{ds}^2 = -r^2 \cos^2{\theta} {dt}^2
+ \frac{{dr}^2}{1-\frac{r_0^2}{r^2}}
+ r^2 {d\theta}^2
+ \biggl(1-\frac{r_0^2}{r^2} \biggr) {d\phi}^2
+ r^2 \sin^2{\theta} {d\psi}^2 \,.
\end{equation}
When we rescale $\phi$ by $r_0$ and set $\sin\theta=\xi$ we recover~\eqref{5dbubble}.

\subsection{From binary black hole to bubble}
\label{subsec:bbhbub}

\begin{figure}
\includegraphics[scale=0.87]{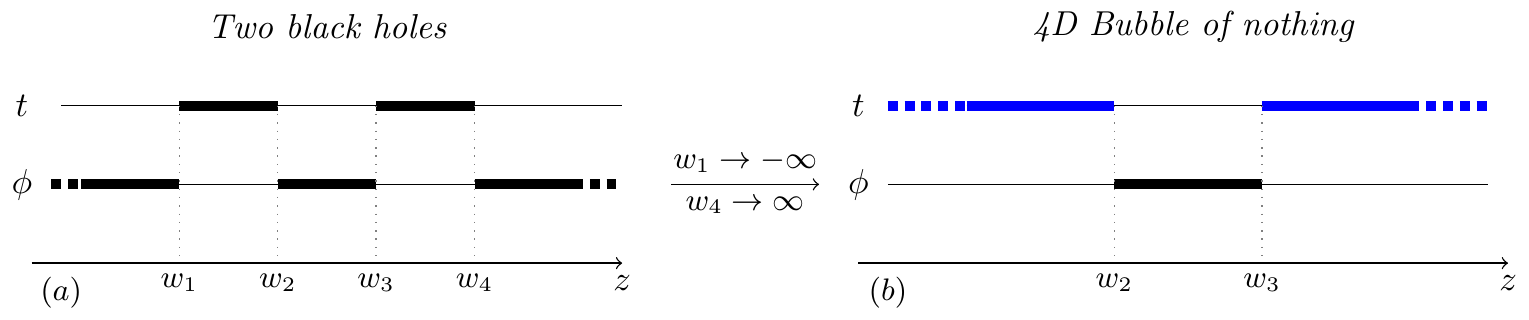}
\caption{{\small $(a)$ Rod diagram for the Bach--Weyl static binary black hole configuration~\cite{Bach1922}.
The thick timelike rods represent the two black hole horizons.
By sending the rod endpoints $w_1\to-\infty$, $w_4\to\infty$, with $w_2$ and $w_3$ fixed, we recover the same diagram as in the bubble of nothing Fig.~\ref{fig:1bh-1bubble}(b).
This limit makes the two black holes infinitely large, while keeping the separation between them finite, as is illustrated in panel $(b)$.}}
\label{fig:bach-weyl}
\end{figure}

Now let us consider the Bach--Weyl solution, with 
\begin{equation}
\label{bach-weyl-metric}
g_{ab} dx^a dx^b =
- \frac{\mu_1\mu_3}{\mu_2\mu_4} {d\tilde{t}}^2
+ \rho^2 \frac{\mu_2 \mu_4}{\mu_1 \mu_3} {d\tilde{\phi}}^2 \ \,, \quad
f = \frac{16 \tilde{C}_f \ \mu_1^3 \mu_2^5 \mu_3^3 \mu_4^5}{\mu_{12} \mu_{14} \mu_{23} \mu_{34} W_{13}^2 W_{24}^2 W_{11} W_{22} W_{33} W_{44}} \,,
\end{equation}
which describes two Schwarzschild black holes aligned along the $z$-axis, and whose rod diagram is pictured in Fig.~\ref{fig:bach-weyl}(a).
By appropriately choosing $\tilde{C}_f$ to be
\begin{equation}
\tilde{C}_f = 16 (w_1-w_2)^2 (w_1-w_3)^2 (w_2-w_4)^2 (w_3-w_4)^2 \,,
\end{equation}
we make the segment $w_2<z<w_3$ of the axis in between the black holes regular, while keeping the standard periodicity of the azimuthal angle $\Delta\phi = 2\pi$.
On the other hand, along the semi-infinite axes from the black holes towards $z\to\pm\infty$ there are conical deficits.
These can be regarded as cosmic strings that keep the black holes apart.

The rod diagram makes manifest how this solution is connected to other black hole/bubble configurations, either via double-Wick rotations that exchange the $t$ and $\phi$ rods\footnote{The Bach--Weyl solution is the double-Wick rotation of the single black hole in the bubble~\cite{Horowitz:2002cx}}, or by taking limits where rod endpoints merge or are sent to infinity.
For our purposes here, we observe that by simply sending the rod endpoints $w_1\to-\infty$ and $w_4\to\infty$, with $w_2$ and $w_3$ fixed, we recover the rod diagram of the 4D bubble of nothing, Fig.~\ref{fig:1bh-1bubble}(b).
When we do so, we make the two black holes infinitely large, while maintaining fixed the separation between them.
This is precisely the type of limit that we discussed in the introduction (see Fig.~\ref{fig:bubblebhs}).
The cosmic strings collapse the space along the outer axes creating a conical deficit angle of $2\pi$, but this is not a problem since this part of the geometry is pushed away to infinity.

To see that the limit works correctly, not only with the rods but also in the entire metric, we conveniently place the bubble poles symmetrically at $w_1=-z_b$ and $w_4=z_b$.
Then, we rescale 
\begin{equation}
\tilde{t} = (2z_b) t \,, \qquad
\tilde{\phi} = \frac{\phi}{2z_b} \,,
\end{equation}
so that the metric~\eqref{bach-weyl-metric} remains finite when we send $z_b \to \infty$. 
One can readily verify that, after rescaling $\tilde{C}_f=C_f/4$ to take into account that in~\eqref{gschbub} $\phi$ has periodicity $4\pi$, the bubble of nothing in the form of~\eqref{gschbub} and~\eqref{fschbub} is recovered.

We have then proven that the gravitational field of the expanding bubble is indeed the same as that between two very large black holes.

\subsection{From black ring to bubble, Weyl style}
\label{subsec:ringtobubble2}

It is now easy to see how the rod diagrams also make transparent the limit from the static black ring to the expanding bubble of nothing, which we discussed using other coordinates in Sec.~\ref{subsec:ringtobubble}. 

\begin{figure}
\includegraphics[scale=0.84]{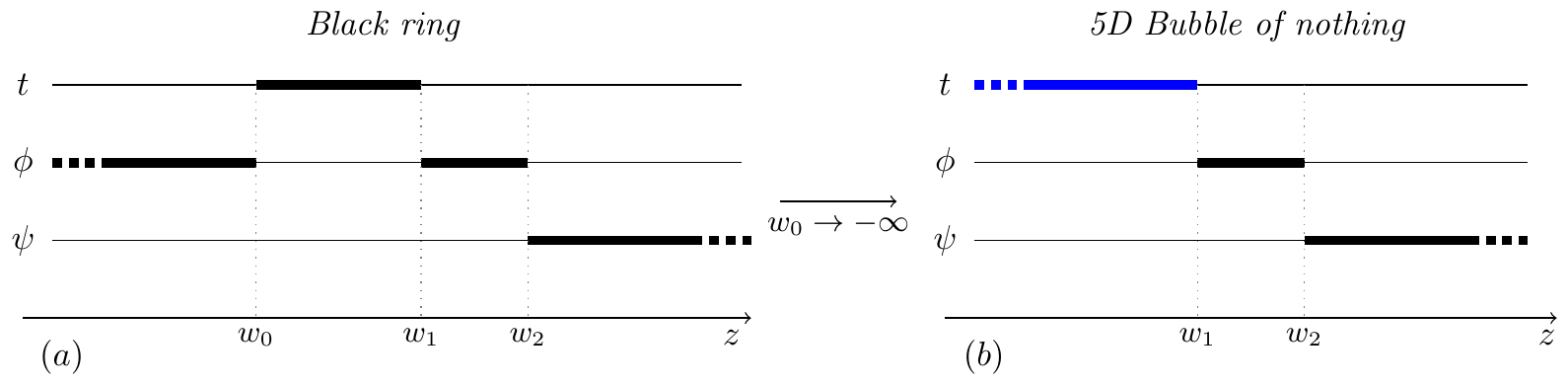}
\caption{{\small $(a)$ Rod diagram for the static black ring.
The thick timelike rod represents the black ring horizon, with topology $S^1\times S^2$.
$(b)$ By sending the rod endpoint $w_0\to-\infty$, with $w_1$ and $w_2$ fixed, we recover the bubble of nothing in Fig.~\ref{fig:5D-1bh-1bubble}(b).
This limit makes the black ring very fat, while keeping its hole finite, as was illustrated in Fig.~\ref{fig:bubblebhs}.}}
\label{fig:statring}
\end{figure}

Fig.~\ref{fig:statring}(a) shows the rod diagram for the static black ring.
The Weyl form of the metric that follows from the diagram is
\begin{equation}
g_{ab} dx^a dx^b =
-\frac{\mu_0}{\mu_1} {d\tilde{t}}^2
+ \rho^2 \frac{\mu_1}{\mu_0\mu_2} {d\tilde{\phi}}^2
+ \mu_2 {d\psi}^2 \,, \qquad
f = C_f
\frac{\mu_2 W_{01}^2 W_{12}}{W_{02} W_{00} W_{11} W_{22}} \,.
\end{equation}
The horizon of the black ring, with topology $S^1\times S^2$, lies at $w_0<z<w_1$, while the ``hole'' of the ring is in the region $w_1<z<w_2$.
If we send $w_0\to -\infty$  keeping all other rod endpoints fixed (hence making the ring very fat while its hole remains finite) we recover the same diagram as for the expanding bubble of nothing in Fig.~\ref{fig:5D-1bh-1bubble}(b).
In the metric, this requires a suitable rescaling of $t$ and $\phi$, similarly to what happens in the 4D case.
The required rescalings are
\begin{equation}
\tilde{t} = \sqrt{2|w_0|} t \,, \qquad
\tilde{\phi} = \frac{\phi}{\sqrt{2|w_0|}} \,.
\end{equation}

\subsection{Five-dimensional black hole binaries and bubbles}
\label{subsec:5dbbh}

We shall briefly mention how a limit can be taken in a five-dimensional black hole binary that is asymptotically flat (save for possible conical defect membranes) to yield a different kind of five-dimensional expanding bubble.

\begin{figure}
\includegraphics[scale=0.84]{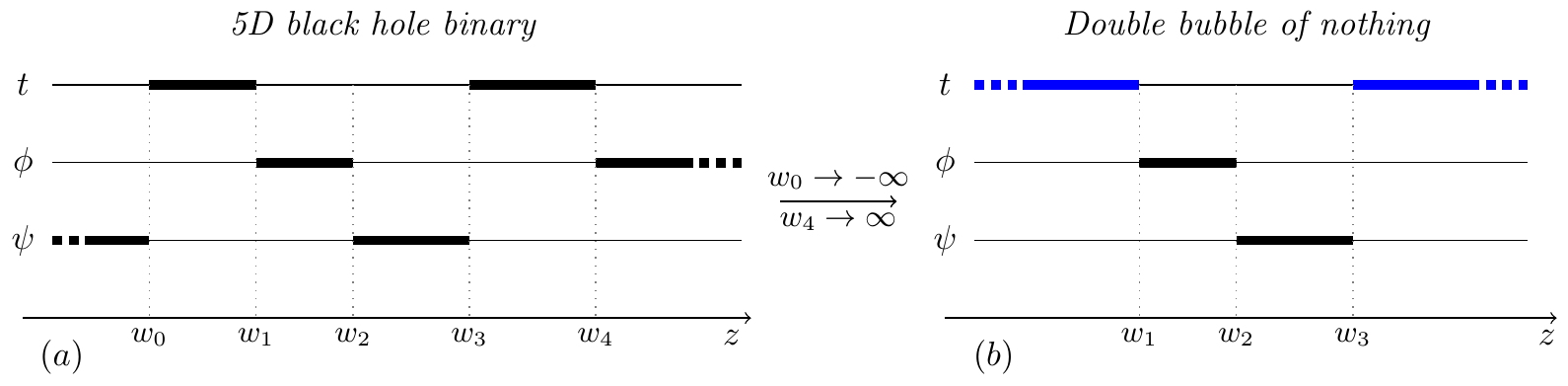}
\caption{{\small $(a)$ Rod diagram for the five-dimensional black hole binary of \cite{Tan:2003jz}.
$(b)$ Limit to the expanding bubble of nothing of Sec.~4.7 in~\cite{Emparan:2001wk}, which is asymptotic to a space compactified on a two-torus.
The space in between the horizons, $w_1<z<w_3$, consists of two topological disks $D_2$, orthogonal to each other and touching at their origins (at $z=w_2$).
In the maximal analytic extension, these become two orthogonal $S^2$ that touch at their poles.}}
\label{fig:5dbbh}
\end{figure}

The Weyl formalism allows to combine two five-dimensional Tangherlini black holes with the rod structure in Fig.~\ref{fig:5dbbh}(a).
This system was studied in~\cite{Tan:2003jz}.
Since the two black holes lie along different axes, they cannot be regarded as collinear.
Nevertheless, the solution is asymptotically flat, as follows from the presence of one semi-infinite rod along $\phi$ and another along $\psi$.
Now take, as in the previous examples, the limit where the two black holes become infinitely large, making their timelike rods semi-infinite.
The result is the system in Fig.~\ref{fig:5dbbh}(b).

This geometry was analyzed in~\cite{Emparan:2001wk}, where one can find the explicit solution (see Sec.~4.7 there).
Here we shall only describe its main properties.
Conical singularities can be avoided along all the axes, and the solution is identified as an expanding bubble of nothing.
In contrast to the simpler five-dimensional bubble of~\eqref{5dbubble} (and~\eqref{g5dbubble}), where the minimal cycle (the bubble) is a sphere $S^2$, in this case it is made of two orthogonal $S^2$, i.e., the meridian lines of one sphere are orthogonal to the meridian lines of the other, and the parallel lines of one lie along $\phi$ and of the other along $\psi$.
The two spheres touch each other at both their north and south poles.
Furthermore, the single bubble in~\eqref{5dbubble} asymptotically has one compact circle, while the double bubble in Fig.~\ref{fig:5dbbh}(b) has two, and therefore represents a Kaluza--Klein compactification from five to three dimensions.
Each of the two $S^2$ is responsible for the compactification of one of the two circle directions.
The solution also differs from the four-dimensional bubble~\eqref{4dbub}, in that the two acceleration horizons here are not symmetric:
the $\phi$ and $\psi$ circles close off at one or the other horizon, and their accelerations can be different.
Indeed, the two $S^2$ can have different sizes.

The two five-dimensional black holes can also be combined in a different fashion with Kaluza--Klein asymptotics~\cite{Emparan:2001wk,Elvang:2002br}.
The configuration has a limit to a ``bubble string'', i.e., the direct product of the 4D expanding bubble and a circle.
Ref.~\cite{Emparan:2001wk} showed that the Weyl formalism allows to generalize all of these solutions to other expanding bubbles in higher dimensions, which compactify spacetime down to three or four dimensions. 

Finally, we could envisage starting from a collinear pair of 5D black holes which lie along a line that is a fixed point of $SO(3)$ rotations (and not $SO(2)$, as above).
In this case, the limit of small separation would result, like in 4D, in a topologically circular $S^1$ bubble.
However, these configurations (and their higher-dimensional counterparts) do not fall within the Weyl class, and they are not known in exact form.

\section{Static black hole binaries and black rings in expanding bubbles}
\label{sec:bhsfrombubbles}

In this section, we will explore some configurations in 4D and 5D that can be regularised by the presence of an expanding bubble of nothing.
First, we will consider a 4D static black hole binary system (a subcase of the Israel--Khan solution~\cite{Israel1964}). As is well known, the Bach--Weyl binary in~\eqref{bach-weyl-metric} necessarily contains conical singularities on the axis $\rho=0$, either in the segment in between the two black holes, or (as we chose above) in the semi-axes towards infinity:
these are, respectively, struts or strings that balance the attraction between the black holes.
We will prove how, by placing the binary within the bubble, we can remove all these singularities and thus obtain a completely regular system on and outside the event horizons.

An analogous construction is possible for the 5D static black ring.
In the manner we presented this solution in Secs.~\ref{subsec:ringtobubble} and~\ref{subsec:ringtobubble2}, the geometry is singular because the tension and self-attraction of the ring, which would drive it to collapse, need to be balanced by a conical-defect membrane.
Again, immersing the ring in an expanding bubble of nothing allows to balance the forces and remove all the conical singularities.

In the following we present the metrics for these systems and prove that it is possible to achieve equilibrium configurations.
A more complete analysis of the physical magnitudes and of the first law of thermodynamics for black hole systems in expanding bubbles will be the subject of future work~\cite{AstorinoBubble}.

\subsection{4D black hole binary in equilibrium inside the expanding bubble}
\label{sec:binary+bubble}

\begin{figure}
\includegraphics[scale=0.98]{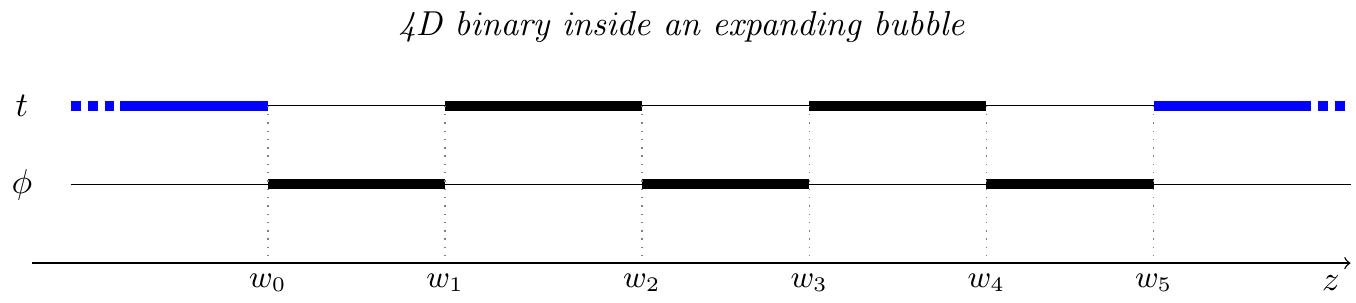}
\caption{{\small Rod diagram for a binary black hole system inside a bubble of nothing.
The black timelike rods (thick lines of the $t$ coordinate) represent the black hole horizons, while the blue timelike semi-infinite rods correspond to the bubble horizon.
This diagram corresponds to the double-Wick rotation of the three-source Israel--Khan solution~\cite{Israel1964}.}}
\label{fig:binary-bubble}
\end{figure}

Superposing the rods of the 4D bubble of nothing and the Bach--Weyl binary, we get the diagram of Fig.~\ref{fig:binary-bubble}.

The solution can be written explicitly in Weyl coordinates~\eqref{static-metric} with
\begin{subequations}
\label{binary-full}
\begin{gather}
\label{gph-binary}
g_{ab} dx^a dx^b =
-\rho^2 \frac{\mu_1 \mu_3 \mu_5}{\mu_0 \mu_2 \mu_4} {dt}^2
+ \frac{\mu_0 \mu_2 \mu_4}{\mu_1 \mu_3 \mu_5} {d\phi}^2 \,, \\
\label{fph-binary}
f = \frac{16 C_f \, \mu_0^5 \mu_1^7\mu_2^5\mu_3^7\mu_4^5\mu_5^7}{\mu_{01} \mu_{03} \mu_{05} \mu_{12} \mu_{14} \mu_{23} \mu_{25} \mu_{34} \mu_{45} W_{02}^2 W_{04}^2 W_{13}^2 W_{15}^2 W_{24}^2 W_{35}^2 W_{00} W_{11} W_{22} W_{33} W_{44} W_{55}} \,.
\end{gather}
\end{subequations}
In the limit in which the bubble horizon is pushed to infinity, for $w_0 \to -\infty$ and $w_5\to \infty$, we recover the standard Bach--Weyl binary~\eqref{bach-weyl-metric}.
On the other hand the limit to the bubble can be obtained in different ways:
by focusing on the bubbles in between black holes as we have done above, e.g..~taking $w_1 \to -\infty$ and $w_4 \to \infty$, or alternatively by eliminating the black holes by collapsing their rods, thus $w_1=w_2=w_3=w_4$.

In general, the geometry contains conical singularities on the $z$-axis in the intervals $(w_0,w_1)$, $(w_2,w_3)$, and  $(w_4,w_5)$, which we eliminate by imposing~\eqref{nocone} on each interval. 
As we mentioned, we can choose $C_f$ (i.e.~a rescaling of $f$) to set $\Delta\phi=2\pi$ without loss of generality.
Then, requiring~\eqref{nocone} on $z\in(w_0,w_1)$ fixes
\begin{equation}
\begin{split}
C_f & = 2^{12}(w_0-w_2)^2 (w_1-w_2)^2 (w_2-w_3)^2 (w_0-w_4)^2 \\
&\quad \times (w_1-w_4)^2(w_3-w_4)^2(w_2-w_5)^2(w_4-w_5)^2 \,,
\end{split}
\end{equation}
while for $z\in(w_2,w_3)$ and $z\in(w_4,w_5)$ we get
\begin{subequations}
\label{equilbinarybubble}
\begin{align}
\frac{(w_0-w_2)(w_2-w_3)(w_1-w_4)(w_2-w_5)}{(w_0-w_1)(w_1-w_3)(w_2-w_4)(w_1-w_5)} &= 1  \,, \\
\frac{(w_0-w_2)(w_0-w_4)(w_2-w_5)(w_4-w_5)}{(w_0-w_1)(w_0-w_3)(w_1-w_5)(w_3-w_5)} &= 1 \,.
\end{align}
\end{subequations}
These can be solved in terms of the bubble parameters $w_0$ and $w_5$, thus leaving the binary parameters $w_{1,2,3,4}$ unconstrained.
To this end, we first choose a convenient parametrization of the rod endpoints in terms of the Komar masses $M_1$, $M_2$ of the two black holes (these are half the coordinate length of the horizon rod), the coordinate distance between them, $d$, and their coordinate distances to the left and right bubble horizons, $\ell_1$ and $\ell_2$, so that
\begin{subequations}
\label{binbubparam}
\begin{gather}
w_0 = -\ell_1 \,, \qquad
w_1 = 0 \,, \qquad
w_2 = 2M_1 \,, \qquad
w_3 = 2M_1 + d \,, \qquad \\
w_4 = 2M_1 + 2M_2 + d \,, \qquad
w_5 = 2M_1 + 2M_2 + d + \ell_2 \,.
\end{gather}
\end{subequations}
We then solve the equilibrium conditions~\eqref{equilbinarybubble} for $\ell_1$ and $\ell_2$, to find
\begin{equation}
\label{solell}
\ell_i = \frac{\sqrt{A_i+B_i^2}-B_i}{2M_1 M_2+d(M_1+M_2)} \,,
\end{equation}
where we have defined
\begin{align}
A_1 & = d (d + 2M_1) (d + M_2) (d + 2(M_1 + M_2)) (2M_1 M_2 + d (M_1 + M_2)) \,, \\
B_1 & = d^2 M_2 + 2 M_1 M_2 (M_1 + M_2) + d (M_1^2 + 3 M_1 M_2 + M_2^2) \,,
\end{align}
and $A_2$, $B_2$ are obtained by changing $1\leftrightarrow 2$.
Since $\ell_1$ and $\ell_2$ in~\eqref{solell} are manifestly positive when $M_1$, $M_2$, $d$ are positive, we have proven that there always exists a unique bubble, with suitably chosen position and size, that provides the necessary expansion to balance an arbitrary binary in static equilibrium (even if unstable).

It is interesting to observe that when the two black holes are very close, $d\ll M_1,M_2$, the bubble distance to them becomes
\begin{equation}
\ell_1, \ell_2 = d + O(d^2) \,,
\end{equation}
i.e., as expected, the bubble snugly hugs the binary.
When the black holes are instead far apart, $d\gg M_1,M_2$, we have
\begin{equation}
\label{farbin}
\ell_1, \ell_2 = \frac{d^{3/2}}{\sqrt{M_1+M_2}} \Bigl( 1+ O(d^{-1/2})\Bigr) \,,
\end{equation}
which we can easily understand.
The Newtonian gravitational potential between the black holes is
\begin{equation}
V_{g} \simeq - \frac{M_1+M_2}{d} \,,
\end{equation}
and the gravitational potential from the de Sitter-like expanding space between them is (for $\ell_{1,2}\simeq \ell$)
\begin{equation}
V_{exp}\simeq -\frac{d^2}{2\ell^2} \,,
\end{equation}
since $1/\ell^2$ acts like a cosmological constant\footnote{The two potentials can be read from $g_{tt}$ in the weak field regime.}.
Then~\eqref{farbin} follows from the equilibrium condition 
\begin{equation}
\frac{\partial(V_{g}+V_{exp})}{\partial d} = 0\,.
\end{equation}

\subsection{Black ring in equilibrium inside the expanding bubble}
\label{sec:ring+bubble}

Now we insert a static black ring inside a five-dimensional bubble of nothing.
Instead of the $(x,y)$ coordinates used in~\eqref{bring}, we will employ Weyl coordinates.
For the black ring, the explicit transformation can be found in~\cite{Emparan:2001wk}.

\begin{figure}
\begin{center}
\includegraphics[scale=1]{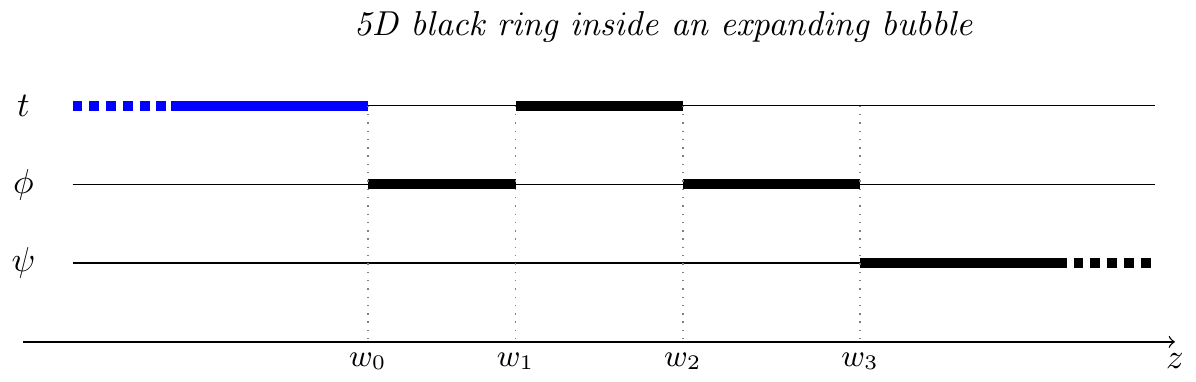}
\end{center}
\caption{{\small Rod diagram for the black ring inside the bubble of nothing.
A semi-infinite timelike rod for $z<w_0$ has been added to the ordinary black ring diagram.}}
\label{fig:ring-bubble}
\end{figure}

The rod diagram for the black ring is represented by the black lines in Fig.~\ref{fig:ring-bubble}, and we add the bubble by putting an extra pole and the blue line representing the bubble horizon.
Incidentally, this diagram is the double Wick-rotated version of the static black Saturn~\cite{Elvang:2007rd} (see also Fig.~\ref{fig:4diagrams}).

The metric corresponding to Fig.~\ref{fig:ring-bubble} is
\begin{subequations}
\begin{align}
g_{ab} dx^a dx^b &=
-\rho^2 \frac{\mu_1}{\mu_0\mu_2} {dt}^2
+ \frac{\mu_0\mu_2}{\mu_1\mu_3} {d\phi}^2
+ \mu_3 {d\psi}^2 \,, \\
f &= C_f
\frac{\mu_3 W_{01}^2 W_{03} W_{12}^2 W_{23}}{W_{02}^2 W_{13} W_{00} W_{11} W_{22} W_{33}} \,.
\end{align}
\end{subequations}
We have to eliminate conical singularities by tuning the parameters of the solution to satisfy~\eqref{nocone} at every spacelike rod.
If we choose $\Delta\psi=2\pi$, we find that~\eqref{nocone} is satisfied along the segment $z\in(w_3,\infty)$ by setting $C_f=1$.
Next, imposing~\eqref{nocone} on the $\phi$ direction along $z\in(w_0,w_1)$ and along $z\in(w_2,w_3)$, we obtain
\begin{subequations}
\label{ring-constraint}
\begin{align}
\frac{(w_1-w_0)^2(w_3-w_0)}{(w_2-w_0)^2} & = \frac{1}{2}\left(\frac{\Delta\phi}{2\pi}\right)^2 \,, \\
\frac{(w_3-w_0)(w_3-w_2)}{w_3-w_1} & = \frac{1}{2}\left(\frac{\Delta\phi}{2\pi}\right)^2 \,.
\end{align}
\end{subequations}
In order to solve these equations, we parametrise the rod endpoints as
\begin{equation}
w_0 = -\ell\,,\qquad w_1 = 0 \,, \qquad
w_2 = 2\mu R^2 \,, \qquad
w_3 = (1+\mu)R^2 \,.
\end{equation}
Here $\ell$ characterizes the bubble size, while $\mu\in [0,1)$ and $R$ are the same parameters for the shape and radius of the ring as in~\eqref{bring}.
Eqs.~\eqref{ring-constraint} are solved with
\begin{equation}
\ell = R^2\Bigl(1-\mu+\sqrt{1-\mu^2}\Bigr) \,,
\end{equation}
and\footnote{We could absorb a scale $\propto R$ in the definition of $\phi$ to make it dimensionless, as we have done before.}
\begin{equation}
\biggl(\frac{\Delta\phi}{2\pi}\biggr)^2 =
2R^2 \Bigl(2+\sqrt{1-\mu^2}\Bigr) \frac{1-\mu}{1+\mu} \,.
\end{equation}
Thus, we can always choose, in a unique way, the bubble size $\ell$ so as to balance into equilibrium an arbitrary static black ring.

To finish this section, we shall mention that, with a straightforward exercise in rodology, which we leave to the reader, one can insert the five-dimensional binary of Fig.~\ref{fig:5dbbh}(a) inside the bubble of Fig.~\ref{fig:5dbbh}(b), and then obtain the corresponding solution (which is a limit of the ones in~\cite{Tan:2003jz}).
Given our previous analyses, it is natural to expect, and consistently with parameter counting, that the bubble parameters can be adjusted to balance an arbitrary binary of this kind.

\section{Other configurations}
\label{sec:otherconfig}

We can extend the discussion of the previous sections to more general configurations, and play with the rods to move from one solution to another.
There are plenty of examples that can be considered, both in four and five dimensions, and even in higher dimensions~\cite{Emparan:2001wk}.
We will consider some of them, just to give a taste of the many possibilities that are offered by the rod diagram machinery.
The limits presented below on the rods diagrams work faithfully on the corresponding metrics.

\subsection{Four dimensions}

\begin{figure}
\includegraphics[scale=0.85]{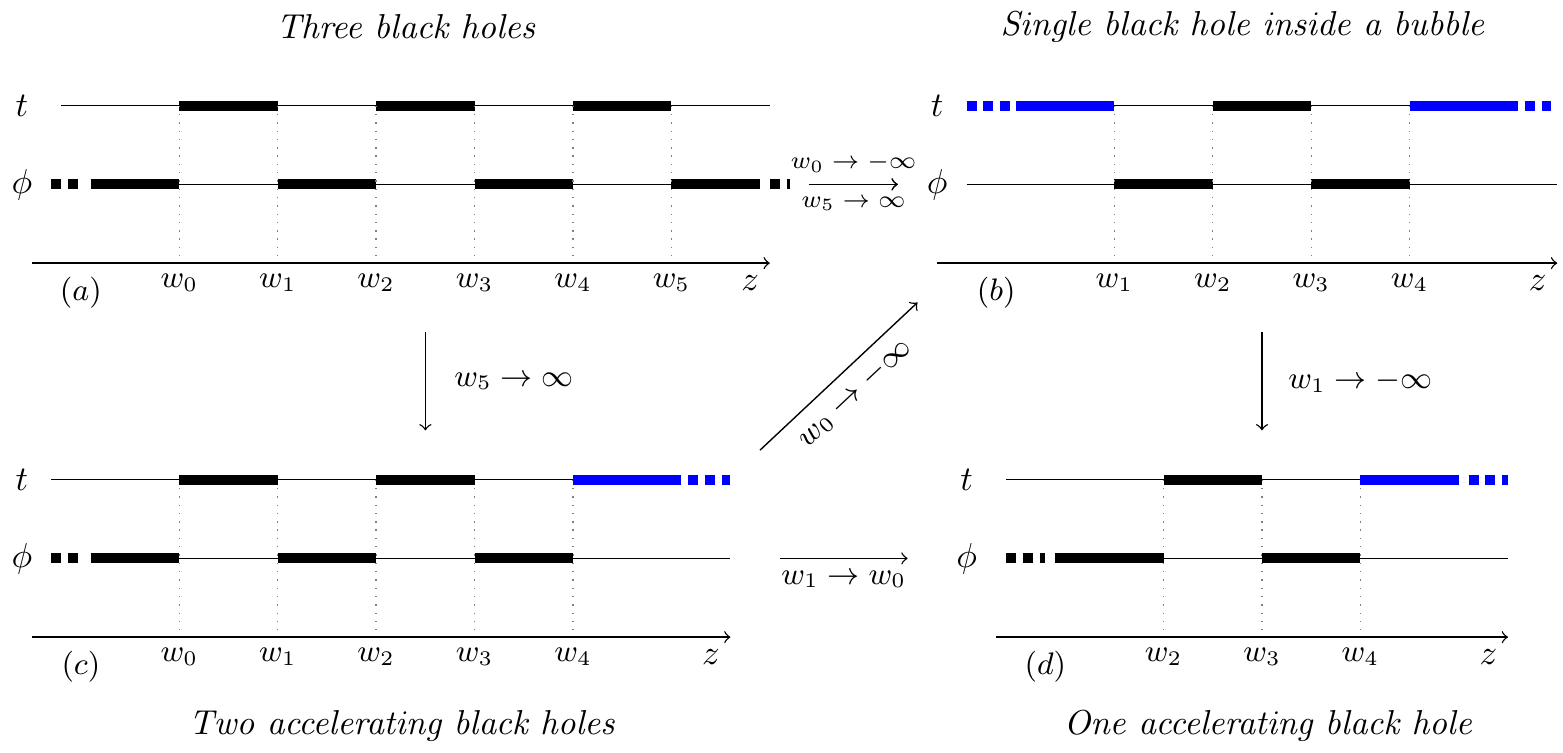}
\caption{{\small $(a)$ Rod diagram for a collinear three-black hole system (an Israel--Khan solution).
The limit $w_0 \to -\infty$ and $w_5 \to \infty$ gives $(b)$ the single black hole in the expanding bubble.
When sending $w_5\to\infty$ in $(a)$ we obtain
$(c)$ two accelerating black holes.
Collapsing one timelike rod in the latter gives $(d)$ the C-metric for a single accelerating black hole.
The rod limits commute, so from the two accelerating black holes in $(c)$, the limit $w_0\to-\infty$ gives $(b)$ a single black hole in a bubble.}}
\label{fig:3bh--1bh+bubble}
\end{figure}

One obvious extension of the binary system studied above is the three-black hole configuration contained in the Israel--Khan solution and represented in Fig.~\ref{fig:3bh--1bh+bubble}(a).
To get the Schwarzschild black hole inside the bubble of nothing, we extend the peripheral timelike rods to infinity, taking the limits $w\to-\infty$ and $w_5\to\infty$.

From the black hole in the bubble we can also generate a metric describing a point-like Curzon--Chazy particle embedded in the bubble.
The procedure is similar to the one used to obtain the Bonnor--Swaminarayan solution from an accelerating binary black hole system~\cite{Astorino:2021rdg}.

Moreover it is very clear, in the 4D setting, how to generate accelerating black hole metrics from the black holes in the bubble, for any number of collinear black holes.
It is sufficient to push away only one of the two poles defining the bubble, for instance $w_0 \to -\infty$ in the binary configuration of Sec.~\ref{sec:binary+bubble}.
In Fig.~\ref{fig:3bh--1bh+bubble} we picture the single black hole case.
The limiting process, however, introduces irremovable conical singularities, unless an external background field is introduced, as in~\cite{Astorino:2021rdg}.
It is clear that this procedure cannot be pursued in 5D.
In that case, there is only a single rod determining the bubble horizon, but more importantly, the five-dimensional C-metric for a uniformly accelerating black hole would have different symmetry ($SO(3)$ rotations, rather than $U(1)^2$) and not be in the Weyl class of solutions.

\subsection{Five dimensions}

\begin{figure}
\includegraphics[scale=0.85]{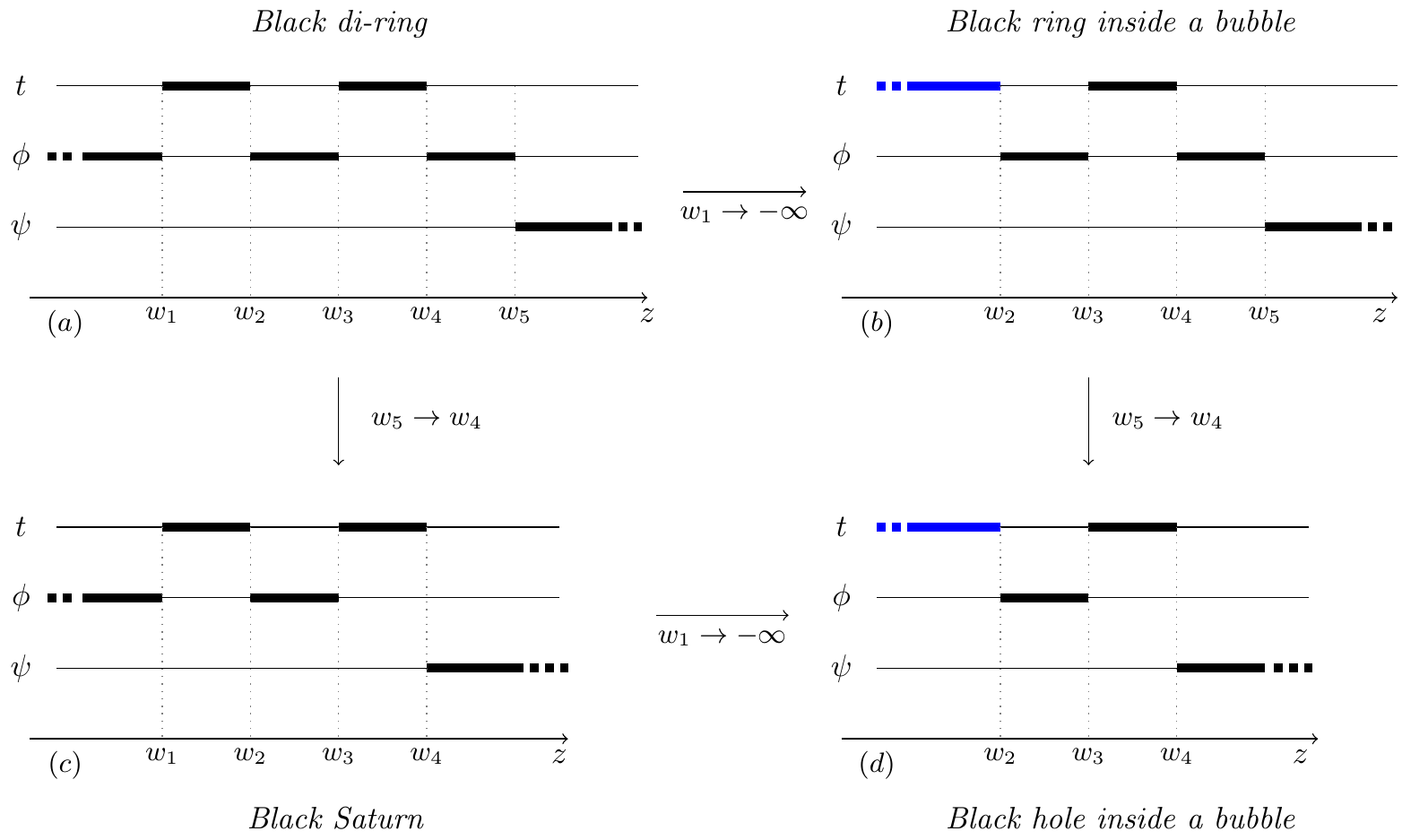}
\caption{{\small $(a)$ Rod diagram for a coaxial double black ring system.
The limit $w_1 \to -\infty$ gives $(b)$ the single black ring in the expanding bubble.
The limit of $(a)$ for $w_5\to w_4$ gives $(c)$ the black Saturn.
Its limit for $w_1 \to -\infty$ gives $(d)$ the five-dimensional black hole in an expanding bubble.
The rod limits commute, so the latter diagram can also be obtained from $(b)$ for $w_5\to w_4$.}}
\label{fig:4diagrams}
\end{figure}

It is interesting that all the five-dimensional configurations studied in this paper can be obtained by performing limits in the black di-ring configuration of Fig.~\ref{fig:4diagrams}(a).

For instance, to recover the black ring-bubble of nothing of Fig.~\ref{fig:4diagrams}(b) (which also corresponds to Fig.~\ref{fig:ring-bubble}), we simply send $w_1\to-\infty$ in the black di-ring diagram.
Furthermore, one can also obtain the black hole inside the bubble from the latter by taking $w_5\to w_4$ to remove a spacelike finite rod.

On the other hand, if we take $w_5\to w_4$ in the di-ring diagram, we recover the black Saturn~\cite{Elvang:2007rd} of Fig.~\ref{fig:4diagrams}(c).
From this diagram, we can send $w_1\to-\infty$ to obtain the 5D black hole-bubble of nothing, which corresponds to the superposition of the two diagrams of Fig.~\ref{fig:5D-1bh-1bubble}.

\section{Outlook}

Black holes and bubbles of nothing are some of the most elementary solutions in General Relativity, and in this chapter we have argued that their properties are closely interrelated.
By revealing how bubbles are present in black hole systems, we have learned that the spacetime expansion in the bubble is driven by the same phenomenon that makes the volume inside a black hole grow.

The basic idea is simple enough to lend itself to easy generalization.
Whenever a small gap region appears between black hole horizons in a maximally extended geometry, it will contain an expanding bubble of nothing.
The bubble is a minimal cycle that links the Einstein--Rosen bridges of the system.
In the simplest instance, namely, the four-dimensional black hole binary in Sec.~\ref{subsec:bbhbub}, the topology of a Cauchy slice is $S^1\times S^2-\{0\}$ (the point at infinity is removed), and the bubble is the minimal $S^1$ in it.
Similarly, for the black ring, the spatial topology is $S^2\times S^2-\{0\}$, and the bubble is the minimal $S^2$.
In the more general ``ringoids'' of~\cite{Kleihaus:2009wh,Kleihaus:2010pr} with spatial topology $S^{d-3}\times S^2-\{0\}$ we find $S^{d-3}$ bubbles.
We have even considered more complex bubbles topologies, such as the double $S^2$ bubble in Sec.~\ref{subsec:5dbbh}, and we have identified that a collinear black hole binary in $d$ dimensions, with Cauchy slices that are $S^1\times S^{d-2}-\{0\}$, must have an $S^1$ bubble.

Such solutions for binary black holes are not known explicitly in arbitrary dimensions, but we can easily find configurations with two disconnected horizons which can be regarded, in the sense explained above, as possessing expanding bubbles of nothing.
The Schwarzschild--de Sitter solution
\begin{equation}
{ds}^2 = -\biggl( 1-\frac{\mu}{r^{d-3}}-\frac{r^2}{L^2}\biggr) {dt}^2 + \frac{{dr}^2}{1-\frac{\mu}{r^{d-3}}-\frac{r^2}{L^2}} + r^2 d\Omega_{d-2} \,,
\end{equation}
with
\begin{equation}
0<\mu<\mu_N\equiv\frac{2}{d-3}\biggl(\frac{d-3}{d-1}\biggr)^\frac{d-1}{2} L^{d-3} \,,
\end{equation}
has a cosmological horizon and a black hole horizon.
In the maximal analytic extension (and identifying regions beyond the horizons to form a spatial circle) the spatial sections have topology $S^1\times S^{d-2}$.
The $S^1$ expands in time, inside the black hole and in the time-dependent region beyond the de Sitter horizon.
The analogue of the bubble limit is the limit $\mu\to \mu_N$, in which (after rescaling $t$) we recover the Nariai solution
\begin{equation}
{ds}^2 = \frac{L^2}{d-1}\biggl[-(1-\xi^2){dt}^2 + \frac{{d\xi}^2}{1-\xi^2} + (d-3) d\Omega_{d-2} \biggr] \,,
\end{equation}
with horizons at $\xi=\pm 1$.
We can change coordinates in this metric (see footnote~\ref{foot:change}) to the form
\begin{equation}
{ds}^2 = \frac{L^2}{d-1}\Bigl[-{dT}^2 +\cosh^2 T {d\chi}^2 + (d-3) d\Omega_{d-2} \Bigr] \,,
\end{equation}
with $0\leq \chi\leq 2\pi$.
Here we recognise the essential features of the $n=1$ circular bubble~\eqref{bon1}, only that now the inhomogeneous, non-compact $(r,\phi)$ cigar is replaced by a round $S^{d-2}$.
Thus, bubble-of-nothing-like expansion is indeed pervasive and connected in wide generality to the phenomenon of spacetime expansion.

It is now clear, given the plethora of black hole topologies and multi-black hole configurations that are possible in higher dimensions~\cite{Emparan:2009vd}, that we can expect a large variety of expanding bubbles of nothing, even in vacuum gravity.
Many of them are unlikely to admit a closed exact solution, but it is intuitively useful to first conceive of them as black hole configurations, as this helps identify new possibilities. 
It would be interesting to know how general the converse is, that is, whether for any expanding bubble one can find a black hole configuration that contains it as a limit.

We have also proven, with explicit examples, that the bubble expansion acts on gravitating systems in much the same way as de Sitter-type inflation:
it counteracts the gravitational attraction between localized objects and allows novel static multi-black hole configurations.
Again, this phenomenon is likely valid for all the expanding bubbles that we have mentioned above, and more generally for other bubbles.
 
Our arguments show that the expanding bubble of nothing is already present in the binary or black ring \emph{even before taking the small-gap limit}, in the sense that there is a minimal cycle that links the system of Einstein--Rosen bridges and which expands because it stretches inside the black holes.
Taking the small-gap limit makes the bubble more symmetric and uniform, and its expansion becomes asymptotically uniform and eternal, since in the limit the interior black hole singularity is pushed away to infinitely late time.
If the black hole system were of finite size, or if it were to merge or collapse, the duration of the expansion would instead be limited, ending on a singularity.
But expanding bubbles of nothing, in the above sense, seem pervasive in black hole systems with multiple or non-spherical horizons\footnote{However, this does not mean that they must admit a good limit to a bubble solution;
we already mentioned in Sec.~\ref{subsec:ringtobubble} that the equilibrium rotating black ring does not admit it, even though at any finite radius it has a bubble in the sense explained above.}.

Does this mean that we should expect bubbles of nothing in astrophysical, dynamical binaries more realistic than the static ones we have studied?
Unfortunately, the answer is no.
The topology of a binary where the black holes formed from collapsing matter is different than in the maximal analytic extensions we have considered.
Collapsing black hole geometries do not have bifurcation surfaces nor Einstein--Rosen bridges.
Even though space expands inside a collapsing black hole, the topology of the Cauchy slices is trivial, and these binaries will not contain any minimal cycle.

However, even if expanding bubbles of nothing may not be present in the sky above, their connection to more conventional black hole systems provides a new, illuminating perspective on their properties and makes them seem more accessible.
Since they behave in many ways like de Sitter space, but without a cosmological constant, and with non-compact horizons, they may provide new venues in which to investigate the holographic description of expanding spacetime, possibly exploiting their relation to Einstein--Rosen bridges and the interiors of black hole systems.

\chapter{Black holes in a swirling universe}
\label{chap:swirl}
\thispagestyle{plain}

In this Chapter we will explore the ``magnetic'' version of the Ehlers transformation presented in Sec.~\ref{sec:ernst} in the context of vacuum General Relativity.
The effect of such a symmetry map has not been investigated until now, at best of our knowledge.
Actually, the spacetime resulting from the application of the modified Ehlers map to the Minkowski metric was presented in an Appendix of~\cite{Gibbons:2013yq}, but it was not studied and its physical properties were not unraveled.

The purpose of this Chapter is to investigate the last unexplored Lie point symmetry of the $SU(2,1)$ group and to explore the properties of the spacetime that it generates.
We will see that such a transformation embeds any given stationary and axisymmetric seed spacetime into a rotating background, which we will dub ``swirling universe'', for its peculiar characteristic.
Indeed, the background can be interpreted as a gravitational whirlpool, and its frame dragging turns a static seed solution into a stationary metric.
The resulting spacetime really looks similar to the Taub--NUT metric, however we will show that it does not share the problematic features of its ``electric'' counterpart:
in fact, our new spacetime does not possess any Misner string, nor is affected by the presence of conical singularities.
The manifold is completely regular, apart from the presence of the usual black hole curvature singularity.

Firstly, we generate the new metric via the ``magnetic'' Ehlers map.
Then, we test the consequences of the transformation on the spherical symmetric black hole, which will be our seed metric, as done with the symmetries previously analysed in the literature.
In this way we generate a novel and analytic exact solution of the Einstein equations, which generalises and deforms the Schwarzschild spacetime.
Of course, because of the well-known no-hair theorems for black hole in four-dimensional General Relativity, the new solution can be a black hole only by renouncing to asymptotic flatness, similarly to black holes embedded in the external electromagnetic field of Melvin universe~\cite{Melvin:1963qx,ErnstMelvin,Gibbons:2013yq}.
We also present the generalisation to the Kerr black hole in the rotating background.
Finally, we conclude the Chapter by establishing a connection (via a double-Wick rotation) between our background metric and the flat Taub--NUT spacetime.
This Chapter is based on~\cite{Astorino:2022aam}.

\section{Generation of the solution}
\label{sec:generation}

As explained above, we consider the magnetic LWP metric of Eq.~\eqref{magnetic-metric},
\begin{equation}
{ds}^2 =
\bar{f} (d\varphi - \bar{\omega} dt)^2
+ \bar{f}^{-1} \bigl[ e^{2\bar{\gamma}} \bigl( {d\rho}^2 + {dz}^2 \bigr)
- \rho^2 {dt}^2 \bigr] \,,
\end{equation}
and the Ehlers transformation~\eqref{ehlers}
\begin{equation}
\label{ehlers-swirl}
\ernst' = \frac{\ernst}{1 + i\jmath\ernst} \,,
\end{equation}
where we dubbed with $\jmath$ the map parameter.

First of all we have to choose the seed.
We start with the Schwarzschild black hole, whose metric, in spherical coordinates, reads
\begin{equation}
\label{schw}
{ds}^2 = - \biggl(1- \frac{2m}{r} \biggr) {dt}^2 + \frac{{dr}^2}{1- \frac{2m}{r}} + r^2 {d\theta}^2 + r^2 \sin^2 \theta {d \varphi}^2 \,.
\end{equation}
The most convenient coordinates for the generating methods, in this case, are the spherical ones $(r,\theta)$, related to the Weyl cylindrical coordinates by
\begin{equation}
\rho = \sqrt{r^2-2mr} \sin\theta \,, \quad
z = (r-m) \cos\theta \,.
\end{equation}
The line element of the LWP metric~\eqref{magnetic-metric} in spherical coordinates reads
\begin{equation}
\label{LWP-m-r-theta}
{ds}^2 =  \bar{f} (d\varphi - \bar{\omega} dt)^2 + \bar{f}^{-1} \biggl[ -\rho^2 {dt}^2 + e^{2\bar{\gamma}} (r^2-2mr+m^2\sin^2\theta) \biggl( \frac{{dr}^2}{r^2-2mr} + {d\theta}^2 \biggr) \biggr] \,.
\end{equation}
Comparing the seed~\eqref{schw} to the above metric we can identify the seed structure functions
\begin{equation}
\bar{f}(r,\theta) = r^2 \sin^2\theta \,, \quad
\bar{\omega}(r,\theta) =0 \,.
\end{equation}
The value of $\bar{\gamma}$ is not fundamental because it is invariant under Ehlers transformations, however we explicit it for completeness
\begin{equation}
\label{gamma0}
\bar{\gamma}_0(r,\theta) = \frac{1}{2} \log \biggl(\frac{r^4 \sin^2 \theta}{r^2 - 2mr + m^2\sin^2\theta} \biggr) \,.
\end{equation}
From definition of $\chi$ in~\eqref{grav-potential}, it is clear that $\bar{\chi}$ is at most constant, but that constant can be reabsorbed via a coordinate transformation.
Therefore without loss of generality we can choose $\bar{\chi}=0$.
Finally, the seed Ernst gravitational potential takes the form
\begin{equation}
\ernst(r,\theta) = \bar{f}(r,\theta) \,.
\end{equation}
The new solution, expressed in terms of the complex potential, is generated via the Ehlers transformation~\eqref{ehlers-swirl}, which gives
\begin{equation}
\label{sol-Er}
\ernst' = \frac{\ernst}{ 1 + i \jmath \ernst}
= \frac{r^2 \sin^2 \theta}{1+ i \jmath\, r^2 \sin^2\theta} \,.
\end{equation}
Note that in case we had used the LWP metric defined in~\eqref{electric-metric}, we would had obtained, via the Ehlers transformation acting on the Schwarzschild seed, the Taub--NUT spacetime, as explained in Chapter~\ref{chap:gentech}.

The solution, in metric form, is extracted from the definition of the transformed Ernst potential $\ernst$.
Hence, according to $\ernst = f + i \chi$, we find
\begin{equation}
f(r,\theta) = \frac{r^2 \sin^2\theta}{1 + \jmath^2r^4\sin^4\theta} \,, \qquad
\chi(r,\theta) = \frac{\jmath \, r^4 \sin^4\theta}{1 + \jmath^2 r^4 \sin^4\theta} \,.
\end{equation}
$\omega$ has to be found from the definition of $\chi$, as in~\eqref{grav-potential}.
The result is
\begin{equation}
\omega(r,\theta) = 4 \jmath (r-2m) \cos\theta + \omega_0 \,,
\end{equation}
where $\omega_0$ is an integration constant related to the choice of reference frame.
Thus, recalling that $\gamma$ is not affected by the Ehlers map, the full new metric is
\begin{equation}
\label{bh-rot-universe}
\begin{split}
{ds}^2 & = F(r,\theta)
\biggl[ - \biggl( 1 - \frac{2m}{r}\biggr) {dt}^2
+ \frac{dr^2}{1 - \frac{2m}{r}} + r^2 d\theta^2 \biggr] \\
&\quad + \frac{r^2 \sin^2 \theta}{F(r,\theta)}
\biggl\{ d\varphi + \big[4\jmath(r-2m)\cos \theta + \omega_0 \big] \ dt \biggr\}^2 \,,
\end{split}
\end{equation}
where we have defined the function
\begin{equation}
F(r,\theta) \coloneqq 1+\jmath^2r^4 \sin^4 \theta \,,
\end{equation}
We can immediately observe that the new metric~\eqref{bh-rot-universe} represents a non-asymptotically flat deformation of the Schwarzschild black hole.
Its structure is quite similar to the Schwarzschild--Melvin spacetime~\cite{ErnstMelvin}, and indeed the magnetic Ehlers map that we have used works in a similar fashion as the Harrison transformation.
For this reason we do not expect that the new parameter can be considered as hair nor as conserved charge of the black hole.  
The physical description of~\eqref{bh-rot-universe} will be analysed in detail in the next Section.

Starting with a more general seed we can obtain generalisations of the metric built above.
In Sec.~\ref{sec:kerr} we embed the Kerr black hole in the swirling background, while in Appendix~\ref{app:zipoy} we generate the Zipoy--Voorhees extension of the spacetime~\eqref{bh-rot-universe}.

\section{Analysis of the background spacetime}
\label{sec:analysis-background}

In the interpretation of the new spacetime~\eqref{bh-rot-universe}, a fundamental point comes from the physical meaning of the new parameter $\jmath$, which defines the behaviour of the gravitational background and, at best of our knowledge, is  unknown.
Thus, we firstly analyse the background metric obtained by turning off the mass parameter $m$ in Eq.~\eqref{bh-rot-universe} and then, in Sec.~\ref{sec:analysis-blackhole}, we study the full black hole solution in his surrounding universe.

When the mass parameter $m$ vanishes the black hole disappears and we are left with the rotating gravitational background only
\begin{equation}
\label{background-metric}
{ds}^2 = F
\bigl( - {dt}^2 + {dr}^2 + r^2 {d\theta}^2 \bigr) + F^{-1} r^2 \sin^2 \theta
( d\varphi + 4\jmath \, r\cos \theta \ dt )^2 \,.
\end{equation}
In cylindrical coordinates
\begin{equation} 
\label{cyly-coord}
\rho = r \sin \theta \,, \quad
z = r \cos \theta \,, 
\end{equation}
the background takes the simpler form
\begin{equation}
\label{background-cyl}
{ds}^2 = \bigl(1+\jmath^2 \rho^4\bigr)\bigl(-{dt}^2 + {d\rho}^2 + {dz}^2\bigr)
+ \frac{\rho^2}{1+\jmath^2\rho^4}(d\varphi + 4\jmath z dt)^2 \,.
\end{equation}
Such a metric has the very same form as the one presented in Appendix C of~\cite{Gibbons:2013yq}:
however, it was not studied in that reference.

The spacetime~\eqref{background-cyl} is characterised by the Killing vectors
\begin{equation}
\partial_t \,, \quad
\partial_\varphi \,, \quad
z\partial_t + t\partial_z - 2\jmath\bigl(t^2+z^2\bigr)\partial_\varphi \,, \quad
\partial_z - 4\jmath t \partial_\varphi \,,
\end{equation}
and by the Killing--Yano form
\begin{equation}
-4\jmath \rho z \, dt \wedge d\rho
+ \jmath \rho^2 \bigl(1 + \jmath^2 \rho^4) \, dt \wedge dz
+ \rho \, d\rho \wedge d\varphi \,.
\end{equation}
It belongs to the Petrov type D class~\cite{Stephani:2003tm}, and its Newman--Penrose spin coefficient is equal to zero:
these features allow us to infer that the metric~\eqref{background-cyl} belongs to the Kundt class (cf. Table 38.9 of~\cite{Stephani:2003tm}).
We can indeed explicitly express the background metric~\eqref{background-cyl} in the standard Kundt form, by performing the rescaling $t\to \jmath t$ and the change of coordinates
\begin{equation}
q = 2\jmath z \,, \quad
p = \rho^2 \,.
\end{equation}
Metric~\eqref{background-cyl} then boils down to (after a rescaling of the conformal factor)
\begin{equation}
\label{kundt}
{ds}^2 = (\gamma^2 + p^2) \bigl( -{dt}^2 + {dq}^2 \bigr)
+ \frac{\gamma^2 + p^2}{\gamma^2 p} {dp}^2
+ \frac{\gamma^2 p}{\gamma^2 + p^2} ( d\varphi + 2\gamma q dt )^2 \,,
\end{equation}
where we have defined $\gamma=1/\jmath$.
This metric is equivalent to (16.27) of~\cite{Griffiths:2009dfa}, once we put
$m=e=g=\Lambda=\alpha=\epsilon_2=k=0$, $\epsilon_0=1$ and $n=\gamma^2/2$.
One can check that the consistency constraints of~\cite{Griffiths:2009dfa} are indeed satisfied.

The metric~\eqref{kundt}, despite being known for a long time (it was discovered by Carter in~\cite{Carter:1968ks}), does not have a clear physical interpretation.
In particular, the physical significance of the parameter $\gamma$ (i.e.~$\jmath$) is unknown:
it has been called ``anti-NUT'' parameter by Pleba\'nski in~\cite{Plebanski:1975xfb} because of its resemblance with the NUT parameter in the Pleba\'nski--Demia\'nski spacetime~\cite{Plebanski:1976gy}\footnote{Actually, there exists a connection between the rotating parameter $\jmath$ and the NUT parameter $\ell$ that will be exploited in Sec.~\ref{sec:double-wick}.}.
A generalisation in the presence of the cosmological constant and some possible interpretations of this background are given in Sec.~\ref{sec:double-wick}.

An interesting limit is given by $p\to\infty$:
by putting $\gamma=0$ after the rescaling $t\to\gamma^{-2/3}t$, $q\to\gamma^{-2/3}q$, $\varphi\to\gamma^{-2/3}\varphi$ and $p\to\gamma^{2/3}p$, one finds
\begin{equation}
{ds}^2 = p^2 \bigl( -{dt}^2 + {dq}^2 \bigr)
+ p\, {dp}^2
+ \frac{1}{p} {d\varphi}^2 \,.
\end{equation}
This is nothing but the Levi-Civita metric~\cite{Levi1919} (cf.~(10.9) of~\cite{Griffiths:2006tk}), in the limiting case when $\sigma=1/4$, which is locally isometric to the asymptotic form of the Melvin spacetime, as shown in~\cite{Bonnor1953,Bonnor1954,Thorne1965};
another connection to the magnetic universe will be explored in Sec.~\ref{sec:double-wick}.
Finally, the above limit can also be expressed in the Kasner-like form~\cite{Kasner1921}.

In order to gain some physical perspective is useful to investigate, in some detail, the properties of the background metric by inspecting its geodesics.

\subsection{Geodesics in the background spacetime}

We define the following geodesic Lagrangian from the background metric~\eqref{background-cyl}
\begin{equation}
\label{lag-background}
\mathscr{L} = \bigl(1+\jmath^2 \rho^4\bigr) \bigl(-\dot{t}^2 + \dot{\rho}^2 + \dot{z}^2\bigr)
+ \frac{\rho^2}{1+\jmath^2\rho^4}\bigl(\dot{\varphi} + 4\jmath z \dot{t}\bigr)^2 \,,
\end{equation}
where the dots stand for the derivatives with respect to an affine parameter $s$.
We can define, via the Killing vectors $\xi = \partial_t$ and $\Phi = \partial_\varphi$, the standard conserved quantities
\begin{equation}
\label{conserved}
-E \coloneqq g_{\mu \nu}u^\mu \xi^\nu \,, \quad
L \coloneqq g_{\mu \nu}u^\mu \Phi^\nu \,,
\end{equation}
where $u^\mu$ is the four-momentum of the test particle, $E$ is the energy and $L$ is the angular momentum.
The explicit definitions for the conserved quantities and the resulting Lagrangian can be found in Appendix~\ref{sec:appback}

The equations of motion derived from the Lagrangian are quite involved.
Qualitative results can be obtained from the normalisation of the four-momentum, i.e.~from equation
$u^\mu u_\mu = \chi$,
with $\chi = -1$ for timelike geodesics and $\chi = 0$ for null geodesics.
The resulting normalisation equation is~\eqref{4mom-BKG}.
For large values of $\rho$ and for fixed $z$, it follows from such equation that
\begin{equation}
\dot{\rho}^2 \approx L^2 \rho^{-2} \,,
\end{equation}
which has solution $\rho(s)\propto \sqrt{2L s}$:
this means that $\rho$ is not limited as the affine parameter grows.

\begin{figure}
\centering
\hspace{-0.2cm}
\includegraphics[width=6cm]{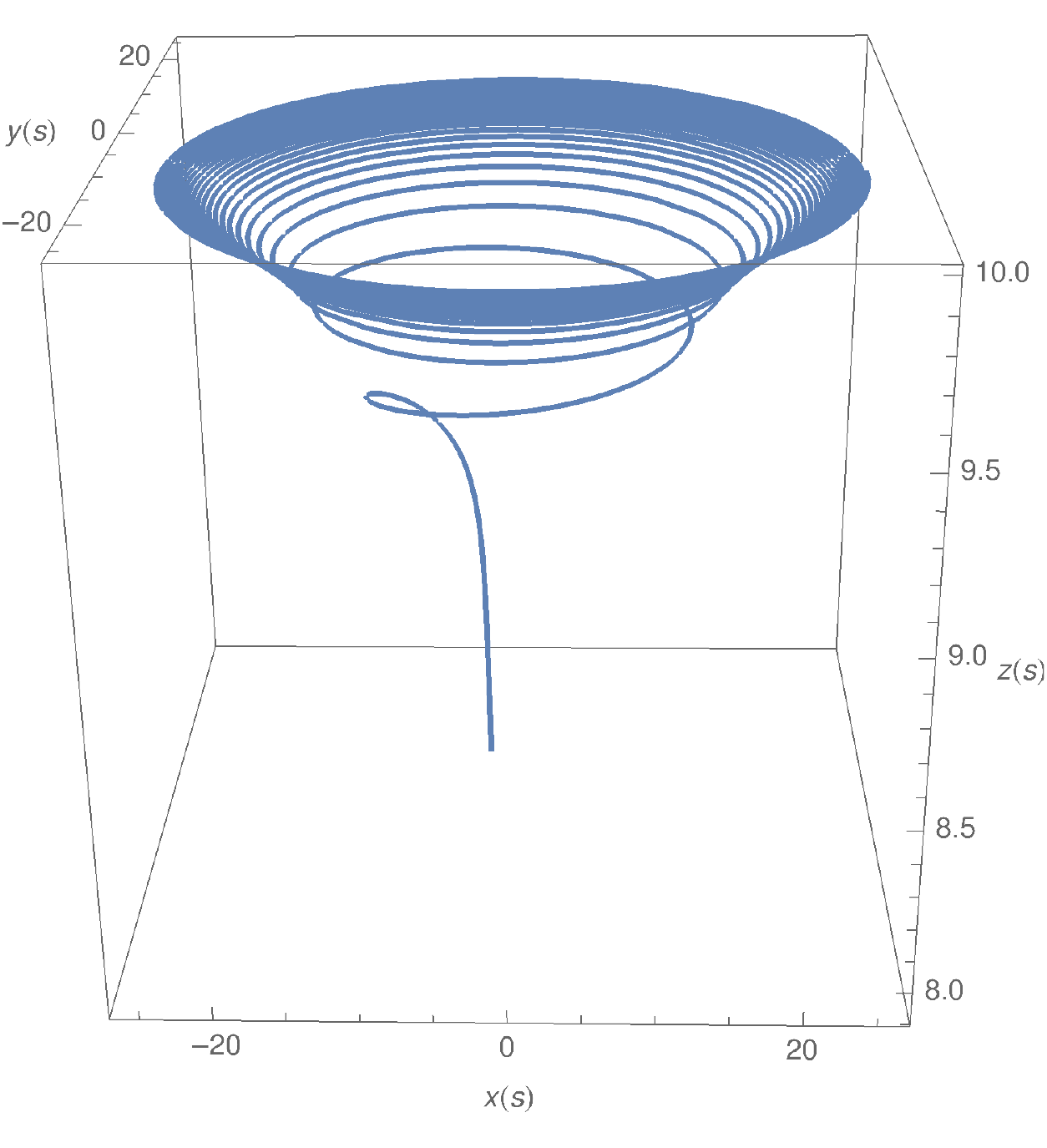}
\caption{\small Geodesic motion of a test particle in the gravitational background for $E=1$, $L=1$, $\jmath=0.1$.}
\label{fig:spiral3d}
\end{figure}

We are interested in analysing the behaviour of $z$ as $\rho$ grows.
We find $\dot{z}^2 = 0$ by letting $\rho\to\infty$ in equation~\eqref{4mom-BKG}, therefore the coordinate $z$ reaches a constant value as $\rho$ approaches infinity.
Moreover, the equation defining $L$, for large values of $\rho$, gives $\phi \approx c^2 L^2 s^2$.
Combining this with the approximate equation for $\rho$, allows one to get the polar equation
$r \approx \sqrt{\frac{2}{c}} \phi^{1/4}$.
Such an equation is the polar form of the generalised Archimedean spiral with exponential 1/4.
Therefore we expect that a geodesic test particle follows a spiral-like path in the $(x,y)$ plane and that it moves toward a constant value of $z$.
These results are in good agreement with the numerical evaluations, as can be observed from the plot in Fig.~\ref{fig:spiral3d}, which shows the trajectory of a test particle in the $(x,y,z)$ space, where $x=\rho\cos\theta$, $y=\rho\sin\theta$.

The statement that $z$ reaches a constant value can be verified by using the equation of motion for $z$
\begin{equation}
4 \jmath^2 \rho^3\dot{\rho} \dot{z} + (\jmath^2\rho^4+1) \ddot{z}
= \frac{4\rho^2 \jmath \dot{t} (\dot{\varphi} +4 \jmath z \dot{t})}{\jmath^2\rho^4+1} \,.
\end{equation}
By inspecting the equations for $\dot{t}$ and $\dot{\varphi}$ in Appendix~\ref{sec:appback}, one can notice that $\dot{t} \propto 1/\rho^4$ and $\dot{\varphi} \propto \rho^2$ as $\rho\to\infty$, therefore the r.h.s.~of the latter equation can be neglected for large values of $\rho$:
\begin{equation}
4 \jmath^2\rho^3 \dot{\rho} \dot{z} +  (\jmath^2\rho^4+1) \ddot{z} = 0 \,.
\end{equation}
Moreover, $1+\jmath^2 \rho^4 \approx \jmath^2 \rho^4$ as $\rho$ approaches infinity.
By using the approximation $\rho(s) \approx \sqrt{2Ls}$ found above, the equation becomes
\begin{equation}
 2 \dot{z} + s \ddot{z} = 0 \,,
\end{equation}
whose solution is
\begin{equation}
z(s) = \frac{D}{s} + C \,,
\end{equation}
where $C$, $D$ are integration constants.
This result clearly shows that $z$ becomes constant as $s$ approaches infinity.

\section{Schwarzschild black hole in a swirling universe}
\label{sec:analysis-blackhole}

\subsection{Physical properties}

The full black hole metric~\eqref{bh-rot-universe}, that we report here for convenience
\begin{equation}
\label{swirling-bh}
\begin{split}
{ds}^2 & = F(r,\theta)
\biggl[ - \biggl( 1 - \frac{2m}{r}\biggr) {dt}^2
+ \frac{dr^2}{1 - \frac{2m}{r}} + r^2 d\theta^2 \biggr] \\
&\quad + \frac{r^2 \sin^2 \theta}{F(r,\theta)}
\biggl\{ d\varphi + \big[4\jmath(r-2m)\cos \theta + \omega_0 \big] \ dt \biggr\}^2 \,,
\end{split}
\end{equation}
with $F(r,\theta) = 1+\jmath^2r^4 \sin^4 \theta$, is a two parameters metric, with $m$ and $\jmath$ related to the mass of the black hole and the angular velocity of the background, respectively.

In view of the previous Section, the spacetime~\eqref{swirling-bh} can be interpreted as a Schwarzschild black hole embedded into a swirling background.
The main casual structure is similar to the Schwarzschild case, as can be readily understood by looking at some $\theta=$ constant slices of the conformal diagram.
For instance, the cases $\theta=\{0,\pm \pi\}$ precisely retrace the static spherically symmetric black hole.

\begin{figure}
\centering
\hspace{-1cm}
\subfloat[\hspace{0.2cm} $m=1$, $\jmath=0.05$]{{\includegraphics[scale=0.2]{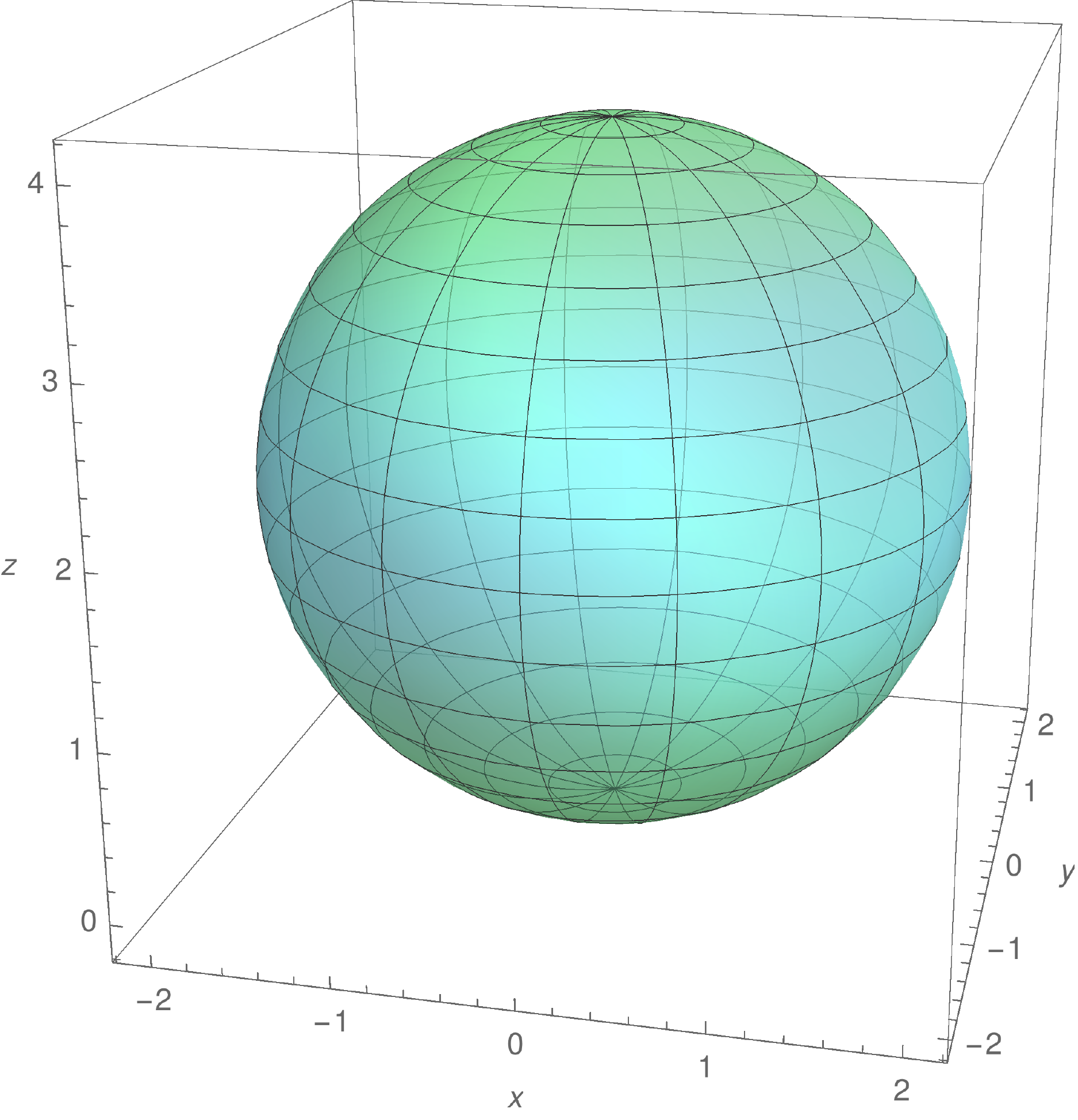}}}
\subfloat[\hspace{0.2cm} $m=1$, $\jmath=0.1$]{{ \hspace{0.5cm} \includegraphics[scale=0.2]{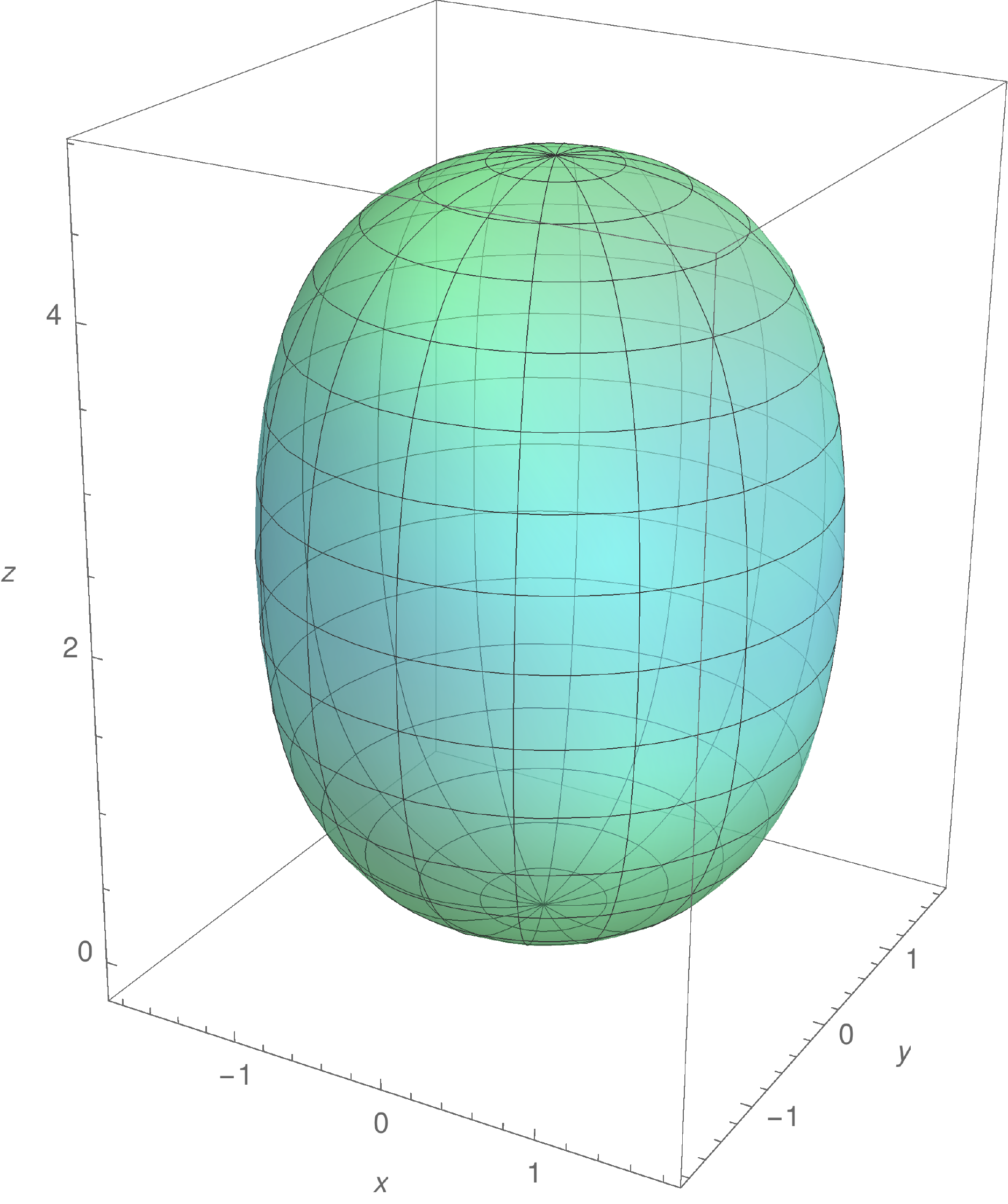}}}
\subfloat[\hspace{0.5cm} $m=1$, $\jmath=0.3$]{{ \hspace{1cm} \includegraphics[scale=0.2]{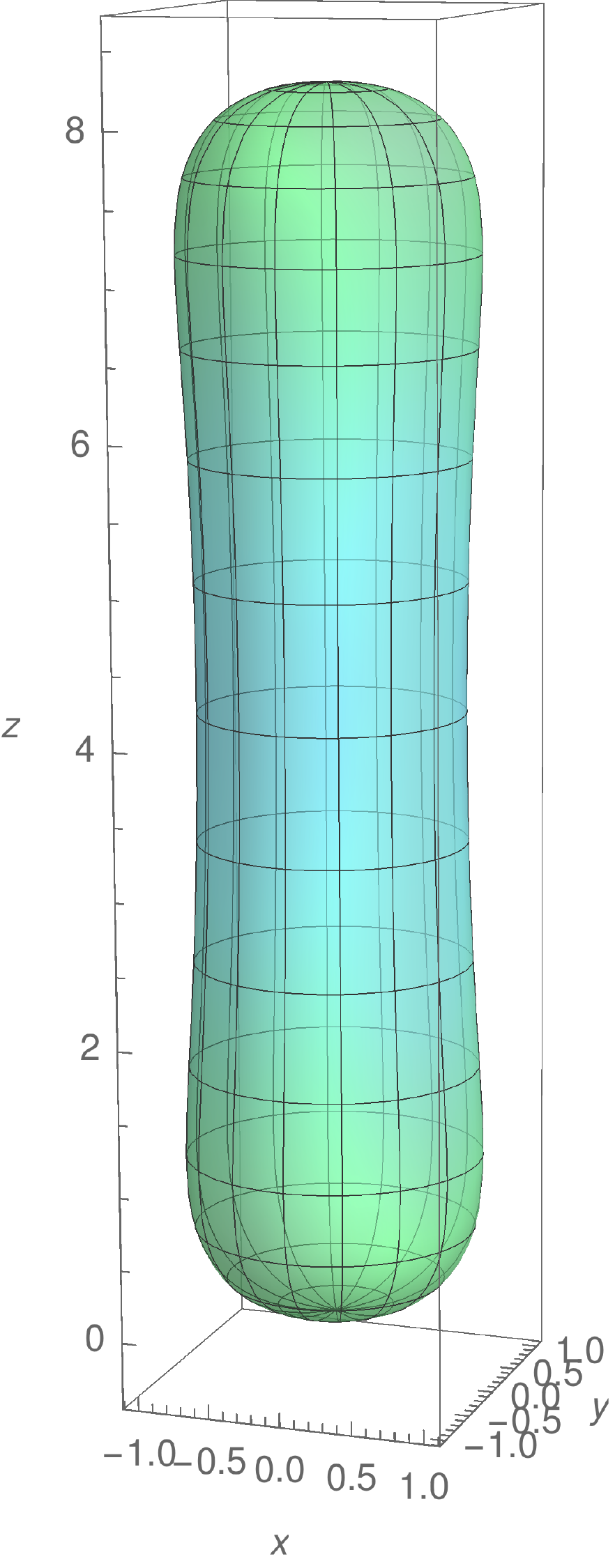}}}
\caption{\small Embedding in Euclidean three-dimensional space $\mathbb{E}^3$ of the event horizon of the black hole distorted by the rotating background, for three different values of the background rotational parameter $\jmath$.}
\label{picture-horizons}
\end{figure}

Indeed the metric~\eqref{swirling-bh} is characterised by a coordinate singularity located at $r=2m$, which identifies the event horizon of the black hole.
This latter is a Killing horizon that has the same significance of the standard Schwarzschild horizon.
The presence of the rotating background deforms the horizon geometry, making it more oblate, while maintaining exactly the same of the Schwarzschild black hole, for the same values of the mass parameter $m$.
In Fig.~\ref{picture-horizons} the deformation is pictured for different intensities of the rotating gravitational background, governed by the new parameter $\jmath$ introduced by the Ehlers transformation.

The solution~\eqref{swirling-bh} is free from axial conical singularities:
to verify this, it is sufficient to consider the ratio between the perimeter of a small circle around the $z$-axis, both for $\theta=0$ and $\theta=\pi$, and its radius.
Such a ratio must be equal to $2\pi$, in case one wants to avoid angular defects.
It turns out that, for the metric~\eqref{swirling-bh}, the ratios in the two limits are equal to $2\pi$
\begin{equation}
\lim_{\theta\to0} \frac{1}{\theta}\int_0^{2\pi} \sqrt{\frac{g_{\varphi\varphi}}{g_{\theta\theta}}
}d\varphi
= 2\pi
= \lim_{\theta\to \pi} \ \frac{1}{\pi-\theta}\int_0^{2\pi} \sqrt{\frac{g_{\varphi\varphi}}{g_{\theta\theta}}}d\varphi \,.
\end{equation}
The metric function $\omega(r,\theta)$ is regular both asymptotically and on the symmetry axis, thus implying the absence of Misner strings or NUT charges.
It is not only continuous, as we can appreciate by the following limits
\begin{equation}
\lim_{\theta\to0} \ \frac{g_{t \varphi}}{g_{tt}}
= \lim_{\theta\to\pi} \ \frac{g_{t\varphi}}{g_{tt}}
= 0 \,,
\end{equation}
but also its first and second derivatives are continuous.

A peculiar characteristic of this metric is that the angular velocity $\Omega$ on the $z$-axis is not constant, and it increases in opposite directions in the two hemispheres
\begin{subequations}
\label{angular-velocity}
\begin{align}
\Omega\big|_{\theta = 0} & = \lim_{\theta \rightarrow 0} \biggl( - \frac{g_{t\varphi}}{g_{\varphi\varphi}} \biggr)
= -4\jmath (r - 2m) + \omega_0 \,, \\
\Omega\big|_{\theta = \pi} & = \lim_{\theta \rightarrow \pi} \biggl( -\frac{g_{t\varphi}}{g_{\varphi\varphi}} \biggr)
= 4\jmath (r - 2m) +\omega_0 \,.
\end{align}
\end{subequations}
This is a feature shared with magnetised Reissner--Nordstr\"om and magnetised Kerr black holes solutions~\cite{Gibbons:2013yq}.

The frame dragging of the whole spacetime is given by~\cite{Poisson:2009pwt}
\begin{equation}
\frac{d\phi}{dt} =
- \frac{g_{t\varphi}}{g_{\varphi\varphi}} = - 4 \jmath ( r -2 m)\cos \theta + \omega_0 \,.
\end{equation}
Hence, outside the event horizon, for $r>2m$, the angular velocity coincides with the asymptotic one $\omega_0$ for $\theta = \frac{\pi}{2}$, while for $\theta \in (\frac{\pi}{2},\pi)$ it is bigger than $\omega_0$ and for $\theta \in (0,\frac{\pi}{2})$ it is smaller than $\omega_0$. 
It is easy to verify that for $r \to \infty$ the angular velocity grows unbounded and that it is equal to $\omega_0$ on the event horizon:
this would lead to the conclusion that superluminal observers exist, since the value of the gravitational dragging can easily exceed 1 (i.e.~the speed of light, in our units) and, then, it would violate causality.
In this perspective, let us study the possible occurrence of closed timelike curves (CTCs):
considering~\eqref{swirling-bh}, curves in which $t$, $r$ and $\theta$ are constants are characterised by
\begin{equation}
{ds}_{t,r,\theta=\text{const}}^2 = F^{-1}(r,\theta) \, r^2 \sin^2\theta {d\phi}^2 \,.
\end{equation}
Such intervals are always space-like since the expression is always positive.
Therefore there are no CTCs and there are no related causality issues:
thus the ``paradox'' of the superluminal observers can be justified with the bad choice of the coordinates\footnote{As it happens, for example, for the Alcubierre spacetime~\cite{Alcubierre:1994tu}}.
A set of coordinates which is adapted to timelike observers does not experience an unbounded growth of the angular velocity, as we will see explicitly when studying the geodesics of the spacetime.

The Kretschmann scalar $R_{\mu\nu\rho\sigma}R^{\mu\nu\rho\sigma}$ suggests that $r=2m$ is a coordinate singularity, while it is divergent for $r=0$, as in the case of the static spherically symmetric black hole in pure General Relativity.
It is indeed possible to find an Eddington--Finkelstein coordinate system that removes the $r=2m$ horizon.
In particular, as $r\to0$ we find
\begin{equation}
R_{\mu\nu\rho\sigma}R^{\mu\nu\rho\sigma} \approx
\frac{48 m^2}{r^6} \,,
\end{equation}
which is exactly the Kretschmann scalar for the Schwarzschild spacetime.
On the other hand, the scalar invariant falls faster than the Schwarzschild metric for large radial distances, indeed one finds, as $r\to\infty$,
\begin{equation}
R_{\mu\nu\rho\sigma}R^{\mu\nu\rho\sigma} \approx
\frac{192}{\jmath^4 \sin^{12}\theta \, r^{12}} \,,
\end{equation}
therefore the solution~\eqref{swirling-bh} is locally asymptotically flat.
We finally notice that for $\theta=0,\pi$ the spacetime has an asymptotically constant curvature:
we find
\begin{equation}
R_{\mu\nu\rho\sigma}R^{\mu\nu\rho\sigma} \big|_{\theta=0,\pi} \approx -192 \jmath^2 \quad
\text{ as } r \to \infty \,,
\end{equation}
thus we see that on the $z$-axis, the spacetime is asimptotically of negative constant curvature.

\subsection{Ergoregions}
\begin{figure}
\centering
\hspace{0.2cm}
\includegraphics[width=6cm]{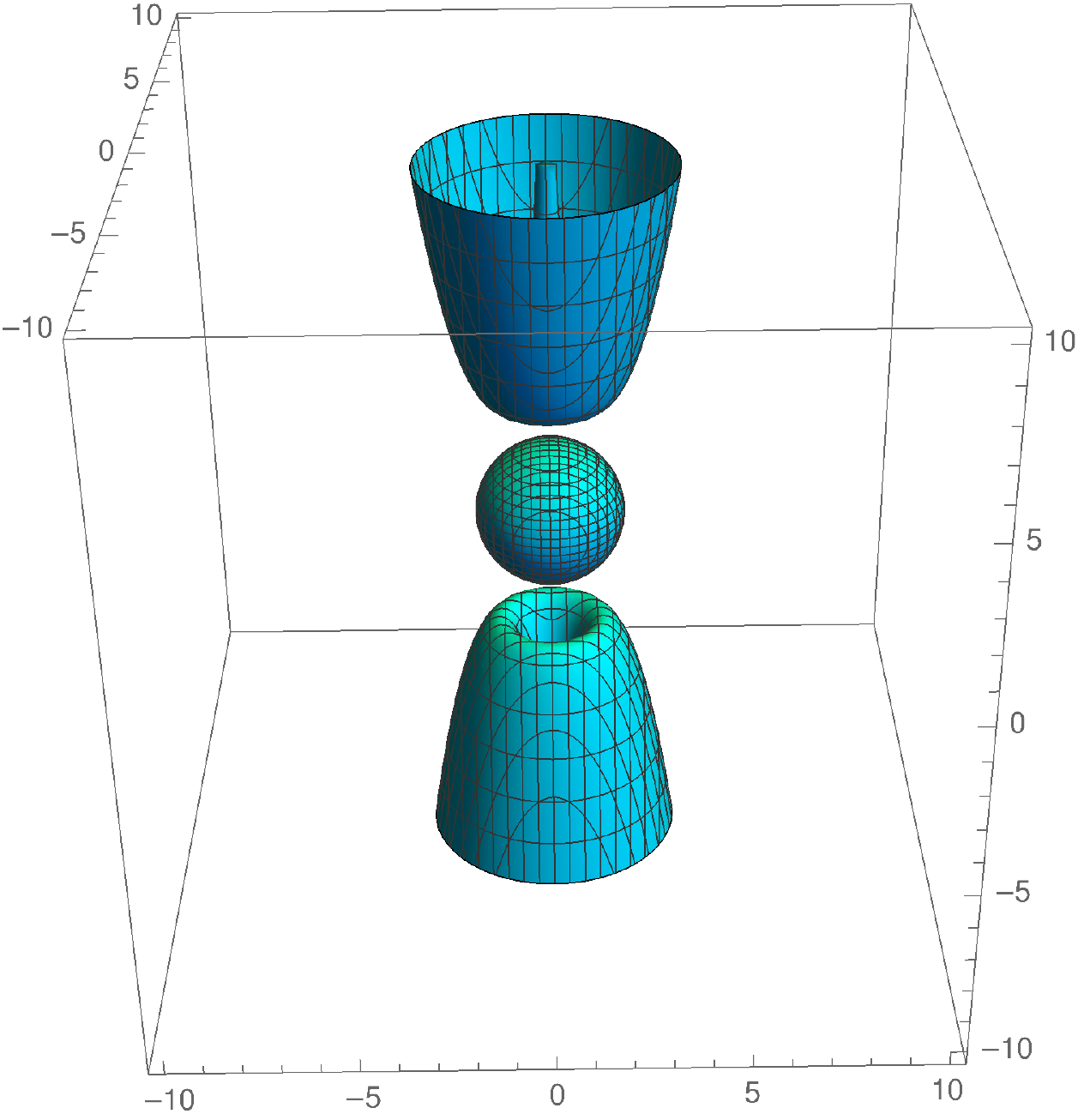}
\caption{\small Ergoregions for the black hole embedded in a rotating universe, with parameters $m=1$, $\jmath=0.3$ and $\omega_0=0$.
The ergoregions extend to infinity in the positive and negative $z$ directions, independently of the choice of the parametrisation for the integrating constants.}
\label{fig:ergoregions3D}
\end{figure}

It is clear, just by inspection, that the $g_{tt}$ component of the metric~\eqref{swirling-bh} becomes null on the event horizon, and that outside the horizon is not everywhere negative. 
Therefore the spacetime presents some ergoregions, analogously to Kerr or magnetised Reissner--Nordstr\"om black holes~\cite{Gibbons:2013yq}.
To analyse these regions it is convenient to use the cylindrical coordinates as defined in~\eqref{cyly-coord} to expand, for large $z$, the $g_{tt}$ part of the metric as follows
\begin{equation}
g_{tt}(\rho,z) \approx \frac{16 \jmath^2 \rho^2 z (z- 4m)}{1+\jmath^2 \rho^4} \,.
\end{equation}
Hence, the ergoregions are located not only in the proximity of the event horizon, as in Kerr spacetime, but also close to the $z$-axis, for large values of $z$.
A numerical analysis of the function $g_{tt}$ is represented in Fig.~\ref{fig:ergoregions3D}:
it shows how the ergoregions extend to infinity around the polar axis, independently of the values of the integrating constants of the solution $(m,\jmath)$.
This behaviour is similar to what happens for magnetised rotating black holes~\cite{Gibbons:2013yq}.

\subsection{Petrov type}

A standard procedure to determine the Petrov type of a spacetime consists in computing the Weyl tensor in a null tetrad basis.
We define a frame by
\begin{subequations}
\begin{align}
e^0 & = F^{1/2} \biggl(1-\frac{2m}{r}\biggr)^{1/2} \, dt \,, \\
e^1 & = F^{1/2} \biggl(1-\frac{2m}{r}\biggr)^{-1/2} \, dr \,, \\
e^2 & = r F^{1/2} \, d\theta \,, \\
e^3 & = r\sin\theta F^{-1/2} \bigl\{ d\varphi + \big[4\jmath(r-2m)\cos \theta + \omega_0 \big] \ dt \bigr\} \,.
\end{align}
\end{subequations}
Given such a frame, the null tetrad is found as
\begin{equation}
k_\mu = \frac{1}{\sqrt{2}} \bigl(e^0_\mu + e^3_\mu\bigr) \,, \quad
l_\mu = \frac{1}{\sqrt{2}} \bigl(e^0_\mu - e^3_\mu\bigr) \,, \quad
m_\mu = \frac{1}{\sqrt{2}} \bigl(e^1_\mu - i e^2_\mu\bigr) \,, \quad
\bar{m}_\mu = \frac{1}{\sqrt{2}} \bigl(e^1_\mu + i e^2_\mu\bigr) \,.
\end{equation}
It is now possible to compute the components of the Weyl tensor in the null basis, as
\begin{subequations}
\begin{align}
\Psi_0 & = C_{\mu\nu\rho\sigma} k^\mu m^\nu k^\rho m^\sigma \,, \\\
\Psi_1 & = C_{\mu\nu\rho\sigma} k^\mu l^\nu k^\rho m^\sigma \,, \\
\Psi_2 & = C_{\mu\nu\rho\sigma} k^\mu m^\nu \bar{m}^\rho l^\sigma \,, \\
\Psi_3 & = C_{\mu\nu\rho\sigma} l^\mu k^\nu l^\rho \bar{m}^\sigma \,, \\
\Psi_4 & = C_{\mu\nu\rho\sigma} l^\mu \bar{m}^\nu l^\rho \bar{m}^\sigma \,,
\end{align}
\end{subequations}
where $C_{\mu\nu\rho\sigma}$ is the Weyl tensor.

One can easily show that $\Psi_1=\Psi_3=0$, while the other components are more involved.
The inspection of the scalar invariants
\begin{equation}
I = \Psi_0\Psi_4 - 4 \Psi_1\Psi_3 + 3\Psi_2^2 \,, \quad
J = \det
\begin{pmatrix}
\Psi_0 & \Psi_1 & \Psi_2 \\
\Psi_1 & \Psi_2 & \Psi_3 \\
\Psi_2 & \Psi_3 & \Psi_4
\end{pmatrix}
\,,
\end{equation}
reveals that $I^2\neq 27J^2$:
this implies that the spacetime is algebraically general~\cite{Stephani:2003tm,Griffiths:2009dfa}.
Thus, the spacetime belongs to the general Petrov type I, contrary to its background~\eqref{background-metric} or its generating seed, which are both type D.
Further, this result shows that the new black hole~\eqref{swirling-bh} does not belong to the Pleba\'nski--Demia\'nski class of spacetimes~\cite{Plebanski:1976gy}.

\subsection{Geodesics}

We follow the same strategy as in the background case and define, from the metric~\eqref{swirling-bh}, the following Lagrangian (dropping the inessential $\omega_0$ term)
\begin{equation}
\label{lag-bh}
\mathscr{L} =  F
\biggl[ - \biggl( 1 - \frac{2m}{r}\biggr) \dot{t}^2
+ \frac{\dot{r}^2}{1 - \frac{2m}{r}} + r^2 \dot{\theta}^2 \biggr] \\
+ F^{-1} r^2 \sin^2 \theta
\Bigl[ \dot{\varphi} + 4\jmath(r-2m)\cos \theta \, \dot{t} \Bigr]^2 \,,
\end{equation}
Proceeding in the same way as the background metric, we obtain the conserved charges equations and the four-momentum normalisation equations, reported in Appendix~\ref{sec:appbh}.

We can extract some qualitative information, especially regarding the quantity $(r-2m)\cos\theta$ that appears in the gravitational dragging.
For stable orbits $r$ is limited, hence the quantity $(r-2m)\cos\theta$ is limited as well.
For unstable orbits we analyse the geodesic motion as $r$ reaches infinity, thus considering large values of $s$.
We notice that $\dot{t} \approx 0$ and $\dot{\varphi} \approx \jmath^2 L r^2 \sin^2\theta$, as $r\to\infty$, and moreover $1- \frac{2m}{r} \approx 1$ and $F \approx \jmath^2 r^4 \sin^4\theta$.
These approximations simplify the Lagrangian~\eqref{lag-bh}, that takes the form
\begin{equation}
\mathscr{L} \approx \jmath^2 r^4 \sin^4\theta
\bigl(
\dot{r}^2 + r^2 \dot{\theta}^2 \bigr)
+ \jmath^2 L^2 \,.
\end{equation}
The constant term is inessential and can be neglected.
By changing to polar coordinates
$x = r \sin \theta$ $y = r \cos \theta$,
the Lagrangian boils down to
\begin{equation}
\mathscr{L} \approx \jmath^2 x^4 \bigl( \dot{x}^2 + \dot{y}^2 \bigr) \,.
\end{equation}
Being the Lagrangian independent of $y$, we find the conserved quantity
\begin{equation}
A = \jmath^2 x^4 \dot{y} \,.
\end{equation}
This result can be plugged into the Lagrangian, and by noticing that it does not depend explicitly on $s$, $d\mathscr{L}/ds=0$, the following equation is derived:
\begin{equation}
\jmath^2 x^4 \dot{x}^2 + \frac{A^2}{\jmath^2 x^4}
= B \,,
\end{equation}
where $B$ is a real constant.
From the last equation we find
\begin{equation}
\dot{x} = \frac{\sqrt{B \jmath^2 x^4 - A^2}}{\jmath^2 x^4} \approx \frac{\sqrt{B}}{\jmath x^2} \,,
\end{equation}
where the numerator $\sqrt{B \jmath^2 x^4 - A^2}$ depends on $x$ which, by our change of coordinate, is proportional to $r$, so when $r$ approaches infinity so does $x$.
Therefore the constant $A^2$ underneath the square root can be neglected, thus justifying the approximation.
Finally we get, by integration,
\begin{equation}
x(s) = \biggl(\frac{3 \sqrt{B}}{\jmath} s
+ C \biggr)^{\frac{1}{3}} \,,
\end{equation}
with $C$ real constant.
So $x \to \infty$ as $s \to \infty$, which means that $\dot{y} \approx 0$ and $y \approx$ constant.
These results can now be plugged into the formula for the gravitational dragging, which gives
\begin{equation}
- \frac{g_{t\varphi}}{g_{\varphi\varphi}}
= - 4 \jmath \biggl(y - 2m \frac{y}{ \sqrt{x^2 + y^2}}\biggr)
\underset{s\to\infty}{\approx} -4\jmath y \,.
\end{equation}
This result shows that, as $s \to \infty$, the angular velocity approaches a constant value.

\begin{figure}
\centering
\hspace{-0.2cm}
\subfloat[\centering $m=1$, $\jmath=0$]{{\includegraphics[width=6.5cm]{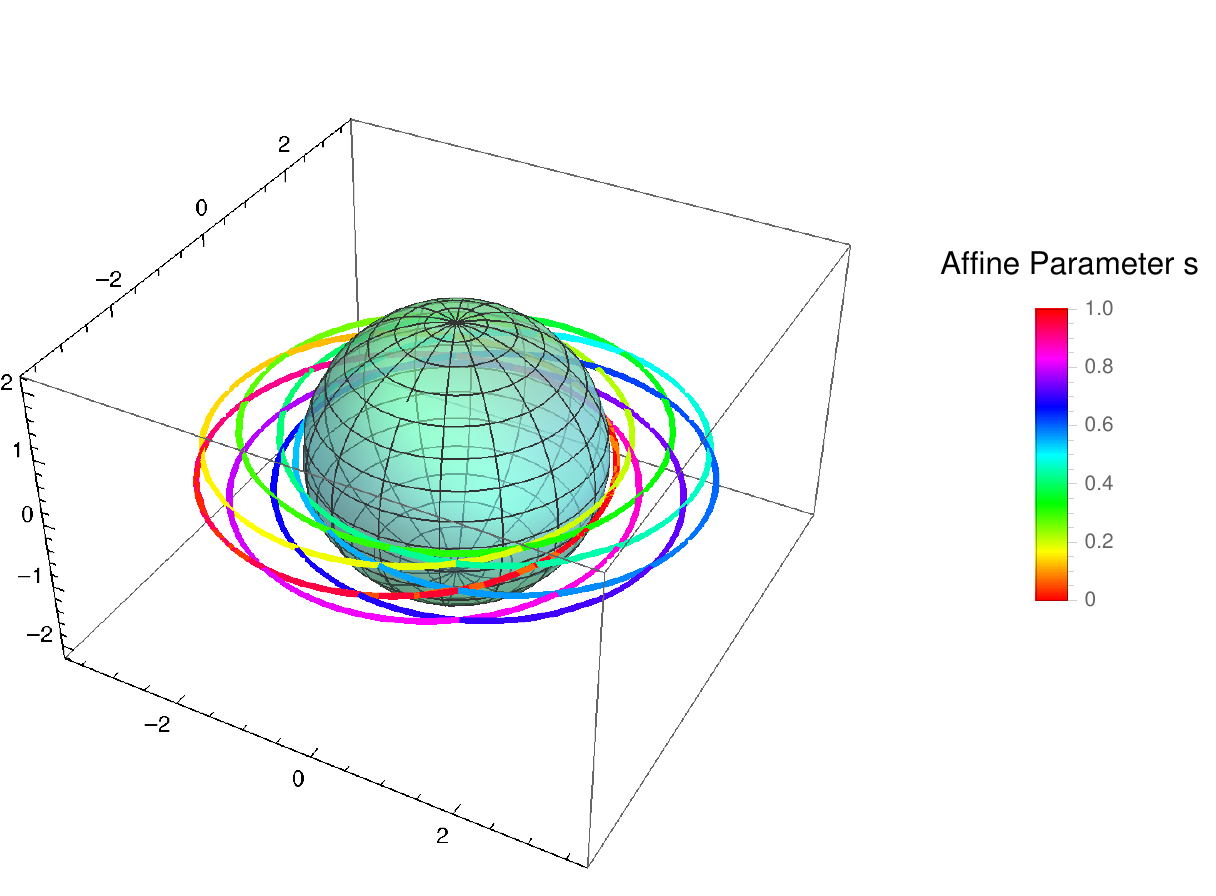}}}
\subfloat[\centering $m=1$, $\jmath=10^{-8}$]{{\hspace{-0.3cm} \includegraphics[width=6.5cm]{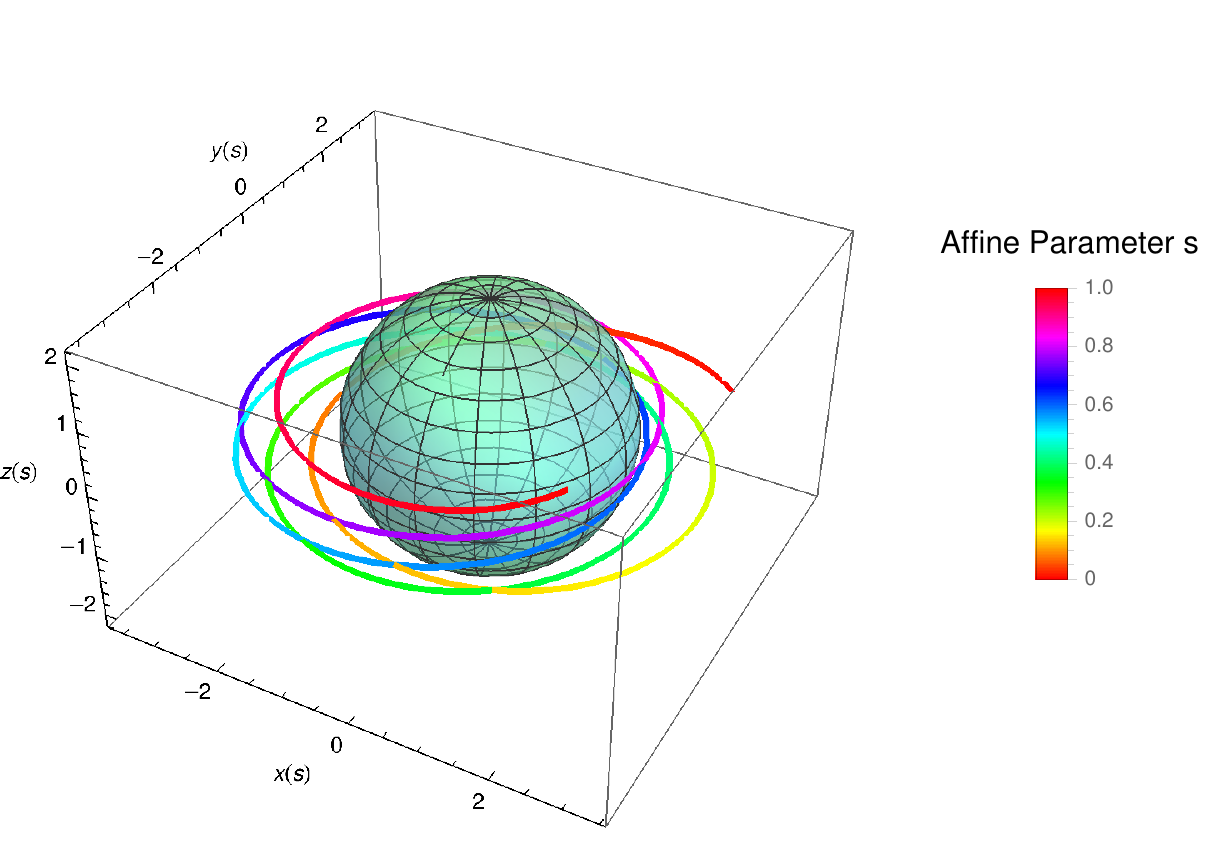}}}
\caption{\small Geodesic motion around the black hole.
The left panel shows the Schwarzschild spacetime, while the right panel shows the new black hole solution~\eqref{swirling-bh}.
The plots share the same initial conditions with $E=1$, $L=12$.}
\label{fig:schwVSswirl}
\end{figure}

\begin{figure}
\captionsetup[subfigure]{labelformat=empty}
\centering
\hspace{-0.2cm}
\subfloat[\centering $m=1$, $\jmath=0.01$]{{\includegraphics[width=6.5cm]{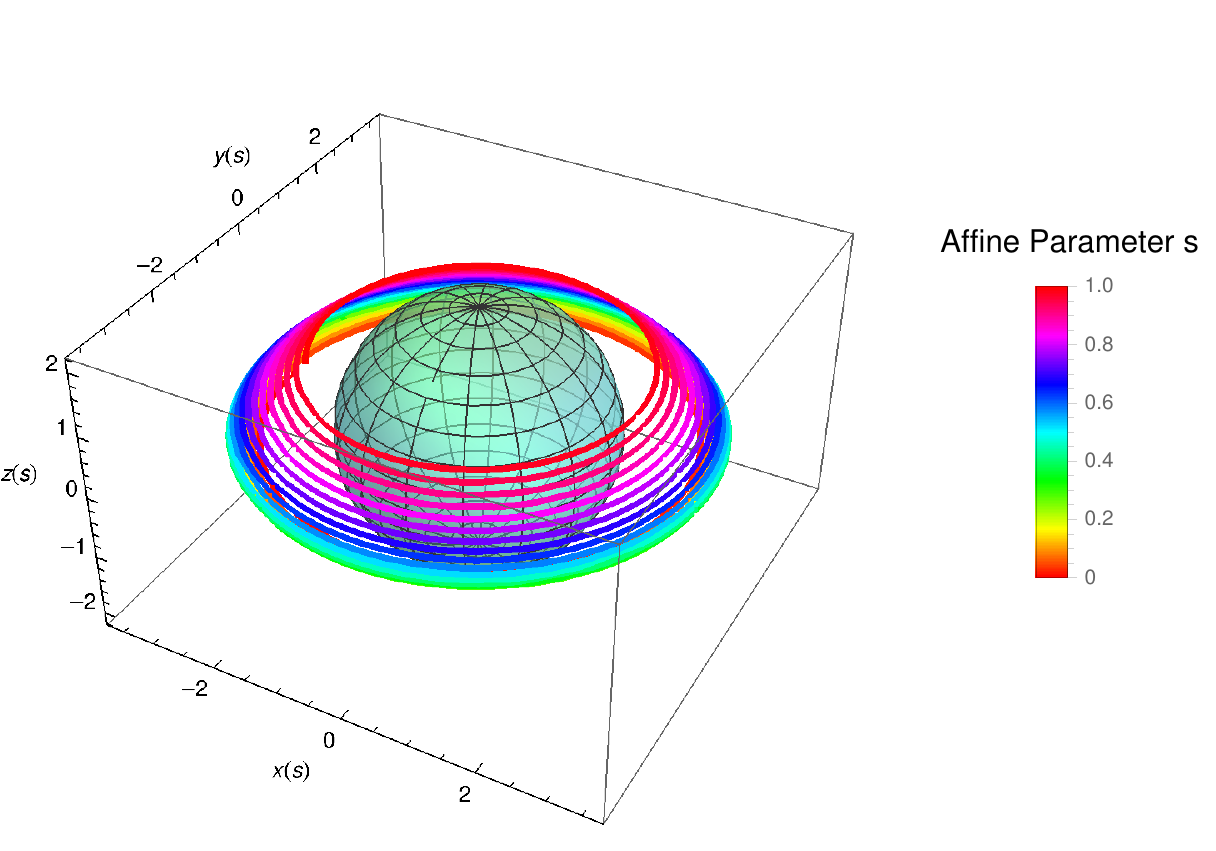}}}%
\subfloat[\centering $m=1$, $\jmath=0.01$]{{\hspace{-0.3cm} \includegraphics[width=6.5cm]{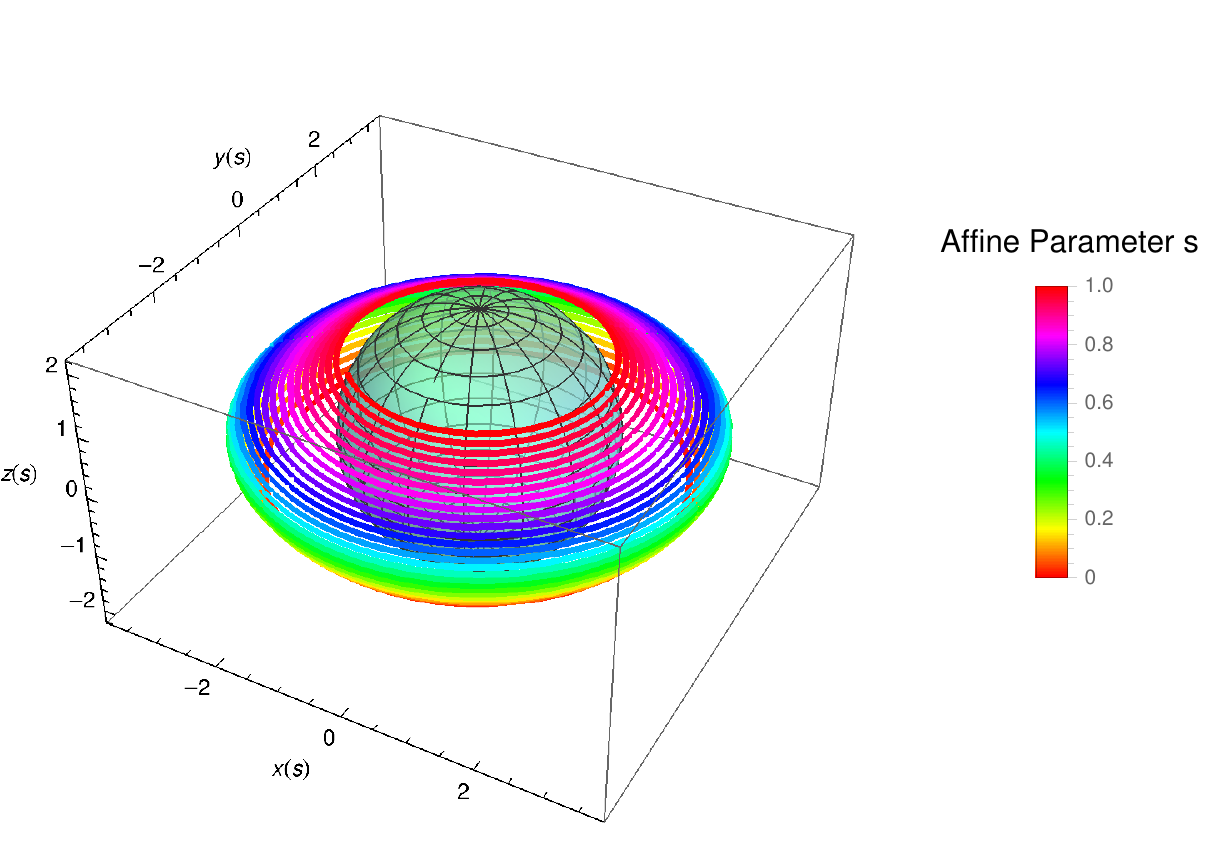}}}
\caption{\small Embedding diagram and geodesics for the new metric~\eqref{swirling-bh} with different initial conditions, $\theta_0 = \frac{\pi}{2}$, $\dot{\theta}_0 = -\frac{1}{2} $, $L=12$ for the l.h.s. diagram and $\theta_0 = \frac{4}{7} \pi$, $\dot{\theta}_0 = -\frac{1}{2} \sin^{-1}(\frac{4}{7}\pi) $, $L=-12$ for the r.h.s. diagram.
Both representations share the following data: $E=1$, $r_0 = 3$, $\dot{r}_0 = 1$ and $\phi_0 = \pi $.}
\label{fig:bh-geod}
\end{figure}

We also plot the geodesic motion in spacetime~\eqref{swirling-bh}:
this amounts to numerically integrate the geodesic equations reported in Appendix~\ref{sec:appbh}, and the results are shown in Fig.~\ref{fig:schwVSswirl},~\ref{fig:bh-geod} and~\ref{fig:unstable}.
More precisely, Fig.~\ref{fig:schwVSswirl} compares geodesics in Schwarzschild spacetime (i.e.~$\jmath=0$) and geodesics in our swirling spacetime~\eqref{swirling-bh}.
Fig.~\ref{fig:bh-geod} shows the geodesics around the black hole for different initial conditions.
Finally, in Appendix~\ref{sec:appbh}, Fig.~\ref{fig:unstable} pictures unstable geodesic motion for two different values of the test particle angular momentum.

\subsection{Charges and thermodynamics}

The total mass of the spacetime can be computed by means of the surface  charges provided by the phase space formalism~\cite{Lee:1990nz,Barnich:2003xg}.
We perturb the metric with respect to the parameters of the solution, and we name that variation $h_{\mu\nu}\coloneqq\delta g_{\mu\nu}$\footnote{In the particular case under consideration the parameter space is spanned by the mass parameter of the black hole $m$ and by the magnitude of the rotational whirlpool dragging, $\jmath$.
Thus the variation takes the form $h_{\mu\nu}= \delta g_{\mu\nu}(m,\jmath) = \partial_m g(m,\jmath) \delta m + \partial_\jmath g(m,\jmath) \delta \jmath$.}.
Then we find the local variation of the charge $K_\xi$ computed along a given Killing direction $\xi^\mu$.

The local variation of the charge must be integrated between the parametric reference background $\bar{\Psi}$ and the actual parametric configuration labelled by $\Psi$, on a $D-2$ dimensional surface $\mathcal{S}$ containing the event horizon $\bigl(d^{D-2}x \bigr)_{\mu\nu} = \frac{1}{2(D-2)!} \epsilon_{\mu\nu\alpha_1...\alpha_{D-2}} dx^{\alpha_1} \wedge \cdots \wedge dx^{\alpha_{D-2}}$.
When the variation of the charge is integrable, all the parametric paths between the reference background and the solution are equivalent\footnote{In case the variation of the charge is not integrable, we still have some gauge degree of freedom in defining the frame of reference, or the normalisation of the time coordinate, to recover integrability.}.

The result gives the total surface charge $Q_\xi$, defined, as in~\cite{Barnich:2003xg,Compere:2009dp}, by
\begin{equation}
Q_{\xi} = \frac{1}{8\pi} \int_{\bar{\Psi}}^{\Psi} \int_\mathcal{S} K_\xi = \frac{1}{8\pi} \int_{\bar{\Psi}}^{\Psi} \int_\mathcal{S} K_\xi^{\mu\nu} \bigl(d^{D-2}x \bigr)_{\mu\nu} \,,
\end{equation}
where
\begin{equation}
K_\xi^{\mu\nu} = \xi^\mu \nabla_\sigma h^{\nu\sigma} - \xi^{\mu} \nabla^\nu h - \xi_\sigma \nabla^\mu h^{\nu\sigma} - \frac{1}{2} h \nabla^\mu \xi^\nu + \frac{1}{2} h^{\sigma \mu} (\nabla_\sigma \xi^\nu - \nabla^\nu \xi_\sigma) \,,
\end{equation}
and where $h\coloneqq h^{\mu}_{\;\;\mu}$.
If we want to compute the mass of the black hole, we have to consider the timelike Killing vector $\xi=\partial_t$, then we find
\begin{equation}
M = Q_{\partial_t} = m \,,
\end{equation}
as for the Schwarzschild black hole.
In this case the presence of the background does not modify the seed black hole mass, similarly to what happens within the context of black holes embedded in an external electromagnetic field~\cite{Astorino:2016hls}.
Following this analogy we expect to observe some stronger coupling with the background in case of more general black hole seeds, as it happens in the next Section.  

The angular momentum can be found analogously, just considering the Killing vector which generates the rotational symmetry $\partial_\varphi$.
In this case one gets null angular momentum 
\begin{equation}
J=Q_{\partial_\varphi} = 0 \,,
\end{equation}
even though the solution is clearly rotating.
In fact the angular momentum refers just to the dipole term in the rotational multipolar expansion of the metric at large distances.
The fact that the metric is rotating, as its non-diagonal form suggests, can be appreciated by the subsequent terms of the multipolar expansion:
the quadrupole, the octupole, etc\footnote{However, note that in contexts where the asymptotia is not of globally constant curvature, the notion of the gravitational multipolar expansion needs some further analysis to be clearly defined.}.

We compute the entropy and the temperature of the event horizon, in order to study the Smarr law and the thermodynamics of the black hole.
The area of the even horizon is found by integrating the $(\theta,\phi)$ part of the metric, hence
\begin{equation}
\mathcal{A} = \int_0^{2\pi} d\varphi \int_0^\theta d\theta
\sqrt{g_{\theta\theta} g_{\phi\phi}} \big|_{r=2m}
= 16 \pi m^2 \,.
\end{equation}
The entropy is given by the Bekenstein--Hawking formula $S = \mathcal{A}/4$.
The validity of the area law also for this unconventional background is confirmed by the conformal field theory dual to the near-horizon geometry of the black hole.
The temperature can be easily obtained via the surface gravity, $\kappa = \sqrt{- (\nabla_\mu\xi_\nu)^2/2}\big|_{r=2m}$, where $\xi=\partial_t$.
We find
\begin{equation}
T = \frac{\kappa}{2\pi} =\frac{1}{8\pi m} \,.
\end{equation}
Note that the entropy and the temperature of the black hole embedded into the swirling background are unaffected by the spacetime rotation:
they remain the same of the \\ Schwarzschild seed.
Again this is peculiar of the Lie point symmetry we used to generate the solution, a general feature shared with the Harrison transformation\footnote{Note that this is true only when the seed does not couple with the background brought in by the transformation.}.

We can easily verify the validity of the Smarr law
\begin{equation}
M = 2 T S \,.
\end{equation}
Further, the conserved charges satisfy the first law of thermodynamics
\begin{equation}
\delta M = T \delta S \,.
\end{equation}

\section{Kerr black hole in a swirling universe}
\label{sec:kerr}

The generating techniques discussed in Sec.~\ref{sec:generation} can be also exploited to embed a rotating black hole in a background endowed with its own rotation.
By using the Kerr metric in Boyer--Lindquist coordinates as a seed, we obtain
\begin{equation}
{ds}^2 = F (d\varphi - \omega dt)^2 + F^{-1}\biggl[-\rho^2 {dt}^2 + \Sigma\sin^2\theta \biggl(\frac{{dr}^2}{\Delta} + {d\theta}^2\biggr)\biggr] \,,
\end{equation}
where the functions $F^{-1}$ and $\omega$ can be expanded in finite power series of $\jmath$
\begin{align}
F^{-1}  & = \chi_{(0)} + \jmath\chi_{(1)} + \jmath^2\chi_{(2)} \,, \qquad
\omega = \omega_{(0)} + \jmath\omega_{(1)} + \jmath^2\omega_{(2)} \,,
\end{align}
with 
\begin{subequations}
\begin{align}
\chi_{(0)} & = \frac{R^2}{\Sigma \sin^2\theta} \,, \\
\chi_{(1)} & = \frac{4 a m \, \Xi \cos\theta }{\Sigma \sin^2\theta} \,, \\
\chi_{(2)} & = \frac{4 a^2 m^2 \Xi^2 \cos^2\theta + \Sigma^2 \sin^4\theta}{R^2 \Sigma \sin^2\theta} \,,
\end{align}
\end{subequations}
and 
\begin{subequations}
\begin{align}
\omega_{(0)} & = -\frac{2 a m r}{\Sigma} + \omega_0\,, \\
\omega_{(1)} & = \frac{4 \cos\theta [-a \Omega (r-m) + ma^4 - r^4(r-2m) - \Delta a^2 r]}{-\Sigma } \,, \\
\begin{split}
\omega_{(2)} & =
-\frac{2m}{\Sigma} \Bigl\{
3ar^5 - a^5 (r + 2m) + 2a^3r^2 (r + 3m) - r^3(\cos^2\theta - 6) \Omega \\
&\quad + a^2[\cos^2\theta(3r-2m) - 6(r-m)] \Omega
\Bigr\}
\,,
\end{split}
\end{align}
\end{subequations}
where
\begin{subequations}
\begin{align}
\Delta & = r^2 - 2mr + a^2 \,, \qquad\qquad\qquad\qquad\qquad\!\!\!\!\!\!\!\!
\rho^2 = \Delta \sin^2\theta \,, \\
\Sigma & = (r^2 + a^2)^2 - \Delta a^2 \sin^2\theta \,, \qquad\qquad\qquad\!\!\!\!\!\!
\Omega = \Delta a \cos^2\theta \,, \\
\Xi & = r^2(\cos^2\theta - 3) - a^2 (1+\cos^2\theta) \,, \qquad
R^2 = r^2 + a^2 \cos^2\theta\,.
\end{align}
\end{subequations}
When $\jmath=0$ we recover the seed metric, i.e.~the Kerr black hole.
For $\jmath\ne0$ we have the direct generalisation of the metric~\eqref{bh-rot-universe}.

However we notice that in this case, because of the spin-spin interaction between the black hole and the background frame dragging, an extra force acts on the axis of symmetry.
But since it is not symmetric on the two hemispheres, the metric is affected by non-removable conical singularities, indeed
\begin{equation}
\lim_{\theta\to0} \frac{1}{\theta}\int_0^{2\pi} \sqrt{\frac{g_{\varphi\varphi}}{g_{\theta\theta}}
}d\varphi
= \frac{2\pi}{(1-4a m \jmath)^2} \neq \lim_{\theta\to \pi} \ \frac{1}{\pi-\theta}\int_0^{2\pi} \sqrt{\frac{g_{\varphi\varphi}}{g_{\theta\theta}}}d\varphi = \frac{2\pi}{(1+4a m \jmath)^2} \,.
\end{equation}
In fact, even though the background spinning parameter $\jmath$ couples to the Kerr angular momentum (for unit of mass) $a$, it is not possible to find a relation among the physical parameters to remove simultaneously both angular defects, unless of course for known subcases such as Kerr, for $\jmath=0$, the spacetime discussed in Sec.~\ref{sec:analysis-blackhole} for $a=0$, or the rotating background for $m=0$.
The presence of a non-removable conical singularity implies that a cosmic string or a strut (with their $\delta$-like stress-energy-momentum tensor on a portion of the $z$-axis) have to be postulated in order to compensate the ``force'' effect induced by the spin-spin interaction of the black hole with the background, which would tent to add acceleration to the black hole\footnote{The metric considered in this Section does not posses the acceleration parameter:
one should work with the rotating C-metric to consistently include the acceleration.
That is why, in this Section, the role of the string uniquely results in the effect of compensating the spin coupling.}.

In the case one wants to immerse the Kerr--Newman black hole into this spinning universe, one has to use the charged generalised version of the Ehlers transformation, as described in Chapter~\ref{chap:gentech}.

\section{Double-Wick rotation of the background: flat Taub--NUT spacetime}
\label{sec:double-wick}

Given the analogies between the rotating background~\eqref{background-cyl} and the Melvin spacetime, and given that the analytical continuation of the Melvin universe corresponds to the Reissner--Nordstr\"om metric with a flat base manifold, it is natural to inquire about an analog analytical continuation for the rotating background.
At this scope, we implement a double Wick rotation between time and the azimuthal angle $t \to i \phi$, $\varphi \to i \tau $ of the metric~\eqref{background-cyl}.
Redefining the integration constant of the rotating background as $\jmath=m/2\ell^3$, changing the coordinate $\rho = \ell \sqrt{2r/m}$ and after the rescaling of the other three coordinates we obtain
\begin{equation}
\label{T-NUT}
{ds}^2 = - \frac{2mr}{r^2+\ell^2} ( dt - 2 \ell \theta d\phi)^2 + \frac{r^2+\ell^2}{2mr} {dr}^2 + (r^2+\ell^2) ({d\theta}^2 + {d\phi}^2 ) \,.
\end{equation}
It is not hard to recognise the Taub--NUT spacetime with a flat, or possibly cylindrical if we keep the azimuthal angle identification, base manifold.
In fact, the flat Taub--NUT metric can be generated via the Ehlers transformation\footnote{In this Section we are referring to transformations applied to the \emph{magnetic} LWP metric, as explained in Sec.~\ref{sec:generation}.} from the Schwarzschild metric, previously composed with a double-Wick rotation\footnote{While this is true for a generic sign of the constant curvature of the seed base manifold, only the metric with positive curvature can be interpreted as a black hole in Einstein gravity.}.
Note that the Ehlers transformation can be used to build, from the Minkowski seed, the rotating background, just considering $m=0$ in the procedure of Sec.~\ref{sec:generation}.
This is analogous to what happens to the Melvin universe, which can be obtained from Minkowski spacetime via the Harrison transformation and whose double-Wick dual corresponds to the flat Reissner--Nordstr\"om metric.
This fact strengthens the link between the Melvin universe and our rotating background.
Actually this correspondence can be summarized by the following proportion:
\begin{equation}
\label{proportion}
\text{Melvin Universe : Harrison transformation = Rotating Universe : Ehlers transformation}
\end{equation}
This formal analogy can be exploited to build new solutions, even outside the range of the generating technique based on the Lie point symmetries of the Ernst equations:
that is important because the symmetry transformations such as the Ehlers and the Harrison maps break in the presence of the cosmological constant.
However, as noted in~\cite{Astorino:2012zm}, the Melvin universe can still be generalised when the cosmological constant is not zero, and it still preserves its relation with the flat Reissner--Nordstr\"om metric with a constant curvature base manifold\footnote{Metrics without a topological spherical base manifold are interpreted as black holes only in the presence of the cosmological constant.}.
Therefore, thanks to the analogy with the Melvin case, we have in our hand a procedure to generalise the rotating background~\eqref{background-cyl} in the presence of the cosmological constant.

It is sufficient to operate a double-Wick rotation of the cosmological version of the flat Taub--NUT metric~\eqref{T-NUT}
\begin{equation}
\begin{split}
{ds}^2 & = - \frac{\frac{\Lambda}{3} r^4 + 2\ell^2\Lambda r^2 + 2mr - \ell^4 \Lambda}{r^2+\ell^2} (dt - 2 \ell \theta d\phi)^2 \\
&\quad + \frac{(r^2+\ell^2)}{\frac{\Lambda}{3} r^4 + 2\ell^2\Lambda r^2 + 2mr - \ell^4 \Lambda} {dr}^2 + (r^2+\ell^2) ({d\theta}^2 + {d\phi}^2) \,.
\end{split}
\end{equation}
Thus, using the same change of coordinates and parametrisation of the case above, we get 
\begin{equation}
\label{kundt-lambda1}
\begin{split}
{ds}^2 & = (1 + \jmath^2 \rho^4) \biggl(-{d\tau}^2 + \frac{\rho^2}{\frac{\Lambda}{4\jmath^2} + \rho^2 - \frac{\Lambda}{2}\rho^4 - \frac{\jmath^2 \Lambda}{12}\rho^8} {d\rho}^2 + {dz}^2 \biggr) \\
&\quad + \biggl(\frac{\Lambda}{4\jmath^2} + \rho^2 - \frac{\Lambda}{2}\rho^4 - \frac{\jmath^2 \Lambda}{12}\rho^8\biggr) \frac{(d\psi + 4\jmath z d\tau)^2}{1+\jmath^2 \rho^4} \,.
\end{split}
\end{equation}
It is not difficult to realise that this metric still corresponds, up to a change of coordinates, to the non-expanding and non-accelerating Kuntdt class of the Pleba\'nski--Demia\'nski family presented in Eq.~(16.26) of~\cite{Griffiths:2009dfa}.
The explicit change of coordinates works as in Sec.~\ref{sec:analysis-background}, namely
$q=2\jmath z$ and $p^2=\rho$, together with the rescaling
$t\to\jmath t$ and the redefinitions $\gamma=1/\jmath$, $\Tilde{\Lambda}=4\Lambda$.
Then, metric~\eqref{kundt-lambda1} becomes
\begin{equation}
\label{kundt-lambda2}
{ds}^2 = (\gamma^2+p^2) \biggl(-{d\tau}^2 + \frac{dp^2}{\mathcal{P}} + {dq}^2 \biggr)
+ \frac{\mathcal{P}}{\gamma^2+p^2} (d\psi + 2\gamma q d\tau)^2 \,,
\end{equation}
where
\begin{equation}
\mathcal{P} = \gamma^4\Tilde{\Lambda} + \gamma^2 p - 2\gamma^2\Tilde{\Lambda} p^2 - \frac{\Tilde{\Lambda}}{3} p^4 \,.
\end{equation}
The latter corresponds to Eq.~(16.26) of~\cite{Griffiths:2009dfa}, where
$m=e=g=\alpha=\epsilon_2=0$, $\epsilon_0=1$, $k=\gamma^4\Tilde{\Lambda}$, $\epsilon=2\gamma^2\Tilde{\Lambda}$ and $n=\gamma^2/2$.

The analogy between the rotating background and the Taub--NUT spacetime can be pushed further:
it is known~\cite{Emparan:2001gm} that the Melvin spacetime corresponds to a couple of magnetically charged Reissner--Nordstr\"om black holes moved towards infinity.
In this sense, the magnetic field which permeates the Melvin spacetime is nothing but the field generated by two black hole sources at infinity.
Thus, it is natural to ask ourselves if a similar construction also holds for the rotating background~\eqref{background-metric}, i.e.~if it can be obtained as a limit of a double black hole metric.

By relying on the above considerations and, more specifically, on the proportion~\eqref{proportion}, the natural candidate for an ``ancestor'' metric is the double-Taub--NUT spacetime with opposite NUT parameters~\cite{Kramer1980}:
the rotation of the two counter-rotating Taub--NUT black holes, once they are pushed at infinity, should produce the rotation of the background that is experienced in the background spacetime.
This interpretation is also consistent with the behaviour of the angular velocity~\eqref{angular-velocity}:
we noticed that the angular velocity increases in opposite directions in the two hemisphere, coherently with the fact that the two black holes rotate in different directions.
Moreover this picture is enforced by the geometry of the ergoregions, since the latter thrive for large values of $z$ on the axis of symmetry.

\section{Outlook}

The transformation we studied in this Chapter, which consists in a proper composition of the Ehlers transformation with a discrete symmetry, allows us to take advantage of the Ernst solution generating technique to non-linearly superpose the Schwarzschild black hole and a swirling universe.
The background geometry can be interpreted as a gravitational whirlpool generated by a couple of counter-rotating sources at infinity.
Its frame dragging transforms the static Schwarzschild metric into a stationary one, removing the asymptotic flatness, but without drastically altering the black hole causal structure nor introducing pathological features.
The analogies between the swirling background and the Melvin universe are numerous, like the metric structure, the ergoregions and the deformations engraved on the event horizon:
in fact the former universe can be considered as the rotating counterpart of the latter.

For this reason we expect that this spinning background can be used as a regularising instrument for metric with conical singularities, exactly as the electromagnetic background brought by the Harrison transformation does.
In the former case to have non-trivial physical effects one needs to exploit the interplay between the coupling of the electromagnetic field of the seed with the one of the background, as suggested by the analysis of the transformed Kerr metric, in Sec.~\ref{sec:kerr}.
Indeed the interaction between the Kerr parameter $a$ and the background parameter $\jmath$ generates an additional ``force'' which impels the system to accelerate.
Unfortunately, the geometry of the spacetime is not general enough to accommodate this physical feature into that metric, yielding a conical singularity which compensates the mutual rotational coupling.
On the other hand we count that the spin-spin interaction between the seed and the background environment can play a relevant role into the regularisation of gravitational models which otherwise would be mathematically defective and physically incomplete.
For instance, in the same way as Ernst showed that the electromagnetic background can remove the conical defect of the accelerating and charged black hole, we foresee that the procedure presented in this paper can remove the axial singularities of the rotating C-metric, providing at the same time a reasonable physical explication for its acceleration~\cite{Astorino:2022prj}.
Also this model furnishes alternative scenarios for black hole nucleation and pair creation, without relying on the electromagnetic field, as discussed in the literature so far~\cite{Garfinkle:1993xk,Hawking:1994ii,Astorino:2013xxa}.

Clearly this procedure may be relevant for other systems, not necessarily accelerating, such as balancing multi-black hole sources to reach an equilibrium configuration.
Also in that case the frame dragging of the background can play a role in removing cosmic strings or strut from the singular spacetime.
On the other hand, preliminary studies suggest that the spin-spin interaction between the swirling universe and a Taub--NUT spacetime are not sufficient to mend also the singular behaviour of that metric, i.e.~to remove the Misner string\footnote{Obviously we are referring to the non-compact time representation of the Taub--NUT metric, because when one considers proper periodic identification of the temporal coordinate the spacetime can be regularised.
Unfortunately, the latter interpretation violates causality because of the appearance of closed timelike curves, which makes this picture nonphysical.}.

From a phenomenological point of view our rotating background might be of some interest in the description of black holes surrounded by interacting matter, which produces intense frame dragging, such as the one caused by the collision of counter-rotating galaxies. 

Since this construction is based on a symmetry transformation of the Ernst equations, it can be directly generalised to the Einstein--Maxwell case, to the minimally and conformally coupled scalar field case and, more generally, to scalar-tensor theories such as Brans--Dicke, just by using the adequate Ehlers transformation as described in~Sec.~\ref{sec:ernst} and in~\cite{Astorino:2013xc} respectively.
The embedding method presented here may reveal useful in estabilishing and improving traversability of wormhole spacetimes.

\chapter*{Conclusions}
\addcontentsline{toc}{chapter}{Conclusions}
\label{chap:conclusions}
\thispagestyle{plain}

Black holes represent one of the most fascinating areas of research in contemporary physics:
they are bizarre objects that challenge our common sense, and whose existence represented a mystery for many years.
They are also the natural candidate for studying quantum gravity, as the region close to a curvature singularity is characterised by strong gravity on very small scales.
As explained in the Introduction, binary and, more generally, multi-black hole systems are relevant for the experiments and for the structure of the gravitational theories.
Thus, it is important to find regular multi-source solutions that can describe actual astrophysical black holes and, from the theoretical point of view, to properly define the thermodynamics of such systems:
indeed, the thermodynamics can give important hints about the microscopic origin of the entropy and the temperature.

In this Thesis, we have studied some configurations that allow one to regularise multi-black hole spacetimes, with the aid of solution generating techniques.

We began by introducing such techniques and explaining the details of their functioning in Chapter~\ref{chap:gentech}, which provided us with the machinery necessary to the construction of our solutions.
We also reported an historical perspective on the development of the various solution generating methods over the years, commenting on the peculiar features and on the pros and cons of the different approaches.

Then, in Chapter~\ref{chap:extfield}, we introduced an external gravitational field as a regularising background, and discussed its properties and multipolar character.
We provided many examples of singularity-free multi-black hole spacetimes:
static, rotating, charged an accelerating systems of black holes were studied and analysed in detail.
They share the common feature of being regularised by an appropriate tuning of the external field parameters, i.e.~by choosing the multipole coefficients in order to balance the attraction among the black holes.
We inspected the action of the external field on the geometry of these ``deformed'' black holes, and we verified that the thermodynamics of these objects was well defined:
this allowed us to conclude that these multi-black hole systems are physically well posed.
Moreover, the external fields have a phenomenological interest since they can model matter surrounding black holes.

Chapter~\ref{chap:bubble} introduced the bubbles of nothing to the business of the multi-black hole regularisation:
the bubbles, which are usually studied in relation to the vacuum stability, were considered as time-dependent expanding spacetimes that can win against the mutual collapse of many black holes.
We showed that bubbles of nothing have a deep connection with black holes, as it can be appreciated by taking some appropriate limits which bring a bubble into a black hole, and viceversa (an analog construction works for bubbles in five dimensions and black rings).
Further, we embedded multi-black holes (in four dimensions) and black rings (in five dimensions) in bubble spacetimes, and explicitly showed that they reach an equilibrium configuration.
We also considered other examples that involve accelerating black holes (in four dimensions) and black Saturn and di-ring (in five dimensions), and how the manipulation of their rod diagrams naturally allows one to move from one configuration to another.
Thus, bubbles revealed to be very powerful solutions, which permit to obtain solutions free of conical singularities, and moreover we proved that their connection to elementary black hole solutions in General Relativity is deeper than one may expect.

Finally, in Chapter~\ref{chap:swirl} we introduced a transformation that add a ``swirling'' background to a given spacetime.
Such a rotational character acts as a sort of whirlpool, which drags the observers that are at rest in the spacetime, as shown by the embedding diagrams of the geodesics.
A static black hole embedded in the swirling geometry is (perhaps surprisingly) not affected by conical singularities:
it is possible to properly define conserved charges and thermodynamics, and to study ergoregions and geometrical properties.
On the other hand, the swirling rotation couples to the angular momentum of a black hole, as shown by the Kerr solution:
in that case, conical singularities appear and the solution is no more regular.
This drawback, however, paves the way to a possible mechanism to regularise a multi-black hole spacetime:
the embedding of a double-Kerr solution into the swirling background and the resulting coupling between the swirl parameter and the spin-spin interaction of the holes, might provide a mechanism to remove the conical singularities.
This is a very promising idea, that can be pursued in a future research.
\vspace{0.2cm}

\addcontentsline{toc}{part}{\large Appendices} \thispagestyle{empty}
\part*{Appendices}
\thispagestyle{empty}

\appendix
\makeatletter
\@addtoreset{equation}{section}  
\makeatother
\renewcommand{\theequation}
  {\thesection.\arabic{equation}}

\renewcommand{\theequation}{A.\arabic{equation}}
\chapter{Conical singularities and energy conditions}
\label{app:conic}
\thispagestyle{plain}

Conical singularities, beyond making the spacetime manifold ill-defined from a mathematical point of view, give also rise to energy issues.
In general, such singularities can be interpreted as strings or struts whose energy-momentum tensor has a $\delta$-like nature.
We show what are the physical issues that the conical singularities bring in when they are present in the spacetime.

Let us consider Minkowski spacetime with an wedge of angle $2\pi\alpha\prime$ artificially removed.
By defining $C=1-\alpha\prime$, we can write the metric as
\begin{equation}
{ds}^2 = -{dt}^2 + {dr}^2 + C^2 r^2 {d\varphi}^2 + {dz}^2 \,,
\end{equation}
where $0\leq\varphi<2\pi$.
One can regard this spacetime as a field sourced by a cosmic string or a strut~\cite{Linet}, whose non-vanishing energy-momentum tensor components are
\begin{equation}
T^t_t = T^\varphi_\varphi = 2\pi \mu \, \delta(x,y) \,,
\end{equation}
where
\begin{equation}
\mu = \frac{1-C}{4C}  \,,
\end{equation}
and $\delta(x,y)$ is the two-dimensional delta function depending on the coordinates orthogonal to the $z$-axis $(x,y)$.
$\mu$ is interpreted as the tension of the filament source.

This result can be generalised to a generic four-dimensional spacetime~\cite{Vickers:1987az}, for which one finds that the Einstein equations
$G_{\mu\nu} = 8\pi T_{\mu\nu}$
give rise to an energy-momentum tensor
\begin{equation}
\label{stress-tensor}
T^0_0 = T^3_3 = 2\pi \mu \, \delta(x,y) \,,
\end{equation}
where now
\begin{equation}
\mu = \frac{2\pi - C}{4C} \,.
\end{equation}
$C$ represents again the angular excess/deficit for the azimuthal angle.

Let us consider, e.g., a two-black hole spacetime from~\eqref{n-bh} ($N=2$):
the result~\eqref{stress-tensor} clearly shows that for $\mu>0$ (positive tension), the source acts as a string that pull a black hole.
This is the behaviour of the conical singularities that one finds at $z<w_1$ and $z>w_4$
There are no negative-energy issues in this case, but the string extends to infinity and the $\delta$ function gives rise to a divergent energy-momentum tensor.

In the case of $\mu<0$ (negative tension), the conical singularity in $w_2<z<w_3$ acts as a strut which pushes apart the two black holes.
The energy density associated to the energy-momentum tensor is negative (i.e.~the strut is composed of anti-gravitational matter) and there is again a divergence due to the $\delta$ function.
\renewcommand{\theequation}{B.\arabic{equation}}
\chapter{Harrison and Kramer--Neugebauer charging trasformations}
\label{app:harrison-kramer-neugebauer}
\thispagestyle{plain}

Both the Harrison and the Kramer--Neugebauer~\cite{Kramer1969} transformations are two symmetries of the Ernst equations for the Einstein--Maxwell theory presented in Chapter~\ref{chap:gentech}.
As it was explained there, the complex Ernst equations enjoy an $SU(2,1)$ Lie-point symmetry group spanned by the finite transformations~\eqref{ernst-group}

The Harrison transformation is given by Eq.~\eqref{harrison}, while the Kramer--Neugebauer one, as defined in~\cite{Manko1992} to charge the Kerr metric embedded in an external gravitational field, is
\begin{equation}
\label{KN}
\ernst' = \frac{\ernst - \zeta^2}{1 - \zeta^2 \ernst} \,, \qquad
\Phi' = \frac{\zeta (\ernst-1)}{1-\zeta^2 \ernst} \,.
\end{equation}
The latter transformation reduces to the one in~\eqref{hat-transf} for static and uncharged seeds.
Since both the Kramer--Neugebauer transformation~\eqref{KN} and the Harrison transformation~\eqref{harrison} have the same physical effects (they add an electric monopole to an uncharged seed), we have the suspect that they are basically the same transformation, up to gauge transformations.
In fact it can be shown that the subsequent composition of transformations~\eqref{gauge1},~\eqref{gauge3} and~\eqref{harrison} to an Ernst seed $(\ernst,\Phi)$ gives
\begin{align}
\ernst' & =
\frac{\lambda \lambda^* \ernst - \beta^*(\beta + 2 \lambda \Phi)}{1 + \alpha^2(-2 \beta + \alpha^* \beta \beta^* - \alpha \lambda \lambda^* \ernst + 2\lambda (\alpha \beta^*-1) \Phi} \,, \\
\Phi' & =
\frac{\beta - \alpha \beta \beta^* + \lambda\lambda^* \alpha \ernst + \Phi - 2\lambda\alpha\beta^* \Phi}{1 + \alpha^2(-2 \beta + \alpha^* \beta \beta^* - \alpha \lambda \lambda^* \ernst + 2\lambda (\alpha \beta^*-1) \Phi} \,.
\end{align}
Then considering a null electromagnetic Ernst potential, $\Phi=0$, the imaginary part of the parameters $\alpha$, $\beta$, $\lambda$ equal to zero and choosing
\begin{equation}
\lambda = \frac{1}{1-\zeta^2} \,, \qquad
\alpha = \zeta \,, \qquad
\beta = - \frac{\zeta}{1-\zeta^2} \,,
\end{equation}
we exactly recover the transformation~\eqref{KN}.
In case of static metrics the latter further simplifies to Eqs.~\eqref{hat-transf}.
Therefore the Kramer--Neugebauer and the Harrison transfomration are basically equivalent, up to gauge transformations, so they might be called collectively Harrison--Kramer--Neugebauer transformation.  

As an explicit example we show the efficacy of the charging transformation~\eqref{hat-transf} on an asymptotically flat, static and discharged metric:
acting on the Schwarzschild metric, we are able to produce the Reissner--Nordstr\"om black hole.

For simplicity we take the seed in spherical symmetric coordinates
\begin{equation}
{ds}^2 = -\biggl(1- \frac{2m}{r} \biggr) {dt}^2 + \frac{{dr}^2}{1-\frac{2m}{r}} + r^2 {d\theta}^2 + r^2 \sin^2 \theta {d\varphi}^2 \,,
\end{equation}
from which we can easily read
\begin{equation}
e^{2\psi} = 1-\frac{2m}{r}  \,, \qquad A_t = 0 \,.
\end{equation}
After the charging transformation~\eqref{hat-transf} we get the new solution
\begin{equation}
e^{2\hat{\psi}} = \frac{r(r-2m)(1-\zeta^2)^2}{[r+(2m-r)\zeta^2]^2} \,, \quad \hat{A}_t = -\frac{2m\zeta}{r+(2m-r)\zeta^2} \,.
\end{equation}
A shift of the radial coordinate
\begin{equation}
r \to \hat{r} - M + \sqrt{M^2-q^2} \,,
\end{equation}
and a rescaling of the parameters
\begin{equation}
\label{zeta}
\zeta \to \frac{M - \sqrt{M^2-q^2}}{q} \,, \qquad
m \to \sqrt{M^2-q^2} \,,
\end{equation}
suffice to recognise the standard Reissner--Norstr\"om spacetime
\begin{align}
{d\hat{s}}^2 & =
- \biggl(1-\frac{2 M}{\hat{r}}+\frac{q^2}{\hat{r}^2} \biggr){dt}^2 + \frac{{d\hat{r}}^2}{1-\frac{2M}{\hat{r}}+\frac{q^2}{\hat{r}^2}} + \hat{r}^2 {d\theta}^2 + \hat{r}^2 \sin^2 \theta {d\varphi}^2 \,, \\
\hat{A} & = -\frac{q}{\hat{r}} \, dt \,.
\end{align}
\renewcommand{\theequation}{C.\arabic{equation}}
\chapter{Geodesics of the swirling spacetime}
\thispagestyle{plain}

We report here the explicit expression for the geodesic equation from Chapter~\ref{chap:swirl}, both for the background and the full black hole metric.

\section{Background geodesics}
\label{sec:appback}

The explicit expressions for the definitions of the conserved quantities~\eqref{conserved} are
\begin{equation}
\dot{t} = \frac{E + 4\jmath L z}{1+\jmath^2 \rho^4} \,, \qquad
\dot{\phi} = \jmath^2 L \rho^2 + \frac{L}{\rho^2}
- 4 \jmath z \frac{E + 4\jmath L z}{1 + \jmath^2 \rho^4} \,.
\end{equation}
By substituting these relations into the Lagrangian~\eqref{lag-background}, we get
\begin{equation}
\mathscr{L} =
(1 + \jmath^2 \rho^2)
\biggl[ \frac{L^2}{\rho^2} - \biggl(\frac{E + 4\jmath L z}{1+\jmath^2 \rho^4}\biggr)^2 + \dot{\rho}^2 + \dot{z}^2 \biggr] \,.
\end{equation}
The equation coming from the normalisation of the four-momentum $u_\mu u^\mu = \chi$ is
\begin{equation}
\label{4mom-BKG}
\begin{split}
\dot{\rho}^2 + \dot{z}^2 & =
\frac{1}{(1 + \jmath^2\rho^4)^2} \Biggl[
E^2 - 8\jmath L z (E + 6 \jmath L z) + \frac{L^2}{\rho^2}
+ 2 \jmath^2 L^2 \rho^2 + \jmath^4 L^2 \rho^6 \\
&\quad + \biggl(4\sqrt{2} \jmath \rho z \frac{E + 4\jmath L z}{1+\jmath^2 \rho^4}\biggr)^2
\Biggr]
+ \frac{\chi}{1+\jmath^2\rho^4} \,.
\end{split}
\end{equation}

\begin{figure}
\centering
\hspace{-0.2cm}
\subfloat[\centering $m=1$, $\jmath=0.01$, $L=1$]{{\includegraphics[width=6.5cm]{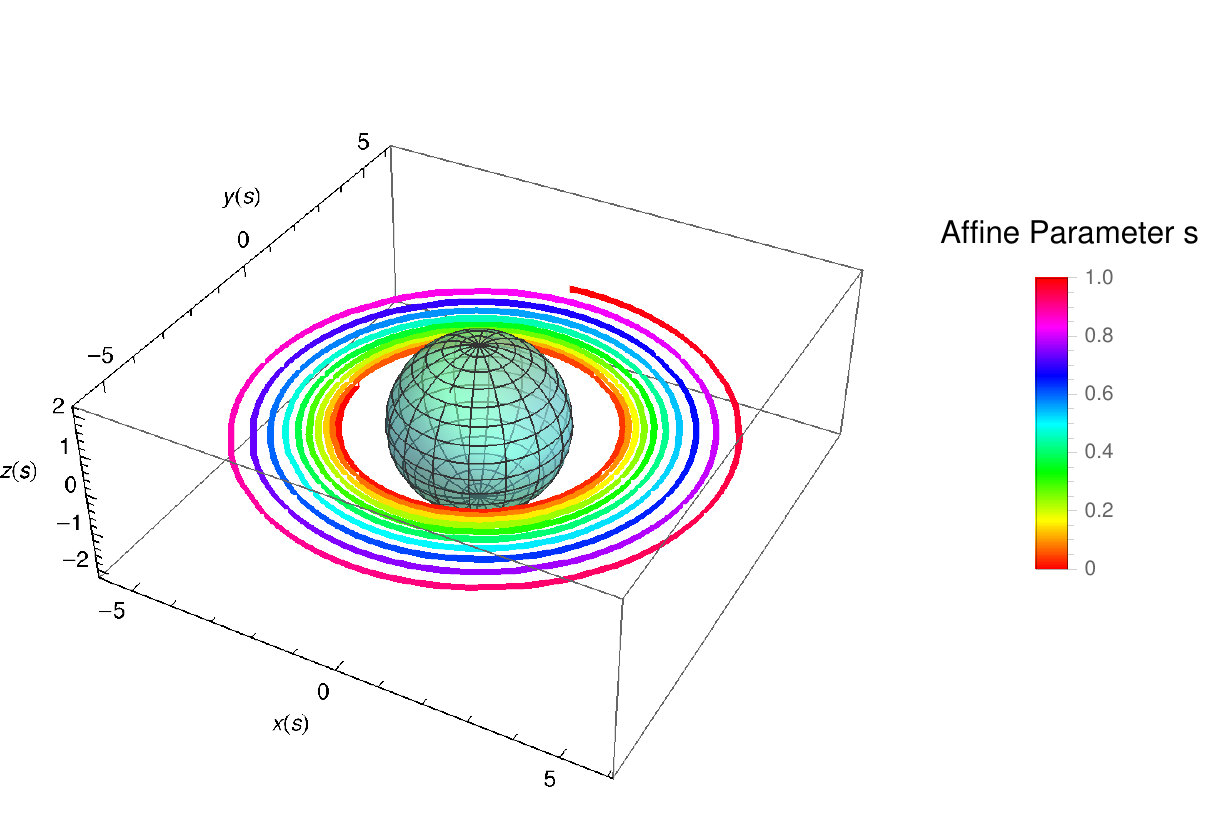}}}
\subfloat[\centering $m=1$, $\jmath=0.01$, $L=36$]{{\hspace{-0.3cm} \includegraphics[width=6.5cm]{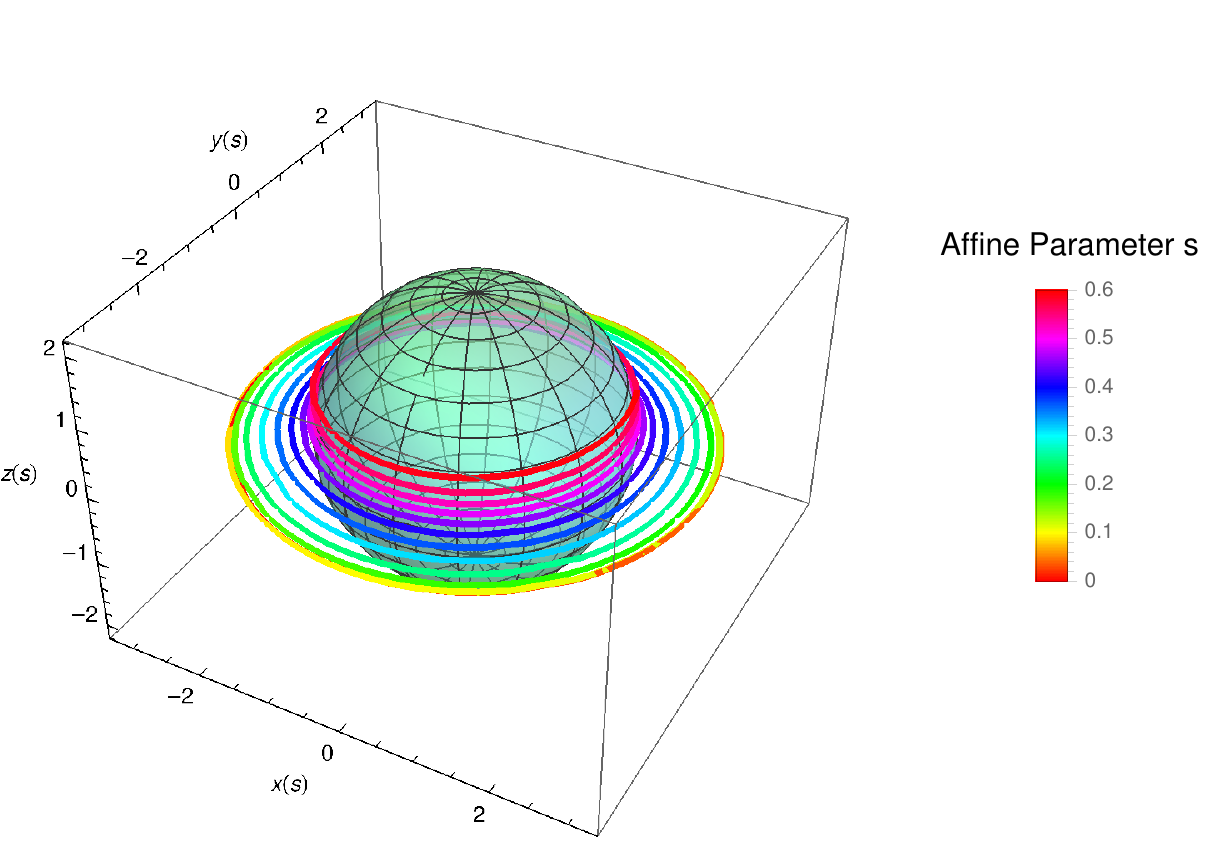}}}
\caption{\small Black hole~\eqref{swirling-bh} with two different unstable orbits.
The first one represents the case where $L$ is small and shows that the orbit approaches an asymptotic value.
The second one shows an orbit with large $L$:
the test particle on the plane defined by $\theta=\pi/2$ is attracted towards the black hole.}
\label{fig:unstable}
\end{figure}

\section{Black hole geodesics}
\label{sec:appbh}

The conserved charges equations are
\begin{subequations}
\label{geod-bh}
\begin{align}
\dot{t} & = \frac{r [ E + 4 \jmath L \cos \theta (r-2 m) ]}{(r-2m) (1 + \jmath^2 r^4 \sin^4\theta)} \,, \\
\dot{\phi} & = \frac{L - \jmath r^3 \sin^2\theta
[ 4 \cos\theta (E-8\jmath L m \cos\theta) - \jmath^3 L r^5 \sin^6\theta + 2\jmath L r (9\cos^2\theta-1) ]}{r^2 \sin^2\theta (1+\jmath^2 r^4 \sin^4\theta)} \,.
\end{align}
\end{subequations}
$L$ is the angular momentum and $E$ is the energy.
From the conservation of the four-momentum it follows
\begin{equation}
\label{4mom-BH}
\begin{split}
& \bigl( 1 + \jmath^2 r^4 \sin^4\theta \bigr)
\biggl[
\biggl(\frac{r ( E + 4 \jmath L \cos\theta (r-2m) )}{(r-2m)(1 + \jmath^2 r^4 \sin^4\theta)}\biggr)^2
+ \frac{\dot{r}^2}{1-\frac{2 m}{r}}+r^2 \dot{\theta}^2\biggr] \\
&\quad + \frac{L^2 (1+\jmath^2 r^4 \sin^4\theta)^2}{r^4 \sin^4\theta}
= \chi \,.
\end{split}
\end{equation}
\renewcommand{\theequation}{D.\arabic{equation}}
\chapter{Zipoy--Voorhees spacetime embedded in the swirling universe}
\label{app:zipoy}
\thispagestyle{plain}

We apply the procedure described in Sec.~\ref{sec:generation} to a slightly more general metric than the Schwarzschild black hole, the Zipoy--Voorhees metric~\cite{Bach1922,Zipoy1966,Vorhees1970}.
This class of spacetime is relevant in General Relativity because, thanks to its rich multipolar expansion, can be used to model the exterior gravitational field of planets or stars.
It is worth to mention, in connection with Chapter~\ref{chap:extfield}, that Kerns and Wild constructed a version of the Zipoy--Voorhees spacetime embedded in the external gravitational field~\cite{Kerns:1982xb}.

The Zipoy--Voorhees metric can be of some interest, when supported by a conformally coupled scalar field, to build hairy black holes or wormholes such as the Bekenstein black hole~\cite{Bekenstein:1975ts} or the Barcelo--Visser wormhole~\cite{Barcelo:1999hq}\footnote{Actually the associated complex Ernst field equations remain the same of the pure general relativistic case, so as the main structure functions in the metric.
Only the decoupled function $\gamma$ has to be slightly modified according to~\cite{Astorino:2014mda}.}.
In particular, the presence of the swirling background might be useful in the wormhole configuration to improve both the stability and the traversability properties of the solution.

We start by casting the Zipoy--Voorhees seed in terms of the magnetic LWP metric~\eqref{magnetic-metric} in prolate spherical coordinates
\begin{equation} 
\label{dsbarxy}
{d\bar{s}}^2 = \bar{f} \bigl( d\phi - \bar{\omega} d\tau \bigr)^2
+ \frac{1}{\bar{f}} \biggl[ -\rho^2 d\tau^2 + \kappa^2 (x^2-y^2) e^{2\bar{\gamma}}  \biggl( \frac{{dx}^2}{x^2-1} + \frac{{d y}^2}{1-y^2} \biggr) \biggr] \,,
\end{equation}
where
\begin{subequations}
\begin{align}
\bar{f}(x,y) & = \kappa^2 \biggl(\frac{x-1}{x+1} \biggr)^{-\delta} (x^2-1)(1-y^2) \,, \\
 \label{gammabar}
\bar{\gamma}(x,y) & = \frac{1}{2} \log \left[ \kappa^2 \biggl(\frac{x-1}{x+1} \biggr)^{-2\delta} (x^2-1)(1-y^2) \biggl( \frac{x^2-1}{x^2-y^2} \biggr)^{\delta^2} \right] \,, \\
\label{rhobar}
\rho(x,y) & = \kappa \sqrt{(x^2-1)(y^2-1)} \,.
\end{align}
\end{subequations}
Clearly, this metric reduces to the static Schwarzschild black hole of Sec.~\ref{sec:generation} when $\delta=1$.
For generic values of $\delta \neq 1$, the metric looses the spherical symmetry and presents naked singularities outside the event horizon, hence it is not suitable for describing legit black holes in pure General Relativity.
However, for $\delta=1/2$, when properly coupled  with a scalar field, it represents the first hairy black hole ever discovered~\cite{Bekenstein:1975ts}.

Thanks to the Ehlers transformation~\eqref{ehlers-swirl}, and following the same procedure illustrated in Sec.~\ref{sec:generation}, we are able to embed the Zipoy--Vorhees metric into the swirling background.
The $\bar{\gamma}$ function remains the same, while
\begin{subequations}
\label{zipoy-metric}
\begin{align}
f(x,y) & = \frac{\kappa^2 (1-y^2)(x^2-1)^{1+\delta}}{(x-1)^{2\delta}+\jmath^2(1+x)^{2\delta}(x^2-1)^2(1-y^2)^2} \,, \\
\omega(x,y) & = 4 \jmath \kappa^2 y (x-\delta) + \omega_0 \,.
\end{align}
\end{subequations}
The metric defined by~\eqref{dsbarxy} and~\eqref{zipoy-metric} represents the $\delta$ extension of the spacetime~\eqref{bh-rot-universe}, therefore the Zipoy--Voorhees spacetime immersed in the rotating background described in Sec.~\ref{sec:analysis-background}.
Further generalisations with angular momentum can be built straightforwardly, starting with seeds of the family of the Tomimatsu--Sato solutions~\cite{Cosgrove1977,Hoenselaers1979}.

\renewcommand{\theequation}{\arabic{chapter}.\arabic{equation}}


\backmatter

\pagestyle{fancy}
\fancyhf{}
\renewcommand{\chaptermark}[1]{\markboth{#1}{}}
\renewcommand{\sectionmark}[1]{\markright{\thesection\ #1}}
\fancyhead[LE]{\bf \thepage}
\fancyhead[RO]{\bf \thepage}
\fancyhead[LO]{{\it Bibliography}}
\fancyhead[RE]{\it Bibliography}
\renewcommand{\headrulewidth}{0.2pt}
\renewcommand{\footrulewidth}{0pt}

\bibliographystyle{unsrturl}
\bibliography{Bibliography}

\end{document}